\documentclass[12pt,notitlepage]{article}
\usepackage{mathrsfs}
\usepackage{latexsym,fullpage}
\usepackage{amssymb}
\usepackage{amsmath}
\usepackage{multirow}
\usepackage[dvips]{graphicx}
\newcommand{\beq}{\begin{equation}}
\newcommand{\eeq}{\end{equation}}
\newcommand{\ie}{{\sl i.e\/}}
\newcommand{\half}{\frac 1 2}
\newcommand{\lag}{{\cal L}}

\newcommand{\act}{\int d^4x\sqrt{-g}\/}

\newcommand{\rta}{\rightarrow}
\newcommand{\bc}{\begin{center}}
\newcommand{\ec}{\end{center}}
\newcommand{\bi}{\begin{itemize}}
\newcommand{\ei}{\end{itemize}}

\newcommand{\Prd}{{Phys. Rev. D}}
\newcommand{\Prl}{{Phys. Rev. Lett.}}

\newcommand{\Plb}{{Phys. Lett. B}}
\newcommand{\Cqg}{{Class. Quantum Grav.}}
\newcommand{\Grg}{{Gen. Rel. Grav.}}
\newcommand{\Np}{{Nuc. Phys.}}
\newcommand{\Fp}{{Found. Phys.}}

\newcommand{\Jpa}{{J. Phys. A}}

\begin{document}
\title{\Large\textbf{{Bouncing Cosmologies
}}}
\author{M. Novello \footnote{
Instituto de Cosmologia Relatividade Astrofisica (ICRA-Brasil/CBPF)
Rua Dr. Xavier Sigaud, 150, CEP 22290-180, Rio de Janeiro, Brazil. E-mail: novello@cbpf.br.}
$\;$,
S. E. Perez Bergliaffa \footnote{Departamento
de Fisica Teorica, Instituto de F\'{\i}sica,
Universidade do Estado do Rio de Janeiro,
R. S\~ao Francisco Xavier, 524,
Maracan\~a - CEP: 20559-900 - Rio de Janeiro,  Brazil. E-mail: sepbergliaffa@gmail.com.}$\;$.}
\maketitle
\begin{abstract}
We review the general features of nonsingular universes ({\em i.e.}
those that go from an era of accelerated collapse to an
expanding era without displaying a singularity) as well as cyclic universes. We
discuss the mechanisms behind the bounce, and analyze examples of
solutions that implement these mechanisms. Observational
consequences of such regular cosmologies are also considered, with
emphasis in the behavior of the perturbations.

\end{abstract}
\maketitle \tableofcontents

\newpage
\begin{flushright}
\emph{The world, an entity out of everything, \\was created by neither gods nor men,\\
but was, is and will be eternally living fire,\\ regularly becoming
ignited \\and regularly
becoming extinguished.}\\
Heraclitus\footnote{The proof of this assertion - which is
still missing - was left by Heraclitus to future
generations.}.
\end{flushright}
\newpage
\section{Introduction}
\label{sint}

The standard cosmological model (SCM) (see for instance \cite{pdg}
for an updated review) furnishes an accurate description of the evolution of
the universe, which spans approximately 13.7 billion years.
The main hypothesis on which the model is based are the following:
\begin{enumerate}
\item Gravity is described by General Relativity.
\item The universe obeys the
Cosmological Principle \cite{coleslucchin}. As a consequence, all the relevant
quantities depend only on global Gaussian time.
\item Above a
certain scale, the matter content of the model is described by a
continuous distribution of matter/energy, which is described by a
perfect fluid.
\end{enumerate}
In spite of its success, the SCM suffers from a series of problems
such as the initial singularity, the cosmological horizon, the
flatness problem, the baryon asymmetry, and the nature of dark matter and
dark energy \footnote{There are even claims that standard cosmology
does not predict the value of the present CMBR temperature
\cite{dac}.}, \footnote{Some ``open questions'' may be added to this list,
such as why the Weyl tensor is null, and what the future
of the universe is.}. Although inflation (which for many is
currently a part of the SCM) partially or totally answers some of
these, it does not solve the crucial problem of the initial singularity
\cite{vile} \footnote{Inflation also presents some problems of its
own, such as the identification of the inflaton with a definite
field of some high-energy theory, the functional form of the
potential $V$ in terms of the inflaton \cite{bennet}, and the need of
particular initial conditions \cite{saka}. See also \cite{narpam}}. The
existence of an initial singularity is disturbing: a
singularity can be naturally considered as a source of lawlessness
\cite{earman}, because the spacetime description breaks down
``there'', and physical laws presuppose spacetime. Regardless of the
fact that several scenarios have been developed to deal with the
singularity issue, the breakdown of physical laws continues to be a
conundrum after almost a hundred years of the discovery of the FLRW
solution \footnote{This acronym refers to the authors that presented for the first time
the
solution of EE that describes a universe with zero pressure
(Friedmann \cite{friedmann}) and nonzero pressure (Lem\^aitre
\cite{lem}), and to those who studied its general mathematical
properties and took it to its current form (Robertson \cite{robe}
and Walker \cite{walker}). For historical details, see \cite{merlau}}
(which
inevitably displays
a past singularity, or in the words of Friedmann \cite{friedmann}, a
\emph{beginning of the world}).

In this review, we shall concentrate precisely on the issue of the
initial singularity \footnote{We shall not analyze the existence of
future singularities, such as the so-called sudden future
singularities \cite{barrowsudden} or the ``Big Rip''
\cite{caldwell}.}. We will see that non-singular universes have been
recurrently present in the scientific literature. In spite of the
fact that the idea of a cosmological bounce is rather old, the first
explicit solutions for a bouncing geometry were obtained by Novello
and Salim \cite{NovelloSalim}, and Melnikov and Orlov \cite{melni}
in the late 70's. It is legitimate to ask why these solutions did
not attract the attention of the community then. In the beginning of
the 80's, it was clear that the SCM was in crisis (due to the
problems mentioned above, to which we may add the creation of
topological defects, and the lack of a process capable of producing
the initial spectrum of perturbations, necessary for structure
formation). On the other hand, at that time the singularity theorems
were taken as the last word about the existence of a singularity in
``reasonable'' cosmological models. The appearance of the
inflationary theory gave an answer to some of the issues in a
relatively economical way, and opened the door for an explanation of
the origin of the spectrum of primordial fluctuations. Faced to
these developments, and taking into account the status of the
singularity theorems at that time, the issue of the initial
singularity was not pressing anymore, and was temporally abandoned
in the hope that quantum gravity would properly address it. At the
end of the 90's, the discovery of the acceleration of the universe
brought back to the front the idea that $\rho + 3p$ could be
negative, which is precisely one of the conditions needed for a
cosmological bounce in GR, and contributed to the revival of nonsingular
universes. Bouncing models even made it to the headlines in the late
90's and early XXI century, since some models in principle embedded
in string theory seemed to suggest that a bouncing geometry could
also take care of the problems solved by inflation.

Perhaps the main motivation for nonsingular universes is
the avoidance of lawlessness, as mentioned above
\footnote{It is worth noting that Einstein was well aware
of the problem of singularities in GR \cite{penrosee}, and he made several attempts
to regularize some solutions of his theory,
such as the so-called Einstein-Rosen bridge, in the early 30´s.
Indeed, he wrote ''The theory (GR)
is based on a separation of the concepts of the gravitational
field and matter. While this may be a valid approximation for weak
fields, it may presumably be quite inadequate for very high densities
of matter. One may not therefore assume the validity of the equations
for very high densities and it is just possible that in a unified theory
there would be no such singularity'' \cite{eins}.}.
Also,
since we do not know how to handle infinite quantities, we would
like to have at our disposal solutions that do not entail
divergencies. As be seen in this review, this can be achieved
at a classical level, and also by quantum modifications. On a
historical vein, this situation calls for a parallel with the status
of the classical theory of the electron by the end of the 19th
century. The divergence of the field on the world line of the
electron led to a deep analysis of Maxwell's theory, including the
acceptance of a cooperative influence of retarded and advanced
fields \cite{rohr} \footnote{In fact, it can be said that the
problem of the singularity of the classical theory of the electron
was transcended, if not resolved, by the quantization of the EM
field.}, and the related causality issues. However, this divergence
is milder than that of some solutions of General Relativity, since it can
be removed by the interaction of the electron with the environment.
Clearly, this is not an option when the singularity is that of a
cosmological model.

Another
motivation for the elimination of the initial singularity is related to
the Cauchy problem.
In the SCM, the structure of spacetime has a natural foliation
(if no
closed timelike curves
are present), from which a global Gaussian
coordinate system can be constructed, with
 $g_{00} = 1$, $ g_{0i} = 0$, in such a way that
$$
ds^{2} = dt^{2} - g_{ij} \, dx^{i} dx^{j}.$$
The existence of a
global coordinate system allows a rigorous setting for the Cauchy
problem of initial data. However, it is the gravitational field
that diverges on a given spatial hypersurface $t=$ const. (denoted by $\Sigma$)
at the
singularity in the SCM. Hence,
the Cauchy problem cannot be well formulated on such a surface: we
cannot pose on $\Sigma$ the initial values for the field to evolve.

There are more arguments that suggest that the singularity should be
absent in an appropriate cosmological model.
According to
\cite{beke2}, the second law of thermodynamics is to be
supplemented with a limit on the entropy of a system of largest
linear dimension $R$ and proper energy $E$, given by
$$
\frac{S}{E}\leq\frac{2\pi R}{\hbar c}.
$$
Currently this bound is known to be satisfied in several physical
systems \cite{schiffer}. It was shown in \cite{beken} that the
bound is violated as the putative singularity is approached in the
radiation-dominated FLRW model (taking as $R$ the particle horizon
size). The restriction to FLRW models was lifted in \cite{schiffer},
where it was shown, independently of the spacetime model, and under
the assumptions that (1) causality and the strong energy condition
(SEC, see Appendix) hold, (2) for a given energy density, the matter
entropy is always bounded from above by the radiation entropy, that
the existence of a singularity is inconsistent with the entropy
bound: a violation occurs at time scales of the order of Planck's
time \footnote{For an updated discussion of the several types of
entropy bounds in the literature, see \cite{entropybounds}.}.

From the point of view of quantum mechanics, we could ask if it is
possible to repeat in gravitation what was done to eliminate the
singularity in the classical theory of the electron. Namely, can the
initial singularity be smoothed via quantum theory of gravity? The
absence of the initial singularity in a quantum setting is to be
expected on qualitative grounds. There exists only one quantity with
dimensions of length that can be constructed
from Newton's constant $G$, the velocity of light $c$, and Planck's constant $\hbar$
(namely Planck's length $\ell _{Pl}=\sqrt{G\hbar/c^3}$). This
quantity would play in quantum gravity a role analogous to that of the energy of the
ground state of the hydrogen atom (which is the only quantity with
dimensions of energy that can be built with fundamental constants)
\cite{bojoi}. As in the hydrogen atom, $\ell _{Pl}$ would imply some
kind of discreteness, and a spectrum bounded from below, hence
avoiding the singularity \footnote{This expectation has received support from
the proof that the spectrum of the volume operator in LQG is discrete, see
for instance \cite{loll}.}. Also, since it is generally assumed that
$\ell _{Pl}$ sets the scale for the quantum gravity effects,
geometries in which curvature can become larger than $\ell
_{Pl}^{-2}$ or can vary very rapidly on this scale would be highly
improbable.

Yet another argument that suggests that quantum effects may
tame a singularity
is
given by the Rayleigh-Jeans spectrum. According to classical physics,
the spectral energy distribution of radiation
in thermal equilibrium diverges like $\omega ^3$ at high
frequencies, but when quantum corrections are taken into account,
this classical singularity is regularized and the Planck
distribution applies \cite{gaspe}. We may expect that QG effects
would regularize the initial singularity.

As a consequence of all these arguments indicating that the
initial singularity may be absent in realistic descriptions of the
universe, many cosmological solutions displaying a bounce were
examined in the last decades.
In fact, the pattern in scientific cosmologies somehow parallels
that of the cosmogonic myths in diverse civilizations, which can be
classified in two broad classes. In one of them, the universe
emerges in a single instant of creation (as in the Jewish-Christian
and the Brazilian Caraj\'as cosmogonies \cite{eli}). In the second class, the
universe is eternal, consisting of an infinite series of cycles (as
in the cosmogonies of the Babylonians and Egyptians)
\cite{primeval}.

We have seen that there are reasons to assume that the
initial singularity is not a feature of our universe. Quite naturally,
the idea of a non-singular universe has been extended to encompass
cyclic cosmologies, which display phases of expansion and
contraction. The first scientific account of cyclic universes is in
the papers of Friedmann \cite{luminet}, Einstein \cite{einstein},
Tolman \cite{tolmanbook}, and Lema\u{\i}tre \cite{lemaitre} and his
Phoenix universe, all published in the 1930's. A long path has been
trodden since those days up to recent realizations of these ideas
(as for instance
\cite{vene}, see Sect.\ref{ekpyrotico}). We shall see
in Ch.\ref{ccyclic} that some cyclic models could potentially solve
the problems of the standard cosmological model, with the interesting
addition that they do not need to address the issue of the
initial conditions.

Another motivation to consider bouncing universes comes from the recognition that
a phase of accelerated contraction can solve some of the problems of the SCM
in a manner similar to inflation. Let us take for instance the flatness problem
(see also Sect.\ref{ccyclic}). Present observations imply
that the spatial curvature term, if not negligible, is at least non-dominant
wrt the curvature term:
$$
r^2=\frac{|\epsilon|}{a^2H^2}\lesssim 1,
$$
but during a phase
of standard, decelerated expansion, $r$ grows
with time. Indeed, if $a\thicksim t^\beta$, then $r\thicksim t^{1-\beta }$.  So we need
an impressive fine-tuning at, say, the GUT scale, to get the observed value of $r$
\footnote{But notice that the flatness problem may actually not be a problem at all
if gravity is not described by GR, see Sect.\ref{riccis}}.
This problem can be solved by introducing an early phase during which the value of
$r$, initially of order 1,
decreases so much in time that its subsequent growth during FLRW
evolution keeps it still below 1 today.
This can be achieved by \cite{vene} power-law inflation ($a\thicksim t^\beta,
\;\beta > 1$), pole inflation ($a\thicksim (-t)^\beta,\;\beta<0, t\rightarrow 0_-$),
and accelerated contraction ($0<\beta<1, t\rightarrow 0_-$)
\cite{deflation}. Thus, an era of
accelerated contraction may solve the flatness problem (and the other kinematical issues
of the SCM \cite{vene}). This property helps in the construction of
a scenario for the creation of the initial spectrum of cosmological perturbations
in non-singular models (see Sect.\ref{cobs}).
%

The main goal of this review is to present some of the many
non-singular solutions available in the literature, exhibit the
mechanism by which they avoid the singularity, and discuss what
observational consequences follow from these solutions and may be
taken (hopefully) as an unmistakable evidence of a bounce. We shall
not pretend to produce an exhaustive list, but we intend to include
at least an explicit form for the time evolution of a representative
member of each type of solution \footnote{The issue of singularities
in cosmology has been previously dealt with in \cite{alere}.}. The
models examined here will be restricted to those close or identical
to the FLRW geometry \footnote{Notice however the solutions given in
\cite{senore}. These are non-singular but do not display the
symmetries of the observed universe, although they are very useful
as checks of general theorems.}. Although theories other than GR
will be examined, we shall not consider multidimensional theories
(exception made for models derived from string theory, see
Sect.\ref{pbb}) or theories with torsion.

We shall start in Sect.\ref{defsing} by stating a working
definition of nonsingular universe, and giving a brief account of
the criteria  that can be used to determine whether a certain
model is singular or not. It will suffice for our purposes in this review to define a
singularity as the region where a physical property of the matter source
or the
curvature ``blows up''
\cite{wald}. In fact, since we shall be dealing almost exclusively with
geometries of the Friedmann type, the singularity is always
associated to the divergence of some functional of the curvature
\footnote{But notice that not all types of singularities have large
curvature, and diverging curvature is not
the basic mechanism behind singularity theorems. If we consider
the problem of singularities in a broad sense,
we seem to be ``treating a symptom rather than the cause'' when addressing exclusively
unbounded curvature \cite{bojobscg}.}.

Let us remark at this point that there are at least two different
types of nonsingular universes: (a) bouncing universes (in which the
scale factor attains a minimum), and (b) ``eternal universes'',
which are past infinity and ever expanding, and exist forever.
Class (a) includes cyclic universes.
The focus of this review are those models in class (a),
although we shall review a few examples of models in class (b)
in Sect.\ref{peu}.

\subsubsection{Notation, conventions, etc}

Throughout this report, the Einstein's equations (EE) are given by
$$
R_{\mu\nu}-\half Rg_{\mu\nu}+\Lambda g_{\mu\nu}=-\kappa T_{\mu\nu},
$$
where $\Lambda$ is the cosmological constant, and
$\kappa=8\pi G/c^4$, which we shall set equal to 1,
unless stated otherwise, while the metric of the FLRW model is
\begin{equation}
ds^2 = dt^2 - a^2(t)\,
\left[\frac{dr^2}{1-\epsilon r^2}+r^2\,(d\theta^2+\sin^2\theta\,d\varphi^2)\right],
\label{frwmetric}
\end{equation}
where $\epsilon=-1,\,0,\,+1$. The 3-dimensional surface of
homogeneity $t=$constant is orthogonal to a fundamental class of
observers endowed with a four-velocity vector field $v^{\mu} =
\delta^{\mu}_{\;0}$.
In the case of a perfect fluid with energy density $\rho$ and
pressure $p$, EE take the form
\begin{equation}
\dot{\rho} + 3(\rho + p) \frac{\dot{a}}{a} = 0, \label{dotRho}
\end{equation}
\begin{equation}
\frac{\ddot{a}}{a} = - \,\frac{1}{6} (\rho + 3p)+\frac{\Lambda}{3}, \label{dotTheta}
\end{equation}
in which $\Lambda$ is the cosmological constant, and the dot denotes
the derivative w.r.t. cosmological time. These equations admit a
first integral given by the so-called Friedmann equation:
\begin{equation}
\protect\label{constraint}
\frac{1}{3}\,\rho=\left(\frac{\dot{a}}{a}\right)^2
+\frac{\epsilon}{a^2}.
\end{equation}
The energy-momentum tensor of a theory specified by Lagrangian
${\cal L}$ is given by
\begin{equation}
 T_{\mu\nu} = \frac{2}{\sqrt {-g}} \frac{\delta(
\sqrt{-g}\:{\cal L})}{\delta g^{\mu\nu}},
\label{emtensor}
\end{equation}
where $g={\rm det}(g_{\mu\nu})$.

\subsection{Singularities, bounces, and energy conditions}
\label{defsing}

The issue of the initial singularity of the
FRLW solution was debated for a long time,
since it was not clear if this
singular state was an inherent trace of the universe or just a
consequence of the high degree of symmetry of the model. This
question was discussed firstly in an analytical manner by Lifshitz and collaborators
in
\cite{lif}, where geometries that are solutions of EE with a maximum number
of allowed functions were analyzed.
The results wrongly suggested that the singularity
was not unavoidable, but a consequence of the special
symmetries of the FLRW solution \footnote{For a reappraisal of the work in \cite{lif},
see for instance \cite{rendall} and references therein.}.

From a completely different point of view, Hawking, Penrose, Geroch
and others developed theorems that give global conditions under
which time and null geodesics cannot be extended beyond a certain
(singular) point \cite{earman}. The goal in this case was not about
proving the existence of a region of spacetime in which some
functional of the metric is divergent. Instead, the issue of the
singularity was considered from a wider perspective, characterizing
a spacetime as a whole, by way of its global properties, such as the
abrupt termination of some geodesics in the manifold.  Let us
present a typical
example of these theorems \cite{haw67}:\\

{\bf Theorem:} The following requirements on space-time
$\mathfrak{M}$ are mutually inconsistent:
\begin{enumerate}
\item There exists a compact spacelike hypersurface (without
boundary) $\mathfrak{H};$

\item The divergence $\theta$ of the unit normals to
$\mathfrak{H}$ is positive at every point of $\mathfrak{H};$

\item $ R_{\mu\nu} \, v^{\mu} \, v^{\nu} \leq 0$ for every non-spacelike
vector $v^{\mu};$

\item $\mathfrak{M}$ is geodesically complete in past timelike
directions.
\end{enumerate}
Notice that the link of this theorem with physics comes through
condition (3) via EE, yielding a statement about the energy-momentum
tensor: \beq T_{\mu\nu}v^\mu v^\nu-\frac T 2 \geq 0, \label{sec1}
\eeq called the strong energy condition (SEC), see the Appendix.
Notice also that, although not explicitly mentioned, this theorem
assume the absence of closed timelike curves \cite{earman}. With
hindsight \footnote{From a mathematical point of view, a negative
energy could also allow for a bounce. We will not examine this
possibility in the present paper.}, it can be said that the strength
of these theorems is the generality of their assumptions (at the
time they were conceived), while their weakness is that they give
little information about how the approach to the singularity is
described in terms of the dynamics of the theory or about the nature
of the singularity. In any case, if we assume that the universe is
nonsingular, a positive attitude regarding the singularity theorems
is to consider that they show the limits of applicability of
``reasonable'' hypothesis (such as GR or the energy conditions, see
the Appendix) \cite{bojobscg}.

A local definition of a bounce can also be
given, in the GR framework, in terms of the so-called Tolman
wormhole \cite{viss1, viss2} (see below). Both in
this case and in that of the above mentioned theorems, the
non-singular behavior in
GR is only possible when the SEC
is violated. The assumption of such a condition seemed reasonable in the
early seventies, but several situations have been examined in
the literature that may be relevant in some epoch of the
evolution of the universe, for which SEC is not fulfilled, such as
curvature-coupled scalar fields and cosmological inflation
\cite{twilight,viss1, rose}.

Next we shall examine in some detail how the singularity can be
avoided.
In the following, we shall use a simple  form of the singularity
theorems \footnote{This will suffice for our goals, more refined
formulations can be found in \cite{senore}.}. Let us first
introduce some definitions (following \cite{ellis}). The covariant
derivative of the 4-velocity $v_\mu$ of the fluid that generates the
geometry can be decomposed as follows
\begin{equation}
v_{\nu ; \mu} =  \frac 1 3 \theta\; h_{\mu\nu} + \sigma_{\mu\nu}
+\omega_{\nu\mu} + v_\mu \dot v_\nu,
\label{veldec}
\end{equation}
where $\theta=v^\mu_{\; ;\mu}$ is the expansion, $h_{\mu\nu} =
g_{\mu\nu} - v_\mu v_\nu$, the trace-free symmetric shear tensor is
denoted by $\sigma_{\mu\nu}$, and $\omega_{\mu\nu}$ is the vorticity
tensor (see Eqns.(\ref{d10}) and (\ref{d11})). Defining $S$ by
\footnote{$S$ corresponds to the scale factor $a$ in the case of the FLRW universe.}
\begin{equation}
\frac{\dot S}{S} = \frac \theta 3,
\end{equation}
the Raychaudhuri equation
\cite{raycha}, which follows from
Eqn.(\ref{veldec}) can be written as \footnote{This equation was
independently obtained by A. Komar \cite{komar}.}
\begin{equation}
3\frac{\ddot S}{S} + 2 (\sigma^2 -\omega^2) - \dot v^{\mu}_{;\mu} = -
\frac 1 2 (\rho + 3p) + \Lambda , \label{ray}
\end{equation}
where $A_\mu = v^\nu v_{\mu ;\nu}\equiv \dot v^{\mu}$ is the acceleration.\\[0.2cm]
{\bf Theorem}\cite{ellisescola}: In a universe where $\rho + 3p\geq
0$ is valid, $\Lambda\leq 0  $, and $\dot v^\mu = \omega^\mu = 0$ at
all times, at any instant when $H = \frac 1 3 \theta
>0$, there must have been a time $t_0< 1/ H$ such that $S\rightarrow 0$
as $t\rightarrow t_0$. A space-time singularity
occurs at $t=t_0$, in such a way that $\rho$
and the temperature $T$ diverge.

Several remarks are in order. First, EE were used to obtain Eqn.(\ref{ray}).
Hence, the consequences of the theorem are only valid
in the realm of GR. Second, the singularity implied in the theorem
is universal: any past-directed causal curve ends at it with a
finite proper length, in line with a coherent
definition of a cosmological singularity \footnote{See
\cite{senore} and \cite{matt3} for a classification of
singularities.}. Third, since there is no restriction on the
symmetries of the geometry, $\theta$ is in
principle a function of all the coordinates,  so that the theorem
applies not only to Friedmann-Lemaitre-Robertson-Walker (FLRW) models, but
also to most of the spatially homogeneous, and to some inhomogeneous models (see examples in
\cite{senore}).
Fourth, as we mentioned before,
the condition $\rho + 3p\geq 0$, or more generally, SEC, is
violated even at the classical level, for instance by the massive
scalar field, and also at the quantum level (as in the Casimir
effect \footnote{In fact, it has been shown in \cite{herd} that
the Casimir effect associated to a massive scalar field coupled to
the Ricci scalar in a closed universe can lead to a bounce.}). So
it would be desirable to have singularity theorems founded on more
general energy conditions, but this goal has not been achieved yet
(see \cite{senore}).

Notice that in the general case, acceleration and/or rotation could in
principle avoid the singularity \cite{senore}, but
high pressure cannot prevent the initial
singularity in the FLRW model. Rather, it accelerates the collapse.
This can be seen as follows. The conservation equations $T^{\mu\nu}_{\;\; ;\mu}=0$
give
$$
v^\mu\rho_{,\mu}+(\rho + p)\theta =0,
$$
$$
(\rho +p)A^\mu =-h^{\mu\nu}p_{,\nu}.
$$
Since $p_{,i}=0$ in the FLRW, there is no acceleration. Furthermore,
the pressure contributes to the the active gravitational mass $\rho + 3p$.
%
Finally, not
even a large and positive $\Lambda$ can prevent the singularity in
the context of the theorem \cite{ellisescola}.

As mentioned before, a bounce can also be defined
locally.
The minimal conditions from a local point of view
for a bounce to happen in the case of a
FLRW universe were analyzed in \cite{viss1}, where a Tolman
wormhole was defined as a universe that undergoes a collapse,
attains a minimum radius, and subsequently expands.
Adopting in what follows the metric Eqn.(\ref{frwmetric}),
to have a bounce it is necessary that
$\dot a_b=0$, and $\ddot a_b\geq 0$. For this to be a
true minimum
of the scale factor (conventionally located at $t=0$)
there must exists a time $\tilde t$ such that $\ddot a>0$
for all $t\in (-\tilde t,0) \cup (0,\tilde t)$. From EE in the FLRW universe we get
$$
\rho =3\left(\frac{\dot a^2}{a^2}+\frac{\epsilon}{a^2}\right),
$$
$$
p=-\left(2\frac{\ddot a}{a}+\frac{\dot a^2}{a^2}+\frac{\epsilon}{a^2}\right).
$$
From these, the combinations relevant for the energy conditions (see Sect.\ref{app1})
are:
$$
\rho + p =2\left(-\frac{d^2\ln a}{dt^2}+\frac{\epsilon}{a^2}\right),
$$
$$
\rho - p =2\left(\frac{1}{3a^3}\frac{d^2(a^3)}{dt^2}+2\frac{\epsilon}{a^2}\right),
$$
$$
\rho + 3p = -6\;\frac{\ddot a}{a}.
$$
From these conditions and $\dot a_b=0$, and $\ddot a_b\geq 0$
it follows that \cite{viss1}
$$
\exists\; {\rm bounce\; and}\;\epsilon =-1\Rightarrow {\rm NEC\;
violated},
\footnote{For the energy conditions,
see the Appendix.}$$
$$
\exists\; {\rm bounce\;and}\; (\epsilon =0; \ddot
a_b>0)\Rightarrow {\rm NEC\; violated},
$$
$$
\exists \;{\rm bounce\;and}\; (\epsilon =1; \ddot
a_b>a_b^{-1})\Rightarrow {\rm NEC\; violated},
$$
The definition of $\rho$ and $p$
and $\ddot a>0$ imply that:
$$
\rho + p <2\;\frac{\epsilon}{a^2},
$$
$$
\rho-p>2\;\frac{\epsilon}{a^2},
$$
$$
\rho-3p<0.$$
It follows that
$$
\exists\;{\rm bounce\;and}\;\epsilon \neq 1\Rightarrow {\rm NEC\;
violated},
$$
$$
\exists\; {\rm bounce} \Rightarrow {\rm SEC\; violated}.
$$
The case that minimizes the violations of the energy conditions can be stated as
$$
\exists\; {\rm bounce\;and} \; (\epsilon =+1;\ddot a_b\leq
a_b^{-1})\Rightarrow {\rm NEC, WEC, DEC\; satisfied; SEC
\;violated}.
$$
This result may be expected since the curvature term with $\epsilon
= +1$ acts like a negative energy density in Friedmann's equation.
Notice that in this analysis, only Einstein's equations and the
point-wise energy conditions were used, without assuming any
particular equation of state. In a certain sense, this is the
inverse of the theorem stated earlier, which assumed the validity of
the SEC \footnote{An analysis along the same lines but with a more
general parametrization for the scale factor was carried out in
\cite{nelson}.}.

The restriction to a FLRW model was lifted in a subsequent paper
\cite{viss2}, and the analysis in a general case was done
following standard techniques taken from the ordinary wormhole
case \cite{viss3}.  It was found that even in the case of a geometry with
no particular symmetries, the SEC must be violated
if there is to be a bounce in GR. Consequently, one can conclude that
the singularity theorems that assume that SEC is valid cannot be
improved. A highlight in these analysis is that only the local
geometrical structure of the bounce was needed; no assumptions
about asymptotic or topology were required, in contrast with the
Hawking-Penrose singularity theorems \cite{sing}.  Equally
important is the fact that, as mentioned above, SEC may not be
such a fundamental physical restriction.

To summarize what was discussed up to now, we can say
that there is a ``window of opportunity'' to avoid the initial
singularity in FLRW models at a classical level
by one or a combination of the
following assumptions \footnote{We shall not consider here the existence of closed timelike
curves as a possible cause of a nonsingular universe.}:
\begin{enumerate}

\item Violating SEC in the realm of GR \footnote{A complete
analysis of the behavior of the energy conditions for different
types of singularities has been presented in \cite{matt3}.};

\item Working with a new gravitational theory, as for instance those that
add scalar degrees of freedom to gravity (Brans-Dicke
theory being the paradigmatic example
of this type, see Sect.\ref{cst}), or by adopting an action
built with higher-order invariants (see Sect.\ref{cho}).
\end{enumerate}
As will be seen below, other ways to avoid the singularity are:
\begin{enumerate}
\item Changing the way gravity couples to matter (from minimal to
non-minimal coupling, see for instance the case of the scalar field in Sect.\ref{cst});

\item Using a non-perfect fluid as a source, see Sect.\ref{visco}.

\end{enumerate}

Finally, quantum gravitational effects also give the chance of
a bounce (see Sect.\ref{slqg}) \footnote{A definition of a nonsingular space
using the so-called principle of quantum hyperbolicity has been given in
\cite{bojobscg}}.

\subsection{Extrema of $a(t)$ and $\rho(t)$}
Let us study the relations imposed by EE between extrema of
the scale factor, the energy density, and the energy conditions, in
the case of one fluid.
Let us recall that the sufficient conditions to
have a bounce are \footnote{We are assuming that $\ddot a\neq 0.$}
$\theta_b=0$ and $\dot\theta_b>0$, where $\theta=3\dot a/a$ , and
the subindex $b$ denotes that the quantities are evaluated at the
bounce. It follows from Raychaudhuri's equation for the
FLRW model (Eqn.(\ref{ray})) with $\Lambda=0$, \beq \dot\theta +
\frac{{\theta}^2}{3}=-\half (\rho + 3p), \eeq that at the bounce we
must have $\left.(\rho + 3p)\right|_b<0$, independently of the value
of $\epsilon$ (as was also shown in the previous section). From the
conservation equation,
$$
\dot \rho = -(\rho + p) \theta,
$$
we see that there may be extrema of $\rho$ when $\theta_e=0$ (as in
the case of a putative bounce) and/or when $\rho_e=-p_e$.
The second
derivative of the energy density is given by
 \beq
 \ddot \rho = -(\dot\rho +
\dot p)\theta -(\rho + p)\dot \theta.
\label{ddrho}
\eeq
Let us assume first that $\theta_e=0$ with $\rho_e + p_e\neq 0$,
which implies that $\dot \rho_e =0$ and
$$
\ddot \rho_e = -(\rho_e+ p_e)\dot \theta_e,\;\;\;\;\;\;\;\;\;\;\dot\theta_e=-\half (\rho_e+3p_e).
$$
The different possibilities, according to the sign of $\dot\theta_e$,
$\rho_e+p_e$, and $\rho+3p$ are displayed in the following table:
\begin{center}
\begin{tabular*}{8.68cm}{|c|c|c|c|c|c|}
\hline
$\rho_e+3p_e $&  $\stackrel{\textbf{.}}{\theta_e}$ & $ \rho_e+p_e$  &  $\ddot\rho_e$ &
  $\rho_e $ & $a_e$   \\ \hline
$<0$ &   \multirow{2}{*}{$>0$}   &  $<0$ &  $>0$  &  min. & min. \\ \cline{3-5}
 & & $>0$ & $<0$ & max. & \\
\hline
$ >0$ &   \multirow{2}{*}{$<0$}   &  $<0$ &  $<0$  &  max. & max. \\ \cline{3-5}
 & & $>0$ & $>0$ & min. & \\
\hline
\label{tabla1}
\end{tabular*}
\end{center}
We see that there are two cases that agree with what may be termed
``normal matter'' (rows 2 and 4), in the sense that maximum (minimum)
compression leads to maximum (minimum) energy density. Notice however that
the case in row 2 violates the strong energy condition (see Appendix).
The other cases are clearly unusual: minimum density
with minimum scale factor (row 1), and the opposite (that is,
maximum density with maximum scale factor, row 3)\footnote{The former
is precisely
the behavior that allows for a bounce in loop quantum gravity
\cite{bojobou}
(see Sect.(\ref{slqg})), while the latter
is what is found in the so-called big-rip
\cite{bigrip}.}.
Notice that it is the null energy condition $\rho +p >0$ (see Appendix) and not the SEC
that
is violated at these unusual cases.
In fact, if the requirement $\rho
+p\geq 0$ is not satisfied, then the equation of energy
conservation for a perfect fluid,
\begin{equation} \dot\rho = -\theta (\rho + p),
\end{equation} says that
compression would entail a decreasing energy density, which is a
rather unexpected behavior for a fluid \footnote{Fluids that violate the
NEC are called \emph{phantom} or \emph{ghost} fluids, and have been studied in
\cite{menace}.}.
Examples of the four behaviors will be found along this review.

When an EOS $p=\lambda \rho$ plus the condition $\rho>0$ are imposed\footnote{Notice
that some models do not satisfy this conditions, see for instance Eqn.(\ref{Neg4}).},
we see that the case in row 1 is permitted for $\lambda <-1$, and
that in row 2, for $\lambda\in (-1,-1/3)$. The case in row
3 is not allowed for any $\lambda$, while
that in row for is permitted for $\lambda >-1/3$.

Notice that all the extrema in $\rho$ in Table \ref{tabla1} are global, since
the other possibility
(given by $\rho_e+p_e=0$) leads to an inflection point in $\rho$,
assuming that
$p=\lambda \rho$.

\subsection{Appendix: Energy conditions}
\label{app1} We shall
give next the general expression of the energy conditions, and
also their form for the particular case
of the energy-momentum tensor given by \beq T^{\mu}_{\;\nu}={\rm
diag}(\rho ,-p,-p,-p). \label{tmunu} \eeq
%
%
\begin{itemize}

\item The null energy condition (NEC) states that for any null
vector, \beq NEC \Leftrightarrow T_{\mu\nu} k^\mu k^\nu \geq 0.
\eeq In terms of Eq.(\ref{tmunu}), \beq NEC \Leftrightarrow \rho +
p \geq 0. \eeq

\item The weak energy condition (WEC) asserts that \beq WEC
\Leftrightarrow T_{\mu\nu} v^\mu v^\nu \geq 0 \eeq for any
timelike vector. In terms of Eqn.(\ref{tmunu}), \beq \rho\geq 0,
\;{\rm and}\; \rho + p \geq 0. \eeq

\item The strong energy condition (SEC) is the assertion that, for
any timelike vector, \beq SEC \Leftrightarrow \left( T_{\mu\nu}-
\frac T 2 g_{\mu\nu} \right)  v^\mu v^\nu \geq 0. \eeq In terms of
Eqn.(\ref{tmunu}), \beq \rho + p \geq 0, \;{\rm and}\; \rho + 3p
\geq 0 .\eeq

\end{itemize}
Each of these three conditions has an averaged counterpart
\cite{libromatt}. There is yet another condition:
\begin{itemize}
\item The dominant energy condition (DEC) says that for any
timelike vector \beq DEC \Leftrightarrow  T_{\mu\nu}   v^\mu v^\nu
\geq 0\; {\rm and}\; T_{\mu\nu}v^\nu \;{\rm is \;not\; spacelike.}
\eeq
\end{itemize}
The different energy conditions are not independent. The following
relations are valid: \beq WEC\Rightarrow NEC, \eeq \beq SEC
\Rightarrow NEC, \eeq \beq DEC\Rightarrow WEC . \eeq Notice that
if NEC is violated then all the other pointwise energy conditions
would be violated \cite{libromatt}.

\newpage

\section{Higher-order gravitational theories}
\label{cho}

Higher-order terms in the action for gravity (such as $R^2,
R_{\mu\nu}R^{\mu\nu}, $ etc.) typically appear due to quantum
effects, either in the case of quantized matter in a fixed
gravitational background \cite{davies}, or in the gravitational
effective action as corrections from quantum gravity \cite{corrqg}
or string theory \footnote{Since in this case the non-linear terms
are always coupled to one or more scalar fields we shall consider it
in Sect. \ref{ssst}.} \cite{corrst}. These terms are expected to be
important in situations of high curvature, when the scale factor is small
\footnote{As opposed to Lagrangians that are negative powers of
$R$, which are currently being considered as candidates to explain the
acceleration of the universe \cite{odin}.}.
The models that are engineered to work in the intermediate regime,
where quantized matter fields evolve on a given classical geometry
(the so-called semiclassical approximation) mirror
the path taken in the early days of quantum field theory,
in which quantum matter was in interaction with a
classical electromagnetic
background field. In the case of
gravity, it is generally agreed that
this approach may be valid for distances above $\ell_{Pl}$,
although this statement can only be verified by a complete quantum
theory of gravitation, not yet available. As we shall see in Ch.\ref{cqc}, some models go
below $\ell_{Pl}$, incorporating effects expected to be present in the
complete theory, but for the time being the
quest of the ''correct theory'' at this energy level seems far
from being settled.

\subsection{Quantized matter on a fixed background}
\label{fb}
Let us start by considering the corrections coming from quantum
matter in a given background. As shown for instance in \cite{uti},
in the models based on the semiclassical approximation
the mean value of the stress-energy tensor $T_{\mu\nu}$ of
a set of quantized fields interacting with a classical geometry is
plagued with infinities.
These divergencies can be removed by a suitable modification of EE
that follows from a renormalization procedure. In order to render
the mean value of $T_{\mu\nu}$ finite, the
cosmological constant $\Lambda$ and Einstein's constant $\kappa$ are
renormalized, and a counterterm of the form

\beq \triangle L = \sqrt{-g} \, ( \alpha \, R^{2} + \beta \,
R^{\mu\nu}\, R_{\mu\nu} ) \label{reno} \eeq must be introduced in
the Lagrangian \footnote{The relevance of this type of series
development was discussed also by Sakharov \cite{sakha}.}.
The corrections arise from the ultraviolet behavior of the field
modes, which only probe the local geometry, hence the appearance
of geometric quantities. After the elimination of the divergences
and with a convenient choice of $\alpha$ and $\beta$, EE with
$<T_{\mu\nu}>$ as a source preserve their form \cite{uti}:
\begin{equation}
G_{\mu\nu} +  \Lambda^{(ren)} \, g_{\mu\nu} = -\kappa^{(ren)} \,
<T^{(ren)}_{\mu\nu}>.
 \label{HOTG2}
\end{equation}
Note that such renormalization does not affect the conservation of
the energy-momentum tensor, that is
\begin{equation}
<{T_{(ren)}^{\mu\nu}}>_{\;;\nu} = 0 \label{23dez1}.
\end{equation}
Since the constants introduced by the counterterm are to be
determined by experiment, instead of fixing their values so as to
eliminate the quadratic contribution to EE (as was done in
\cite{uti}), we can shift them as $\alpha \rightarrow \alpha +
\eta$
 and
 $ \beta   \rightarrow \beta + \gamma \, \eta$. The new equations are
\begin{equation}
G_{\mu\nu} + \eta( \chi_{\mu\nu} + \gamma \, Z_{\mu\nu} )  +
\Lambda^{(ren)} \, g_{\mu\nu} = -\kappa^{(ren)} \,<T^{(ren)}_{\mu\nu}>,
 \label{HOTG3}
\end{equation}
where
 \begin{equation}
\frac{1}{2} \,\chi_{\mu\nu} \equiv  R (R_{\mu\nu} - \frac{1}{4} \,
R g_{\mu\nu} ) + R_{\;;\mu;\nu} - g_{\mu\nu}\Box R  \label{HOTG4},
\end{equation}
and
\begin{equation}
Z_{\mu\nu} \equiv R_{\;;\mu;\nu} - \Box R_{\mu\nu} - \frac{1}{2}
\,( \Box R + R_{\alpha\beta} \, R^{\alpha\beta}) g_{\mu\nu} + 2
R^{\alpha\beta}\, R_{\alpha\mu\beta\nu}.
 \label{HOTG5}
\end{equation}
Cosmological solutions of Eqn.(\ref{HOTG3}) in the
case of the FLRW metric were studied in \cite{nato}. For a flat universe,
the equations
take the form
\beq 3\left(\frac{\dot a}{a}\right)^2 + 3 t_c^2
\left\{ \left(\frac{\ddot a}{a} + \left(\frac{\dot
a}{a}\right)^2\right) \left(\frac{\ddot a}{a} - \left(\frac{\dot
a}{a}\right)^2\right) - 2 \left(\frac{\dot a}{a}\right)
\left(\frac{\ddot a}{a}+\left(\frac{\dot
a}{a}\right)^2\right)\right\} = \rho, \eeq \beq
\dot\rho+3\left(\frac{\dot a}{a}\right)(\rho+p)=0, \eeq where
$t_c\equiv 1/(c\mu)$, and $\mu \equiv -2\eta (\gamma +3)$.
For the case of radiation ($\rho=\rho_ca_c^4/a^4$), we get
\beq H^2+t_c^2\left\{\left(\frac{\ddot
a}{a}-H^2\right)^2-2H\left(\frac{\stackrel{...}{a}}{a}-H^3
\right)\right\}= \frac{\rho_c}{3}\left(\frac{a_c}{a}\right)^4, \eeq
where $H=\dot a/a$, and $\rho_c=\rho(t_c)$.
If we impose the
existence of a bounce by the conditions  $a_b>0, \dot a_b=0, $ and
$\ddot a_b>0$, it follows from this equation that $\mu>0$. It as also shown in
\cite{nato} $t_c\leq 3.33\times
10^{-4}$ sec. in order that the theory does not conflict with the
three classical tests of GR.


Vacuum solutions of Eqn.(\ref{HOTG3}) in the FLRW geometry were
studied in \cite{mul1}. Notice that taking the trace of
Eqn.(\ref{HOTG3}) in the absence of matter we obtain
$$
\ddot R + h \dot R + \sigma R = 0,
$$
where $\sigma = 1/(2\eta(1+\gamma))$,
$h=d[\ln(-g)^{1/2}]/dt$. This equation is analogous to that
of a damped harmonic oscillator. Depending on the sign of the
parameter $\sigma$, there may be damped oscillations for $R$ around
$R=0$, or exponentially decaying or growing solutions \cite{mul1}.

Corrections coming from one-loop contributions of
conformally-invariant matter fields on a FLRW background were
studied in \cite{starobinsky} (see also
\cite{fischetti}). They allow for nonsingular solutions
that are not of the bouncing type since they describe a universe
starting from a deSitter state. A thorough analysis of this
setting was given in \cite{ander1}, where the back-reaction
problem for conformally invariant free quantum fields in FLRW
spacetimes with radiation was studied, for both zero \cite{ander1} and non-zero
\cite{ander2} curvature and/or $\Lambda$. It was found that depending
on the values of the regularization parameters, there are some
bouncing solutions that approach FLRW at late times.

\subsection{Lagrangians depending on the Ricci scalar}
\label{riccis}
On approaching the singularity, powers of the curvature
may be expected to play an important dynamical role, hence other
possible nonlinear Lagrangians are those belonging to the class
defined by
\begin{equation}
S = \int \sqrt{-g} \, f(R)\, d^{4}x,
\end{equation}
where $f(R)$ is an arbitrary function of the curvature scalar,
encompassing polynomials as a particular case. The problem of the
singularity using this type of Lagrangians has been repeatedly
discussed in the literature (see for instance \cite{buch,bao}).
The EOM that follows from this action is
\begin{equation}
f^{'} \,R_{\mu\nu} - \frac{1}{2} \, f \, g_{\mu\nu}   - \Box f \,
g_{\mu\nu}   + {f^{'}}_{,\mu;\nu}   = 0, \label{24dez1}
\end{equation}
where $f^{'} \equiv df/dR.$
This equation can be expressed in $f$ and its derivatives as
\begin{equation}
f^{'} \,R_{\mu\nu} - \frac{1}{2} \, f \, g_{\mu\nu} + f^{''} \,
(R_{,\mu;\nu} - \Box R \, g_{\mu\nu} )  + f^{'''}\, (R_{,\mu}
R_{,\nu} - R_{,\lambda} R^{,\lambda} \, g _{\mu\nu}) = 0,
\label{23dez4}
\end{equation}
or, using the trace,
\beq f^{'} \,\left(R_{\mu\nu}-\frac 1 4 R
g_{\mu\nu}\right)+ f^{''}\left(R_{,\mu;\nu} - \frac 1 4 \,
g_{\mu\nu} \Box R \right) +f^{'''}\, \left(R_{,\mu}R_{,\nu} -\frac
1 4 R_{,\lambda} R^{,\lambda} \, g _{\mu\nu}\right) = 0.
\label{fr} \eeq

The particular example given by \beq
 f(R)  = R + \alpha \, R^{2}
\label{f(R)} \eeq
was studied by many authors \cite{nariai, ruz, gies, mac}.
In principle a term
$R_{\alpha\beta\gamma\delta}R^{\alpha\beta\gamma\delta}$ should be
included in the action, but the existence of a topological
invariant yields
$$
\delta\int(R_{\alpha\beta\gamma\delta}R^{\alpha\beta\gamma\delta}-4R_{\alpha\beta}
R^{\alpha\beta}+R^2)\;\sqrt{-g}\;d^4x = 0,
$$
in such a way that the Riemann-squared term can be omitted. The
equations of motion for the Lagrangian introduced in Eqn.(\ref{f(R)}) in
the presence of matter are
\begin{equation}
(1 + 2\alpha R )\, R_{\mu\nu} - \frac{1}{2} ( R + \alpha R^{2} )\,
g_{\mu\nu} + 2\alpha ( R_{,\mu;\nu} - \Box R \, g_{\mu\nu} ) = -
T_{\mu\nu}. \label{Nariai2}
\end{equation}
If we restrict
ultra-relativistic matter, the $0-0$ component of this
equation yields
\begin{equation}
\rho = \frac{1}{3} \, \theta^{2} + \frac{3\epsilon}{a^{2}} -
2\alpha \dot{\theta} \left(\dot{\theta} + \frac{2}{3} \, \theta^{2} \right) +
\frac{18 \epsilon^{2} \, \alpha}{a^{4}} + \frac{4 \epsilon \,
\alpha}{a^{2}} + 2\alpha\theta \, \dot{R}, \label{Nariai3}
\end{equation}
where $ R = 2 \dot{\theta} + 4\, \theta^{2}/3 +
{6\epsilon}/{a^{2}}.$ At the point where the bounce occurs,
$\theta_{b} = 0$ and $\dot\theta_{b} > 0$, and Eqn.(\ref{Nariai3})
reduces to
\begin{equation}
\rho_{b} = - \, 2 \alpha \dot{\theta_{b}}^{2} +
\frac{3\epsilon}{a_{b}^{2}} \, \left( 1 +
\frac{6\alpha\epsilon}{a_{b}^{2}}+ \frac{4\alpha}{3}\right).
\label{Nariai1}
\end{equation}
Let us take as an example the case in which the section is
Euclidean. If we want to have
a minimum with positive energy density,
it follows from Eqn.(\ref{Nariai1}) that
$ \alpha < 0$.
As shown in \cite{ruz}, such a choice
for the action of the gravitational field admits
solutions in the FLRW framework that allow a
regular transition from a contracting to an expanding phase.
Although negative values of $\alpha$ remove the initial singularity, it was
shown in \cite{ruz,gies} that the solutions with $ \alpha < 0$ do not go to
the corresponding FLRW
solution ($a\propto t^{1/2}$) for large $t$.

A theory that generalizes that defined by Eqn.(\ref{f(R)}), namely
$$
f(R)=R+\alpha R^n
$$
was studied in \cite{ruz2}. It was found that the FLRW solution for $n=4/3$
and $p=\rho/3$ is regular for all values of $t$, and
tends to the radiation solution for large values of $t$. Later,
solutions of this theory with dust as a source were found to have
similar properties in \cite{guro}.

Another type of corrections, given by
the Lagrangian
\beq
{\cal L}=R + \Lambda +BR^2+CR^2\ln|R|,
\label{guro}
\eeq
were studied in \cite{guro1} (with
$B$ and $C$ constants). The quadratic and logarithmic terms are
consequences of vacuum polarization \cite{wit}.
Although this form
of the Lagrangian does not eliminate the singularity in the FLRW solutions, addition of
particle creation effects through a viscosity term does (see
Ch.\ref{visco}) \footnote{A Bianchi I solution of this theory with and
without self-consistent particle production
was considered in \cite{gusta}. It was shown that
particle production quickly isotropizes the
model.}.

The stability analysis of the FLRW solution in theories with ${\cal
L}=f(R)$ was performed in \cite{bao}, along with necessary and
sufficient conditions for the existence of singularities.
Eqn.(\ref{23dez4}) in the case of a FLRW geometry in the presence of
matter reduces to \cite{kerner}
\beq f^{''}\dot a (a^2
\stackrel{...}{a}+a \dot a \ddot a -2\dot a ^3 - 2\dot a \epsilon )
+ \frac 1 6 f^{'} a^3 \ddot a +\frac{1}{36}fa^4 + \frac{1}{18}
a^4T_{00} = 0. \label{eqkerner} \eeq The argument of the function
$f$ is given by \beq R = \frac{6}{a^2}(a\ddot a + \dot a^2
+\epsilon).
\eeq
Assuming that near the bounce the scale factor can
be developed in a power series as \beq \label{ps} a(t)=a_0+\half\;
a_1t^2+\frac 1 6\; a_2 t^3 +..., \eeq a necessary condition for the
bounce was given \cite{bao}:
\beq f_0a_0+6a_1f'_0\leq 0, \eeq
where
$f_0=f(R_0)$, and $R_0=-6a_0^{-2}(a_0a_1+\epsilon)$, and it was
assumed that $T_{00}>0$. In the quadratic case given by
Eqn.(\ref{f(R)}), this condition takes the form \beq
6\alpha\epsilon^2-a_0^2\epsilon-6\alpha a_1^2a_0^2<0. \eeq When
$\epsilon=0$, the condition $\alpha >0$ is regained, but there are
other possibilities when $\epsilon=1,-1$ \cite{bao}.
In the same vein, but without using a series development,
conditions for a bounce in $f(R)$ theories were studied
in \cite{carloni}
\footnote{Bounce solutions were also shown to exist in
orthogonal spatially homogeneous Bianchi cosmologies in $f(R)=R^n$ in \cite{goheer}.}. The basic equations are, that follow from
Raychaudhuri's equation and the Gauss-Codazzi equation are
$$
\frac{\ddot a_b}{a_b}=-\frac{\rho_b}{f'_b}+\frac{f_b}{f'_b},
$$
$$
R=6\left(\frac{\ddot a_b}{a_b}+\frac{\epsilon}{a_b^2}\right).
$$
These equations were used in \cite{carloni} to analyze a possible
bounce in the theories given by $f_1(R)=R^n$, $f_2(R)=R+\alpha R^m$,
$f_3(R)=\exp (\lambda R)$. Bounces for $\epsilon =\pm 1$ are possible in the
case of $f_1$. This case can describe an ``almost-FRLW'' phase
folowed by an accelerated phase
if $n > 1$ and $n$ is odd for $\epsilon =-1$
and $R > 0$. The same happens with $n$ even
and $n < 0$ with $R > 0$ or $0 < n < 1$ with $R < 0$, where in the second case $n$ can be
only rational. For $f_2$, closed bounces
are allowed for every integer value of $m$ (often together with open bounces). For $m$
rational, closed bounces are not allowed in general for $0 < m < 1$. For $m$ rational
with even denominator there is no closed bounce for $(m > 1, \alpha < 0)$ and no bounce
at all for negative $m$ and $\alpha$.
In the case of $f_3$, one of the following two conditions must be satisfied in order
to have a bounce:
$\lambda >0$ and $R_b>\ln (2\rho_b)/\lambda$, or
$\lambda <0$ and $R_b<\ln (2\rho_b)/\lambda$.

Some exact solutions have been recently found in \cite{clifton2}
for the theory defined by $f(R)=R^{1+\delta}$. For the vacuum case
with $\epsilon=0$, there is bouncing (entirely due to the dynamics
of the theory), for $0<\delta<1/4$. There are vacuum solutions for
$\delta =1/2$ and $\epsilon\neq 0$, are given by
$$
ds^2=dt^2-(\kappa-\kappa t^2\pm t^4)\left(\frac{dr^2}{1-\epsilon
r^2}+r^2d\Omega^2\right).
$$
This solution exhibits a bounce for $\kappa>0$. Bouncing solutions
were also obtained for a perfect fluid with $p=(\gamma -1)\rho$ in
the case $\delta = 1/(3\gamma-1)$ \footnote{Cyclic solutions were
obtained in the case $\delta=(3\gamma-4)/(2(7-3\gamma))$ for a
convenient choice of the integration constants.}.

We would like to close this section by pointing out that Eqn.\ref{eqkerner}
illustrates the fact that the flatness problem is not a priori a problem in
theories other than GR (no definite
behavior of $|\Omega -1|$ with time follows from \ref{eqkerner}).

\subsubsection{Saturation}

An interesting idea was proposed in \cite{kerner} to limit
the curvature by adding terms in the Lagrangian,
following
the lines that Born and Infeld \cite{bi} devised to avoid
singularities in electromagnetism. The Born-Infeld Lagrangian,
given by
\begin{equation}
{\cal L}_{BI} = \beta^{2} \left[\sqrt{ 1 -
\frac{\mathscr{H}^2-\mathscr{E}^2}{\beta^4}}-1 \right] \label{lagbi}
\end{equation}
is such that
the invariant $\mathscr{H}^2-\mathscr{E}^2$ cannot take values higher
than $\beta^4$. The fact that it takes more and more energy to
increment the field when it takes values near $\beta^2$ is a
phenomenon called saturation \footnote{This is analogous to the
fact that it takes an infinite amount of energy to accelerate a
mass moving with $v\approx c$ in special relativity.}. A similar cutoff may be postulated
for the curvature tensor
when quantum gravitational fluctuations become non-negligible,
that is (presumably), when
$$
R\approx \ell_{Pl}^{-2}\approx 10^{66}{\rm cm}^{-2}.
$$
In \cite{kerner}, non-polynomial
Lagrangians $f(R)$ were considered
such that they reduce to $R$ when $R<<\ell_{Pl}^{-2}$, and
required that $f(R)\rightarrow $ constant for $R\rightarrow
\infty$. This condition is of course not enough to determine the
Lagrangian, but a qualitative guess can be made. A typical
Lagrangian that fulfills the above given conditions is
\begin{equation}
f(R) = \frac{R}{1 -\ell_{Pl}^{2}R}. \label{lagkerner}
\end{equation}
An
approximate solution of the EOM (\ref{eqkerner})
for
(\ref{lagkerner}) by a development as a power series of $t$ for $\epsilon =0$
was built in \cite{kerner2}, the solution
being non-singular though strongly dependent on the
non-linearities of the chosen Lagrangian.
%

The idea of saturation was subsequently explored in \cite{kerner3}, where an
explicit nonsingular solution given by
\beq
a(t) = \sigma \left( 1 + \frac{\beta^4
t^2}{\sigma^4}\right)^{1/4},
\label{ans}
\eeq
was inserted in Eqn.(\ref{eqkerner}),
where
$\sigma$ is a small parameter. This expression tends to the
radiation-dominated scale factor for $\beta^4 t^2/\sigma^4>>1$.
With this $a(t)$ and using that $R=-3\beta^4\sigma^4/a^8$,
Eqn.(\ref{eqkerner}) can be rewritten as an ordinary linear
second-order differential equation for $f(R)$. This equation was
integrated for all the values of the 3-curvature. The dependence
of the resulting $f(R)$ on the chosen form of $a(t)$ was tested
in the case $\epsilon =0$ with that obtained from
$a^8(t)=1+2(1+\alpha)t^2+ t^4$, which has the same asymptotic limit
of Eqn.(\ref{ans}). The result in this second case is not distinguishable from the
first.

A related analysis was carried out in \cite{biswas}, where it was asked that the theory
defined by $f(R)$
be asymptotically
free (implying that gravity becomes weak at short distances, in such a way that
pressure may counteract the gravitational attraction, thus avoiding the singularity), and
also ghost-free (so that the bounce is not caused by negative-energy-density matter)
\footnote{For the relation between $f(R)$ theories and ghosts, see
\cite{ghosts}.}.
The actions studied in \cite{biswas} that satisfy these requirements
were specified by
\footnote{It was shown in \cite{biswas}
that polynomial actions in $R$ do not satisfy these
requirements.}
\beq
f(R)=R+\sum_{n=0}^\infty\;c_nR\Box^nR,
\label{bisw}
\eeq
and can be rewritten in terms of a higher-derivative scalar-tensor action:
$$
S=\int d^4x \sqrt{-g}\left(\Phi R + \psi \sum_1^\infty c_i\Box^i\psi-(\psi(\Phi-1)-c_0\psi^2\right),
$$
from which it follows that $\psi = R$ (from the EOM of $\Phi$).
After a conformal transformation and linearization
it follows that the EOM for the scalar fields are \cite{biswas}
$$
\psi=3\Box\phi,\;\;\;\;\;\;\;\;\;\;\phi=2\left(\sum_1^\infty c_i\Box^i\psi+c_0\psi\right)
$$
with $\Phi=e^\phi$.
From these we get
$$
\left(1-6\sum_0^\infty c_i\Box^{i+1}\right)\phi\equiv\Gamma(\Box)\phi=0,
$$
and the scalar propagator is
$$
G(p^2) \propto \frac{1}{\Gamma(-p^2)}.
$$
It is precisely the function $\Gamma$ that controls the absence of ghosts
and the asymptotic properties of the theory,  which was parameterized
in \cite{biswas} as $\Gamma(-p^2)=e^{\gamma(-p^2)}$, with $\gamma$ analytic.
To actually show the existence of bouncing solutions with the properties mentioned above,
the scale factor
$$
a(t)=a_0\cosh\left(\sqrt{\frac{\omega}{2}}\;t\right),
$$
was imposed in the equation for $G_{00}$ written in terms of $\Gamma$ and its derivatives,
and compared with the
r.h.s. composed of radiation and cosmological constant, thus yielding the following
constraints on $\Gamma$:
$$
\Gamma'(\omega) =\frac 2 3 \Gamma'(0) - \frac{1}{3\omega},
$$
$$
2\omega\Gamma'(\omega)-1\geq 0
$$
(the latter coming from demanding that the bounce be caused by the nonlinearities, and not
by the radiation energy density). The authors go on to show that the
kinetic operator defined by
$$
\gamma(\omega)=k_1\omega-k_2\omega^2+k_4\omega^4,
$$
where $k_i$ are constants, satisfies the constraints
and has the correct
Newtonian limit. So a bouncing solution that is ghost and asymptotically
free exists for the theory defined by
Eqn.(\ref{bisw}), although
the Lagrangian in the original variable $R$ was not exhibited.
%
%
%

\subsection{The limiting curvature hypothesis (LCH)}
\label{slch}

A different proposal to deal with the singularity problem in the higher-order-curvature
scenario is to
adopt the limiting curvature hypothesis, introduced by M. Markov
\cite{maxcurv} as the limiting density hypothesis
\footnote{For boucing solutions that implement this hypothesis through modifications
of the EOS, see \cite{rosen}.}. The LHC
postulates the existence of a maximum value for the
curvature, in such a way that
$$
R^2<\ell_{Pl}^{-4},\;\;\;\;\;R_{\mu\nu}R^{\mu\nu}<\ell_{Pl}^{-8},\;\;\;\;\;\;
W_{\alpha\beta\gamma\delta}W^{\alpha\beta\gamma\delta}<\ell_{Pl}^{-8},
$$
etc, and that any geometry must approach a definite nonsingular
solution (typically the de Sitter solution) when the limiting
curvature is reached. This automatically guarantees that all curvature
invariants are finite \cite{mamu}. A nonsingular higher order theory was
constructed in \cite{mukh} in which every contracting and
spatially flat, isotropic universe avoids the big crunch by ending
up in a deSitter state enforced by the LCH, for all initial
conditions and general matter content \footnote{Note that the LFH
furnishes in this case a nonsingular
universe without bounce.}.
The action used in \cite{mukh} was the linear action plus
a
non-linear term $I_2$ with the property that \beq
I_2(g_{\mu\nu})=0\Leftrightarrow g_{\mu\nu}=g_{\mu\nu}^{DS},
\label{cond} \eeq
and enforced that $I_2\rightarrow 0$ for large
curvatures using an auxiliary field (see below).
In a subsequent paper \cite{bran1}, the method was applied to an
isotropic, homogeneous universe, both in vacuum and in the
presence of matter. The solutions corresponding to $\epsilon =1$ display a
deSitter bounce. In the case in which matter is present, it is
shown that its coupling to gravity is asymptotically free.
Later, the model was generalized to include a dilaton field
\cite{bd1}, in which case it admits flat bouncing solutions. The
starting point is the dilaton gravity action with an added
non-linear term ($I_2$) times a Lagrange multiplier $\psi$ subject
to a potential $V(\psi)$: \beq S =
-\frac{1}{2\kappa^2}\act\left(R-\half\;
(\nabla\phi)^2+\frac{1}{\sqrt{12}}\;
\psi\;e^{\gamma\phi}I_2+V(\psi)\right). \eeq The potential is to
be tailored from the EOM and the constraint equations in such a
way that $I_2$, given by
$$
I_2 = \sqrt{4R_{\mu\nu}R^{\mu\nu}},
$$
goes to zero for large curvatures. Notice that this form of $I_2$
satisfies condition (\ref{cond}), so all the curvature invariants
are automatically bounded. Restricting to an FLRW metric with $k=0$,
the EOM are
\beq \dot\psi = -3H\psi + 6H - \frac 1 H \left(\half
\chi^2+V(\psi)\right), \label{b1} \eeq
\beq \dot H = - V'(\psi),
\label{b2} \eeq
\beq \dot\chi = -3H\chi, \label{b3} \eeq
with
$\chi = \dot\phi$, and a prime denotes derivative wrt $\psi$.
An example was given in \cite{bd1}, where
\beq V(\psi) =
\frac{\psi^2-\frac{1}{16}\psi^4}{1+\frac{1}{32}\psi^4}. \eeq
was chosen.
This potential yields the dilaton
gravity action at low curvatures, enforces that $I_2$ go to zero
at large curvatures, and enables a bounce. By
means of a phase space analysis of Eqns.(\ref{b1})-(\ref{b3}), it
was shown \cite{bd1} that all the solutions are non-singular,
and that some of them display a bounce either with or without the
dilaton. In particular, the flat bouncing solutions with a
non-zero dilaton interpolate between a contracting
dilaton-dominated phase and an expanding FLRW epoch, thus avoiding
the graceful exit problem of pre-big-bang cosmology (see below).

One obvious drawback of the LCH is that the non-linear terms are
not dictated by first principles: they are chosen in such a way as
to render the theory finite.

\subsection{Appendix: $f(R)$  and scalar-tensor theories}

Higher-order Lagrangians can be related to
scalar-tensor gravity (see for instance \cite{tey}).
Let us start with the function $f(R)$ is
given by
\beq f(R)=R+\alpha R^2. \eeq The EOM that follow from this
Lagrangian is \beq 2\alpha R_{;\mu\nu} -(1-2\alpha R) R_{\mu\nu}
+g_{\mu\nu} \left(\half \alpha R^2 +\half R -2\alpha \Box R
\right) = 0, \label{4o} \eeq the trace of this equation being \beq
\Box R -\frac{R}{6\alpha} = 0. \eeq It
was shown in \cite{tey} that
this theory is equivalent to the one given by the action
\begin{equation}
S = \int \sqrt{-g} \, d^{4}x \, \left[ (1 + 2\alpha \varphi) R -
\alpha \varphi^{2} \right]. \label{24dez17}
\end{equation}

Varying independently $g_{\mu\nu}$ and $\varphi$ in the action given
in Eqn.(\ref{24dez17}), one obtains
\begin{equation}
(1 + 2\alpha) ( R_{\mu\nu} - \frac{1}{2} R g_{\mu\nu}) +
\frac{\alpha}{2} \, \varphi^{2} g_{\mu\nu} - 2\alpha
(\varphi_{,\mu;\nu} - \Box \varphi \, g_{\mu\nu}) = 0,
\label{24dez171}
\end{equation}
and
\begin{equation}
 2 \alpha (R - \varphi) = 0.
 \label{24dez172}
 \end{equation}
In turn, as shown in \cite{whitt} the conformal transformation \beq
\widetilde{g}_{\mu\nu} = (1+2\alpha \phi)\;g_{\mu\nu}, \eeq takes
this theory to Einstein gravity with a massive scalar field.

Except in the case in which $\alpha$ vanishes (which is precisely
the case in general relativity) the second equation yields that the
scalar field is nothing but the scalar of curvature. Inserting this
result into Eqn.(\ref{24dez171}) one arrives precisely at
Eqn.(\ref{4o}).
The equivalence can be generalized to functions $f(R)$ (see
\cite{wands2}) \footnote{It was later proved that
all higher order, scalar-tensor and string actions
are conformally equivalent to general relativity with additional
scalar fields which have particular (different in each case)
self-interaction potentials \cite{confor}.}.
Based on the equivalence, the singularity problem in fourth order
theories was analyzed in \cite{zia} for homogeneous cosmological
models with a diagonal metric.

\section{Theories with a scalar field}
\label{cst}
\subsection{Scalar field in the presence of a potential}

Violations to some of the energy conditions are produced even at
the classical level by some scalar field theories. From the singularity
theorems discussed in Ch. 1, we can expect the existence of bouncing solutions in this
scenario \footnote{The role of scalar fields in Cosmology has been examined for instance
in \cite{kamen}.}. We shall see next examples of avoidance of the singularity
in scalar field models that violate some of the energy conditions,  as well as
theories with nonminimal coupling.

A universe filled with radiation and pressureless matter
coupled to a classical conformal massless scalar field was studied
in \cite{bekesc}. The coupling was provided by the action
\beq
S=-\half \int (\psi_{,\alpha}\psi^{,\alpha} + \frac 1 6 R
\psi^2)\sqrt{-g} - \int (\mu +f\psi)d\tau, \label{act} \eeq where
$\mu$ is the mass of the particle, and
$$
-f\int \psi d\tau = -f\int d^4x\left[\sqrt{-g}\psi\int
(-g)^{-1/2}\delta^4(x^\mu - x^\mu (\tau)) d\tau\right],
$$
(this interaction was suggested in \cite{bekesc} as a classical
analog of the pion-nucleon coupling). Assuming that we have a FLRW
universe filled with a uniform distribution of identical $\mu$
particles,  in the continuum approximation, the field equation for
$\psi$ takes the form
\beq F_{,\eta,\eta}+kF=-fN,
\label{efe}
\eeq
where $F=a\psi$ and $N=na^3$=constant. The calculation of the
trace of the
total stress-energy tensor from Eq.(\ref{act}) yields
$$
T_\alpha^{\;\alpha}= - \mu n,
$$
so we get for the trace of EE
\beq
a''+\epsilon a = \frac{4\pi}{3}N\mu.
\label{trace}\eeq
Finally the Friedmann equation is given by
\beq
a'^2+\epsilon a^2 = \frac{4\pi}{3} (F'^2+\epsilon F^2 +
2N a\mu + 2 N f F + 2 B),
\label{fri}
\eeq
where $B$ is a constant that gives the amount of radiation. The
system composed of Eqns.(\ref{efe}-\ref{fri}) was solved in
\cite{bekesc} for all values of $\epsilon$, and it was shown that
a bounce is possible for the three cases when some relations
between the integration constants are fulfilled. However, physical
requirements show that only the $\epsilon = + 1$ solution can
bounce provided $N^2f^2>2B$. A nice feature of this solution is
that it satisfies the weak energy condition.

Another non-singular universe based on a scalar field was
presented in \cite{turco}. A closed FLRW model was considered,
with a conformally coupled scalar field $\phi$ as matter content,
which can be thought as a perfect fluid with comoving velocity
defined by
$$
v^\mu = \frac{\phi^{,\mu}}{(\phi_{,\alpha} \phi^{,\alpha})^{1/2} }.
$$
In this case, the energy density and the pressure are given by
$$
\rho = \half \dot\phi^2 +\half \phi^2\left[\left(\frac{\dot
a}{a}\right)^2+ \frac{1}{a^2}\right]+\frac{\dot
a}{a}\phi\dot\phi+V,
$$
$$
p=\frac 1 6 \dot\phi^2 + \frac 1 3 \phi \frac{dV}{d\phi} + \frac 1
6  \phi^2\left[\left(\frac{\dot a}{a}\right)^2+
\frac{1}{a^2}\right]+\frac 1 3 \frac{\dot a}{a}\phi\dot\phi-V.
$$
EE were written as
$$
\left( \frac{\dot a}{a}\right)^2+ \frac{1}{a^2}=
\frac{\rho}{6},
$$
$$
\frac{\ddot a}{a} + \half \left(\gamma - \frac 2 3 \right) \rho =
0,
$$
with $p=(\gamma -1)\rho$. From these equations we get
$$
\frac{\ddot a}{a} + \left( \frac 3 2 \gamma -1\right)
\left(\frac{\dot a^2+1}{a^2}\right)=0.
$$
Introducing the conformal time through $dt = a(\eta) d\eta$, and
with the changes of variables $u=a'/a$, and $u=w'/(cw)$, with $c=3\gamma/2-1$, the solution for $a(\eta)$ is
\cite{turco}
$$
a(\eta) = a_0[\cos(c\eta +d)]^{1/c},
$$
where $a_0$ and $d$ are integration constants, which were fixed
resorting to the limiting curvature hypothesis (see Sect. \ref{slch}). The
potential $V$ was then reconstructed in terms of the scale factor
(assuming that the EOS changes inthe different eras of the universe)
and $\phi$ from $\gamma = 1+ p/ \rho$, and the evolution of
$\phi$ was obtained by numerical integration.

More general models, given by solutions of the theory
$$
S=\int d^4x
\sqrt{-g}\{F(\phi)R-\partial_\alpha\phi\partial^\alpha\phi-2V(\phi)\},
$$
in which $\phi$ is nonminimalluy coupled to gravity through $F$,
were studied in \cite{gunzig}, where it was shown that there are
bouncing solutions, which were later proved to be unstable
\cite{abramo}. A phase-space analysis of the models given by $F(\phi )=\xi \phi^2$
showed the
existence of bouncing solutions, under certain
restrictions on the constants of the potential $V(\phi )=\alpha \phi^2 +\beta \phi^4 +\Lambda$ \cite{gunzig2}.

Nonsingular solutions for a scalar field in the presence of a potential
were also studied in \cite{altshu}, for theories defined by
$$
{\cal L} = \half \omega \dot\phi^2-U(\omega ),
$$
where $\omega$ is determined by
$ dU/d\omega = \half \dot\phi^2$.
The existence of a bounce was shown for
a tailored potential given by
$$
U(\omega )=\lambda\left(\omega^{-1}+\frac{1-\alpha}{\alpha}\omega^{\alpha/(1-\alpha )}-\frac 1
\alpha\right ),
$$
where $\lambda$ is a constant with dimensions of energy density,
and $\alpha$ is a number parameterising the classes of theories
\footnote{This potential interpolates between $p=\rho$ for $\rho<<\lambda$,
and $p<0$ for high densities.}.
The bounce exists for $\alpha <1/3$, and $\epsilon = +1$.
Later, this approach was generalized to Bianchi I cosmologies in
\cite{frogun}.

So far we have examined a classical scalar field on a given
background. A quantum scalar field $\phi (x)$ in a classical
geometry was studied in \cite{melni,melni2} where, inspired by the features
of the mechanism of spontaneous symmetry breaking, the authors seek a
solution in which the fundamental state of $\phi$ is given by
\begin{equation}
\langle 0|\phi |0\rangle = \sqrt{\frac{3}{\lambda}}\ \frac{f(\eta
)}{A(\eta )} \label{31dez1400},
\end{equation}
where $\eta$ is the conformal time of an open Friedmann geometry
given by
\begin{equation}
ds^2=a^2(\eta )\left[d\eta^2-d\chi^2-\sinh^2 \chi \left(d\theta^2+
\sin^2\theta d\phi^2\right)\right].
\end{equation}
For a massless field the equation of motion for the scale factor
reduces to
\begin{equation}
\frac{a''}{a}=1.
\end{equation}
From the
Lagrangian
$$
{\cal L}=\half \partial_\mu\phi\partial^\mu\phi - \half \sigma
\phi^4
$$
we obtain the equation of the scalar field $\phi$, given by
\begin{equation}
\phi ''+2\phi '\frac{a'}{a}+2\sigma a^2\phi^3=0.
\end{equation}
Compatibility of these two equations with the assumption in
Eqn.(\ref{31dez1400}) yields the relation
\begin{equation}
\sigma =\frac{\lambda}{6}\ .
\end{equation}
For the scale factor as function of the Gaussian time $t$ we obtain
\begin{equation}
a(t)=\sqrt{t^2-L^2} \label{31dez1402},
\end{equation}
where $L$ is a constant and
\begin{equation}
f''-f+f^3=0.
\end{equation}
It was shown in \cite{melni} that the solution $f=0$ is unstable,
while the solutions $f^2=1$ are stable.
%
%
%
%
%
From the equation for $g_{\mu \nu}$ and specializing for $\mu =\nu
=0$ we obtain the value of the constant $L$ in
Eqn.(\ref{31dez1402}):
\begin{equation}
L^2=\frac{\kappa}{24\sigma}
\end{equation}
which represents the minimum allowable value of the scale factor.
%
%
%
From standard quantum field theory in curved spacetime,
\[
G_{\mu \nu}=-\kappa_{(ren)}T_{\mu \nu},
\]
it follows that $E_{|0\rangle}=-\frac{3L^2}{a^4}< 0$, which shows
explicitly the expected violation of the weak energy condition that
causes the absence of a singularity in this model. Note that the
gravitational constant in the vacuum state is renormalized:
\[
\frac{1}{\kappa_{(ren)}}=\frac{1}{\kappa}-\frac{\phi^2}{6}=
\frac{12\sigma t^2-\kappa /2}{12\sigma \kappa a^2}.
\]
It follows that $\kappa_{ren}<0 $ for $
t^2<\frac{\kappa}{24\sigma}$ and $\kappa_{ren}>0$ for $
t^2>\frac{\kappa}{24\sigma}$, thus showing that a change in the
sign of the gravitational constant can be induced by the
non-minimal coupling of scalar field with gravity, yielding repulsive gravity.

The phenomenon of repulsive gravity can also be generated at a classical level
by means of a non-minimally coupled complex scalar field \cite{indianos}.
The Lagrangian is given by
$$
{\cal L} = \partial_\mu\phi\partial^\mu\phi^*-\sigma(\phi^*\phi)^2-\frac 1 6
R(\phi^*\phi)+\kappa^{-1}R+{\cal L}_m,
$$
where $\sigma$ is the constant that measures the auto-interaction of $\phi$, and
${\cal L}$ is the matter Lagrangian. The EOM following from this Lagrangian are
$$
\Box\phi+2\sigma\phi^*\phi^2+\frac 1 6 R\phi = 0,
$$
$$
G_{\mu\nu}=-\tilde\kappa(\theta_{\mu\nu}+T_{\mu\nu}),
$$
where
\beq
\tilde\kappa=\kappa\left(1-\frac\kappa 6 \phi^*\phi\right),
\label{repgrav}
\eeq
$$
\theta_{\mu\nu}=\half\left(\partial_\mu\phi^*\partial_\nu\phi+\partial_\nu\phi^*
\partial_\mu\phi-g_{\mu\nu}(\partial_\rho\phi^*\partial^\rho\phi-\sigma(\phi^*\phi)^2)
+\frac 1 3 g_{\mu\nu}\Box(\phi^*\phi)-\frac 1 3 (\phi^*\phi)_{;\mu\nu}\right),
$$
and $T_{\mu\nu}$ is the energy-momentum tensor associated to matter.
From Eqn.(\ref{repgrav}) we see that the gravitational constant is renormalized
at the classical level by the scalar field.
In fact, as shown in \cite{indianos}, for the open FLRW metric \footnote{This
scenario does not work for the closed case.} the scalar field has
three vacuum solutions: $\phi=0$, and $\phi=\pm \gamma/a(t)$, where $\gamma$ is a constant.
Only the nonzero solutions are stable, and they are also more favorable from the point of
view of energy \cite{indianos}. Since they are inversely proportional to $a$,
it may be argued that the scalar field was in a nonzero vacuum in the early universe. Hence,
$$
\tilde\kappa = \kappa\left[1-\frac{a_c^2}{a^2}\right]^{-1},
$$
where $a_c=(\kappa/12\sigma)^{1/2}$ signals the
change of sign of the gravitational interaction. Nonsingular solutions were
obtained in \cite{indianos} for matter given by radiation ($\rho = \epsilon/a^4$):
$$
a(t) = \frac{\varpi}{\sqrt 2}\cosh t,
$$
where $\varpi^2=a_c^2-\frac 2 3 \kappa \epsilon$. This case reduces to
the case without matter for $\epsilon = 0$.

\subsection{Dynamical origin of the geometry}
\label{dynor}

We shall see in this section that a cosmological scenario displaying
a bounce arises in an extension of Riemannian geometry called Weyl
Integrable Space-Time (WIST) \cite{NovelloS2}.

Let us begin by recalling that one of the central hypothesis of
General Relativity is that gravitational processes occur in a
Riemannian space-time structure. This means that there exists a
metric tensor $g_{\mu\nu}$ and a symmetric connection
${\Gamma^{\alpha}}_{\mu\nu}$ related by
\begin{equation}
g_{\mu\nu;\alpha} \equiv  g_{\mu\nu,\alpha}  -
\Gamma^{\epsilon}_{\alpha\mu} \,  g_{\epsilon\mu} -
{\Gamma^{\epsilon}}_{\alpha\nu}\, g_{\mu\epsilon} = 0.
\end{equation}
\protect\label{rs1} In other words, the connection is metric and
can be written in terms of the metric tensor as follows
\begin{equation}
{\Gamma^{\alpha}}_{\mu\nu} = \left\{
 ^{\alpha}_{\mu\nu} \right\} \equiv  \frac{1}{2} g^{\alpha\beta} [
g_{\beta\mu,\,\nu} + g_{\beta\nu,\,\mu} - g_{\mu\nu,\,\beta} ].
\end{equation}
\protect\label{rs2} A direct method to deduce such metricity
condition
is given by the first order Palatini variation (in which the variation of the metric tensor
and of the connection are independent). The starting point is
the
Hilbert action:
\begin{equation}
{\cal S} = \int \sqrt{-g}  R d^{4}x.
\end{equation}
\protect\label{rs3}
In a local
Euclidean coordinate system,
\begin{equation} \delta R_{\mu\nu} =
\delta {\Gamma^{\alpha}}_{\mu\alpha;\nu} - \delta
{\Gamma^{\alpha}}_{\mu\nu;\alpha},
\end{equation}
\protect\label{rs4}
where the covariant derivative represented
by a semicolon must be taken in the non-perturbed background
geometry. From this equation it follows that
 \begin{equation}
 \delta {\cal L} =  ( R_{\mu\nu} - \frac{1}{2} R  g_{\mu\nu} )
\sqrt{-g} \delta  g^{\mu\nu} + \sqrt{-g}  g^{\mu\nu} \delta
R_{\mu\nu}.
\end{equation}
\protect\label{rs5} Correspondingly
\begin{eqnarray}
\delta {\cal S}  &=& \int \sqrt{-g}  ( R_{\mu\nu} -
\frac{1}{2} R  g_{\mu\nu} )  \delta  g^{\mu\nu} \nonumber \\
 &+& \int
\left\{ ( \sqrt{-g} g^{\mu\epsilon} )_{;\alpha}  - \frac{1}{2} (
\sqrt{-g} g^{\mu\nu} )_{;\nu}   \delta^{\epsilon}_{\alpha} -
\frac{1}{2} ( \sqrt{-g} g^{\nu\epsilon} )_{;\nu}
\delta^{\mu}_{\alpha} \right\} \delta
{\Gamma^{\alpha}}_{\mu\epsilon}.
\end{eqnarray}
\protect\label{rs6} Hence,
\begin{equation}
( \sqrt{-g} g^{\mu\epsilon} )_{;\alpha}  - \frac{1}{2} ( \sqrt{-g}
g^{\mu\nu} )_{;\nu}   \delta^{\epsilon}_{\alpha} - \frac{1}{2} (
\sqrt{-g} g^{\nu\epsilon} )_{;\nu} \delta^{\mu}_{\alpha} = 0,
\end{equation}
\protect\label{rs7} and we obtain
\begin{equation}
(\sqrt{-g} g^{\mu\epsilon})_{;\alpha} = 0. \end{equation}
\protect\label{rs8} After some algebra it can be shown that
space-time has a Riemannian structure, that is, it obeys the
metricity condition,
\begin{equation}
g_{\mu\epsilon;\alpha} = 0.
\end{equation}
\protect\label{rs9} The other equation that follows from the
variational principle yields Einstein's equations.
The lesson we learn from this calculation is that the structure of
the manifold associated to space-time is not given {\emph a
priori}, but may depend on the dynamics. Surely, we should
check whether the addition of matter alters this feature. The
answer is not unique: it depends crucially on the way matter
couples to gravity. There will be no modification to the precedent
structure if we adopt the minimal coupling (that is, if the strong
equivalence principle is valid). However, when the interaction is
non-minimal, the geometrical structure obtained by the Palatini
variation is not Riemannian in general. The simplest way to show
this is with an example. Let us take the Lagrangian which
describes the non-minimal interaction of a scalar field with
gravity in the form:
\begin{equation}
L_{int} = \sqrt{-g} \, R \, f(\varphi) \protect\label{rs10}.
\end{equation}
Following the procedure sketched above we get:
\begin{eqnarray}
\delta S_{int} &=& \int \sqrt{-g}  \, f\, ( R_{\mu\nu} -
\frac{1}{2} R  g_{\mu\nu} )  \delta  g^{\mu\nu} \nonumber \\
&+& \int \left\{( \sqrt{-g} \, f\, g^{\mu\epsilon}
)_{;\alpha}  - \frac{1}{2} ( \sqrt{-g} \, f \, g^{\mu\nu}
)_{;\nu} \delta^{\epsilon}_{\alpha} -  \frac{1}{2} ( \sqrt{-g}\,
f\, g^{\nu\epsilon} )_{;\nu}   \delta^{\mu}_{\alpha}
\right\} \delta {\Gamma^{\alpha}}_{\mu\epsilon},
\end{eqnarray}
\protect\label{rs11} and it follows that
\begin{equation}
 \left\{ \sqrt{-g} f(\varphi)\, g^{\mu\nu} \right\}_{;\epsilon}
= 0.
\end{equation}
\protect\label{rs12} This equation shows that the covariant
derivative of the metric tensor is not zero but
\beq
g_{\mu\nu ;\alpha}  =  Q_{\mu\nu\alpha},
\end{equation}\protect\label{rs13}
where
$
Q_{\mu\nu\lambda} = - (\ln f )_{,\lambda} \, g_{\mu\nu}.
$
Taking the cyclic permutation of Eqn.(\ref{rs13}) yields
\begin{equation}
  {\Gamma^{\lambda}}_{\mu\alpha} = \left\{ ^{\lambda}_{\mu\alpha}
  \right\} - \frac{1}{2} [ {{Q_{\mu}}^{\lambda}}_{\alpha} +
{Q^{\lambda}}_{\alpha\mu} - {Q_{\alpha\mu}}^{\lambda} ].
\end{equation}
\protect\label{rs15} The equation
\begin{equation}
g_{\mu\nu;\alpha} = -  (\ln f )_{,\lambda}\, g_{\mu\nu}.
\protect\label{rs17}
\end{equation}
shows that the structure generated by the Lagrangian (\ref{rs10})
using the Palatini variation
is not Riemannian but, as we shall see in the next section, a
special case of Weyl geometry.

\subsubsection{WIST (Weyl Integrable Space Time)}

A Weyl geometry is defined by the relation \cite{weylbook}
\begin{equation}
g_{\mu \nu ;\alpha}=W_{\alpha}g_{\mu \nu}. \label{26dez1}
\end{equation}
This equation implies that there is a variation of the length
$\ell_{0}$ of any vector under parallel transport, given by
\begin{equation}
\Delta \ell =\ell_0 W_{\mu}\Delta x^{\mu}. \label{deltal}
\end{equation}
This property has the undesirable consequence that the measure of
length depends on the previous history of the measurement
apparatus, as pointed out by Einstein in the beginning of the past century
in a
criticism against Weyl's proposal for the geometrization of the
electromagnetic field \cite{pauli}. Einstein�s remark led to the
abandonment of this type of geometry. However, there is just one
particular case in which this problem disappears: the so-called
Weyl integrable spacetime (WIST). By definition, a  WIST is a
particular Weyl spacetime in which the vector $W_{\mu}$ is
irrotational:
$$ W_{\mu} \equiv \partial_{\mu} \varphi .$$
It follows that in a closed trajectory
\begin{equation}
\oint \Delta \ell =0,
\end{equation}
which solves the critic raised by Einstein. From the definition
given in Eqn.(\ref{26dez1}) it follows that the associated
connection is given by
\begin{equation}
C^{\alpha}_{\mu \nu}=\left\{{\alpha}\atop {\mu \nu}\right\}-
\frac{1}{2}
\left(W_{\mu}\delta^{\alpha}_{\nu}+W_{\mu}\delta^{\alpha}_{\nu}-
W^{\alpha}g_{\mu \nu} \right).
\end{equation}
Using this equation we can write the contracted curvature tensor
$R^{(W)}_{\mu \nu}$ in terms of the tensor
$R_{\mu\nu}$ of the associated Riemann space constructed
with the Christoffel symbols $\left\{{\alpha}\atop {\mu
\nu}\right\}$. We obtain
\begin{equation}
R^{(W)}_{\mu \nu}= R_{\mu \nu} - \varphi_{,\mu ;\nu} - \frac{1}{2}
\,\varphi_{,\mu} \varphi_{,\nu}  + \frac{1}{2} \,
\varphi_{,\lambda} \varphi^{,\lambda} g_{\mu \nu}-\frac{1}{2} \,
\Box \varphi \, g_{\mu \nu} \label{z1}
\end{equation}
where the covariant derivatives are taken in the associated
Riemannian geometry and $ \Box$ is the d'Alembertian in the
Riemannian geometry. Thus, for the curvature scalar,
\begin{equation}
R^{(W)}= R - 3 \Box \, \varphi +\frac{3}{2} \, \varphi_{,\lambda}
\varphi^{,\lambda} \label{H1}
\end{equation}
in which $R$ is the curvature scalar of the associated Riemannian
spacetime.

The expressions in Eqns.(\ref{z1}) and (\ref{H1}) are very
similar to those obtained by a conformal mapping of a Riemannian geometry
as shown in Sec.\ref{app2}.

\subsubsection{WIST duality: the Weyl map}

A Weyl integral spacetime is determined by both a metric tensor
and a scalar field. In \cite{weylbook}, Weyl introduced
a generalization of the conformal mapping, which he called a gauge transformation,
given by
\beq
g_{\mu \nu} \rightarrow \tilde{g}_{\mu \nu} = e^{\chi} g_{\mu
\nu},\;\;\;\;\;\;\;\;\;\;\;\varphi \rightarrow \tilde{\varphi}= \varphi +
\chi ,
\label{gauge1}
\eeq
in which $\chi$ is an arbitrary function. Under such
transformations the affine connection and the curvature and Ricci
tensors are invariant:
\begin{eqnarray*}
\tilde C^{\alpha}{}_{\mu \nu} &=& C^{\alpha}
{}_{\mu \nu},  \nonumber \\
{\tilde{R}^{(W)\, \alpha}}{  }_{\beta\mu\nu} &=& {R^{(W)\,\alpha}}_{\beta\mu\nu},\\
\nonumber
 \tilde{R}^{(W)}_{\mu\nu} &=& R^{(W)}_{\mu\nu}.
\end{eqnarray*}
Note however that this is not the case for the scalar of
curvature,  which changes as
$$ \tilde{R}^{(W)} = e^{-\chi}R^{(W)}.$$
This property has been used to construct gauge-invariant theories, as we shall see next.
\subsubsection{Invariant Action in WIST}

From the behavior of the geometric quantities under a Weyl map, it
is not difficult to write an action that is invariant under the
transformation given by Eqns.(\ref{gauge1}):
\begin{equation}
S_{W} = \int \sqrt{-g} \, e^{-\varphi}\, R^{(W)}.
\label{action1}
\end{equation}
This Lagrangian can be rewritten in terms of the associated
Riemannian quantities as follows:
 \begin{equation}
S_{W} = \int \sqrt{-g}e^{-\varphi} \left(R - 3 \Box \, \varphi
+\frac{3}{2} \, \varphi_{,\lambda} \varphi^{,\lambda}\right).
\end{equation}
After some algebra, we arrive (up a total divergence)
to the result
\begin{equation}
S_{W} = \int \sqrt{-g}e^{-\varphi} \left( R - \frac{3}{2}
\varphi_{,\mu} \varphi^{,\mu} \right).
\end{equation}
Note that the kinematical term of the scalar
field for the scalar field
appears with the ``wrong'' sign.
This can be interpreted as a ghost field term hidden in the WIST structure.

\subsubsection{A particular case of WIST Duality}

Let us go one step further and add to the above Lagrangian a
kinematical term:
\begin{equation}
S_{K} = \int \sqrt{-g}e^{-\varphi} \varphi_{,\mu} \varphi^{,\mu}.
\end{equation}
If we restrict to the case in which $\chi$ (given in Eqn.(\ref{gauge1}))
is a functional of $\varphi$, it follows that the
complete action
\begin{equation}
S = \int \sqrt{-g}\;e^{-\varphi} ( R^{(W)} + \beta \,
\varphi_{,\mu} \varphi^{,\mu} )
\end{equation}
is invariant under the restricted map
\begin{eqnarray}
g_{\mu \nu} &\rightarrow& \tilde{g}_{\mu \nu} = e^{-2\varphi} g_{\mu \nu},\\
\nonumber \varphi &\rightarrow& \tilde{\varphi}= - \, \varphi,
\end{eqnarray}
which is a special case of the general transformation (\ref{gauge1}). In terms of Riemann variables,
\begin{equation}
S = \int \sqrt{-g}\;e^{-\varphi} \left[ R + \left(\beta - \frac{3}{2}\right)
\varphi_{,\mu} \varphi^{,\mu}  \right].
\end{equation}

There are three invariants of dimension (length)$^{2}$ that can be
constructed with the independent quantities of a WIST geometry:
$R^{(W)}$, $\varphi^{\alpha} \, \varphi_{\alpha}$, and
$\varphi^{\alpha}{}_{;\alpha}$, where $\varphi_{\alpha}\equiv
\varphi_{,\alpha}$. Now, since the covariant derivative ``;'' in the
WIST spacetime can be written in terms of the Riemann covariant
derivative (denoted by ``$||$'') as
$$
\varphi^\alpha_{\; ;\alpha} = \varphi^\alpha_{\; ||\alpha}
-2\varphi^{,\alpha}\varphi_{,\alpha},
$$
the three invariants reduce to two.
The most general action can then be written as
\begin{equation}
S = \int \,\sqrt{-g} \, [ R^{(W)} + \xi \, \varphi^{\alpha} _{\;
;\alpha}] \protect\label{WIST1}.
\end{equation}
Independent variation of the metric tensor and the WIST field
$\varphi$ yields
\begin{equation}
\Box \,\varphi = 0 \protect\label{WIST3},
\end{equation}
(the operator $\Box$ is calculated in the Riemannian spacetime)
and
\begin{equation}
{R^{(W)}}{}_{\mu\nu} - \frac{1}{2} \,{R^{(W)}} \,g_{\mu\nu} +
\varphi_{,\mu ;\nu} - 2(\xi - 1) \varphi_{,\mu}
\varphi_{,\nu} + (\xi - \frac{1}{2}) g_{\mu\nu}
\varphi_{,\alpha} \varphi^{,\alpha} = 0.
\protect\label{WIST2}
\end{equation}
This equation can be rewritten exclusively in
terms of the associated Riemannian structure
\begin{equation}
R_{\mu\nu} - \frac{1}{2} \, R \,g_{\mu\nu} - \lambda^{2} \,
\varphi_{,\mu}\, \varphi_{,\nu} + \frac{\lambda^{2}}{2}\,
\varphi_{,\alpha}\,
 \varphi^{,\alpha} \, g_{\mu\nu} = 0,
\protect\label{WIST4}
\end{equation}
where
\beq
\lambda^{2} = \frac{1}{2} \, (4 \xi - 3).
\label{eqlambda}
\eeq

\subsubsection{A nonsingular cosmological model in WIST}

Let us now show how a nonsingular cosmological scenario in the WIST
framework can be constructed, following \cite{NovelloS2}. We shall work with
the standard form of the FLRW
metric:
\begin{equation}
ds^2 = dt^2 - a^2(t)\,
\left[\frac{dr^2}{1-\epsilon r^2}+r^2\,(d\theta^2+\sin^2\theta\,d\varphi^2)\right],
\end{equation}
As in the case of a standard scalar field, the
WIST configuration can be represented by a perfect fluid, so that
Eq.(\ref{WIST4}) becomes Einstein�s equation for a perfect fluid
with $v^{\mu} = \delta^{\mu}_{\;0}$, energy density
$\rho_{\varphi}$ and pressure $p_{\varphi}$, given by
\beq
\rho_{\varphi} = p_{\varphi} = -\frac{1}{2} \,  \lambda^{2} \,
\dot{\varphi}^{2}
\label{pr}
\eeq
In this interpretation, the WIST structure is
equivalent to a Riemannian geometry, satisfying the
equations of General Relativity with a perfect fluid having
negative energy density as a
source.
The gauge vector $\varphi_{\lambda}$ for this geometry becomes
\begin{equation}
\varphi_{\gamma}=\partial_{\lambda}\varphi
(t)=\dot{\varphi}\;\delta^0_{\lambda}\ ,
\end{equation}
where the dot denotes differentiation with respect to the time
variable.
Use of Eq.(\ref{WIST3}) yields a first
integral for the function  $\varphi (t)$:
\begin{equation}
\dot{\varphi}=\gamma a^{-3},
\label{firsti}
\end{equation}
where $\gamma =$ constant. In turn, EE (\ref{WIST4}) for
the Friedman scale factor $a(t)$ are
\begin{equation}
\dot{a}^2+\epsilon +\frac{\lambda^2}{6}\ (\dot{\varphi}a )^2=0,
\label{fr1}
\end{equation}
\begin{equation}
2a\ \ddot{a}+\dot{a}^2+\epsilon -\frac{\lambda^2}{2}\
(\dot{\varphi}a )^2=0\ ,
\end{equation}
where $\epsilon$ is the 3-curvature parameter of the FLRW
geometry. From Eqns.(\ref{firsti}) and (\ref{fr1}) we see that
$(3-4\xi )=-2\lambda^2<0$, and an open Universe is obtained (i.e.,
$\epsilon =-1$). Combining Eqns.(\ref{pr}) and (\ref{fr1}) we get
the fundamental dynamical equation
\begin{equation}
\dot{a}^2=1-\left[\frac{a_0}{a}\right]^4\ , \label{sf}
\end{equation}
with $a_0=[\gamma^2\lambda^2/6]^{1/4}$.
Before entering into the details of the solution of the system of
structural and dynamical equations (\ref{fr1}) and (\ref{firsti}), let us comment
some of the consequences of this cosmological model and list some
interesting results.\\[0.3cm]
{\bf Features of the model}\\[0.2cm]
An immediate consequence of Eq.(\ref{sf}) is that the scale factor
$a(t)$ cannot attain  values smaller than $a_0$. Let us consider a
time reversal operation and run backwards into the past of  the
cosmic evolution.  As the cosmic radius $a(t)$ decreases, the
temperature of the material medium grows. In Hot Big Bang models such
increment is unlimited; in the present theory, on the other hand,
there is an epoch of greatest condensation in the vicinity of the
minimum radius $a_0$. Close to this period, there occurs a
continuous  ``phase transition'' in the geometrical background: a
Weyl structure is activated, according to Eq.(\ref{firsti}):
the Universe attains the minimum radius $a_0$ at $(t=0)$, and
consequently an unbounded growth of the temperature is inhibited.
%
Notice that since the Universe had
this infinite collapsing era to become homogeneous, in the
present scenario the horizon problem of standard cosmology
does not arise.

For very large times, the scale factor behaves as
$a\sim t$. Thus, asymptotically, the geometrical configuration
assumes a Riemannian character (since $\dot{\varphi}\rightarrow
t$) in the  form of a flat Minkowski space (in Milne's coordinate
system). Consequently, in the present model the evolution of the
universe may be started by a primordial instability of Minkowski
spacetime at the remote past, due to Weyl perturbations of the
Riemann structure through Eq.(\ref{rs17}). In order to prescribe
the behavior of these perturbations, knowledge of the
time dependence of the gauge vector $\varphi_{\lambda}$
is required. Since the
WIST function $\dot{\varphi}$ has a maximum at $t=0$, the largest
deviation of the Riemannian configuration corresponds to the epoch
of greatest contraction near to the value $a_{0}.$\\[0.3cm]
{\bf Stability of the solution}
\\[0.2cm]
Among the difficult questions concerning bouncing
Universes, one may count the problem of their survival with
respect to eventual metric perturbations (see Sect.(\ref{cobs})).
We shall show that
during the stage of greatest
condensation the WIST model of the Universe is stable. Applying the perturbative
scheme
$$
\varphi \rightarrow \varphi +\delta \varphi ,
$$
$$
a\rightarrow a+\delta a,
$$
to Eqs.(\ref{firsti}) and (\ref{fr1}), one obtains
$$
\delta \dot\varphi  =-\frac{3\gamma}{a^4}\;\delta a,
$$
$$
\delta \dot{a}\sim 2 \frac{a^4_0}{a^5\dot{a}}\; \delta
a.
$$
Hence,
$$
\delta\dot \varphi =-\frac{9\dot{a}}{\gamma \lambda^2}\;
a\delta \dot{a},
$$
$$
\frac{\delta \dot{a}}{\delta a}\sim a^{-3}
[a^4-a^4_0]^{-\frac{1}{2}}\ .
$$
Far from $a_{0}$ (\emph{i.e.}, for large $t$) we have $a>>a_0$;
then,
$$
\frac{\delta \dot{a}}{\delta a}\sim a^{-5} ,$$
$$
da\sim dt\ ,
$$
so with $(\delta a)_i$ being the initial spectrum of
perturbations, one obtains
\[
\delta a\sim (\delta a)_i\exp[a^{-4}].
\]
The solutions of the system Eqs. (4.12) and (4.15) are therefore
stable against metric perturbations in the course of the infinite
collapsing phase.\\[0.3cm]
{\bf The exact solution}\\[0.2cm]
No closed solution can be obtained in terms of the cosmological time,
so it is convenient to move to conformal time $\eta$, in which case
the solution is easily shown to be
\beq
a(\eta) = a_0\sqrt{\cosh 2 (\eta - \eta_0)},
\label{wistb}
\eeq
where $\eta_0$ is an integration constant. The following qualitative plot shows
the difference between this bouncing solution and the radiation-dominated
model in standard cosmology.
\begin{figure}[htp]
\begin{center}
\includegraphics[angle=-90,width=0.5\textwidth]{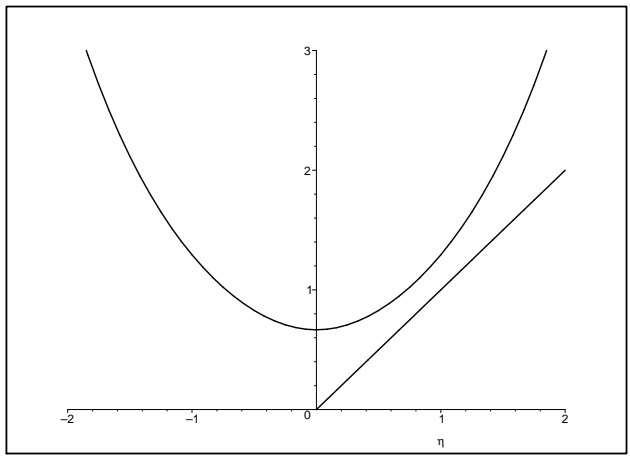}
\caption{The qualitative plot shows (in conformal time)
the scale factor for the bouncing model
given by Eqn(\ref{wistb}, and the scale factor for radiation in
the SCM, $a\propto \eta$.}
\label{bouncesing}
\end{center}
\end{figure}
The scale factor has a minimum for $a=a_0$, which corresponds to
$\eta=\eta_0$. Thus the Universe had a collapsing era for $\eta<\eta_0$, attained
its minimum dimension at $\eta=\eta_0$, and thereafter initiated an
expanding era. Both the collapse and the expansion run
adiabatically, i.e., at a very slow pace.

\begin{figure}[htp]
\begin{center}
\includegraphics[angle=-90,width=0.5\textwidth]{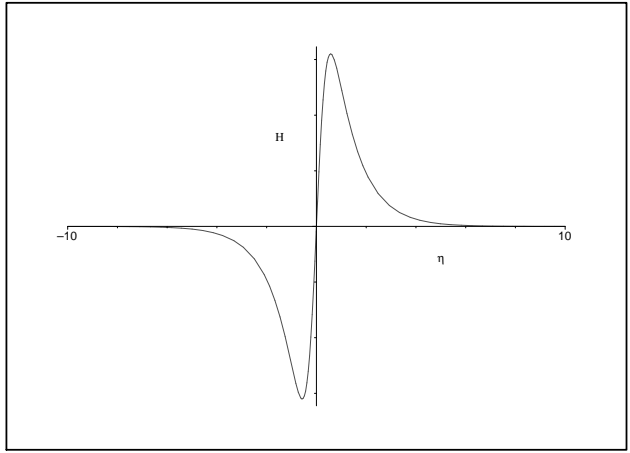}
\caption{Plot of the Hubble parameter in conformal time for $a_0=1$.}
\label{hwist}
\end{center}
\end{figure}

The correlate behavior of the Hubble expansion parameter $H \equiv
(\dot{a}/a)$ helps to understand the
model (Fig. \ref{hwist}).
Indeed, the Hubble parameter
$H$ is a smooth function of the conformal time $\eta$ and does not
diverge at the origin of the expanding era; quite on the contrary,
it vanishes at $\eta =\eta_0$. The corresponding evolution of the Cosmos
may be outlined as follows: the Universe stays for a long period
in a phase of slow adiabatic contraction, until $H$
attains its minimum value.
Then an
abrupt transition occurs: a fast compression turns into a fast
expansion up to the maximum of $H$,
and afterwards the expansion proceeds in an adiabatic slow pace again.
%
While this image supplies a picture of the behavior of an Universe
driven by $\varphi(t)$, it is however incomplete, due to the fact that
the production of large amounts of matter and
entropy has been neglected. This topic will be discussed in Sect.\ref{matcre}).

\subsubsection{The WIST function $\varphi (t)$: structural transitions}

\begin{figure}[htp]
\begin{center}
\includegraphics[angle=-90,width=0.5\textwidth]{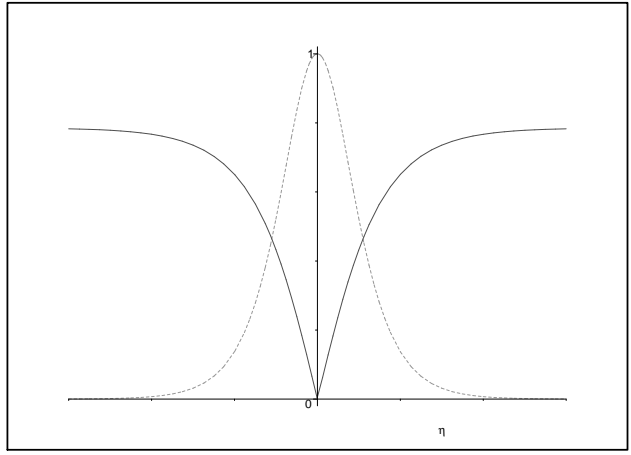}
\caption{Plot of $\varphi$ (full line)
and $\dot\varphi$ (dotted line)
in conformal time for $a_0=1$.}
\label{fvarphi}
\end{center}
\end{figure}

According to the basic conception of the scenario
presented above, the WIST function $\varphi (t)$ governs the
cosmic evolution. Taking into account
the solution Eq.(\ref{sol}) for the scale factor $a(t)$, the first
integral equation (\ref{firsti}) yields for $\varphi (t)$ the expression
\begin{equation}
\varphi = \frac{\gamma}{2a^2_0}\arccos
\left[\frac{a_0}{a}\right]^2\ .
\label{varphi}
\end{equation}
The behavior of $\varphi(t)$ is qualitatively portrayed in Fig. \ref{fvarphi},
along with $\dot\varphi$.
Note that when $a\rightarrow \pm \infty$ (i.e., for large times),
$\varphi \rightarrow \pm \gamma \pi/4a^2_0=$constant, which is
consistent with the assumption that the Universe originated from a
Minkowskian ``nothing'' state.
The behavior of the time
derivative $\dot{\varphi}=\gamma /a^3$, which appears in
Eq.(\ref{pr}) of the energy density $\rho_{\varphi}$ of
the ``stiff matter'' state associated to the WIST field is also shown in Fig.\ref{fvarphi}.
Since this function has a strong peak in the neighborhood of the
minimum radius $a_0$, the greatest deviation from the Riemannian
configuration happens at this point. In this sense, a sort of
''structural phase transition'' takes place when the Universe
approaches its maximally condensed state. The increase of the
(negative)
energy of the WIST ``fluid'' precludes the collapse to a singularity, reversing
the
cosmic evolution into an expansion. Note that the ``kinky''
aspect of the behavior
of the WIST function $\varphi (t)$ in Fig.\ref{fvarphi} suggests a similarity
between the Weyl structural transition described above and the
propagation of instantons in Euclideanized models of quantum
creation (see Eqn.(\ref{sf})).

\subsubsection{WISTons and anti-WISTons: On the geometrization of instantons}

\vspace{3mm} In the derivation of the solution of the
WIST structural function $\varphi (t)$ (given by Eq.(\ref{varphi})),
no attention was paid
to the sign of the constant $\gamma$. Since the only
information we have about $\gamma$ is that
$\gamma^2=6a^4_0/\lambda^2$, according to Eqs.(\ref{firsti}) and (\ref{sf}),
$\gamma$ can be either positive or negative:
$$
\gamma^{(\pm )}=\pm \sqrt{6}\ \frac{a^2_0}{|\lambda |}.
$$
Hence, Eqns.(\ref{firsti}) and (\ref{varphi}) actually yield two
equations, as follows:
\begin{equation}
\varphi^{(\pm )}=\varphi_0^{(\pm )}\arccos
\left[\frac{a_0}{a}\right]^2,
\end{equation}
\begin{equation}
\dot{\varphi}^{(\pm )}=\frac{\gamma^{(\pm )}}{a^3}\ ,
\end{equation}
in which $\varphi^{(\pm )}_0=\gamma^{(\pm )}/2a^2_0=\pm
\sqrt{3/2}\ |\gamma |^{-1}$. Thus the amplitude of the solutions
$\varphi^{(\pm )}$  depends exclusively on the dimensionless
parameter $\xi$ (see Eqn.(\ref{eqlambda})).
The plot
of the WIST functions $\varphi^{(-)}(t)$ and
$\dot{\varphi}^{(-)}(t)$
is given by the mirror image of Fig.\ref{fvarphi} with respect to the horizontal
axis.
Note, however, that the energy density $\rho_{\varphi}$ of the
``stiff matter'' state associated with the WIST field $\varphi
(t)$ is the same in both cases, since from Eqns. (\ref{pr}) and (\ref{firsti})
we have
\begin{equation}
\rho_{\varphi}=-\frac{\lambda^2}{2}\ \dot{\varphi}^2=-3
\left[\frac{a^4_0}{a^6}\right] \ .
\end{equation}
Thus, in spite of the fact that the pairs of WIST functions
$(\varphi^{(+)},\dot{\varphi}^{(+)})$ and
$(\varphi^{(-)},\dot{\varphi}^{(-)})$ have different
characteristics, they induce the same type of cosmological
evolution. Their only
distinction, in fact, is connected to length variations, since
according to Eq.(\ref{deltal}) one now has $\Delta L^{(\pm
)}=L\dot{\varphi}^{(\pm )}\Delta t$.

It is interesting to observe that the system is invariant with
respect to the time reversal operation $t\rightarrow (-t)$ if
$\varphi^{(+)}$ is concurrently mapped into $\varphi^{(-)}$  and
reciprocally. In this sense, the WIST instanton-like functions
$\varphi^{(+)}$ and $\varphi^{(-)}$ may be called ``WISTon''
and ``anti-WISTon'' solutions, respectively, since an anti-WISTon
may be described as a WISTon running backwards in time. According
to Eq.\ref{WIST3}, WISTons are defined up to an additive constant.

\begin{figure}[htb]
\begin{center}
\includegraphics[angle=-90,width=0.5\textwidth]{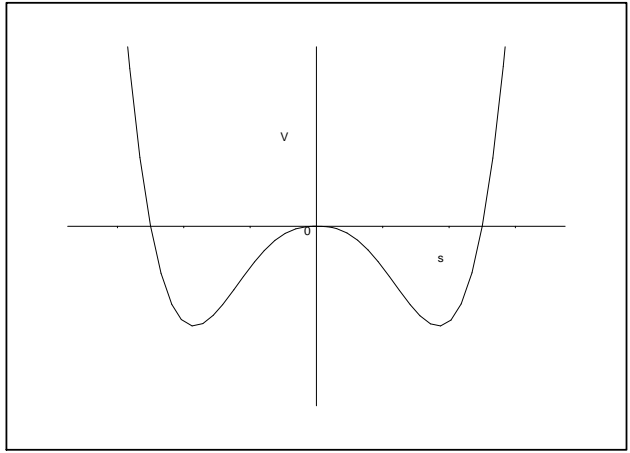}
\caption{Qualitative plot of $V(s)$.}
\label{potwist}
\end{center}
\end{figure}

A closer inspection of the equations governing the behavior of
$\varphi (t)$ reveals an instanton-like behavior typical of
nonlinear theories of self-interacting scalar fields.  Of course,
the root of such nonlinearity is the fact that $\varphi (t)$ is
taken as the actual source of the curvature of the metric
structure, which in turn modifies the D'Alembertian operator
$\Box$ due to the introduction of $\varphi$-dependent terms. A
direct way to clarify this issue is to make explicit, by means of
a change of variables, the hidden nonlinearity of the system of
equations of motion involving the scale
factor $a(t)$ and the WIST function $\varphi (t)$. Define the new
variable $s (t) \equiv \dot{\varphi} (t)$. Using Eqns. (\ref{WIST3})
and (\ref{firsti}), we
have
\begin{equation}
\left\{
\begin{array}{c}
\dot{s}+3\gamma a^{-4}\dot{a}=0\ ,\\
a^3-\gamma s^{-1}=0.
\end{array}
\right.
\end{equation}
Taking $s (t)$ to represent a generalized coordinate
associated with a one-particle dynamical system yields the
conservation equation
\begin{equation}
\frac{1}{2}\ \dot{s}^2+V(s )=0 \ ,
\end{equation}
in which the associate potential $V(s )$ is given by
\begin{eqnarray}
V(s ) &=& \frac{9}{2\gamma^2}\ \left[a^4_0\; s^4-
\gamma^{\frac{4}{3}} s^{\frac{8}{3}}\right]\nonumber \\
&=& \frac{3\lambda^2}{4}\
\left[s^4-b^2 s^{\frac{8}{3}}\right] \ ,
\end{eqnarray}
with $b^2 = 6\lambda^{-2}\gamma^{2/3}$. Thus the evolution of
field $s$ is equivalent to a unit mass particle moving in a
potential with vanishing total energy. Due to the nonlinear
character of this potential, the instanton-like aspect of
functions $\varphi ^{(\pm )}(t)$ is not surprising. Figure \ref{potwist}
shows
the behavior of $V(s )$. The potential vanishes at $s =
0$ and at $s^{(\pm )}_B=\gamma^{(\pm )}a_0^{-3}$ its extrema
are at $s =0$, and at $s^{(+)}_m=(2/3)^{3/4}\gamma^{(\pm
)}a^{-3}_0$ (which are minima). However, the system cannot
remain at the stable states $V(s_m^{(\pm )})=\left(- \
\frac{2}{3}\right)\gamma^2a^{-8}_0$, since in this case
$\dot{s} \neq 0$; this in turn implies, of course, a
nontrivial, evolving cosmic configuration.
This nonlinear scheme provides a succinct picture of the evolution
of the Universe: its development is initiated at $s =0$
(which corresponds to Minkowski space time at $t\rightarrow
-\infty$), attains its minimum radius $a(t=0)=a_0$ at either
$s_B^{(+)}$  or $s^{(-)}_B$ and returns back to $s= 0$
(which now corresponds to a Minkowski spacetime at
$t\rightarrow +\infty$). According to whether the system proceeds
along the right or the left branches (i.e., from $s =0$ to
$s^{(+)}_B$ or $s^{(-)}_B$) of the figure, the cosmic
evolution is driven by a WISTon or an anti-WISTon, respectively.

The appearance of instanton-like configurations is
a direct consequence of the fundamental dynamical equation
(\ref{sf}), in combination with the ``structural'' equation
(\ref{firsti}) which prescribes the degree of ``Weylization'' of
space time.

\subsubsection{Weylization}

We shall see next that
the ``structural
transitions'' discussed above are equivalent to a quantum tunnelling process
in models of quantum
creation from ``nothing''. Consider a generic Einstein equation
for a Friedman scale factor,
\begin{equation}
\dot{a}^2=-\epsilon +\frac{1}{3}\ \rho a^2\ ,
\label{acima}
\end{equation}
It was shown in \cite{coleman}
that a semiclassical description of a
quantum tunnelling process is given by the bounce solutions of
Euclideanized field equations, \ie, of field equations in which
the time parameter t is changed into ($-it$). Applying such an
Euclideanization procedure to Eq.(\ref{acima}), one obtains
\begin{equation}
\dot{a}^2=+\epsilon -\frac{1}{3}\ \rho a^2\ .
\label{oneo}
\end{equation}
In the case of an $\epsilon =+1$ universe driven by a
(positive) cosmological constant $\Lambda = 3\varsigma^2$
this approach was used in \cite{vile2} to obtain, instead of the
classical de Sitter solution, namely
$$
a(t) = \frac{1}{\varsigma}\cosh(\varsigma t),
$$
the
solution
\begin{equation}
a_E(t)=\left(\frac{1}{\varsigma}\right)\cos (\varsigma t)\ ,
\end{equation}
corresponding to a de Sitter instanton -- a ``kink''
configuration-- propagating with  negative classical energy, which
bounces at the classical turning point $a=a_0=(1/\varsigma )$
interpreted as representing the tunnelling to classical de Sitter
space from ``nothing.''

Now consider Eqn.(\ref{acima}) in the case of a closed
Universe driven by the energy density $\rho =3[a^4_0/a^6]$. The
euclideanized  version of Eq.(\ref{oneo}) gives
$$
\dot{a}^2=1-\left[\frac{a^4_0}{a^4}\right]\ .
$$
But this is precisely the fundamental dynamical equation (\ref{sf}) of
the WIST cosmological scenario. In this way, an equivalence is
established between the Euclideanization of a closed Universe
model driven by a positive energy density and a ``structural
transition'' to a Weyl configuration which results in an {\it
open} Universe model driven by a ``stiff matter'' state of
negative energy. Just as in models of quantum creation the
propagation of an instanton is seen to represent the tunneling of
the Universe from a primordial quantum ``nothing'' state, in the
present scenario  the propagation of a WISTon (i.e., a deviation
of the Riemannian structure) is tantamount to the development of
the Universe from a primordial empty Minkowski space.

It has been argued that
solutions obtained through Euclideanization are in fact
non-realistic, since they are to be interpreted as instantons,
field configurations which tunnel across a classically forbidden
region.  Other authors endorse the view that such solutions
correspond to an actual primordial phase of the cosmic evolution
in which the basic Lorentzian nature of spacetime is changed into
an Euclidean one. According to the present model, a different
interpretation may be ascribed to these solutions, since an
enlargement of the spacetime structure to a Weyl configuration --
in which the geometry is characterized by the pair ($g_{\mu
\nu},\varphi_{\lambda}$) of fundamental variables -- supplies, at
least in a particular case, the same basic behavior. It then
becomes possible to conciliate the opposing interpretations of an
``abstract soliton configuration'' \cite{duncan} and of a truly
observable Euclidean cosmic phase \cite{hawking5}. The WIST solution is
observable, whereas its basic nature is always Lorentzian. It is
the Riemannian character of spacetime structure that results
altered; allegorically, the choice is no
longer Euclid or Lorentz, but rather Riemann or Weyl.

\subsubsection{Solution with matter generation}
\label{matcre}
\begin{figure}[htb]
\begin{center}
\includegraphics[angle=-90,width=0.5\textwidth]{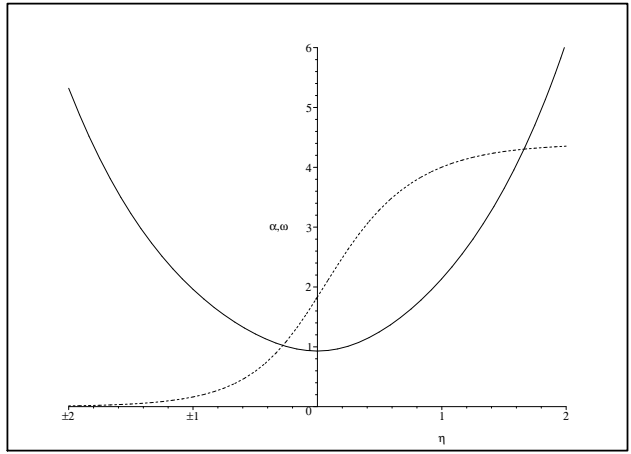}
\caption{Plot of $a$ and $\varphi$ for $k_0 = 1/7$ and $a_0 = 0.93$, values
chosen by imposing that the solution in Eqn.(\ref{sol}) enters the
radiation era for $t \approx 10^{-8}$ seg.}
\label{aomegamatter}
\end{center}
\end{figure}

We have mentioned above that the
the model must be improved by taking into account matter creation.
A non-singular solution in WIST that incorporates the effect of
the creation of matter on the geometry was studied in
\cite{nilton}. Friedman equation in conformal time is given by
\beq a'^2-a^2=-\frac{\lambda^2}{6}(\varphi
'a)^2+\frac{a^4}{3}\rho_m, \label{fct} \eeq while the second EE is
\beq
-3\left(2\frac{a''}{a^3}-\frac{a'^2}{a^4}-\frac{1}{a^2}\right)=\rho_m+3\rho_\varphi.
\label{atdcf} \eeq The conservation of the stress-energy tensor in
the case of ultra-relativistic matter is \beq
(a^4\rho_m)'+\frac{1}{a^2}(a^6\rho_\varphi)'=0. \label{ultra} \eeq
A particular solution to these equations that describes creation of
relativistic matter only around the bounce, and enters a radiation
phase with a constant scalar field in a short time is given by the
expression \cite{nilton}
\beq a(\eta ) = \beta\sqrt{\cosh (2\eta )
+k_0\sinh (2\eta ) - 2k_0(\tanh \eta +1)}, \label{sol} \eeq with
$\beta = a_0/\sqrt{1-k_0}$, and $0<k_0<1/7$. The dependence of
$\varphi$ with $\eta$ can be obtained from Eqns.(\ref{fct}) and
(\ref{atdcf}). An asymmetry is to be expected both in the scale
factor and in $\varphi$, since the evolution of this universe
starts from the vacuum and enters a radiation dominated epoch.
This is pictured in Fig.\ref{aomegamatter}.
Notice that since the scalar field tends rapidly to a constant value, the production
of matter (controlled by $\varphi '$, see Eqn.(\ref{ultra})) stops
soon, and the model enters a radiation phase without the
need of a potential. In this sense, this solution describes a
hot bounce, as opposed to cold bouncing solutions, which
do not enter the radiation era unless they are heated up
\cite{massimo}. Another nice feature of this solution is that the scalar field
(formally equivalent to the dilaton of string theory)
goes automatically to a constant value for $\eta\rightarrow\infty$,
in such a way that the solution could be taken as the
leading order of a perturbative development (as is the
case in string theory). Again, no potential was needed
in order to display this feature.

\subsection{Scalar-tensor theories}

Scalar-tensor theories are a generalization of the Brans-Dicke
Lagrangian \cite{bd}, in which the constant appearing in the
kinetic term of the scalar field $\phi$ becomes a function of
$\phi$. Among the possible Lagrangians to describe these theories, one
possibility is \cite{stei}
\beq {\cal L} = -f(\phi ) R + \half
\phi_{,\mu}\phi^{,\mu} + 16\pi {\cal L}_{\rm matter},
\eeq where
the scalar field $\phi$ couples non-minimally with the curvature through
$f(\phi)$.
With the redefinition $\varphi = f(\phi)$, the Lagrangian becomes
\beq
\lag = -\varphi R + \frac{\omega (\varphi)}{\varphi}
\varphi_{,\mu}\varphi^{,\mu} + 16\pi\lag_{\rm matter},
\label{lagrst} \eeq
with $\omega (\varphi ) = \half
f/f_\varphi^2$ and $f_\varphi\equiv df/d\varphi$. Brans-Dicke
theory is a special case of this Lagrangian, $f(\phi) \propto
\phi^2$ which entails $\omega = $const. This Lagrangian also
describes the gravity-dilaton sector of low-energy string theory
for $\omega = -1$ \cite{corrst}. The differences between the two
Lagrangians have been analyzed in \cite{liddle}. Following the
results of the discussion presented there, we shall use
Eqn.(\ref{lagrst}) as the definition of scalar-tensor theories.

The equations of motion corresponding to Eqn.(\ref{lagrst}) are
\beq
R_{\mu\nu} = -\frac{1}{\varphi} (T_{\mu\nu} -
\half g_{\mu\nu} T) - \frac{\omega(\varphi)} {\varphi^2}
\varphi_{,\mu} \varphi_{,\nu} - \frac{1}{\varphi} \varphi_{,\mu
;\nu} - \frac{1}{2\varphi} g_{\mu\nu} \Box \varphi \label{dst},
\eeq
\beq [3+2\omega (\varphi) ] \Box\varphi =  T -
\omega_\varphi \varphi_{,\mu}\varphi^{,\mu} \label{sf2}. \eeq
Eqn.(\ref{dst}) suggests that it may
be possible to find solutions in which matter satisfies SEC, but
the whole r.h.s. is such that $R_{\mu\nu} v^\mu v^\nu \geq 0$
\footnote{The same happens in some wormhole configurations in
Brans-Dicke theory. See \cite{bdwh}.}. This implies, via the
singularity theorem given in Sect.(\ref{defsing}) that nonsingular solutions
may exist in scalar-tensor theories. Using Eqn.(\ref{dst}), the
inequality $R_{\mu\nu} v^\mu v^\nu \geq 0$ translates for the flat
FLRW case and EOS $p=\lambda \rho$ to \beq - \frac{1}{\varphi}
(1+3\lambda) \rho\;\frac{\omega+2}{2\omega+3}
-\frac{\dot\varphi^2}{\varphi}\;\left(\frac{\omega}{\varphi}
-\frac{\omega'}{2(2+3\omega)}\right)-\frac{\ddot\varphi}{\varphi}\geq
0. \eeq Solutions satisfying this constraint, and hence exhibiting
a bounce, have been presented in \cite{barrow1}, for $\epsilon=0$ in the
cases of vacuum and radiation (for which $T=0$, see r.h.s. of
Eqn.(\ref{sf2}))
\footnote{A shadow of doubt has been cast on these
results in \cite{kalo}, where it was shown that gravitons would
still see a singularity, even if the rest of matter does not.
}. With these restrictions,
Eq.(\ref{sf2}) written in conformal time takes the form \beq
\varphi'' + \frac{2a'}{a}\varphi'= -
\frac{\varphi�^2\omega_{\varphi}}{3+2\omega}, \eeq which
integrates to \beq \varphi' a^2 = \frac{\sqrt 3
A}{\sqrt{2\omega+3}}, \label{int} \eeq where $A$ is a constant.
Introducing the variable $y=\varphi a^2$ and using Eq.(\ref{int}), the
Friedmann equation takes the form \beq y'^2 = 4\Gamma y + A^2,
\eeq ($\Gamma\geq 0$ is a constant coming from energy
conservation) yielding for $y(\eta)$, \beq y(\eta) = A(\eta +
\eta_0) \eeq in the case of vacuum, and \beq y(\eta) = \Gamma
(\eta + \eta_0)^2 - \frac{A^2}{4\Gamma} \eeq in the case of
radiation. Dividing now Eq.(\ref{int}) by $y=\varphi a^2$ we
obtain \beq \int\frac{\sqrt{2\omega(\varphi)+3}}{\varphi}
\;d\varphi = \sqrt 3 A \int \frac{d\eta}{y(\eta)}. \label{invert}
\eeq If this equation is invertible, we could obtain from it
$\varphi = \varphi(\eta)$, and then $y = \varphi a^2$ yields
$a(\eta)$. To integrate Eq.(\ref{invert}), we need to specify the
function $\omega(\varphi)$. The choice in \cite{barrow1} was \beq
2\omega(\varphi)+3 =
2\beta\left(1-\frac{\varphi}{\varphi_c}\right)^{-\alpha}, \eeq
where $\alpha$, $\beta>0$ and $\varphi_c$ are constants. With this
choice of $\omega$, Eq.(\ref{invert}) can be solved for
$\varphi(\eta)$ in the cases $\alpha =0$ (which corresponds to
Brans-Dicke theory), $\alpha = 1$ and $\beta = -\half$ (which
defines a theory introduced by Barker \cite{barker}), and $\alpha
= 2$. The latter was studied in \cite{barrow1}. The solutions for
the vacuum case are given by \beq a(\eta)^2 = \frac
{A(\eta+\eta_0)(1+(\eta+\eta_0)^\lambda)}{\varphi_c
(\eta+\eta_0)^\lambda} \eeq \beq \varphi(\eta) =
\frac{\varphi_c(\eta+\eta_0)^\lambda}{1+(\eta+\eta_0)^\lambda},
\eeq with $\lambda = \sqrt{3/2\beta}$. These solutions were shown
to be nonsingular for $\beta<3/2$. Hence the radiation solutions
(which approach those for the vacuum for $\eta\rightarrow 0$
\cite{barrow1}) are also nonsingular. All the solutions for
$\alpha = 2$ approach the FLRW radiation regime at late times
because $\varphi$ tends to a constant, and then $\omega (\varphi)
\rightarrow\infty$, but in order to be in agreement with solar
system experiments, $\alpha$ must be greater than $1/2$
\cite{barrow1}.

The case of stiff matter (defined by $\rho = p$) sourcing the
scalar field was studied in \cite{mimoso}. Since the density of a
barotropic fluid ($p=(\gamma-1)\rho$) evolves as $\rho\propto
a^{-3\gamma}$, this kind of matter is expected to dominate at
early times, and the associated solutions give information about
the early evolution of the universe. One of the results in
\cite{mimoso} is that a necessary condition for $\dot a = 0$
when spatial curvature is negligible
is
$\omega = -6M\varphi/A$, where $A$ and $M$ are positive constants, yielding
a negative
kinetic term
for $\varphi$ (see Eqn.(\ref{lagrst}).
A thorough qualitative study of the case in which
$\omega(\varphi)$ is a monotonic but otherwise arbitrary function
of $\varphi$ was presented in \cite{serna}, where the existence of
nonsingular solutions in theories which agree with GR in the weak
field limit was proved.

The first term on the left hand side of Eqn.(\ref{dst}) suggests
that the gravitational constant is not actually a constant but
varies with $\varphi^{-1}$. Based on this idea, a generalization
of scalar-tensor theories (the so-called hyper-extended
scalar-tensor) was advanced in \cite{diego}.
The Lagrangian
associated to these theories is given by \beq \lag =
-G(\varphi)^{-1} R + \frac{\omega (\varphi)}{\varphi}
\varphi_{,\mu}\varphi^{,\mu} + 16\pi\lag_{\rm matter},
\label{lagrhypst} \eeq which reduces to Eqn.(\ref{lagrst}) when
$G(\varphi)  = 1/\varphi$. Sufficient conditions on $G(\varphi)$,
$\omega(\varphi)$, and their derivatives in order to have bouncing
cosmological solutions were given in \cite{fay}, generalizing the
work of \cite{kal} for the case of ST theories.

Another descendant of the original ST theory are the
multiscalar-tensor (MST) theories \cite{copenu}, which are the generic product
of a compactification process of a higher-dimensional theory. The
scalar content of a given MST theory depends on the details of the
internal manifold that results from the compactification (usually
gauge fields are set to zero in cosmological applications).
Typically, one or more fields are associated to the size of
the extra dimensions. In string theory, the coupling constants
depend on the expectation value of massive scalar fields (called
moduli fields) also associated with the size and shape of the
extra dimensions, the most popular example of them being the
dilaton. The moduli are an inescapable ingredient of string
theory, hence several problematic issues raised by them must be
confronted, such as stabilization, overcritical density, and
violations of the Equivalence Principle. Cosmological solutions of
low-energy string theories have been extensively studied (see
\cite{reviewssc} for a review). Needless to say, the results
depend on the field content, which in turn depends on the given
string theory under scrutiny.

A possible way to parameterize an action of a MST theory is
\cite{fabris}
\beq
{\cal L} = \sqrt{-g}\left[\phi R - \omega
\frac{\phi_{,\rho}\phi^{,\rho}}{\phi}-\phi^n\psi_{,\rho}
\psi^{,\rho}-\chi_{,\rho}\chi^{,\rho}\right]+{\cal L}_{\rm matter}.
\label{julio}
\eeq
This Lagrangian represents pure multidimensional theories when
$\psi=$constant, $\chi=$ constant, and $\omega=(1-d)/d$, where $d$
is the number of compactified dimensions (assuming that they have
the topology of a torus). The same case but with $\psi\neq$
constant and $n=-2/d+1$ corresponds to a two-form gauge field in
higher dimensions. If this field is conformal, it is associated to
a $(d+4)/2$-form, leading to $n=-2/d$. In the case of string
theory, $\omega = -1$, and the field $\psi$ is associated to a
three-form field $H_{\mu\nu\lambda}$, leading to $n=-1$. The
scalar $\chi$ is related to another three-form field coming from
the R-R sector of type IIB superstring theory.

The existence of bouncing solutions for this Lagrangian in vacuum
and in the presence of radiation for the FLRW geometry for all
values of the three-curvature
and for arbitrary values of $\omega$ and $n$
has been studied in \cite{fabris}. The results show
that generically there is bounce for $n<1$ and $\omega<0$.

\subsubsection{Corrections coming from String Theory}
\label{ssst}

Superstring theory is a candidate for a unified theory of the
fundamental interactions, including gravity \cite{sstrings}.
Since the fundamental objects in this theory
are at least one-dimensional, geodesics of point
particles are replaced by world-volumes.
It is a valid question then to ask whether string theory has anything
to say about the singularity problem. In this regard, it must be noted that
in string theory, the gravitational
excitations are defined on a fixed metric background.
Since singularities in general relativity are boundaries of space-time, which are
a consequence
of the dynamics governing its structure, a fixed manifold is certainly
a restriction. Yet another difficulty is the
breakdown of string perturbation
theory in the regime of interest \cite{bojobscg}. However, we have seen in the previous section that
the incorporation of the massless degrees of freedom (correspongind to the lowest order
EOM), which
applies on scales below the
string scale and above those where the string symmetries are broken, may smooth out the singularity.
One could go further and include higher-order corrections in the
action of string theory. There are two types of corrections.
First,
there are
the classical corrections arising from
the finite size of the strings, when the fields vary over the string length scale, given by
$\lambda_s = \sqrt{\alpha '}$. These terms
are important in the regime of large curvature, and lead to a series in $\alpha'$ (the inverse of
the tension of the string).
Then there are the loop (quantum)
corrections.
The loop expansion is
parameterized by powers of the string coupling parameter
$e^\phi=g^2_{\rm string}$,
which is a time-dependent quantity in cosmological models. In the so-called strong coupling regime,
the dilaton becomes large and quantum corrections are important.

The effective action at the one-loop level is  given by (see for
instance \cite{anto})
\beq S = \int d^4x \sqrt{-g} \left\{ \frac R
2 + \frac 1 4 (\nabla\phi)^2 + \frac 3 4 (\nabla \sigma)^2 +
\frac{1}{16}[\lambda e^\phi - \delta \xi(\sigma)]R^2_{GB}\right\},
\eeq where $\phi$ is the dilaton, $\sigma$ is a modulus field, and
$\lambda=2/g^2$ ($g$ is the string coupling), $\delta$ is
proportional to the 4-d trace anomaly, and $\xi(\sigma) =
\ln(2e^\sigma\eta^4(ie^\sigma))$, where $\eta$ is the Dedekind function.
The correction to the gravitational term is given in terms of the
Gauss-Bonnet invariant,
$$
R_{GB}^2 = R_{\mu\nu\kappa\lambda}R^{\mu\nu\kappa\lambda} - 4
R_{\mu\nu}R^{\mu\nu} +R^2.
$$
The EOM that follow from this action in the case of a FLRW flat
spacetime with the metric $g_{\mu\nu} = {\rm
diag}(1,-e^{2\omega}\delta_{ij})$ are
\cite{anto} \footnote{See
\cite{easther} for the case of nonzero spatial curvature.}
\beq 3\dot\omega^2 - \frac 3
4 \dot\sigma^2-\frac 1 4 \dot\phi^2 + 24 \dot f\dot\omega^3=0,
\eeq \beq 2\ddot \omega + 3 \dot\omega^2+\frac 3 4 \dot\sigma^2 +
\frac 1 4 \dot\phi^2 + 16 \dot f\dot\omega^3 + 8 \ddot
f\dot\omega^2 + 16\dot f \dot \omega\ddot \omega = 0, \eeq \beq
\ddot\sigma + 3 \dot\omega\dot\sigma + \delta
\frac{\partial\xi}{\partial\sigma} \dot\omega^2 (\dot\omega^2 +
\ddot \omega)=0, \eeq \beq \ddot\phi +
3\dot\omega\dot\phi-3\lambda e^\phi\dot\omega^2 (\dot\omega^2 +
\ddot \omega)=0, \eeq where $f = \frac{1}{16}(\lambda e^\phi -
\delta \xi(\sigma))$. These equations are not linearly independent
due to the conservation of $T_{\mu\nu}$.

It
was shown in
\cite{anto} that there are solutions with bounce for $\delta <0$,
which interpolate between an asymptotically flat and a
slowly expanding universe with a period of rapid expansion. The
bounce is essentially due to the violation of the strong energy
condition by the modulus field (the dilaton playing an unimportant
role). In a subsequent paper \cite{rizos} it was shown that
non-singular solutions can be obtained under the assumptions that
$\xi$ is a smooth function that has a minimum at some point
$\sigma_0$, and grows faster than $\sigma^2$ for
$\sigma\rightarrow\pm\infty$, and $\delta>0$. However, these
solutions were later shown to be generically unstable for tensor
perturbations \cite{soda}. Less symmetric models (Bianchi I
\cite{kawai} and Bianchi IX \cite{yajima}) were also studied for
this action, confirming the findings of \cite{soda}.

Another attempt to avoid the singularity is to consider the effect
of matter terms to the action of string theory. In
\cite{shinji} an action including dilaton, axion and one modulus
field was considered along with matter (radiation or a ``stringy''
gas) and higher-order dilaton corrections in a flat FLRW
background in $d$ dimensions. In this case, the results of \cite{shinji} show that
the energy densities of matter, axion and modulus are strongly
suppressed in the inflationary phase driven by the dilaton, and
hence the higher-order corrections coming from this field take the
system through the graceful exit.

Yet another model inspired in string theory is the so-called
ekpyrotic universe and its extension, the cyclic universe which
will be discussed in Sect.\ref{ekpyrotico}.

\subsubsection{String Pre-Big Bang}
\label{pbb}

\begin{figure}[htb]
\begin{center}
\includegraphics[width=0.5\textwidth
]{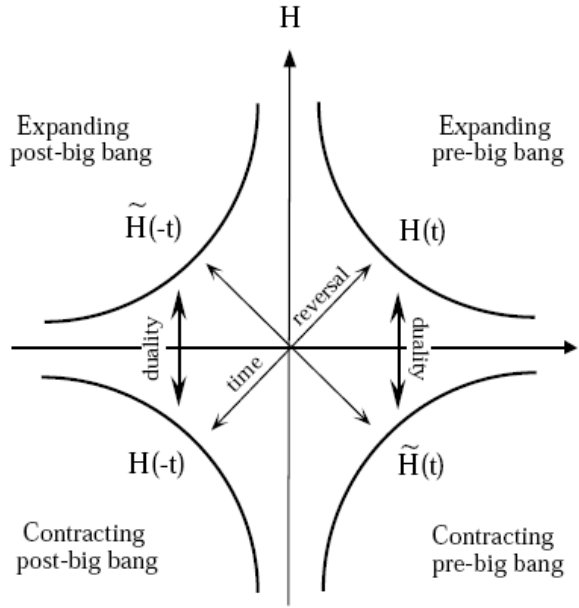}
\caption{The four branches of the low-energy string cosmology backgrounds.
Taken from \cite{vene}.}
\end{center}
\label{fdual}
\end{figure}

A very-well developed example of the string cosmology
approach is the so-called ``pre-big bang'' \cite{vene}, which we
shall call ``string pre-big bang'' (SPBB), to differentiate it
from similar models not coming from string theory (see Sect.\ref{dynor}).
There are two properties of string theory that can be expected to
play an important role in cosmology \cite{vene2}. First, in the
short-distance regime, a fundamental length $\lambda_s$ is
expected to arise, thus introducing an ultraviolet cut-off and
bounding physical quantities such as $H^2$ and $a$. Hence a bounce
may be expected. Second, as we discussed before, at lower
energies, the action of string theory is not Einstein's but a
(multi)scalar-tensor theory, where one of the scalar fields is the
dilaton, which controls the coupling constants. If these are
really constant today (see \cite{constants}), the dilaton must be
seated at the bottom of its potential, but it may have evolved in
cosmological times. The idea of the SPBB is that during the
cosmological evolution, the kinetic term of the dilaton drove a
period of deflation (or inflation, depending on
whether we consider the Einstein frame or the string frame)
``before the big bang''(that is, in the
contracting phase)\footnote{This idea was also suggested in
\cite{NovelloS2}.}, which can solve the horizon and flatness problems
\cite{deflation}.
In this approach, the universe starts from a
perturbative state, passes through a high-curvature and
high-coupling stage, and then (hopefully) enters the radiation-dominated FLRW
evolution. Duality symmetries present in the low-energy action of
string theory are invoked to support this line of reasoning \cite{dualities}: in
the isotropic case, the gravidilaton EOM in the FLRW setting are
invariant under a time inversion,
$$ t\rightarrow -t \Rightarrow
H\rightarrow -H,
$$
$$\dot\phi \rightarrow -\dot\phi, $$
and under the duality transformation
$$a\rightarrow \tilde a =
a^{-1},
$$
$$\phi\rightarrow\tilde\phi = \phi-6\ln a.$$
(compare with the Weyl transformation, Eqn.(\ref{gauge1})).
These transformations relate four branches of the solution (PBB,
and post-big-bang expansion and contraction). In particular, to
any expanding solution with decreasing curvature (such as those in
the standard cosmological model), duality associates an
accelerated contracting solution (see Fig.\ref{fdual}). It is this pairing (which is possible
only in the presence of the dilaton) that
supports the whole idea of the SPBB.
One of the issues of this idea is
the joining of the two phases through the
putative singularity (the graceful exit problem).
It has been proved in
\cite{kaloper} that
the graceful exit transition from the initial
phase of inflation to the subsequent standard
radiation dominated evolution
must take place
during
a ``string phase'' of high curvature or strong coupling is actually required.
The corrections to the lowest-order lagrangian can be parameterized as
\cite{madden}
$$
{\cal L}_c = {\cal L_{\alpha '}}+{\cal L}_{q},
$$
where
\beq
\half{\cal L_{\alpha '}}=e^{-\phi}\left(\frac 1 4 R_{GB}^2-\half (\nabla\phi)^4\right),
\label{hoc}
\eeq
and ${\cal L}_{q}$ designates the quantum loop corrections.
Several forms of ${\cal L}_{q}$ were studied in
\cite{madden}. The existence of a bounce in the Einstein frame, yielding a
solution to the graceful exit problem,
was shown by numerical integration of the EOM
in \cite{madden} for the case
${\cal L}_{q}=-2(\nabla\phi)^4$,
${\cal L}_{q}=-2(\nabla\phi)^4+ R^2/3$, and for the two-loop
correction
${\cal L}_{q}=2e^\phi R_{GB}^2$,
in all cases by choosing the appropriate sign for the correction.

An even more general form of the corrections was studied in
\cite{cartier}, where ${\cal L}_c$ was given by
$$
{\cal L}_c=-\frac 1 4 e^{-\phi} \left(a R^2_{GB}+ b \phi (\nabla\phi)^2+ c  G^{\mu\nu}\partial_\mu\phi\partial_\nu\phi+
d(\partial_\mu\phi)^4\right),
$$
and $4b+2c+d=-4a$ ($G^{\mu\nu}$ is the Einstein tensor).
The quantum corrections were included by adding a suitable power of
the string coupling, so the total effective Lagrangian is given by
$$
{\cal L}=R+(\partial_\mu\phi)^2+ {\cal L}_c+Ae^{\phi}{\cal L}_c+Be^{2\phi}{\cal L}_c,
$$
and the parameters $A$ and $B$ set the scale for the loop corrections.
Solutions with graceful exit were found in \cite{cartier}
for a large range of parameters, but it is very hard to obtain
the transition in the weak coupling regime, whilst keeping the loop corrections
small.

A problem that remains to be solved is the stabilization
of the dilaton to a constant value (otherwise there would be violations to the Equivalence
Principle and to the observed ``constancy of the coupling constants''). This
was achieved in the previously mentioned articles in a number of ways: 1)
by introducing by hand a friction term
in the equation of motion of the dilaton,
and then coupling it to radiation in such
a way as to preserve overall conservation, 2) by
``turning off'' by hand the
quantum Lagrangian by means of a step function, and 3) by the manipulation of the sign
and size
of the higher-loop corrections.

\subsection{Appendix:Conformal Transformation}
\label{app2}
Consider the map
\begin{equation}
 \tilde{g}_{\mu\nu}\,(x) = \Omega^{2}\,(x)\, g_{\mu\nu}\,(x).
 \label{CT1}
 \end{equation}
Then, for the contravariant components:
\begin{equation}
 \tilde{g}^{\mu\nu}\,(x) = \Omega^{-\,2}\,(x)\, g^{\mu\nu}\,(x).
 \label{Pescara14,1450}
 \end{equation}
The conformal transformation of the connection is provided by
\begin{equation}
\widetilde{\Gamma}^{\alpha}_{\mu \nu}= \Gamma^{\alpha}_{\mu \nu} +
\frac{1}{\Omega} \left(\Omega_{,\mu} \, \delta^{\alpha}_{\nu} +
\Omega_{,\nu} \, \delta^{\alpha}_{\mu} - \Omega_{,\epsilon} \,
g^{\epsilon\alpha} \, g_{\mu\nu}  \right) \label{Pescara14,1451},
\end{equation}
and for the curvature tensor:
\begin{equation}
 {\tilde{R}^{\alpha\beta}}_{\;\;\mu\nu} = \Omega^{-\,2}
\,{R^{\alpha\beta}}_{\mu\nu} - \frac{1}{4}
{\delta^{[\,\alpha}}_{[\,\mu} \, {Q^{\beta\,]}}_{\nu\,]},
\label{Pescara21}\end{equation}
 where
$$ {Q^{\alpha}}_{\beta} \equiv 4 \, \Omega^{-\,1} \,
(\Omega^{-\,1})_{,\,\beta;\,\lambda} g^{\alpha\lambda} - 2 \,
(\Omega^{-\,1})_{,\,\mu} \,(\Omega^{-\,1})_{,\,\nu} \,g^{\mu\nu}
\delta^{\alpha}_{\beta}. $$ Contracting Eqn.(\ref{Pescara21}) we
get
\begin{equation}
 {\tilde{R}^{\alpha}}_{\;\mu} = \Omega^{-\,2} \,{R^{\alpha}}_{\mu}
- \frac{1}{2}\,   {Q^{\alpha}}_{\mu} - \frac{1}{4} Q
\delta^{\alpha}_{\mu} \label{Pescara22}, \end{equation} and
contracting again,
\begin{equation}
\tilde{R} =  \Omega^{-\,2} [\,R + 6\,\Omega^{-\,1} \, \Box\,\Omega
\,]. \label{Pescara23}
\end{equation}
A direct comparison of this conformal scalar of curvature  and the
Weyl scalar equation (\ref{H1}) shows that they coincide (up to a
multiplicative factor) if we set
$$ \Omega = \exp\left(- \, \frac{1}{2} \, \varphi\right),$$ and Eqn.(\ref{Pescara23})
takes the form
$$
\tilde{R} =  e^{\varphi} [\,R -3\Box\varphi+\frac 3 2
\varphi_{,\mu}\varphi^{,\mu}],
$$
which is exactly the transformed of the Ricci scalar for the WIST:
$$
\tilde{R} =  e^{\varphi}R^{(W)}.
$$

\section{Maxwellian and Non-Maxwellian Vector Fields}

\subsection{Introduction}

The model described by the FLRW geometry
with Maxwell's electrodynamics as its source displays a
cosmological singularity at a finite time in the past \cite{Kolb}.
However, this is not an intrinsic property of the combined
electromagnetic and gravitational fields. Indeed,
modifications of Maxwell electrodynamics (or, generically,
massless vector field dynamics) can generate non-singular
spatially homogeneous and isotropic (SHI) solutions of classical
GR. We shall examine here two modifications that are relevant to the
singularity problem:
\begin{itemize}
\item The non-minimal coupling of the EM field
with gravity, and
\item the self-interaction of the EM field.
\end{itemize}
These
modifications will be introduced by means of Lagrangians which depend
nonlinearly on the field invariants or on the space-time
curvature. In both cases, the singularity theorems (see Ch.\ref{sint})
are circumvented by the appearance of a large, but nevertheless
finite, negative pressure in an early phase of the SHI geometry.

\subsection{Einstein-Maxwell Singular Universe}
\label{emsu}
The fact that Maxwell electrodynamics minimally coupled to gravity
leads to singular models for the  universe in the FLRW framework is a
direct consequence of the singularity theorems (see Ch.\ref{sint}).
Essentially, this can be understood from the examination of the
energy conservation law and Raychaudhuri equation, as
follows.
To be consistent with the symmetries of the SHI metric,
an averaging procedure must be performed if electromagnetic fields
are to be taken as a source for the EE
\cite{tolmanbook}. As a consequence, the components of the electric
$\mathscr{E}_{i}$ and magnetic $\mathscr{H}_{i}$ fields must satisfy the following
relations:
\begin{eqnarray}
\overline{\rule{0pt}{2ex}{\mathscr{E}}_i} = 0,\qquad
%
\overline{\rule{0pt}{2ex}{\mathscr{H}}_i} &=& 0,\qquad
%
\overline{\rule{0pt}{2ex}{\mathscr{E}}_i\, {\mathscr{H}}_j} = 0,
\label{meanEH}\\[1ex]
\overline{\rule{0pt}{2ex}{\mathscr{E}}_i\,{\mathscr{E}}_j} &=& -\, \frac{1}{3} \mathscr{E}^2
\,g_{ij},
\label{meanE2}\\[1ex]
\overline{\rule{0pt}{2ex}{\mathscr{H}}_i\, {\mathscr{H}}_j} &=&  -\, \frac{1}{3} \mathscr{H}^2
\,g_{ij}.
\label{meanH2}
\end{eqnarray}
The symmetric energy-momentum tensor associated with Maxwell
Lagrangian is given by
\begin{equation}
E_{\mu\nu} = F_{\mu\,\alpha}\,F^{\alpha}\mbox{}_{\nu} +
\frac{1}{4} \,F \,g_{\mu\nu}, \label{Maxwell}
\end{equation}
in which $F \equiv F_{\mu\nu}\, F^{\mu\nu}=2({\mathscr{H}}^2-{\mathscr{E}}^2)$.
Using the above average values it follows that the $T_{\mu\nu}$
reduces to a perfect fluid configuration with energy density
$\rho_\gamma$ and pressure $p_\gamma$ given by
\begin{equation}
\overline{\rule{0pt}{2ex}E_{\mu\nu}} = (\rho_\gamma + p_\gamma)\,
v_{\mu}\, v_{\nu} - p_\gamma\, g_{\mu\nu}, \label{Pfluid}
\end{equation}
where
\begin{equation}
\label{RhoMaxwell} \rho_\gamma = 3p_\gamma = \frac{1}{2}\,(\mathscr{E}^2 +
\mathscr{H}^2).
\end{equation}
The fact that both the energy density and the pressure in this case are positive
definite for all values of $t$ implies the singular nature of FLRW universes. In
fact, the solution of EE for the above
energy-momentum configuration gives for the scale factor the
singular form \cite{Robertson}
\begin{equation}
a(t)=\sqrt{a_0^2t-\epsilon t^2},
\label{A(t)Maxwell}
\end{equation}
where $a_0$ is an arbitrary constant. We conclude that the
space-time singularity in the Einstein-Maxwell system is
unavoidable.

\subsection{Non-minimal interaction}

Most of the articles concerning the interaction of Electrodynamics
with Gravitation assume the principle of minimal coupling, which is a
direct application of the strong form of the Equivalence Principle.
In the absence of stringent limits from observation, ideally we should keep an open
mind and consider other possibilities.
Non-minimal coupling of the EM field with gravity
has recently been applied in cosmology, following the trend initiated by
scalar field theories interacting conformally with
gravitation. These opened the way to the exam of more general theories,
such as those in which curvature is directly coupled with the
fields.

There are seven possible Lagrangians for the interaction of the EM field with Gravity
which can be
constructed as linear functionals of the curvature tensor. They are
divided in two classes. Class I is given by:
\begin{eqnarray}
{\cal L}_1 &=& R\ A_{\mu}A^{\mu}\nonumber, \\
{\cal L}_2 &=& R_{\mu \nu}A^{\mu}A^{\nu}.\nonumber
\end{eqnarray}
These two Lagrangians are gauge dependent but no dimensional
constant must be added since they already have the right
dimensionality. As shown in \cite{NovelloSalim} the EOM obtained from ${\cal L}_2$
in Einstein's gravity with the addition of a kinetic term for $A^\mu$ do not admit
a FLRW solution. Thus, in the following
we shall limit our analysis to ${\cal L}_1$.

In Class II, there are five Lagrangians :
\begin{eqnarray}
{\cal L}_3 &=& R\ F_{\mu \nu}F^{\mu \nu}\nonumber ,\\
{\cal L}_4 &=& R\ F_{\mu \nu}F^{\mu \stackrel{\ast}{\nu}}\nonumber ,\\
{\cal L}_5 &=& R_{\mu \nu}\ F^{\mu}_{~\alpha}F^{\alpha \nu}\nonumber, \\
{\cal L}_6&=& R^{\alpha \beta \mu \nu}\ F_{\alpha \beta}\ F^{\nu
\mu}\nonumber ,\\
{\cal L}_7 &=& \stackrel{\ast}{W}_{\alpha \beta \mu \nu}\ F^{\alpha
\beta}\ F^{\mu \nu},
\end{eqnarray}
where $W^{\alpha \beta \mu \nu}$ is the Weyl tensor and the star in the Weyl tensor
means
$$
\stackrel{\ast}{W}_{\alpha \beta \mu \nu}\ = W_{\alpha \beta \stackrel{\ast}{\mu \nu}}
= W_{\stackrel{\ast}{\alpha \beta} \mu \nu} =\half \eta_{\alpha\beta}^{\;\;\rho\sigma}
W_{\rho\sigma\mu\nu}.
$$
These
Lagrangians are gauge independent but they all need the introduction
of a length $\ell_0$ in order to have the correct dimensionality.

Another Lagrangian sometimes studied in the literature that is not
explicitly contained in this list is
\[
{\cal L}_{8}=R_{\stackrel{\ast}{\alpha}\beta\mu\stackrel{\ast}{\nu}}\,
F^{\alpha \beta}F^{\mu \nu}.
\]
However, ${\cal L}_{8}$ is not independent of $({\cal L}_1,...{\cal L}_7)$. Indeed,
the double dual
$R_{\stackrel{\ast}{\alpha}\beta\mu\stackrel{\ast}{\nu}}$ satisfies
the identity
\begin{equation}
R_{\stackrel{\ast}{\alpha}\beta\mu\stackrel{\ast}{\nu}}=
R_{\alpha\beta\mu\nu}-2W_{\alpha\beta\mu\nu}-\frac{1}{2}\ R\
g_{\alpha\beta\mu\nu},
\end{equation}
or, equivalently,
\begin{eqnarray}
R_{\stackrel{\ast}{\alpha}\beta\mu\stackrel{\ast}{\nu}}&=& -
W_{\alpha\beta\mu\nu}+\frac{1}{2}\left(R_{\alpha \mu}g_{\beta
\nu}+ R_{\beta \nu}g_{\alpha \mu}-R_{\alpha \nu}g_{\beta \mu}-
R_{\beta \mu}g_{\alpha \nu}\right)-\nonumber \\
&-& \frac{1}{3}\ Rg_{\alpha\beta\mu\nu}.
\end{eqnarray}
Thus,
\begin{eqnarray}
{\cal L}_8 &=& -{\cal L}_6-\frac{1}{3}\ R\left(g_{\alpha \mu}g_{\beta \mu}-
g_{\alpha \nu}g_{\beta \mu}\right)F^{\alpha \beta}F^{\mu \nu}+
\frac{1}{2}\left(R_{\alpha \mu}g_{\beta \nu}+g_{\alpha
\mu}\right.-\nonumber
\\
&-& \left.R_{\alpha \nu}g_{\beta \mu}-R_{\beta \mu}g_{\alpha
\nu}\right)F^{\alpha \beta}F^{\mu \nu}\nonumber.
\end{eqnarray}
Hence, ${\cal L}_8 = -{\cal L}_6-\frac{2}{3}\ {\cal L}_3-2{\cal L}_5$.

\subsection{An example of a non singular universe}
\label{examplensu}

The first example of a nonsingular universe driven by the nonminimal
coupling of EM and gravity was presented in
\cite{NovelloSalim}, using the ${\cal L}_1$ of the previous
section:
\begin{equation}
{\cal L} = \,R - \frac{1}{4} \,F^{\mu\nu}\,F_{\mu\nu} +
 \beta \, R\,A_{\mu}\, A^{\mu}.
 \label{15fev10}
\end{equation}
As mentioned in Sect.\ref{emsu},  in order to obtain a SHI geometry in the
realm of General Relativity having a vector field as a source, an
average procedure is needed. In the present non-minimal case there
is another possibility, which we shall now explore. Since this theory is
not gauge-invariant, it is possible to find a non-trivial solution
for $A_\mu$ such
that $F^{\mu\nu} $ vanishes.

The equations of motion that follow from the Lagrangian (\ref{15fev10})
are:
\begin{equation}
( 1 + \beta \, A^{2} ) \, ( R_{\mu\nu} - \frac{1}{2} \, R
g_{\mu\nu} ) - \beta \, \Box A^{2} \, g_{\mu\nu} + \beta \, (
A^{2} )_{;\mu;\nu} + \beta \, R A_{\mu} A_{\nu} = -  E_{\mu\nu} -
, \label{15fev101}
\end{equation}
\begin{equation}
F^{\mu\nu}_{;\nu} = - 2\beta \, R \, A^{\mu}.
\label{15fev102}
\end{equation}
From the trace of (\ref{15fev101}) it follows
$$R = - \, 3\beta \, \Box A^2, $$ which
when inserted in the equation of evolution of the electromagnetic
field yields a nonlinear equation:
\begin{equation}
{F^{\mu\nu}}_{;\,\nu} - 6 \beta^2 \,(\Box \,A^2) \, A^{\mu} = 0.
\label{B4}
\end{equation}
The non-linearity induced by the non-minimal
coupling with gravity is a generic feature for any field. To
obtain a solution in which the geometry is nonsingular for a SHI
geometry without imposing an average on the fields \cite{NovelloSalim} we
can consider the case in which $F_{\mu\nu}$ is zero. This is
possible due to the explicit dependence of the dynamical equations
on the vector $A_{\mu}$. We take the vector field $A_{\mu}$ of
the form
\begin{equation}
A_{\mu} = A(t)\,\delta^{0}_{\mu}. \label {B5}
\end{equation}
Defining the quantity $\Omega$ by
\begin{equation}
 \Omega(t) \equiv 1 + \beta \,A^2,
 \label{B6}
 \end{equation}
the set of equations
(\ref{15fev101}-\ref{15fev102}) in a FLRW geometry  reduces to the following:
\begin{equation}
3 \,\frac{\ddot{a}}{a} = - \,\frac{\ddot{\Omega}}{\Omega},
\label{eq1}
\end{equation}
\begin{equation}
\frac{\ddot{a}}{a} + 2 \,\left(\frac{\dot{a}}{a}\right)^2 +
\frac{2\,\epsilon}{a^2} = - \frac{\dot{a}}{a}
\,\frac{\dot{\Omega}}{\Omega}, \label{eq2}
\end{equation}
\begin{equation}
\Box \,\Omega = 0. \label{eq3}
\end{equation}
The last equation implies that $ a^{3} \, d\Omega/dt$ is a
constant. Thus we set $ d\Omega/dt = b \,a^{-\,3}$. A particular solution
of this set of equations for $\epsilon =-1$  is given by
\cite{NovelloSalim}
\begin{eqnarray}
A^2(t) &=& 1-\frac{t}{a(t)} \\
a(t)   &=& \sqrt{t^2 + \alpha_0^2}
\label{aex}
\end{eqnarray}
where $\alpha_0$ is a constant that measures the minimum possible
value of the scale factor. When $\alpha_0=0$ the system reduces to
empty Minkowski space-time in Milne coordinates.
%
%
For $\alpha_0\neq
0$ this model represents an eternal universe without singularity and
with a bounce \footnote{This form of the scale factor is similar
to Melnikov-Orlov geometry \cite{melni}, the difference being in the
interpretation of the minimum radius $a_0$ and the source of the
curvature.}. The system (\ref{eq1}-\ref{eq3}) can be written as
%
a planar autonomous system, which was examined in \cite{romero}\textbf{....}
Notice that in recent years theories with negative energies have been examined
in a cosmological context \cite{makler}. One way to achieve this goal is by
introducing an \emph{ad-hoc} term in the Lagrangian with the wrong sign. In
the case of a scalar field this is given as
\begin{equation}
S = \int \, \sqrt{-g} \, ( R - \frac{1}{2} \, \partial_{\mu}
\varphi \,
\partial^{\mu} \varphi).
\label{NE1}
\end{equation}
A fluid with this odd feature can also be obtained by the
non-minimal interaction of a vector field with gravity. Indeed,
the solution presented in the precedent section can be interpreted
as a perfect fluid with negative energy. The equations of motion
presented in \cite{NovelloSalim} can be re-written in the form:
\begin{equation}
R_{\mu\nu} = \frac{\Omega_{, \mu;\nu}}{\Omega}, \label{Neg1}
\end{equation}
were  $\Omega$, given by Eqn.(\ref{B6}), depends only in time .
 The structure of the corresponding system of equations
 is equivalent to the equations of General Relativity in the
 SHI geometry having as its source the energy-momentum tensor of a
perfect fluid with negative energy density and pressure given by
\begin{equation}
p = \frac{1}{3} \, \rho = - \, \frac{a_0^{2}}{a^{4}}   \label{Neg4}
\end{equation}
In this way, fluids with the "wrong"  sign in Einstein\rq s equation
can be interpreted as vector fields with non-minimal interaction
with gravity.

\subsection{Nonlinear electrodynamics}

\label{snled}

As pointed out in the introduction of this Chapter,
linear electromagnetism unavoidably
leads to
a singularity. This situation changes drastically in
the case of non-minimal coupling. In this section, we shall deal
with another type of theories, in which it is the nonlinearity of
the self-interaction of the EM field that provides the necessary
conditions for a cosmological bounce to occur.
The theories that will be examined
are described by Lagrangian which are
arbitrary functions of the invariants $F$ and $G$
that is $ {\cal L} = {\cal L}(F,G) $,
where $ F=F_{\mu\nu}F^{\mu\nu}, G \stackrel{.}{=} 
\frac{1}{2}\eta_{\alpha\beta\mu\nu}F^{\alpha\beta}F^{\mu\nu}$.
Their corresponding energy momentum
tensor, computed from Eqn.(\ref{emtensor})
yields
\begin{equation}
\protect\label{Tmunu} T_{\mu\nu}=-4\,{\cal L}_F\,F_\mu\mbox{}^\alpha
F_{\alpha\nu} + (G{\cal L}_G-{\cal L})\,g_{\mu\nu},
\end{equation}
where ${\cal L}_{A}\equiv d{\cal L}/dA$, with $A=F,G$. It follows that
\beq
\rho = -{\cal L}+G{\cal L}_G-4{\cal L}_F{\mathscr{E}}^2,
\label{eden}
\eeq
\beq
p={\cal L}-G{\cal L}_G-\frac 4 3 (2{\mathscr{H}}^2-{\mathscr{E}}^2){\cal L}_F.
\label{pre}
\eeq

We shall start our analysis by studying a toy model generalization
of Maxwell\rq s electrodynamics generated by a Lagrangian quadratic
in the field invariants as in \cite{klippert}, that is:
\begin{equation} L = -\frac{1}{4}\,F + \alpha\,F^2 + \beta\,G^2,
\label{Order2}
\end{equation}
where $\alpha$ and $\beta$ are dimensionfull constants \footnote{If
we consider that the origin of these corrections come from quantum
fluctuations then the value of the constants $\alpha$ and $\beta$
are fixed by the calculations made by Heisenberg and Euler.}.

\subsubsection{Magnetic universe}
\label{magu}
In the early universe, matter behaves to a good approximation
as a primordial plasma %
\cite{Tajima,
Campos}. Hence, it is natural to limit our considerations to the case in which only
the average of the squared magnetic field $\mathscr{H}^2$ survives %
\cite{Dunne,Tajima}. This is formally equivalent to put $\mathscr{E}^2=0$ in
(\ref{meanE2}), and physically means to neglect bulk viscosity terms
in the electric conductivity of the primordial plasma.

\begin{figure}[htb]
\begin{center}
\includegraphics[width=0.7\textwidth
]{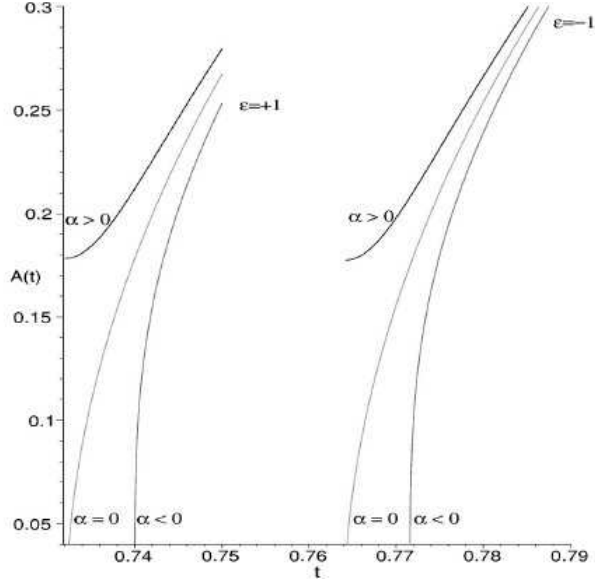}
\caption{Plot of the scale factor as a function of $t$ for
different values of $\epsilon$ and $\alpha$. Taken from \cite{klippert}.}
\label{sf1}
\end{center}
\end{figure}

\begin{figure}[htb]
\begin{center}
\includegraphics[width=0.7\textwidth
]{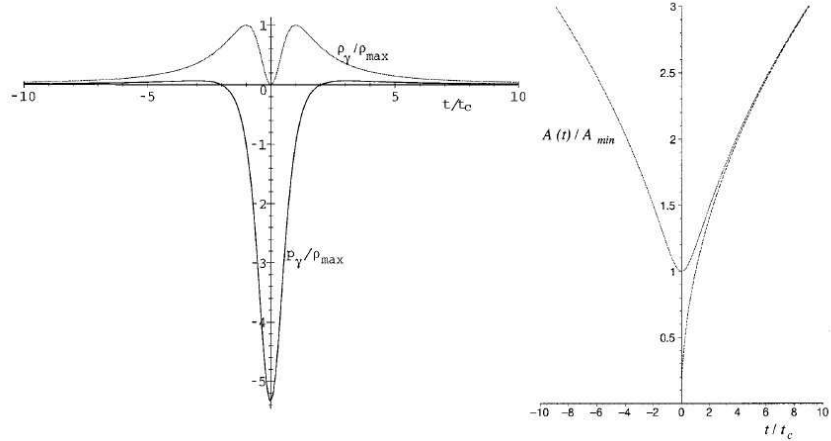}
\caption{Left panel: time dependence of $\rho$ and $p$ for $\epsilon =0$
and $\alpha >0$. Right panel: scale factor for $\epsilon =0$
and $\alpha >0$ (full line), and for Mawxell's case (dashed line).
Taken from \cite{klippert}.}
\label{fk2}
\end{center}
\end{figure}

The Lagrangian (\ref{Order2}) requires some spatial averages over
large scales, such as the one given by equations
(\ref{meanEH})--(\ref{meanH2}). If one intends to make similar
calculations on smaller scales then either more involved
Lagrangians should be used,
or some additional magnetohydrodynamical effect \cite{Thompson} 
should be devised in order to achieve correlation \cite{Jedamzik} at
the desired scale. Since the average procedure is independent of the
equations of the electromagnetic field we can use the above formulae
(\ref{meanEH})--(\ref{meanH2}) to arrive at a counterpart of
expression (\ref{Pfluid}) for the non-Maxwellian case. The average
energy-momentum tensor is identical to that of a perfect fluid
(\ref{Pfluid}) with modified expressions for the energy density
$\rho$ and pressure $p$, given by
\begin{eqnarray}
\rho &=& \frac{1}{2} \, \mathscr{H}^2 \,(1 - 8\,\alpha\,\mathscr{H}^2),
\label{rho}\\[1ex]
\protect\label{P} p &=& \frac{1}{6} \,\mathscr{H}^2 \,(1 -
40\,\alpha\,\mathscr{H}^2).
\end{eqnarray}
Inserting expressions (\ref{rho})--(\ref{P}) in the conservation equation
(\ref{dotRho}) yields
\begin{equation}
\mathscr{H}=\frac{\mathscr{H}_0}{a^2}, \protect\label{H->A}
\end{equation}
where $\mathscr{H}_0$ is a constant.
With this result, equation
(\ref{constraint}) leads to
\begin{equation}
\label{eqA2} \dot{a}^2=\frac{\mathscr{H}_0^2}{6\,a^2}
\left(1-\frac{8\alpha \mathscr{H}_0^2}{a^4}\right)-\epsilon.
\end{equation}
Since the right-hand side of equation (\ref{eqA2}) must not be
negative it follows that,  for $\alpha>0$ the scale factor $a(t)$
cannot be arbitrarily small regardless of the value of $\epsilon.$
The solution of Eqn.(\ref{eqA2}) is implicitly given as
\begin{equation}
\label{solution} t=\pm\int_{a_0}^{a(t)}\frac{\textstyle dz}
{\sqrt{\textstyle\frac{\mathscr{H}_0^2}{\textstyle6z^2}
-\frac{\textstyle8\alpha
\mathscr{H}_0^4}{\textstyle6z^6}-\epsilon}},
\end{equation}
where $a(0)=a_0$. The linear case described by Eqn.(\ref{A(t)Maxwell}) can be
regained from Eq.(\ref{solution}) by setting $\alpha=0$.
For the Euclidean section, expression (\ref{solution}) can be
solved as \footnote{Nonsingular solutions in Bianchi universes
with nonlinear electrodynamics as a source were studied in
\cite{breton}.}
\begin{equation}
\label{A(t)} a^2 = \mathscr{H}_{0} \,\sqrt{\frac{2}{3} \,(t^2
+12\,\alpha)}.
\end{equation}
From Eqn.(\ref{H->A}), the average strength of the magnetic
field $\mathscr{H}$ evolves in time as
\begin{equation}
\label{Ht} \mathscr{H}^2 = \frac{3}{2}\,\frac{1}{{t^2} +
12\,\alpha}.
\end{equation}
Expression (\ref{A(t)}) is singular for $\alpha<0$, as there exist
a time $t=\sqrt{-12\alpha}$ for which $a(t)$ is arbitrarily small.
Otherwise, for $\alpha>0$ at $t = 0$ the radius of the universe
attains a minimum value (see Fig.\ref{sf1})
$a_{0}$, given by
\begin{equation}
\label{Amin} a^2_{0} = \mathscr{H}_{0} \, \sqrt{8\,\alpha},
\end{equation}
which depends on $\mathscr{H}_0$.
The energy density $\rho_\gamma$ given by Eqn.(\ref{rho}) reaches its
maximum value $\rho_{max}=1/64\alpha$
at the instant $t=t_c$, where
\begin{equation}
\label{tc} t_{c} =  \,\sqrt{12\,\alpha}.
\end{equation}
For smaller values of $t$ the energy density decreases, vanishing at
$t = 0$, while the pressure becomes negative (see Fig.\ref{fk2}, left panel).
Notice that we have a minimum of $a(t)$ along with a minimum of the energy density,
entailing a violation of the NEC condition, in accordance with the first row of
Table\ref{tabla1}.

Only for times
$t\,\raisebox{-0.5ex}{$\stackrel{<}{\scriptsize\sim}$}
\,\sqrt{4\alpha}$ the non-linear effects are relevant for the
normalized scale-factor, as shown in
Figure \ref{fk2}, left panel. Indeed, the solution (\ref{A(t)}) yields the
standard expression (\ref{A(t)Maxwell}) of the Maxwell case at the
limit of large times.
Notice that the energy-momentum tensor (\ref{Tmunu}) is not trace-free for
$\alpha\neq0$. Thus, the equation of state
\begin{math}
p_\gamma = p_\gamma(\rho_\gamma)
\end{math}
is no longer that of Maxwell's; it has instead a
term proportional to the
constant $\alpha,$ that is
\begin{equation}
\label{newstate} p = \frac{1}{3} \,\rho -
\frac{16}{3}\,\alpha \,\mathscr{H}^4.
\end{equation}
%
This scenario has been generalized in several ways in
\cite{ademir}. First, the general expression for the scale factor was shown to be
\beq
a(t) = a_0(4\alpha_0^2t^2+4\alpha_0\beta_0 t+1)^{1/4},
\label{gen}
\eeq where
$$
\alpha_0=\sqrt{\frac 2 3}\mathscr{H}_0,\;\;\;\;\;\beta_0=\pm\sqrt{1-8\alpha\mathscr{H}_0}.
$$
Eqn.(\ref{A(t)}) follows as a particular
case from Eqn.(\ref{gen}), which describes a bounce with
$$ a_{\rm
min} = a_0(8 \omega \mathscr{H}_0^2)^{1/4},\;\;\;\;\;\;t_{\rm min} =
-\beta_0/(2\alpha_0),\;\;\;\;\;\;\mathscr{H}_{\rm
min}=\frac{1}{2\sqrt{2\alpha}},\;\;\;\;\;\;\rho_{\rm
min}=0.
$$

Solutions of this model with the addition
of a cosmological constant $\Lambda$ were also discussed in \cite{ademir}.
It was shown that nonsingular solutions are  possible both
for a constant $\Lambda$, and for certain choices
of $\Lambda
= \Lambda (t)$.

\subsubsection{Born-Infeld electrodynamics}

A widely studied EM theory is that proposed by Born and
Infeld, with Lagrangian
\begin{equation}
L_{BI} = \beta^{2} \left( 1 - \sqrt{X} \right) \label{BI1}
\end{equation}
where
\begin{equation}
X \equiv 1 + \frac{1}{2\beta^{2}} \, F  - \frac{1}{16\beta^{4}}\,
G^{2} \label{BI2}
\end{equation}
Note that, following Born-Infeld's original work, a constant
term has been added in the Lagrangian in order to eliminate a cosmological
constant and to set the value of the Coulomb-like field to be zero
at the infinity. Using equation (\ref{eden}) for the energy density we
obtain
\begin{equation}
\rho =  \frac{\beta^{2}}{\sqrt{X}} \, \left(  1 - \sqrt{X} +
\frac{{\mathscr{H}}^{2}}{\beta^{2}} \right) \label{BI3}
\end{equation}
and for the pressure
\begin{equation}
p =  \frac{\beta^{2}}{\sqrt{X}} \, \left(  \sqrt{X} - \beta^{2} +
\frac{2}{3} \, \frac{{\mathscr{E}}^{2}}{\beta^{2}} - \frac{1}{3} \,
\frac{{\mathscr{H}}^{2}}{\beta^{2}} \right) \label{BI4}
\end{equation}
A straightforward calculation of $\rho+3p$
shows that this theory cannot yield a nonsingular universe.

\subsubsection{Bouncing in the Magnetic Universe}
\label{bmu}
The ``magnetic universe'' displays a
very interesting property due to the nonlinear dynamics:
its energy density can be
can be interpreted as composed of $k$
non-interacting fluids, in
the case in which the dynamics is provided by the
polynomial
\begin{equation}
{\cal L} = \sum_k \, c_{k} \, F^{k} \label{19dez1},
\end{equation}
where $k\in \mathbb{Z}$.
The conservation of the energy-momentum tensor projected in the
direction of the co-moving velocity $v^{\mu} = \delta^{\mu}_{\; 0}$
yields
\begin{equation}
\dot{\rho} + (\rho + p) \theta = 0 \label{M1}.
\end{equation}
From the expression for the energy density and pressure given
in Eqns.(\ref{eden}) and (\ref{pre}) with ${\mathscr{E}}=0$
we get that
$ \rho = \sum_k \rho_{k}$, and $p = \sum_k p_{k}$
where
\begin{eqnarray}
\rho_{k} &=&- c_{k} 2^{k} {\mathscr{H}}^{2k} \nonumber \label{den1}\\
p_{k} &=& c_{k}  \, 2^{k} {\mathscr{H}}^{2k} \left(1 - \frac{4k}{3}  \label{pre1}\right),
\end{eqnarray}
in such a way that we can associate to each power of $k$ an
independent fluid characterized by $\rho_k$ and $p_k$,
with an EOS
$$
p_k=\left(\frac{4k}{3}-1\right)\rho_k.
$$
Inserting the total energy density and pressure
(from the sum of $\rho_k$ and $p_k$ in Eqns.(\ref{den1}) and (\ref{pre1}))
in the conservation equation (\ref{M1}) we obtain
\beq
{\cal L}_F\left[({\mathscr{H}}^2)^. +4{\mathscr{H}}^2\frac{\dot a}{a}\right]=0.
\label{factor}
\eeq
The important result that this equation shows is that each $k$-fluid is separately conserved,
since the dependence of the conservation equation on the specific form
of the Lagrangian factors out, in such a way that
${\mathscr{H}}$ evolves with the scale factor as
\beq
{\mathscr{H}}=\frac{{\mathscr{H}}_0}{a^2}
\label{hevol}
\eeq
for any ${\cal L}$ of the form given in Eqn.(\ref{19dez1}).

\subsubsection{Two-fluid description}
\label{tfd}
It follows from equations (\ref{rho}), (\ref{P}) and
(\ref{H->A}) that in the case of the nonlinear Lagrangian
given by Eqn.(\ref{Order2}
it is not possible to write an equation of state
relating the pressure to the energy density.
This is a drawback if we want to use a fluid
description of the averaged electromagnetic field. In order to
circumvent such difficulty a two-fluid description can be adopted,
because of the remarkable fact that there exists a separate law of
conservation for each component of the fluid, as we saw above. The
fact that the dynamical equation for ${\mathscr{H}}$ factors (see Eqn.(\ref{factor})
means that
the fluids are conserved independently: the energy-momentum tensor
can be separated into two pieces, each representing a perfect
fluid which is conserved independently. In other words, there is
no interaction between fluids $1$ and $2.$ We shall see in Section
\ref{spqm} that the analysis of the stability of the non-singular
universe described in this section is more transparent when using
the two-fluid description. This case can be generalized to a
multi-component fluid, but we shall restrict here to the 2-fluid application
for a pure magnetic field.

In order to get a better understanding
of the properties of the cosmic
geometry controlled by the magnetic field let us analyze
the case in which the spatial section is closed $(\epsilon = 1).$
The crucial equations for such analysis are the
conservation law, the Raychaudhuri equation for the expansion and
the Friedman equation, that is:
\begin{equation}
\dot{\rho} + (\rho + p) \, \theta  = 0
 \label{11abril4},
\end{equation}
\begin{equation}
\dot{\theta} + \frac{1}{3} \, \theta^{2} = - \frac{1}{2} \, (\rho +
3p).
 \label{11abril5}
\end{equation}
\begin{equation}
\rho = \frac{1}{3} \, \theta^{2} + \frac{3}{a^{2}} \label{11abril3},
\end{equation}
In the magnetic universe we have
\begin{equation}
\rho = \frac{{\mathscr{H}_0}^{2}}{2 a^{4}} \, \left( 1 - 8 \alpha \,
\frac{{\mathscr{H}_0}^{2}}{a^{4}}\right) \label{11abril6}.
\end{equation}
A necessary condition for the existence of a bounce is given
by the vanishing of the
expansion factor for a given value of $t$. This leads
to an algebraic equation of third
order in $x \equiv a_{b}^{2} $:
\begin{equation}
x^{3} - \frac{{\mathscr{H}}_0^{2}}{6} \, x^{2} + \frac{4}{3} \, \alpha {{\mathscr{H}}_0}^{4} = 0.
\label{11abril7}
\end{equation}
Using the fact that $\alpha$ is a very small
parameter, it can be shown that
this equation has three real
solutions. Two of them are positives and the third is negative.
Thus we retain only the positive solutions which will be called
$X_{1}$ and $X_{2}.$ The important quantity for our analysis is
contained in the expression
\begin{equation}
\rho_b + 3 p_b
= \frac{{\mathscr{H}}_0^{2}}{x^{4}} \, ( x^{2} - 24 \alpha {{\mathscr{H}}_0}^{2} ).
 \label{11abril8}
\end{equation}
Thus, at one of the points, say $X_{1}$ there is a local maximum for
the scale factor; and at the other,  $X_{2}$ there is a minimum for
$ x^{2} < 24 \alpha {{\mathscr{H}}_0}^{2}.$ Note that at the bounce (where
$\theta = 0$), there is an
extremum of the total energy: $\dot{\rho}_{b} = 0$.
The analysis of the second derivative in the bounce depends on the
location of $X_{2}$ through the equations:
\begin{equation}
\ddot{\rho_b} = \frac{1}{3}\, \frac{{{\mathscr{H}}_0}^{4}}{x^{8}} \, \left( x^{2} - 16
\alpha {{\mathscr{H}}}_0^{2} \right) \, \left( x^{2} - 24 \alpha {\mathscr{H}}_0^{2} \right)
\label{11abril9}.
\end{equation}
At $x=X_{1}$ the density is a minimum. For $x=X_{2}$ the extremum
depends on the location of the bounce respect to the point in
which the quantity $ \rho + p$ changes sign. For the case in which
$16 \alpha {{\mathscr{H}}_0}^{2} < X_{2} < 24 \alpha {{\mathscr{H}}_0}^{2}$, it follows that
the density has a maximum
at $X_{1}$. On the other hand if $X_{2} < 16
\alpha {{\mathscr{H}}_0}^{2}$ it is a minimum. To understand completely the behavior
of the energy density the existence of other
critical points for $\rho$ must be addressed. This is controlled by
equation (\ref{11abril4}).
Thus, the extra extremum (which are not bounce or  turning
points) occur at $x$ such that
\begin{equation}
\rho +  p  = \frac{2}{3} \, \frac{{{\mathscr{H}}}_0^{2}}{x^{4}} \, (x^{2} - 16
\alpha {\mathscr{H}}_0^{2} )=0,
 \label{11abril10}
\end{equation}
that is, at points in which the scale factor takes the value $\sqrt{16
\alpha {{\mathscr{H}}_0}^{2}}$. Direct inspection shows that these are points of
maximum density.

Another consequence of nonlinear electromagnetism in cosmology is the
occurrence of cyclic universe, as will be discussed in Sect.\ref{cunlem}.

\subsection{Appendix}
\subsubsection{Repulsive gravity}

A peculiar result which provides a
framework to generate cosmological scenarios without singularity
comes from the nonminimal interaction of EM with gravity, rendering
gravity repulsive. The theory is defined by
\begin{equation}
L=\sqrt{-g}\ \left\{
R-\frac{1}{4}\ F_{\mu \nu}
F^{\mu \nu}+\beta R\ A_{\mu}A^{\mu}\right\} \label{cot1},
\end{equation}
where $\beta$ is a dimensionless constant.
This Lagrangian is not gauge-invariant and can be interpreted in
terms of a photon having a mass which depends on the
curvature of the geometry.

Variation of $g_{\mu\nu}$ and $A_{\mu}$
yield the equations of motion:
\begin{eqnarray}
\left(\frac{1}{\kappa}+\beta A^2\right)G_{\mu \nu} &=& \beta g_{\mu \nu}\Box
A^2-
\beta A^2_{~,\mu ;\nu}-\beta RA_{\mu}A_{\nu}-E_{\mu \nu}, \label{Pescara153} \\
F^{\mu \nu}_{~;\nu} &=& -2 \beta RA^{\mu} \label{Pescara154},
\end{eqnarray}
where $E_{\mu \nu}$ is Maxwell's energy-momentum tensor given by
equation (\ref{Maxwell}).
As will be shown next, this set of equations allows a renormalization of the
gravitational constant. Consider for instance the case in
which $A_{\mu}A^{\mu}=Z= $ constant$\neq 0$. Then
\begin{equation}
\left(\frac{1}{\kappa}+\beta Z \right) \,G_{\mu\nu}=-\beta R \,
A_{\mu}A_{\nu}- E_{\mu\nu} .\label{Pescara151}
\end{equation}
Taking the trace of this equation we obtain $ R=0, $ and inserting
this result back into Eqn.(\ref{Pescara151}) we get
\[
R_{\mu \nu}=-\tilde\kappa E_{\mu \nu},
\]
where the renormalized constant $ {\tilde\kappa}$ is given by
\[
\frac{1}{\tilde\kappa}=\frac{1}{\kappa}+\beta Z.
 \label{Pescara155}
\]
Thus, Eqns.(\ref{Pescara153}) and (\ref{Pescara154}) can be
written as
\begin{equation}
R_{\mu \nu}= -\tilde\kappa E_{\mu \nu}, \;\;\;\;\;\;\;\;\;\;\;\;\;
F^{\mu\nu}_{\;\; ;\nu} = 0, \label{A1}
\end{equation}
which are nothing but Maxwell's electrodynamics minimally coupled to
gravity with a re-normalized gravitational coupling plus the
condition $ A_{\mu}A^{\mu}=constant = Z. $

The addition of
other forms of neutral matter, such that the corresponding
energy-momentum tensor is traceless, takes the Lagrangian to
\begin{equation}
L=\sqrt{-g}\ \left\{\frac{1}{\kappa}\ R-\frac{1}{4}\ F_{\mu
\nu}F^{\mu \nu}+ \beta \, RA_{\mu}A^{\mu}+L^{(m)}\right\},
\end{equation}
where $L^{(m)}$ represents the Lagrangian for all other kinds of
matter such that $ T^{(m)}_{\mu \nu}g^{\mu \nu}\equiv T^{(m)}= 0. $ The
equations of motion in this case are given by
\beq
\left(\frac{1}{\kappa}+ \beta \, A^2 \right) G_{\mu \nu}=\beta
\Box A^2g_{\mu \nu}- \beta A^2_{,\mu ;\nu}-\beta
RA_{\mu}A_{\nu}-E_{\mu \nu}-T_{\mu \nu}^{m}
\label{e1}
\eeq
\beq
F^{\mu
\nu}_{\;\; ;\nu}=-2\beta RA^{\mu} \label{A2}.
\eeq
Taking again the case $A_{\mu}A^{\mu}=$ constant, yields $R=0$. Then
Eqns.(\ref{e1}-\ref{A2}) take the reduced form
\begin{eqnarray}
R_{\mu \nu} &=&-\tilde\kappa E_{\mu
\nu}-\tilde\kappa T^{(m)}_{\mu \nu}, \nonumber \\
 F^{\mu\nu}_{\;\; ;\nu}&=& 0\nonumber \label{A3},
\end{eqnarray}
where
$\tilde{\kappa}$ was given above. Thus, the renormalization of
the gravitational constant by the non-minimal coupling represented
by the presence of the term $RA_{\mu}A^{\mu}$ in the Lagrangian in
the state where $A^{\mu}A_{\mu}$ is constant is still valid in the presence
of matter with null trace.



\subsubsection{Global Dual invariance}

While observation must be the ultimate judge of the choice among the
possible couplings, if it scarce or not available, we can resort to
criteria coming from theoretical considerations. One of them is
related to the invariance of the Lagrangian under a given
transformation, such as the dual rotation. A dual map is a
transformation on the set of the bi-tensors $F_{\mu \nu}$ such that
\begin{equation}
F_{\mu \nu}\rightarrow F'_{\mu \nu}=\cos \theta F_{\mu \nu}+\sin
\theta F^{*}_{\mu\nu} \label{29dez1725}.
\end{equation}
Classical Maxwell's electrodynamics is invariant under such
transformation only if the angle $\theta$ is constant. In a Minkowskian
background it is not possible to implement such invariance for a
local map $\theta = \theta(x).$ However, this can be
achieved in the case of a non-minimal coupling of the
electromagnetic field with the metric of a non-flat geometry.
In fact, using the identities
\begin{eqnarray*}
F_{\mu\alpha} \, F^{\alpha\nu} &-& F^{*} _{\mu\alpha} \, F^{* \,
\alpha\nu} = - \, \frac{F}{2} \, \delta^{\nu}_{\mu}
\nonumber \\
F_{\mu\alpha} \, F^{* \, \alpha\nu} &=& - \, \frac{G}{4} \,
\delta^{\nu}_{\mu}
\end{eqnarray*}
it can be shown that the combined Lagrangian:
\begin{equation}
L_{\rm DI}=L_5 - \frac{1}{4}\ L_3=\left(R_{\mu \nu}-\frac{1}{4}\ R\
g_{\mu \nu}\right)F^{\mu}_{~\alpha}F^{\alpha \nu}.
\end{equation}
is invariant under local
dual rotations: $ \tilde{L}_{\rm DI} = L_{\rm DI}.$
This is a remarkable
property which has no counterpart in the flat space limit.



\section{Viscosity}
\label{visco}

A full knowledge of the global properties of the universe
cannot be achieved without giving a description of the
thermodynamics of the cosmic fluid. In the last decades, this task was addressed
in three distinct periods. In the first period the
universe was treated as a system in equilibrium in which all global
processes were described by classical reversible thermodynamics, in
such a way that total entropy was conserved. The salient
feature of this phase was the development
of the standard
cosmological model, which comprises the homogeneous and isotropic
FLRW geometry, and the characterization of the matter content of the
universe as a one-component perfect fluid in equilibrium. In order
to solve the EE, the energy density $\rho$ and the
pressure $p$ were considered functions of the cosmological
time only, and they were related by a linear EOS $ p = \lambda\rho.$ The
FLRW models generated in this way share the common property of having an
initial singularity (with $\lambda > -1/3$).

Later, it was realized \cite{dresden} that the validity of thermal
equilibrium near the initial singularity is perhaps too strong an
assumption. A second phase then started, in which the
description of the cosmic fluid was improved by allowing viscous processes.
Some of the motivations for this alteration are the following:
\begin{itemize}
\item The examination of the possible role of viscosity in the
dissipation of eventual primordial anisotropies (chaotic
cosmologies),
\item The effect on the existence and/or the form of
the singularity,
\item The application in cosmology of results
obtained from non-equilibrium thermodynamics.
\end{itemize}
In 1973 a FLRW cosmological model without singularity was
presented \cite{murphy} (see also \cite{kli}), using a viscous fluid as a source. The
energy-momentum tensor was given by
$$ T_{\mu\nu} = (\rho + p) \, v_{\mu} \, v_{\nu} - p \, g_{\mu\nu},
$$
in which $ p = p_{\rm th} - \zeta \, \theta;$ where $p_{\rm th}$ is the
thermodynamical pressure,  $\zeta$ is a viscous coefficient and
$\theta$  is the Hubble parameter, which
is exactly the case of the
energy-momentum tensor representing particle creation \cite{belinsky}.
The SEC in this case is
given by the inequalities
$$ \rho +  p_{th}   > 0 $$
and
$$ \rho + 3 p_{th}   > 0, $$
which are weaker than the correspondent ones in the
case of perfect fluid, hence allowing for the absence of singularity.
The solution found in \cite{murphy} is nonsingular, and past-eternal.

More general forms for the dependence of viscous quantities have
been investigated for arbitrary Stokesian regimes in which the
fluid parameters become more general (for instance nonlinear)
functions of the expansion. With these modifications, there are
non-singular cosmological solutions, but they may suffer from a possibly
worse disease than the initial singularity: they are unstable and
display non-causal propagation. In fact,
the instability of the model in
\cite{murphy} was proven by the analysis made
in \cite{belinsky}. It was also proved in \cite{belinsky} that the avoidance
of the singularity is not generic. In other words, the singularity
is not avoided for any type of viscosity (that is, for any dependence of
the coefficients of viscosity on the expansion factor).

In this
second phase, local equilibrium \cite{prigo2}
is still imposed,  in such a way that
the thermodynamical variables are described as if the
dissipative fluxes - e.g. heat flux - do not influence local
variables like for instance the entropy, although as a whole the
system is not in equilibrium.  As another example, a fluid
in the regime
$$
\tilde p = p+\alpha \theta + \beta \theta ^2
$$
was analyzed in \cite{novaraujo}, both for
$\alpha = \beta = $ constant, and $\alpha =0$, $\beta = M\rho^m$, with $M$ and $m$
constants. In the second case, nonsingular solutions were found using
tools from dynamical systems analysis.


Let us remark that in  general, the imposition of local equilibrium leads
to
causal difficulties, allowing dissipative signals to travel with
infinite velocity of propagation.
These causal problems were the focus
of the third phase, where
extended irreversible thermodynamics was used
\cite{causal}. In this theory,
the basic
quantities become dependent not only on local
variables of classical thermodynamics but also on the dissipative
fluxes. This has very important consequences, the most important one
being the preservation of causal connections for the whole system.
In \cite{hensalim}, a FLRW universe was studied in this context,
the net consequence of the assumption of
extended irreversible thermodynamics
being to provide an
additional equation of motion for the
non-equilibrium pressure $ \pi$, with
$p = p_{\rm th} +  \pi$, given by
\beq
\tau_0\dot\pi +\pi=-\xi \theta.
\label{vis}
\eeq
(where $\tau_{0} $ is the relaxation time) which preserves the causal
structure.  Thus, contrary to the previous case in which the viscous
term is assumed to be a polynomial on
$\theta$, here it must obey Eqn.(\ref{vis}). The other quantities relevant
to thermodynamics (that is, the entropy flux $s^\alpha$ and the
particle flux per unit of proper volume $n$) are determined by
$$
n\dot s=\frac{\pi ^2}{\xi T},\;\;\;\;\;\;\theta = -\frac{\dot n}{n}.
$$
Assuming an EOS given by $p_{\rm th}=\lambda \rho$, the cases
$\xi=$ constant, and $\xi = \beta \rho$, (with $\beta =$ constant)
were analyzed in \cite{hensalim}, always with $\tau_0=$ constant,
and nonsingular solutions were discovered in both cases, for $\lambda =0$
and $\lambda =1/3$. The
relevant equations of this system can be put in the form
of an autonomous planar system:
\begin{eqnarray}
\frac{d \theta}{dt}  &=& - 3/2 ( 1 + \lambda) \theta^{2} -
\frac{\pi}{2}  + \frac{( 1 + \lambda)}{2} \, \Lambda \nonumber , \\
\frac{d \pi}{dt}  &=& - \frac{1}{\tau_{0}} ( 1 + 3 \zeta \, \theta),
\end{eqnarray}
where $\Lambda$ is the cosmological constant.
The set of integral curves
of this system was studied in
\cite{joao}, where it was shown that the solution found in
\cite{hensalim} is stable.
\\[0.3cm]
\textbf{Bifurcations in the early cosmos}

Quadratic dissipative processes were analyzed from a new perspective
in \cite{ligia}, where it was shown that dissipative
processes may lead to the appearance of bifurcations.
This is a consequence of the application of a theorem due to
Bendixson \cite{bendix}
to the system of EE that describes a universe with curvature
controlled by a dissipative fluid. Indeed, let us
consider a planar autonomous system that contains a parameter, say
$\sigma$, of the form
\begin{eqnarray}
 \dot{x} &=& F( x, y; \sigma)  \nonumber \\
 \dot{y} &=&  G( x, y; \sigma),
 \label{2junho900}
 \end{eqnarray}
where the functions $F$ and $G$ are non-linear and the parameter $\sigma$
has a domain $\cal{D}.$ Applying methods of qualitative analysis to
this system and restricting to
the two-dimensional plane $ \Gamma$ of all integrals
of this system, one arrives to the notion of "elliptical" and
"hyperbolic" sectors, that characterize, as the names indicates, the
behavior of the integral curves in the neighborhood of a multiple
equilibrium point (that is, an isolated points that
is a zero of both $F$ and $G$). Let us
call ${{\cal E}}$ and ${\cal H}$ the number of elliptical and hyperbolic sectors of
a given equilibrium point $M \equiv (x_{0}, y_{0})$ of $\Gamma,$
respectively. Then the Poincar\'e index is defined by the formula
$$
I_{P} = \frac{{\cal E} - {\cal H} }{2} + 1.$$ This is a measure of
the topological properties of the integral curves in the phase plane
$\Gamma.$
If above a certain value $\sigma_{c}$ of $\cal{D}$ the topological
properties of the system (\ref{2junho900}) change, then
there is
an abrupt change of behavior of the physical system
in the vicinity of the unstable equilibrium point. The crucial consequence of
the above-given theorem is the appearance of indeterministic features.
In \cite{ligia}
this theorem was applied to
spatially homogeneous and isotropic cosmological models, whose dynamics is
described by a planar
autonomous system, given by
 \begin{eqnarray}
 \dot{\rho} &=& - \gamma \, \rho \, \theta + \alpha \, \theta^{2}
 + \beta \, \theta^{3} \nonumber ,\\
 \dot{\theta} &=& - \frac{ 3 \gamma - 2}{2}  \, \rho + \frac{3 \alpha}{2}
 \, \theta + \left(\frac{3 \beta}{2} - \frac{1}{3} \right) \, \theta^{2},
 \label{ver}
 \end{eqnarray}
where $\sigma$ (referred to in the theorem)
can be either $\alpha$, $\beta$ or $\gamma$, and the energy-momentum tensor is
$$ T_{\mu\nu} = \left( \rho + \tilde{p} \right)\, v_{\mu} \, v_{\nu} - \tilde{p} \,
g_{\mu\nu}, $$
where
$$ \tilde{p} = p_{th} + \alpha \, \theta + \beta \, \theta^{2}, $$
with $ p_{\rm th} = ( \gamma - 1 ) \, \rho.$

The viscous terms (parameterized by $\alpha$ and $\beta$)
can be a phenomenological description of particle creation in a nonstationary
gravitational field as proposed in \cite{veresh} and
\cite{z1}. Applying the methods of qualitative analysis
to the system given in Eqn.(\ref{ver})
it was shown in \cite{ligia}
that for
$\gamma-3\beta<0$, the Poincar\'e index $I_P({\cal B})=-1$ (saddle point); for $\gamma-3\beta\geq0$,
$I_P({\cal B})=1$ (two-tangent node). This situation characterizes a bifurcation
in the singular point,
when $\rho =\theta =\infty$.
This bifurcation, caused by
dissipative processes involving quadratic viscous terms generates a high degree
of indeterminacy in the development of the solution of EE, which
enshrouds the past of this
model of the universe. In this case, nothing can be stated about the existence of the
initial cosmological
singularity.

\section{Bounces in the braneworld}
\label{bbw}
Theoretical developments coming from string theory have revived
the idea that our universe may have more than 4 dimensions (first
considerated by Kaluza in the context of unification of gravity and
electromagnetism). Among the multidimensional models, those with one or more branes
that live in a bulk space
have been thoroughly studied recently (see for instance \cite{maartens}).
In these models, the matter fields are typically confined
to a 3-brane in $1+3+d$ dimensions, while the gravitational field
can propagate also in the $d$ extra dimensions, which need not be
small, or even finite, as shown in one of the models introduced by
Randall and Sundrum \cite{rs}, where for $d = 1$, gravity can be
localized on a single 3-brane even when the fifth dimension is
infinite. The Friedmann equation on the brane is
modified by high-energy matter terms and also by a term which
incorporates the nonlocal effects of the bulk onto the brane
\cite{bine,maartens}:
\beq
H^2 = \frac\Lambda 3 + \frac{\kappa^2}{3}\rho -
\frac{\epsilon}{a^2}+\frac{\overline\kappa^4}{36} \rho^2+\frac 1 3
\left(\frac{\overline \kappa}{\kappa}\right)^4{\cal
U}_0\left(\frac{a_0}{a} \right)^4,
\label{fotb}
\eeq
where $\epsilon$ is the 3-curvature, $H=\dot a/a$,
$\rho$ is the energy density of the matter on the brane,
$\overline\kappa^2=8\pi/\overline M_{Pl}^3$, $\overline M_{Pl}^3$ is the
fundamental 5-dimensional Planck mass, $\kappa^2=8\pi/
M_{Pl}^2$, and
$$
\Lambda = \frac{4\pi}{\overline M_{Pl}^3}\left[ \overline\Lambda +
\left(\frac{4\pi} {3\overline M_{Pl}^3}\right)\lambda^2\right],
$$
where $\lambda$ is the tension of the brane, and
$\overline\Lambda$ is the 5-dimensional cosmological constant.
Finally, ${\cal U}_{\;0}$ is the constant corresponding to the
non-local energy conservation equation. This term comes from the
projection of the Weyl tensor of the bulk on the brane
\cite{maartens}. From Eqn.(\ref{fotb}) we see that a necessary condition
to have a bounce with $\rho>0$ in the $\epsilon=0,-1$ cases
is that either $\Lambda<0$ or ${\cal U}<0$, or both.

The case that includes matter in the bulk, without cosmological constant
for a flat
FLRW $d+1$-dimensional was studied in \cite{foffa}. A neceessary condition in order to have a bounce
is that $dH/dt>0$, with
\beq \frac{dH}{dt} =
\frac{\kappa^2}{d} (R+P)-\left(\frac{8\pi
G_N}{d-1}+\frac{\kappa^4}{4d}\;\rho\right) (\rho + p) -
\frac{d+1}{d(d-1)} E^0_0, \eeq where
(in a notation slightly different from that used in Eqn.(\ref{fotb}))
$\kappa$ is the bulk
gravitational coupling, $G_N$ the effective Newton constant on the
$(d+1)$-dimensional brane, $E$ is the projection of the bulk brane
Weyl tensor on the brane, and $T^\mu_{\;\nu} = (-R, \vec P)$ is the
projection of the bulk energy-momentum tensor on the
brane. It follows from this equation that a necessary condition to have a bounce without
resorting to exotic forms of matter (that is, matter that violates $\rho>0$
or $\rho _p>0$) is a negative $E^0_0$
\cite{foffa}. This is precisely the approach taken in \cite{tamva,
peloso}, where a brane evolving in a charged AdS black hole
background was studied. Bouncing solutions were found for both
critical ($\Lambda =0$) and non-critical ($\Lambda \neq 0$) branes, the bounce generically depending
on the parameters of the black hole, and on the matter content of the brane
\footnote{The bounce in the model presented in \cite{peloso} was analyzed from the point of view
of the causal entropy bound in \cite{medved}, and its stability was
put in doubt in \cite{myers}.}.

The abovementioned necessary condition was explicitly checked in
the case of the dilaton-gravity braneworld \cite{foffa}, and
bouncing solutions were obtained for a a flat FLRW brane in a static spherically symmetric
bulk
\footnote{Bouncing solutions for a domain wall in the presence
of a Liouville potential were found in \cite{chamb}.}.
This solution describes (in the string frame)
a pre-big bang model where the transition between the branches
is realized at low curvature and weak coupling, thus providing an example of succesful
graceful
exit without resorting to quantum or ``stringy'' corrections.

Notice that the extra dimension(s) could be spacelike or
timelike. The latter case was analyzed in \cite{sahni}.
The usual incantations \cite{maartens}
for the
case of an extra timelike dimension and an homogeneous and
isotropic brane lead to \cite{sahni} \beq
H^2+\frac{\epsilon}{a^2}=\frac{\Lambda}{3}+\frac{8\pi
G\rho}{3}-\frac{\rho^2}{\bar M_{\rm Pl}^6}+\frac{C}{a^4}, \eeq where $G$ and
$\Lambda$ are the effective gravitational and cosmological
constant, respectively, and $M$ is the 5-dimensional Planck mass.
Notice that the minus sign in front of $\rho^2$ may lead to a bounce
instead of a singularity, since this term grows faster than the
others, leading to $H=0$, this feature being independent
of the equation of state and also of the spatial curvature of the
universe. The simplest of these bouncing universes, described by
\beq H^2=\frac{8\pi G}{3}\rho -
\frac{\rho^2}{\bar M_{\rm Pl}^6} \label{free}
\eeq
will be discussed in Sec.\ref{cms}, since it may lead to
a cyclic universe.

The case with an extra timelike dimension in this scenario
was
also extended to Bianchi I universes \cite{sahni}, which exhibit
an anisotropic bounce as long as the shear scalar
$\sigma_{\alpha\beta}\sigma^{\alpha\beta}$ does not grow faster
than $a^{-8}$ as $a$ goes to zero at the end of the contraction
phase. All these results were obtained by neglecting the induced
curvature on the brane, which can trigger the formation of a
singularity at the beginning or at the end of the evolution
\cite{sahni}.

\begin{figure}[htb]
\begin{center}
\includegraphics[width=0.7\textwidth
]{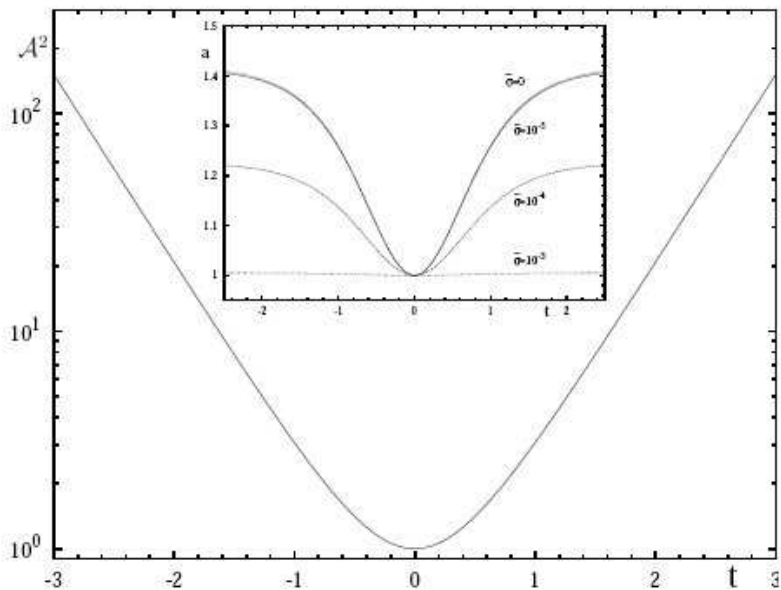}
\caption{Effective scale factor $\mathcal{A}(t)$ plotted for
$\bar\sigma = 4\pi\sigma/(L^{(3-d)}_p(d-1))=10^{-4}$. The insert displays the dependence of
$\mathcal A$
with $\bar\sigma$. Taken from \cite{siete}.}
\end{center}
\label{seven1}
\end{figure}

Another model along these lines was introduced in \cite{siete},
where a ``test brane'' (\emph{i.e.} one that does not modify the ambient geometry)
moves in a higher-dimensional gravitational background.
Using the thin-shell formalism,
in which the field equations
are re-written as junction conditions relating the discontinuity in the brane extrinsic
curvature to its vacuum energy,
the motion of domain walls in de Sitter and
anti-de Sitter (AdS) time-dependent bulks was discussed. This motion induces a dynamical law for
the brane scale factor, and it was shown in \cite{siete}
that in the case of a clean brane the scale factor may
describe a non-singular universes. In order to build the class of geometries of interest,
two copies of (d+1)-dimensional dS (AdS) spaces $M1$ and $M2$ undergoing expansion were considered.
From each of them,
one identical $d$-dimensional region  $\Omega_i$ ($i=1,2$) was removed.
yielding two
geodesically incomplete manifolds with boundaries given by the hypersurfaces $\partial\Omega_1$
and
$\partial\Omega_2$. Finally, the boundaries were identified up to
an homeomorphism $h: \partial\Omega_1
\rightarrow\partial\Omega_2$. Hence, the resulting manifold that is defined by the connected sum $M1\#M2$
is
geodesically complete.
The starting point is the action
$$
S=\frac{{\ell}_{\rm Pl}^{(3-d)}}{16\pi}\int_M\;d^{d+1}x\;\sqrt{g}\;(R-2\Lambda)+\frac{{\ell}_{\rm Pl}^{(3-d)}}{8\pi}
\int_{\partial\Omega}d^dx\;\sqrt{\gamma}\;\mathcal{K}+\sigma\int_{\partial\Omega}d^dx\;\sqrt\gamma,
$$
where
the
first term is the usual Einstein-Hilbert action with a cosmological constant $\Lambda$,
the second term is the Gibbons-Hawking boundary term,
$\mathcal{K}_{MN}$ is the extrinsic curvature, and $\sigma$ is the intrinsic tension
of the $d$-dimensional brane.
The spatial coordinates on $\partial\Omega$
can be taken to be the angular variables
$\phi_i$, which for a spherically symmetric configuration are always well defined up to
an overall rotation. Generically, the line element of each patch can be written as
$$
ds^2=-dt^2+A^2(t)[r^2d\Omega^2_{(d-1)}+(1-kr^2)^{-1}dr^2],
$$
where $\epsilon$ takes the values 1 (-1) for dS (AdS), $\Omega^2_{(d-1)}$
is the corresponding metric on
the unit $d-1$-dimensional sphere, and $t$ is the proper time of a clock measured in the
higher-dimensional spacetime.
In order to analyze the dynamics of the system, the brane is allowed to move radially.
Let the position of the brane be described by $x_\mu(\tau , \phi_i) \equiv (t(\tau ), a(\tau ), \phi_i)$, with $\tau$ the
proper time (as measured by co-moving observers on the brane) that parameterizes
the motion, and the velocity of a piece of stress-energy at the brane satisfying $ u^Mu_M =
-1$.
With these assumptions the brane will have an effective scale factor
$\mathcal{A}^2(t) = a^2(t)A^2(t)$.
The constraint
$$
\frac{d\tau}{dt}=\pm\sqrt{1-\frac{(A\dot a)^2}{1-\epsilon a^2}}
$$
along with the result of the
integration of EE across the boundary (done with the junction conditions)
\cite{siete} yields two differential equations for $A$ and $a$.
For the case of a background composed by two patches of dS
undergoing expansion, $A(t)={\ell}\cosh(t/\ell)$, and $\epsilon =1$, where
$\ell^2=d(d-1)/|\Lambda|$ is the dS radius.
In this case the EOM for the brane is
$$
\frac{4\pi}{L_p^{(3-d)}(d-1)}\;\sigma = \frac{\pm \dot a \sinh (t/\ell)+[a\ell \cosh(t/\ell)]^{-1}(1-a^2)}{(1-a^2-[\ell \dot a\cosh(t/\ell)]^2)^{1/2}}.
$$
Nonsingular
analytical
solutions of this equation for $\sigma =0$ can be obtained, while for
$\sigma\neq 0$, numerical methods must be used. This latter case
also yields bouncing solutions (see Fig.\ref{seven1}).

The motion of a test brane in a background produced by a collection of branes
was discussed in \cite{kiritsis} (the so-called mirage cosmology).
Adopting spherically-symmetric backgrounds, it was shown that although there is
a singularity
in the evolution of the 4-d brane, the higher-dimensional geometry is regular.
The origin of the singularity on the brane is actually the
embedding of the brane in the bulk, in such a way that the
singularity is smoothed out when the solution is lifted to higher dimensions.

The effect of inflation on a bouncing brane was used in
\cite{aranha} to set limits on the parameters of the braneworld.
Specifically, the model consists of closed a FLRW metric embedded
in a 5-d conformally flat bulk with one extra timelike dimension,
containing a conformally coupled scalar field (the inflaton field)
and a radiation fluid, evolving on the brane with corrections due
to the bulk. The non-singular bouncing solutions considered were
oscillatory and bounded, or initially bounded. They are in
principle stable and would never enter an inflationary phase with
an exponential growth of the scale factor since they correspond to
periodic orbits of the integrable dynamics in the gravitational
sector. The introduction of a massive scalar field, even in the
form of small fluctuations, turns non-integrable the dynamics of
the system \cite{aranha}. As a consequence, non-linear resonance
phenomena are present in the phase space dynamics for certain
domains of the parameter space of the models, and the associated
dynamical configurations become metastable, allowing the orbits
escape to the de Sitter infinity in a finite time. From the
conditions for these orbits to happen, limits on the parameters
$(\sigma, m, E_0)$ are set, where $\sigma$ is the brane tension,
$m$ is the mass of the scalar field, and $E_0$ is a constant
proportional to the total energy of the fluid.

Yet another turn in the mirage model was introduced in
\cite{sling}, where the brane moves in an open orbit around a
non-trivial spherically-symmetric background. In this model, the
brane is moving on a Calabi-Yau manifold generated by a heap of
D3-branes, and the mirage effects dominate the evolution of the
Universe only at early time, i.e. when the brane moves in the
throat of the background manifold. The new feature is the
influence of the angular momentum of the test brane on its motion
in the higher-dimensional space. In fact, the effective 4-d metric
has two parameters: the energy $U$ and the angular momentum $L$
of the 4-d brane, which determine the form of the orbit. In
particular, to have an open orbit in an asymptotically Minkowskian
background,
$$
L ^4-4(U+2)U^3\geq 0.
$$
As discussed in \cite{sling}, the effective metric corresponding to
orbits satisfying this constraint display
cosmological contraction during the ingoing part of the orbit, expansion
during the outgoing part, and a bounce at the turning point
\footnote{Further effects of the angular momentum on the motion of the brane,
including cyclic universes, were studied in \cite{cycling}.}.

Another model based on the brane scenario is the ekpyrotic
universe \cite{ek}, the cyclic
version of which shall be considered in Sect.\ref{ccyclic}.

\section{Variable cosmological constant}
\label{cvcc}
General Relativity allows for the introduction of only one
arbitrary
constant, the so-called cosmological constant $\Lambda$. At least two
attitudes can be taken regarding $\Lambda$ \cite{padma}.
The first one is to consider it as a
derived quantity, that emerges from vacuum fluctuations
(see for instance \cite{zeld}). One way out of
the huge disagreement between theory and observation in this case
\cite{caldwell} is to assume that
$\Lambda$ is actually time-dependent.
The second attitude that can be adopted is that $\Lambda$ is, along with $G$, a fundamental
parameter of the theory, to be determined by observation
\footnote{Notice that this second attitude is somewhat different from
Einstein's original ideas leading to GR, since there would be curvature even in the absence of matter,
caused by $\Lambda$.}
\cite{novellocc}.
 In fact, from a gravitational point of view what matters
is the ``effective'' cosmological constant, since
the matter Lagrangian can sometimes contribute with a $\Lambda$-like term, as in the case of
the scalar
field in the presence of a potential with a minimum:
$$
\Lambda_{\rm eff} = \Lambda +  V(\phi_{\rm min}),
$$
where $\Lambda$ is the ``bare'' cosmological constant.
Any change in $\phi_{\rm min}$ during the evolution leads to changes in
the value of $\Lambda_{\rm eff}$. In fact, the effect of the
evolution of the universe on the ground state is to add a
temperature dependence, which can be translated into a time
dependence \cite{kir}. A model along these lines based on a gauge
field (instead of a scalar field) was presented in \cite{barcelos}
\footnote{In fact, any classical nonlinear field theory (such as
nonlinear electromagnetism) admits a fundamental state that
generates a cosmological constant \cite{nosso}.}.
This is another motivation to consider a variable
$\Lambda$,  that is not a constant but a function of spacetime
coordinates, in such a way that its value is determined by the
dynamics of the theory under scrutiny (following the line of
reasoning of other ``variable constant'' theories, see
Section \ref{ccyclic}.). In fact,
a time-dependent cosmological constant has also been called
upon to explain the current accelerated expansion and
the fact that this phase started in the recent past.

In the case of $\Lambda = \Lambda (t)$, EE for the
FLRW metric take the form
\beq
\frac{\dot{a}^2}{a^2}=\frac{1}{3}\rho + \frac{\Lambda
(t)}{3}-\frac{\epsilon}{a^2}, \label{lv1} \eeq \beq
\frac{\ddot{a}}{a} = \frac{\Lambda (t)}{3}-\frac{1}{6}
(\rho + 3p), \label{lv2}
\eeq
and the continuity equation is given by
\beq
\dot\rho+3\frac{\dot a}{a}\;(\rho +p)=-\dot \Lambda
. \label{lv3} \eeq As seen from Eqn.(\ref{lv3}),
$\Lambda$ can supply or absorb energy from ordinary matter and
radiation. In fact, it follows from this equation that \beq
TdS=-Vd\Lambda. \eeq Hence, $\Lambda$ is a source
of entropy. Requiring that $dS/dt>0$ implies $d\Lambda/da<0$
through cosmic expansion.

Assuming that only radiation is present, Eqn.(\ref{lv3}) gives
$$
\frac{d\rho}{da}+\frac{d\Lambda}{da}+\frac{4\rho}{a}=0,
$$
which can be integrated to \beq
\rho=\rho_0\left(\frac{a_0}{a}\right)^4-\frac{1}{a^4}\int^a_{a_0}A^4\frac{d\Lambda}{dA}dA,
\label{key} \eeq where $\rho=\rho_0$ when $a=a_0$, and the
subindex 0 denotes quantities evaluated at $t=0$. Notice that the
model is completely determined in this case by providing the
function $\Lambda = \Lambda (a)$, since Eqn.({\ref{key})
then yields $\rho=\rho (a)$, and
$a=a(t)$ follows from Eqn.(\ref{lv1}).
A cosmological model based on this scenario was
discussed in \cite{taha2}, where the dependence of $\Lambda$ on $a$
was fixed by imposing that $\rho = \rho_c$ for all values of $t$,
where $\rho_c=3H^2$ is the critical density. It follows from
Eqn.(\ref{lv1}) that \beq \Lambda = \frac{\alpha\epsilon}{a^2}.
\label{lambda} \eeq The conditions $\dot\Lambda\geq 0$ and $\dot
a\geq 0$ give $\epsilon>0$, hence $\epsilon = 1$. In the model
presented in \cite{taha2}, at $t=0$ the universe had only a nonzero
cosmological constant. With $\rho_0=0$,
Eqn.(\ref{key},\ref{lambda}) give
\beq \rho (a)=
\frac{\alpha}{a^2}\left( 1-\frac{a_0^2}{a^2}\right).
\label{i1}
\eeq
Note that $\rho_0=0$ implies that $a_0\neq 0$, in such a way
that the singularity at $t=0$ is absent. An estimation of $a_0$
was made in \cite{taha2} by assuming that the maximum temperature reached is
$T_{max}\sim M_{Pl}$,
which gives
$$
a_0\sim\frac{2.5}{\sqrt N}\times 10^{-20} ({\rm GeV})^{-1},
$$
where $N=N(T)$ is the effective number of degrees of freedom at
temperature $T$.

The fact that this model does not display a horizon problem was
also shown in \cite{taha2}. In fact, the time $t_c$ at which global
causality is established is given by
$$
t_c=a_0\sinh \frac\pi 2 \sim 2.3 a_0,$$ which indicates that
global causal connection was established at a very early time. The
model is also free of the monopole problem, but it is worth noting
that there is an inflationary period. From
Eqn.(\ref{lv1}) we get
\beq a^2=a_0 ^2+t^2. \eeq
A peculiarity of
this model is that $a\rightarrow\infty$ for $t\rightarrow\infty$,
even though $\epsilon =1$. Needless to say, other choices of $\Lambda$
would give a different asymptotic behavior.

The same form of $\Lambda$, namely \beq \Lambda (t) =
\frac{\gamma}{a(t)^2}, \label{l1} \eeq where $\gamma$ is a
constant to be determined by observations, was studied in
\cite{wei}, but without the assumption that $\rho=\rho_c$. The
conservation equation (\ref{lv3}) can be solved for dust and
radiation. Inserting the solution in Eqns.(\ref{lv1}) and
(\ref{lv2}) we get
$$
\frac{\dot{a}^2}{a^2}+\frac{\Upsilon}{a^2}=\frac{1}{3}\rho^{(i)},
$$
$$
\frac{\ddot{a}}{a} = -\frac{1}{6}\rho^{(i)},
$$
where $\Upsilon=\epsilon-2\gamma/3$ for radiation, and
$\Upsilon=\epsilon-\gamma$ for dust,
and $\rho^{(i)}$ is the
energy density of dust or radiation for the case $\Lambda =0$.
These equations show that the effect of assuming that
$\Lambda\propto a^{-2}$
is to shift the curvature parameter $\epsilon$ by a
constant value.
A nonsingular cosmological model based on the model presented in
\cite{wei}
has been analyzed in \cite{abdel}. Notice that Eqn.(\ref{l1})
along with condition $d\Lambda/da<0$ require that $\gamma$ be
positive. A positive $\Lambda$ for all $t$ implies, through
Eqn.(\ref{lv2}) that there may be a zero in $\dot a$, and hence
the possibility of a bounce. For this to happen we need that $\dot
a $ be zero at the putative bounce. Supposing there is a bounce,
it follows from Eqn.(\ref{lv1}) evaluated at the bounce that
$$
\alpha^{-1} \rho_0a_0^2=\epsilon-\gamma.$$ Hence, $\rho_0>0$
implies that $\epsilon>\gamma>0$, and so $\epsilon=1$. Introducing
the \emph{Ansatz} (\ref{l1}) in the Friedmann equation, we get \beq
a^2\dot a^2=(2\gamma-1)(a^2-a_0^2), \eeq so it follows that
$\gamma>1/2$. Hence, $1/2<\gamma\leq1$. This equation can be
integrated to get
$$
a^2=(2\gamma -1)t^2+a_0^2,
$$
which leads to bounded-from-above densities and temperature
\footnote{The evolution of perturbations in this model was studied
in \cite{abdalla}.}.

Yet another form for the dependence of $\Lambda$,  given by
$$
\Lambda = \Lambda_1+\Lambda_2\;a^{-m},
$$
where $\Lambda_1$, $\Lambda_2$ and $m$ are constants (with $\Lambda_2>0$), was
studied in \cite{maty}. The analysis of the dynamics was
carried out using the analog of the one-dimensional problem of the particle
under the influence of the potential $V(a)$ given by
$$
V(a) = -\Lambda_1\delta\frac{a^2}{\alpha +2}-\Lambda_2\delta \frac{a^{2-m}}{\alpha -m+2}+ba^{-\alpha},
$$
where $\alpha = 1+3\lambda$, $\delta=1+\lambda$, $b$ is a positive integration constant,
and $p=\lambda \rho$. Denoting by $r$ the maximum of the potential,
cyclic solutions are obtained for the cases $\epsilon=1$ with
$\Lambda_1$, $\Lambda_2>0$,
and $r>-1$, and for $\Lambda_1<0$, $\Lambda_2>0$, and $m\leq 2$, regardless of the sign of
$\epsilon$.

The proposal in Eqn.(\ref{l1}) was later generalized in
\cite{arbab} to
\beq \Lambda = 3\beta H^2+\frac{3\gamma}{a^2},
\eeq
where $\beta$ and $\gamma$ are dimensionless numbers, and $H=\dot a/a$.
following \cite{carv}.
With this \emph{Ansatz}, the Friedmann
equation for a radiation-dominated phase
can be rewritten as
\beq \dot a
^2=\frac{2\gamma-\epsilon}{1-2\beta}+A_0a^{-2+4\beta},
\eeq
which allows a bouncing solution at $t=0$ for
$A_0<0$, $\beta<1/2$, $\epsilon=1$ (with $\rho_0>0$).
The value $\gamma=1$ was chosen in \cite{arbab} so that
$dS/da$ is always greater than zero, thus solving the entropy problem.
In this case, the model gives $\Omega<1$ for all $t$.

A thorough review of variable-$\Lambda$ models
has been presented in \cite{over}. The models analyzed were power-laws
of the different relevant parameters, namely
$$\Lambda_1={\cal A}t^{-{\ell}},\;\;\;\;
\Lambda_2={\cal B}a^{-{m}},\;\;\;\; \Lambda_3={\cal C}H^{{n}},\;\;\;\;
\Lambda_4={\cal D}q^{r},$$
where
${\cal A}$, ${\cal B}$, ${\cal C}$, ${\cal D}$,
${\ell}$, $m$, $n$, and $r$ are constants. Let us state from
\cite{over}
the relevant results for this review: (1) no bouncing models were found for $\Lambda_1$ with $k=0$
and ${\ell}=1,2,3,4$, irrespectively of the sign of ${\cal A}$.
(2) For $\Lambda_2$, it was shown (numerically) that there are nonsingular models
for dust, $\epsilon=1$,
with $m=1$, $\Omega_0=0.34$,  and $0.68<\Omega_{0\Lambda}<0.72$, and also with
higher values of $m$ and $\Omega_0$.
(3) For $\Lambda_3$, the value $n=2$ admit analytical solution.
For this $n$, there are bouncing solutions for $\gamma>2/3$ and $\epsilon= 1$
with ${\cal C}>3(3\gamma/2-1)\Omega_0$, and also for $\gamma>2/3$ and $\epsilon= -1$,
for ${\cal C}<3(3\gamma/2-1)\Omega_0$.
(4) Only the value $r=1$ was explored for $\Lambda_4$. Defining
$\lambda_0=-{\cal D}q_0/3$, there are closed bouncing solutions for
$\lambda_0>-\Omega_0$, and open bouncing solutions for $\lambda_0<-\Omega_0$.

The examples given above show that
varying-$\Lambda$ scenarios
are worth examining because they address
a number of pressing problems in cosmology (horizon problem, entropy, initial
singularity)\footnote{Nonsingular cosmological
solutions for the case in which the cosmological constant
is replaced by a second-rank tensor $\Lambda^\mu_{\;\nu}$ were
studied in \cite{bronnikov}.}. Furthermore, many of them are simple enough to draw
definite conclusions about their viability. One of the drawbacks
is perhaps the lack of strong motivation for
choosing any given form of $\Lambda$. In this regard, let us remember that
many of the varying-$\Lambda$ models can be reverse-engineered
to scalar-field models with a potential. Unfortunately, in most cases the
corresponding
models lack predictive power or clear particle physics motivation \cite{padma}.

\section{Past-eternal universes}
\label{peu}

In this section, we shall examine some models which are nonsingular but do not exhibit a
bounce. Historically, perhaps the most important example of these
is the Steady-State model \cite{ssm} \footnote{For an updated version,
see Sect.\ref{oqss}.}. As mentioned in Sect.\ref{fb},
nonsingular solutions
that start from a deSitter state were discussed in
\cite{starobinsky, fischetti}.
Another example
is that discused in \cite{mukh} in which every contracting and
spatially flat, isotropic universe avoids the big crunch by ending
up in a deSitter state enforced by the limiting curvature hypothesis.

\subsection{Variable cosmological constant}

As noted in \cite{ademir2}, in all the articles mentioned in Sect.\ref{cvcc},
the dependence of $\Lambda$ on $a$ and $\dot a$ was set either
from ''first principles'' (for instance quantum gravity, as in
\cite{wei}), or by extrapolating backwards current cosmological
data, including the current value of $\Lambda$. However, another
view can be taken. Since $\Lambda$ can be considered as a remnant
of a period of inflation, a complete model should also describe the
era of inflationary expansion. This is precisely the proposal in \cite{ademir2}, where
$\Lambda$ was taken as \beq \Lambda (H) = 3\beta H^2+3(1-\beta)
\frac{H^3}{H_\ell}, \eeq where ${H_\ell}$ is the timescale of
inflation, and $\beta$ is a parameter. Note that when
$H={H_\ell}$, $\Lambda = 3{H_\ell}^2$, as required by inflation,
while $\Lambda\sim 3\beta H^2$ for large cosmological times. In
the case of $\epsilon=0$, and for
$$
p=(\gamma -1) \rho,
$$
an equation for the Hubble parameter follows \cite{ademir2}:
$$
\dot H + \frac{3\gamma
(1-\beta)}{2}\;H^2\left(1-\frac{H}{H_{\ell}}\right)=0,
$$
whose solution is
$$
H=\frac{H_{\ell}}{1+Ca^{3\gamma (1-\beta )/2}},
$$
where $C$ is a $\gamma$-dependent integration constant
\footnote{Here the value $\epsilon = 0$ was chosen, but this restriction was lifted in
\cite{trodden}.}.
This
equation can be integrated to yield
$$
H_{\ell}t=\ln\left(\frac{a}{a_*}\right)+\frac{2C}{3\gamma (1-\beta
)}\;a^{3\gamma (1-\beta )/2},
$$
where $a_*$ is an arbitrary value of the scale factor.
It follows from this equation that the evolution of the
universe starts from a deSitter stage $a\sim e^{H_\ell t}$ for
$Ca^{3\gamma (1-\beta )/2}<<1$, and evolves towards a FLRW phase,
$a\sim t^{2/3\gamma (1-\beta )}$ for $Ca^{3\gamma (1-\beta
)/2}>>1$.

\subsection{Fundamental state for $f(R)$ theories}

A novelty in some theories described by Lagrangians that depend only on $R$
is the possibility of the
emergence of an intrinsic cosmological constant.  This is not the
case, however, in theories generated by Lagrangians
that are a linear combination of $R^2$ and $R_{\mu\nu} R^{\mu\nu}$
as can be seen by a direct
inspection of the EOM (\ref{23dez4}).
The proof of this assertion follows from the fact that the tensors
$\chi_{\mu\nu}$ and $Z_{\mu\nu}$ appearing in the EOM \ref{HOTG3}
are traceless in the case of a constant curvature scalar.
($R_{\mu\nu} = \Lambda \, g_{\mu\nu}$).
However, restricting to the $f(R)$ case,
Lagrangians that are not linear in $R^2$ can bypass
such prohibition.
The existence of a deSitter solution in the absence of
matter occurs when the function obeys the condition
\begin{equation}
\frac{f^{'}}{f} = {\rm constant}. \label{25dez1750}
\end{equation}
A typical example is provided by the exponential Lagrangian
$$ f(R) = \exp{\left(\frac{R}{2\Lambda}\right)}. $$
It follows straightforwardly from Eqn.(ref{fr})
that $R_{\mu\nu} = \Lambda \, g_{\mu\nu} $ is a possible state of
the system.

\subsection{The emergent universe}
\label{eu}
Another example of past eternal universe was given in
\cite{ellismaartens}. This model uses general relativity plus a scalar field with a potential,
and matter. The relevant equations are
$$
\ddot\phi+3H\dot\phi+V'(\phi )=0,
$$
$$
\frac{\ddot a}{a}=-\left[ \half (1+3\omega )\rho+\dot\phi^2-V(\phi )\right],
$$
$$
H^2=\rho+\half \dot\phi+V(\phi )-\frac{\epsilon}{a^2}.
$$
From these, it follows that
to have a minimum of the scale factor we need to impose the conditions
$$
\half (1+3\omega )\rho+\dot\phi^2<V(\phi ),
$$
and
$$
\half \dot\phi_i^2+V_i+\rho_i=\frac{\epsilon}{a_i^2},
$$
where the subindex $i$ means that the quantities are evaluated at $t_i$, the time
at which $a$ is minimum. Assuming positive potentials and energy density, it follows that
only $\epsilon =+1$ is allowed.
It follows that
$$
\half (1-\omega_i)\rho_i+V_i=\frac{2}{a_i^2},
$$
where $V_i=\Lambda_i$, and
$$
(1+\omega_i)\rho_i+\dot\phi^2=\frac{2}{a_i^2},
$$
so a model can be constructed with
$\rho_i=0$ and constant $\dot\phi^2$.
This can be achieved in the limit $t\rightarrow\infty$
with the potential \cite{ellismaartens}
$$
V(\phi ) = V_f+(V_i-V_f)\left[\exp\left(\frac{\phi-\phi_f}{\alpha}\right)-1\right]^2,
$$
where $\phi_f$ is the value of the field for which $V$ is minimum,
and $\alpha$ is a constant energy scale.
In order to achieve the Einstein universe state in the
far past,  some fine-tunning on $a_i$ and $\dot\phi_i$ is needed, which is not
necessarily a hindrance \cite{ellismaartens}\footnote{In particular,
the initial scale factor could be chosen in such a way to avoid the quantum gravity regime.}.
The choice of such a highly-symmetric state
as the initial state is supported by various arguments: it is stable agains some types of
inhomogeneous linear perturbations,
it has no horizon problem, it maximizes the entropy within the family of FLRW radiation
models, and it is the unique highest symmetry non-empty FLRW model (with a
7 dimensional group of isometries). The model was elaborated further in
\cite{murugan}, where it was shown that an explicit form for the
potential can be found such that the model leaves the inflationary stage and enters a reheating
phase, followed by standard evolution.

\section{Quantum Cosmology}
\label{cqc}

As discussed in Sect.\ref{sint}, there are reasons to suppose
that at very
high energies some of the hypotheses of the singularity theorems
are rendered invalid: if the universe ever attains this regime,
an important role is to be played
by quantum gravitational effects, in such a way that a
quantum theory of gravitation is needed to have a proper
description.

Although there is yet no complete realization of quantum gravity,
there are some attempts to tackle the singularity problem in a
quantum framework. A standard method of quantizing General
Relativity is canonical quantization \cite{hartle} where the
momentum and Hamiltonian constraint equations are interpreted as
operators, and it is required that they annihilate the quantum
state. The Hamiltonian constraint gives the Wheeler-DeWitt (WdW) equation
\cite{wdw}, which depends on the choice of the factor ordering in the products
of generalized momenta and ``velocities''. For some choices of the
ordering, the
WdW equation turns it into a Klein- Gordon
equation on an indefinite DeWitt metric in the infinite-dimensional
superspace (space of three-metrics), with a potential term
\cite{wdw}. In addition to the WdW equation, initial
conditions must be specified,
the two most popular being the
``no-boundary'' \cite{hartlehaw}, and the ``tunnelling''
condition \cite{Vilenkin}.

In practice, the infinite degrees of
freedom of the superspace are truncated to obtain a minisuperspace
model, usually under the assumptions of isotropy and homogeneity.
Once a solution to the WdW equation has been found, there
is the question of how to interpret it and extract probabilities from it.

Among other issues related to the WdW equation,
there is the fact that a suitable
initial condition must be chosen to get a solution. It would be desirable that
the initial condition be
somehow determined by the dynamical law (see for instance \cite{bojoic}). In fact, the
most well-accepted proposals mentioned above
do not solve the singularity problem \cite{asht}.
Moreover,
in the quantization following the ADM procedure, time is fixed by a gauge choice, and the
results are dependent of this choice \cite{nelson}
\footnote{In this regard, it was shown in \cite{demaret}
that a Bianchi I universe, quantized following the ADM
recipe with a particular choice of the time coordinate
\cite{lund} in the presence of dust is nonsingular.}.

As we shall see below, there are other approaches to
Quantum Cosmology which may yield a nonsingular universe in the
regime where the WdW equation is valid.
We shall discuss
two possibilities: the Bohm-de Broglie interpretation of QM, and
Loop Quantum Cosmology (LQC).

\subsection{The ontological (Bohm-de Broglie) interpretation}
\label{onto}
If the universality of quantum mechanics is assumed, the Universe
must be describable by a wave function (furnished by a
yet-to-be-discovered quantum theory of gravity and matter fields) in every step of
its evolution. Moreover, this description must have a well-defined
classical limit. The orthodox interpretation of Quantum Mechanics
(the so-called Copenhagen interpretation) \cite{copen} is ill-suited for
the task of describing the universe, since it assumes
the existence of a ``classical apparatus'' external to the system
to solve the measure problem by forcing the collapse of the wave
function. Clearly, there is no classical apparatus outside the
universe. Therefore, the least we can say is that an alternative to
the Copenhagen interpretation is needed. One such alternative that
has received some attention recently is that of Bohm and de
Broglie (BdB)\cite{holland}
\footnote{Other possibilities (not free of problems,
though) are the many-worlds interpretation \cite{manyw}, non-linear quantum mechanics
\cite{nlqm}, and
decoherence \cite{decoh}.}. In
classical physics, the dynamics of a point in configuration space is
determined by the principle of extremal action, yielding the classical
EOM. According to the BdB interpretation,
in quantum physics the evolution of the configuration
variables is guided by a quantum wave which obeys Schr\"odinger's
equation. The associated Hamilton-Jacobi equation displays a new term (of quantum origin, see below),
that can be
interpreted as part of the potential.
It should be emphasized that the BdB interpretation furnishes a framework to
make predictions based on the wave function of the system, which must be
obtained by some means (for instance, through the WdW equation).

Let us briefly review first the quantum mechanics of a single particle
in the BdB interpretation, and afterwards the results will be translated,
\emph{mutatis mutandis}, to the context of FLRW cosmology.
The Schr\"odinger equation for a non-relativistic particle in a potential $V$ is given by
$$
i\hbar\; \frac{d\psi (x,t)}{dt}=\left(-\frac{\hbar^2}{2m}\nabla^2+V(x)\right)\psi (x,t).
$$
With the replacement $\psi=R\exp (iS/\hbar)$, this equation becomes
\beq
\frac{\partial S}{\partial t}+\frac{(\nabla S)^2}{2m}+V-\frac{\hbar ^2}{2m}
\frac{\nabla^2R}{R}=0,
\label{bdb1}
\eeq
\beq
\frac{\partial R}{\partial t}+\nabla . \left(R^2 \frac{\nabla S}{m}\right)=0.
\label{bdb2}\eeq
This last equation suggests that $\nabla S/m$ can be interpreted as a velocity field,
leading to the identification $p=\nabla S$, in such a way that
Eqn.(\ref{bdb1})
is the Hamilton-Jacobi equation for the particle in the classical potential $V$
plus a ``quantum
potential'' $Q=-\hbar^2\nabla^2R/2mR$. The BdB interpretation
argues that a
quantum system is composed of a particle \emph{and} a field, and that
quantum particles follow trajectories $x(t)$,
independent on the existence of an ouside observer.
These trajectories can be determined from
$$
m\frac{d^2x}{dt^2}=-\nabla V-\nabla Q,
$$
or from
$p=m\dot x=\nabla S$,
after $S$ and $R$ are determined using Eqns.(\ref{bdb1}) and (\ref{bdb2}).
In practice, since $S$ is the phase of the wave function, it can be
read off from the explicit solution of Schr\"odinger's equation.

Let us analyze an example developed in \cite{colis}, where the Lagrangian
was given by
$$
{\cal L} = \sqrt{-g}\;\left(R-C_\omega
 \phi_{,\mu}\phi^{,\mu}
\right),
$$
where $C_\omega =
(\omega+\frac 3 2)$. From the metric
$$
ds^2= -N^3dt^2+\frac{a(t)^2}{1+(\epsilon/4)r^2}\;\left(dr^2+r^2d\Omega^2\right),
$$
and the definitions $\beta^2=4\pi{\ell}_{Pl}^2/3V$, $\bar\phi=\phi\sqrt{C_\omega/6}$,
we get
$$
{\cal H}=N\left(-\beta^2\frac{p_a^2}{2a}+\beta^2\frac{p_{\bar\phi}^2}
{2a^3}-\epsilon\frac{a}{2\beta^2}
\right),
$$
with $p_a=-a\dot a/(\beta^2N)$, $p_{\bar\phi}=a^3\dot{\bar\phi}/(\beta^2N)$.
Defining $\tilde a=a/\beta$, setting $\beta=1$ and $\alpha\equiv\ln \tilde a$, we get
\beq
{\cal H}=\frac{N}{2\exp(3\alpha)}\left(-p_\alpha^2+p_\phi^2-\epsilon\exp(4\alpha)\right),
\label{ham}
\eeq
where
$$
p_\alpha=-\frac{\dot\alpha e^{3\alpha}}{N},\;\;\;\;\;\;p_\phi=\frac{\dot\phi e^{3\alpha}}{N}.
$$
Notice that $p_\phi=\bar k$ is a constant of the motion.
We shall restrict to the case $\epsilon =0$ since it is analytically tractable.
The classical solutions are given by
$$
a=3\bar k t^{1/3},\;\;\;\;\;\;\phi=\frac 1 3 \ln t+c_2,
$$
where $c_2$ is an integration constant.
Depending on the sign of $\bar k$,
this solution contracts to or expands from a singularity.

The Wheeler-DeWitt equation corresponding to the Hamiltonian given in Eqn.(\ref{ham})
is given by \cite{colis}
$$
-\frac{\partial^2\Psi}{\partial\alpha^2}+\frac{\partial^2\Psi}{\partial\phi^2}+
\epsilon\;e^{4\alpha}\Psi=0.
$$
The solution, obtained by separation of variables, reads
$$
\Psi(\alpha , \phi)=\int F(\kappa)A_\kappa(\alpha)B_\kappa(\phi)d\kappa,
$$
where $\kappa$ is a separation constant, $F(\kappa)$ is an arbitrary function of $\kappa$,
$$
A_\kappa(\alpha)=a_1\exp(i\kappa\alpha)+a_2\exp(-i\kappa\alpha),
$$ (for $\epsilon =0$),
and
$$
B_\kappa(\phi)=b_1\exp(i\kappa\phi)+b_2\exp(-i\kappa\phi).
$$
A direct application of the formalism sketched for the case of a one-particle system,
generalized to several degrees of freedom yields from the Hamiltonian
(\ref{ham}) \cite{colis}
$$
Q(\alpha,\phi)=\frac{e^{3\alpha}}{2R}\left(\frac{\partial^2R}{\partial\alpha^2}-
\frac{\partial^2R}{\partial\phi^2}\right),
$$
with the ``guidance relations''
$$
\frac{\partial S}{\partial\alpha}=-\frac{e^{3\alpha}\dot\alpha}{N},\;\;\;\;\;\;
\frac{\partial S}{\partial\phi}=\frac{e^{3\alpha}\dot\phi}{N}.
$$
A state is now needed to read off from it $S$ and $R$.
A Gaussian superposition was chosen in \cite{colis},
given by
$$
\Psi(\alpha, \phi)=\int F_\kappa B_\kappa(\phi)[A_\kappa (\alpha)+
A_{-\kappa}(\alpha)]\;d\kappa,
$$
with
$$
F(\kappa)=\exp\left(-\frac{(\kappa-d)^2}{\sigma^2}\right).
$$
and $a_2=b_2=0$. Performing the integration in $\kappa$, we can extract from the result
the phase $S$ which, when inserted into the guidance relations (in the $N=1$ gauge)
furnishes
a planar system:
\beq
\dot\alpha=\frac{\phi\sigma^2\sin(2d\alpha)+2d\sinh(\sigma^2\alpha\phi)}
{\exp{3\alpha}\left(2(\cos(2d\alpha)+\cosh(\sigma^2\alpha\phi))\right)},
\label{ps1}
\eeq
\beq
\dot\phi=\frac{-\alpha\sigma^2\sin(2d\alpha)+2d\cos(2d\alpha)+2d\cosh
(\sigma^2\alpha\phi)}
{\exp{3\alpha}\left(2(\cos(2d\alpha)+\cosh(\sigma^2\alpha\phi))\right)}.
\label{ps2}
\eeq
The plot of this system (see Fig.\ref{pps}) shows that there are bouncing trajectories
for $\alpha >0$, and also oscillating universes near the centre points (white
points in the plot).
\begin{figure}[htb]
\begin{center}
\includegraphics[width=0.7\textwidth
]{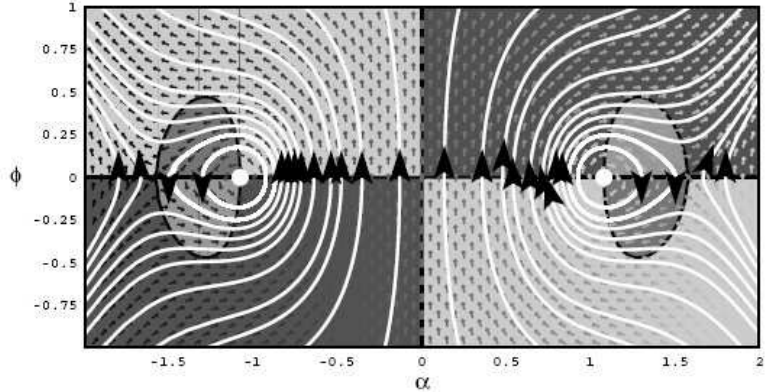}
\caption{Field plot of the planar system (\ref{ps1})-(\ref{ps2}) for
$\sigma = d = 1$.
Taken from \cite{colis}.}
\label{pps}
\end{center}
\end{figure}
The BdB interpretation has been applied to mini-superspace models
in Quantum Cosmology (see for instance \cite{acacio, nelson}), and
non-singular solutions have been found for models with scalar
fields or radiation \cite{felipe}. The bounce is due to the action
of the quantum potential, which generates a repulsive ``quantum
force'', large enough to reverse the collapse.

One of the advantages of this
formulation is that, starting from WdW equation, it yields a
dynamics that is invariant under time re-parameterizations.
Notice however that the results are dependent on the state chosen to
represent the system.

\subsection{Loop Quantum Gravity}
\label{slqg}

Loop Quantum Gravity is a background-independent, non-perturbative
canonical quantization of gravity in which the classical metric
and the extrinsic curvature are turned into operators on a Hilbert
space \cite{rovelli}. The classical description of space-time is
replaced by a quantum counterpart, in such a way that quantum
effects are important at very short scales, for instance near
putative singularities. In this scenario, the evolution of the
universe is divided in three epochs. First there is a quantum
epoch with high curvature and energy, described by difference
equations for the wave function of the universe. These are a
direct consequence of the discreteness of space and time, the step
size being dictated by the lowest non-zero eigenvalue of the area
operator (see \cite{bojoic}). It is this discreteness that modifies the
behavior near the singularity, leading to a theory that
is not equivalent to the WdW description (even in the isotropic case), which
furnishes a continuous spectrum for the scale factor.
A semiclassical epoch follows, with
differential equations for matter and geometry modified by
non-perturbative quantization effects. Finally, a classical phase
is reached, described by the usual cosmological equations.

Since difference equations are often difficult to analyze or
to solve explicitly, and at such a fundamental level, the emergence
of space-time in
inhomogeneous models with many degrees
of freedom from
the underlying quantum state is hard to understand, a suitable strategy is to
use special models allowing exact solutions.
Care must be taken in the extension of results from particular examples
to more general cases. In any case, it may be instructive to have
a detailed understanding of how the singularity is resolved
in some instances.

Yet another convenient simplification is to work
in an effective semiclassical theory, which takes
into account only some quantum effects. This theory can be
understood as governing the motion of a wave packet
that solves the difference equation
\cite{bojo1}, and can be obtained as an asymptotic
series of correction terms to the equations of motion in the
isotropic case \cite{bojo}. For instance, in the case of a matter
term generated by a scalar field under the influence of a
potential, the effective Klein-Gordon equation is
\cite{topo}
\beq
\ddot \phi = \dot\phi
\left(-3H+\frac{\dot D}{D}\right) - D V'(\phi),
\label{effkg}
\eeq
where
$$
D(q)=\left(\frac{8}{77}\right)^6\;q^{3/2}
\{7[(q+1)^{11/4}-|q-1|^{11/4}]-11q[(q+1)7/4\}-|q-1|^{7/4}{\rm
sgn}(q-1)]\}^6,
$$
with $q=a^2/a^2_*$ and $a_*^2=\gamma {\ell}_{Pl}^2 j /3$, where
$\gamma\approx 0.13$, and $j$ is
a quantization parameter, which takes half-integer values. This
equation represents an approximate expression for the eigenvalues
of the inverse volume operator \cite{bojo5}. The function $D$
varies as $a^{15}$ for $a\ll a_*$, has a global maximum at
$a\approx a_*$, and falls monotonically to $D=1$ for $a> a_*$. In
turn, the effective Friedmann equation is given by
\beq
\frac{\dot a^2}{a^2} + \frac{\epsilon}{a^2} = \frac 1 3 \left(\frac{\dot\phi^2}{2D}+
V(\phi)\right),
\label{efffri}
\eeq
and the effective Raychaudhuri equation is
\beq
\frac{\ddot a}{a} = -\frac 1 3 \dot\phi^2\left( 1-\frac{\dot
D}{4HD}\right)+ \frac 1 3 V(\phi).
\label{effray}
\eeq
These approximations are valid for $a_i< a< a_*$, where
$a_i=\sqrt\gamma\: {\ell}_{Pl}$. Below  $a_i$ the quantum nature of
spacetime cannot be replaced by an effective theory, while above
$a_*$ we recover classical cosmology.
It was
shown in \cite{topo} that a closed universe with a minimally
coupled scalar field will bounce (avoiding the so-called big
crunch) as soon as
$a\approx a_*$
for any choice of the initial conditions.
The
bounce in this case is due to the change of sign of the
``friction'' term in Eqn.(\ref{effkg}), which becomes frictional for
$a<<a_*$, freezing the field $\phi$ in some constant value, and
turning the effective EOS into a cosmological constant EOS
\cite{topo}. Similar results were obtained in the case of
anisotropic models \cite{bojo3}.

The previous example incorporated quantum gravitational effects on the
matter (represented by a scalar field) Hamiltonian, but there may also
be modifications
of the gravitational Hamiltonian due to quantum
geometry. Recently, some calculations
illustrating the effects of quantum geometry on both
the gravitational and matter Hamiltonians were carried out in the case of
a spatially homogeneous, isotropic
$\epsilon=0$ universe with a massless scalar field (a system which is singular
both classically and according to the WdW formalism
in the Copenhagen interpretation of QM). It was shown in
\cite{asht}
that the
singularity is resolved in the sense that a complete set of Dirac
observables on the physical Hilbert space remains well-defined
throughout the evolution;
the big-bang is replaced by a big-bounce in the quantum theory due to the quantum corrections
to the geometry;
there is a large classical universe on the "other side", and the
evolution bridging the two classical branches is deterministic,
thanks to the background independence and non-perturbative
methods \footnote{In a subsequent paper the Hamiltonian was modified
to forbid the bounce at low densities \cite{assingh}.}.
Notice also that no boundary condition was imposed (it was
asked instead that the quantum state be semiclassical at late times)
\footnote{An analysis along the same lines was carried out in
\cite{vander} for the case $\epsilon=-1$, and it was shown that the singularity is
avoided too.}.

Surely the major
limitation in all the analysis of LQC
is that, since
a satisfactory quantum gravity theory
which can serve as an unambiguous starting point is not available yet,
the theory is not developed by a systematic truncation of full quantum
gravity. Another limitation is
the restriction to
isotropy and homogeneity.

\subsection{Stochastic approach }
%
%
%
A different approach was introduced in
\cite{marioluiz}, which starts form the observation made in
\cite{markov} that the universe could be enlarged through
an ``analytic extension''.
In \cite{markov}, such an extension is achieved from
the geometrical construction of
a semiclosed universe, namely a closed Friedmann model extended by gluing
a given geometry to the FLRW before the maximum expansion. This gluing can be done in different ways,
through the junction conditions. In \cite{markov} an asymptotically flat geometry was chosen.
A collection of this configuration (called
\textit{friedmon} in \cite{marioluiz})
was considered in \cite{marioluiz}, in such a way that each member of the collection
perceives the remaining systems as a
perturbative effect of random character, as in a stochastic process.
Noting that in the case of an open universe,
the Friedman equation takes the form of
the energy conservation for a harmonic oscillator, namely
$$
\dot a^2+ \frac 1 3 \Lambda a^2=1,
$$
a Hamiltonian can be defined by setting
$q=a$, $p=\dot a$, and the quantum theory of the harmonic oscillator can be developed according to
\cite{nelson4}. A straightforward calculation leads to the result
$$
E[a^2(t,W)]=a^2_{Cl}+\half \sqrt 3\frac{\hbar}{\Lambda},
$$
where
$E$ is the expectation value, $a_{Cl}$ is the classical value of $a$,
and $W$ is the white noise associated to the stochastic process.
One arrives at the result that the net effect of the environment is to preclude the
collapse of the model, the minimum of the radius being large if $\Lambda$ is
small.

\section{Cyclic universes}
\label{ccyclic}

Oscillating universes
have been explored in several
contexts in an attempt to solve some problems of the standard
cosmological model. The first example of such universes
was that presented in the
seminal paper by Lema\^itre \cite{lemaitre}, who stated that
``The solutions where the universe successively expands and
contracts, periodically reducing to an atomic system with the
dimensions of the solar system, have an incontestable poetic
charm, and bring to mind the Phoenix of the legend'' \cite{lemaitre}
\footnote{Note however that Lema\^itre did not produce an explicit solution
for the cyclic universe.}.
Let us briefly recall some of the issues of the
standard model and the solution that oscillating models can provide:
\begin{itemize}
\item The flatness problem. The Friedmann equation can be written
as
$$
|\Omega_{\rm tot}(t)-1|=\frac{|\epsilon|}{a^2H^2},
$$
As already discussed in Section \ref{sint},
in a situation in which the universe is dominated by matter
or radiation, the difference $|\Omega_{\rm tot}(t)-1|$ grows as a
power of $t$. Since present data indicate that $\Omega_{\rm tot}$ is
very close to 1, it must have been incredibly close to one far in
the past, if $\Omega_{\rm tot}\neq 1$ initially. This is the so-called
flatness problem. As we shall see below, in a cyclic universe
$\Omega_{\rm tot}$ starts deviating from
1 only when $a$ approaches its maximum. Since the maximum grows
with the number of cycles, in a sufficiently old cyclic universe
it may take a long time for $\Omega_{\rm tot}$ to deviate from 1
\cite{durrer}.

\item The horizon problem. In the SCM, light signals can propagate
only a finite distance between the initial singularity and a given time $t$,
provided the energy density changes faster than $a^{-2}$. Hence, microphysics
would not have enough time to take the universe to
its high degree of homogeneity.
In the cyclic model the age of the universe is given by the sum of the duration of
all the previous cycles. This would solve the horizon problem, provided
correlations safely traverse the bounce.

\end{itemize}
Some implementations of the cyclic model
may also solve the so-called ``coincidence problem'' (why did the universe
begin its accelerated expansion only recently?). The model
in \cite{wang} has its parameters tuned in such a way that the
fraction of time that the universe spends in the coincidence state is
comparable to the period of the oscillating universe.

Oscillating models have been also used to explain the observed values of the
dimensionless constants of nature. In \cite{smolin},
the value of these constants is randomly
set after a bounce (see also \cite{mtw}).
In order to see whether cosmological evolution establishes any trend in the behaviour of the
``constants'',
cyclic models were studied in
\cite{magueijo} as solutions of varying-constants theories, such as the
varying $\alpha$ theory presented in \cite{sandvik}, the Brans-Dicke theory, and
the variable-speed-of-light theory \cite{magueijo2}
\footnote{A word of caution regarding this latter type of theory was issued in
\cite{ellisc}.}.
The cyclic solutions were studied both for non-interacting and interacting scalar field
(which models evolution of the ``constant'')
 plus radiation, and the bounce was caused by negative-energy scalar fields. In all three theories, the models showed monotonic changes in the constants from cycle to cycle (the scale factor qualitatively behaving as explained in
\cite{dabro2}).

\subsection{Thermodinamical arguments}
\label{targ}

The existence of oscillatory solutions
in the FLRW model was shown by Tolman (see \cite{tolmanbook} and references therein).
His argument can be understood from a purely mechanic point of view, by
modelling the Friedmann equation as a one-particle system:
examination of the effective potential for a closed universe shows
that there are oscillatory solutions for some values of the
parameters of the model (assuming that there is a mechanism to
revert the contraction
into expansion before the singularity).
These solutions are permitted from a
thermodynamical point of view, since the matter term in the FLRW
model is a perfect fluid, whose entropy is constant. Hence the
expansion is reversible, although at a finite rate.
In more realistic models however, entropy generation is
inevitable, arising from various sources (such as viscosity
effects from particle creation). However notice that, as discussed
in \cite{tolmanbook}, the entropy of each element of the fluid
need not attain a maximum, as would be the case in an isolated
thermodynamical system, because the energy of the fluid element is
not constant. In fact, each time a given element of fluid returns
to the same volume, its energy density is higher than in the
previous passage through the same volume, due to a lag behind
equilibrium conditions. The increment in the entropy leads to
non-reversibility, which forbids identical oscillations. As a
consequence of the raising energy density, the maximum value for
$a$ grows in each cycle
\footnote{Notice that in these
considerations neither the mechanism that allows safe passage
through the singularity nor the details of the entropy generation
are given.}. This can be easily seen from Friedmann equation,
taking the case $\Lambda = 0$, $\epsilon = 1$ as an example:
$$
\dot a ^2 + 1 = \frac{1}{3} \rho a^2.
$$
After one cycle, the 3-volume goes back to a value it had before
when $a$ does. Since $\rho$ grows with the number of cycles, this
growth can only be attributed to an increment in $\dot a$. Hence a
sufficiently ``old´´ cycle is strongly peaked, and $\Omega_{tot}$
remains close to 1 until $a$ is very near the maximum,
thus yielding a solution to the flatness problem.\\
Starting from the fact that the entropy of the universe today is
finite, and making the reasonable hypothesis that the increment in
the entropy through each bounce shares this property, Zeldovich
and Novikov \cite{novi} among others (see \cite{dicus}) have
estimated the number of cycles back to an initial state (which
should not be singular, to keep the idea of a cyclic universe
attractive).

To move from qualitative arguments to actual calculations, the key
issue is the production of entropy. The irreversible energy
transfer from the gravitational field to particle generation was the
source of entropy considered in \cite{prigo}, while it was
suggested in \cite{gunzig,de} that black hole evaporation could be
responsible for the entropy growth. An analytical study that
showed the correctness of Tolman's arguments was presented in
\cite{dabro2}, where closed Friedmann universes with $\Lambda \neq
0$ were scrutinized, including an {\it ad-hoc} mechanism of
entropy generation, and assuming that there is a bounce, without
entering in the details of its realization. The entropy growth was
implemented by relating the constant coming from the conservation
laws
$$
\rho_i{a^\alpha}={\rm const.}= C_i,
$$
where $i$ denotes radiation or dust, and $\alpha = 4$ or 3
respectively, to the expression for the entropy in each case. Let
us take the case of radiation, in which
$$
S_r={\rm constant} = \frac{8}{3}\pi^2\beta T^3a^3,
$$
so we can set $T^3a^3={\rm const.}=\gamma$.
From this equation and the conservation law it follows that
$$
C_r=\frac{G\gamma^{1/3}}{\pi c^4}\;S_r,
$$
thus linking the increment in entropy to the change in the
constant appearing in the solution. In the same way it is shown
that $C_m$ is related to $S_m$ through a similar expression. In
\cite{dabro2} it was assumed that the entropy is constant within a
cycle, but increases at the beginning of each cycle through the
increment in the constants $C_r$ and $C_m$. The behaviour of
models with different combinations of matter, radiation and
cosmological constant were studied for positive and negative
$\Lambda$. The results show that for $\Lambda
>\Lambda_c$ (where $\Lambda_c =\Lambda_c (C_r,C_m)$) the universe
stops its oscillations with increasing maximum and starts an
ever-expanding phase (see Fig. \ref{dabro}). In other words, when the oscillations become
large enough the cosmological constant dominates over the matter
and radiation terms, the oscillations cease, and the universe
enters a deSitter regime. If $\Lambda <\Lambda_c$, the
oscillations are not interrupted.
\begin{figure}[htb]
\begin{center}
\includegraphics[width=0.7\textwidth
]{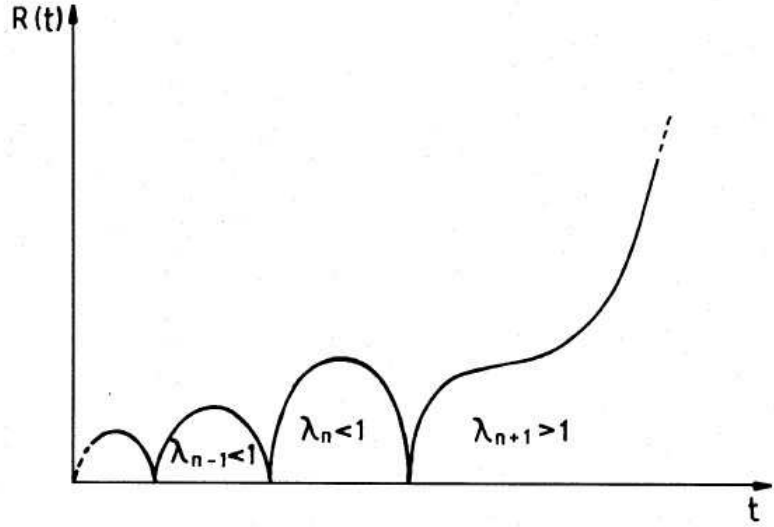}
\caption{The plot shows that transition of a $\Lambda >0$,
radiation-filled universe
from the oscillating phase to the ever-expanding phase, due to
the growth of entropy (given by the
increment in $\lambda_n=\sqrt{4C_{rn}\Lambda/3})$.
Taken from \cite{dabro2}.}
\label{dabro}
\end{center}
\end{figure}
Oscillations in anisotropic models were also studied in
\cite{dabro2}, paying attention to the question of isotropization
after a large number of oscillations.
As the entropy
increases,
the volume of Bianchi I
universes with $\Lambda <0$ oscillates with growing maximum
amplitude, while the shear anisotropy vanishes
\footnote{Axisymmetric Bianchi type IX, dust
Kantowski-Sachs, Bianchi IX, and some features of inhomogeneous
cyclic cosmological models were also studied in
\cite{dabro2}.}.

A more sophisticated model was studied in
\cite{clifton},  where FLRW two-fluid out of equilibrium
models were considered. Exact solutions were found for a particular cases
of the energy exchange, conserving the total energy.
In the case of nonzero spatial curvature,
cyclic models were shown to exist.
The energy exchange between the fluids was modelled by
a function $s$
such that
$$
\dot\rho+3H\gamma\rho=s,\;\;\;\;\;\;\;\;\;\;\dot\rho_1+3H\Gamma\rho_1=-s,
$$
where $\gamma-1$ and $\Gamma-1$ are the EOS parameters of each fluid .
Solutions of these equations along with
$$
H^2=\rho+\rho_1-\frac{\epsilon}{a^2}
$$
were found in \cite{clifton} for different forms of $s$, for the cases radiation and dust,  radiation and scalar field, and radiation and negative vacuum energy. In the second case,  a new feature appears (as well as the ``runaway stage'' mentioned in
\cite{dabro2}): the increment in magnitude of the minima in the scale factor
as time increases. This was interpreted by the authors as a consequence
of the energy exchange: the scalar field reached negative energy values after transferring energy to radiation. Surely this behaviour depends on the specific form of the function $s$.
The examples studied in \cite{clifton} suggest that caution is needed when it is said that cyclic models can solve the flatness problem, since in some of them the
cycles cannot become indefinitely large and long-lived, while in others the minimum of the expansion increases.

\subsection{Realizations of the cyclic universe}

We present in this section some concrete examples of theories that
yield cyclic regular solutions ({\em i.e.} which actually bounce at the minimum of the
expansion without presenting singularities),
along with some of its successes and
conundrums.

\subsubsection{Changes in the matter side of EE}
\label{cms}
One way to generate a cyclic universe is to add matter that will
certainly produce a bounce, and consider next what conditions are to be imposed on
it to produce oscillations. A necessary condition that the extrema of the
expansion factor must satisfy is given by $H=0$, with
$$
H^2= \frac{8\pi}{3M_{\rm Pl}^2}(\rho-f(\rho)).
$$
This amounts to $\rho-f(\rho)=0$, where the
function $f(\rho)$ is positive.
A cyclic universe has been generated along this line in \cite{dabro}, where
``wall-like'' and ``string-like'' matter (whose energy
scales as $a^{-1}$ and $a^{-2}$ respectively) generate the
required $f(\rho)$
\footnote{Earlier attempts along these lines, imposing that
$p\propto -a^{-n}$, and $\rho = p \propto -a^{-6}$ are respectively given
in \cite{pachner} and \cite{rosen2}. For a somewhat different
approach, see \cite{hoyleold}. }. These rather exotic sources can be also
thought as originating from scalar fields under the influence of a
potential, using the procedure presented in \cite{barrow}.
A modification of the Friedman
equation coming from brane models was used to fix the form of
$f(\rho)$ in \cite{freese}, where
\beq
H^2=\frac{8\pi}{3M_p^2}\left(\rho -
\frac{\rho^2}{2|\sigma|}\right),
\label{free2}
\eeq
see Sect.\ref{bbw}.
The dominant component in this model is the so-called ``phantom'' matter,
which has
an energy-conditions-violating equation of state characterized by
$$
\omega_Q = \frac{p_Q}{\rho_Q}<-1.
$$
Since the energy
density of matter with state parameter $\omega$ scales with the
expansion as
$$
\rho = a^{-3(1+\omega )},
$$
we see that $\rho$ grows with the expansion. Surely before
reaching an infinite energy density, quantum gravity effects will
take over the evolution. The somewhat paradoxical situation arises
in which very high-density effects must be incorporated in the
description of the universe for both very small and very large
values of the scale factor. The central idea in \cite{freese} is
that the same physics causes then the bounce and the turnaround,
both governed by Eqn.(\ref{free2}). After a bounce, the universe
follows the standard evolution until the phantom energy dominates.
This energy may erase every trace of structure \cite{caldwell}, and
dominates the evolution until high-density effects are again
important, producing the turnaround.
As will be discussed in Sect.\ref{contraarg}, one of the problems to be faced in
the collapsing phase is the merging of
black holes into a ``monster black hole´´.
The energy density the universe must reach in order that
black holes are torn apart was shown in \cite{freese} to be
$$
\rho_{\rm br}\propto
M_P^4\left(\frac{M_P}{M}\right)^2\frac{3}{32\pi}\frac{1}{|1+3\omega_Q|}.
$$
This energy density must be reached before the turnaround,
characterized by $\rho_{\rm ta}=2|\sigma|$. The value $\sigma\approx
m_{GUT}$ is enough for all but Planck-mass black holes to be torn
apart (some of them evaporate before the universe enters the
phantom energy stage). These Planck-mass remnants may
help in explaining the dark matter puzzle \cite{freese}
\footnote{Details about the evolution
of this model and its relation with the so-called coincidence
problem can be found in \cite{wang}.}. Some
problems still remain in this model. First, the generation of
structures in the contracting phase needs to be addressed, to see
that the black hole problem does not recur. Second, as stated
before, entropy production would lead actually to quasi-cyclic
evolution.

A similar model has been studied in \cite{marek}, given by
\beq
H^2=\frac{1}{3} \rho +\nu \rho^2 +\frac{\Lambda}{3},
\label{freem}
\eeq
where $\nu$ is a real constant. Analytical solutions of this
equation have been found in the case of dust, and their generic
feature seems to be the replacement of the initial singularity by
a bounce, some solutions displaying also a cyclic behaviour (those for $\Lambda\leq0$
and $\nu<0$).

An interesting twist to the entropy problem in cyclic universes was
introduced in \cite{frampton}, where a model described by Eqn.(\ref{freem})
was studied, with the cosmological constant replaced by a dark energy component
with EOS $p=\omega \rho$ and $\omega<-1$,
matter and radiation as normal components, and
$\nu<0$. The  model takes advantage of the Big Rip
phenomenon, where bound systems become unbound and their constituents causally disconnected
as a result of the increasing value of the dark energy density. As a consequence of the
Big Rip, the universe would disintegrate in a huge number of disconnected patches.
The new ingredient of the model is that the turnaround is placed an instant before the
``total
Big Rip'', when each patch would contain almost no matter at all, and  only a small
amount of radiation \cite{frampton2}
and dark energy. Due to the Big Rip, the huge entropy
of the universe is
distributed between the enormous number of patches, hence leaving each patch
with very low entropy. The subsequent contraction of each patch
is free of ``formation of structure'' problems, and proceeds until a bounce occurs. After
the bounce, a normal inflationary phase follows (vastly increasing the entropy),
and the cycle starts again.

\subsubsection{Cyclic universes in nonlinear electrodynamics}
\label{cunlem}

As discussed in Sect.\ref{snled},
nonlinear electrodynamics can describe a nonsingular universe.
Here it will be shown how a cyclic model arises from the theory given by
the Lagrangian \cite{nossou}
\begin{equation}
{\cal L} = -\frac{1}{4}\,F + \alpha\,F^2 - \frac{\gamma^{2}}{F}
\label{11jan1}
\end{equation}
where $\alpha$ and $\gamma$ are constants,
with the dependence of the magnetic field on the scale factor
given by $\mathscr{H} = \mathscr{H}_0 / a^2$ (see Eqn.(\ref{hevol}).
The time-evolution of the scale factor can be qualitatively
described by the effective potential,
which arises from Friedmann equation written as a ``one-particle'' system. For the case at hand, the effective potential is
given by
\beq
V(a)=\frac{A}{a^6}-\frac{B}{a^2}-Ca^6
\label{pot}. \eeq
The constants in $V(a)$ are given by
$$
A=4\alpha \mathscr{H}_0^4,\;\;\;\;\;\;
B=\frac{1}{6}\mathscr{H}_0^2,\;\;\;\;\;\;C=\frac{\gamma^2}{2\mathscr{H}_0^4},
$$
and are all positive.
The analysis of $V(a)$ and its derivatives implies
solving polynomial equations in $a$, which can be reduced to
cubic equations through the substitution $z=a^4$.
The existence and features of the roots of such
equations are discussed in \cite{bir}.
A key point to the analysis is the sign of $D$, defined as follows.
For a general cubic equation
$$x^3+px=q,
$$
the discriminant $D$ is given by
$$
D=\left(\frac p 3 \right)^3+\left(\frac q 2 \right)^2.
$$
We will denote by $D_V$ the discriminant corresponding to the
potential and $D_{V'}$ that of the derivative of V. From the
behaviour of the potential and its derivatives for $a\rightarrow 0$
and $a\rightarrow \infty$ we see that only one or three zeros of the
potential are allowed. The case of interest here (given by
$D_V>0$, $D_{V'}=0$) is plotted in Fig. \ref{potencials}, which
shows the qualitative behavior of the potential for
typical values of the parameters.
\begin{figure}[htb]
\begin{center}
\includegraphics[angle=-90,width=0.5\textwidth]{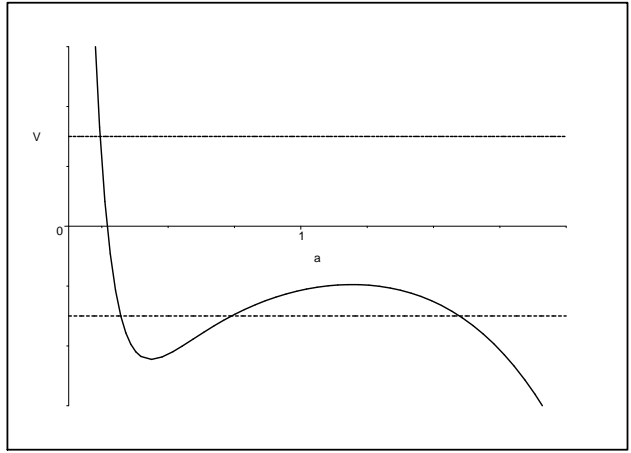}
\caption{Qualitative plot of the effective potential for $D_V>0$, $D_{V'}=0$. The lower
dotted line
corresponds to $\epsilon =1$.}
\label{potencials}
\end{center}
\end{figure}
The model is nonsingular
for any value of $\epsilon$,
and a cyclic model is obtained for $\epsilon =1$.

This setting was generalized in \cite{aline}, where the
Lagrangian
\beq {\cal L}_{T} = \alpha^{2} \, F^{2}
-\frac{1}{4} \, F - \frac{\mu^{2}}{F} + \frac{\beta^{2}}{F^{2}}.
\label{laga} \eeq
was considered, with
$\alpha, \beta$ and
$\mu$ constants. As shown in \cite{aline},
four distinct phases can be described
with this Lagrangian: a bounce, a
radiation era, an acceleration era and a turnaround.
This unity of four stages, christened \emph{tetraktys} in \cite{aline},
constitutes an eternal
cyclic configuration.
The cyclic behavior is
a manifestation of the invariance under the dual map of the scale
factor $ a(t) \rightarrow 1/ a(t),$ a consequence of the
corresponding inverse symmetry of the Lagrangian (\ref{laga}) wrt
the electromagnetic field ($ F
\rightarrow 1/ F,$ where $F \equiv F^{\mu\nu}F_{\mu\nu}$).
Restricting to a magnetic universe, as defined in Sect.\ref{magu},
the Lagrangian ${\cal L}_{T}$
yields for the energy density and pressure given in equations
(\ref{eden}-\ref{pre}):
\begin{equation}
\rho = - \, \alpha^{2}\, F^{2} + \frac{1}{4} \, F +  \,
\frac{\mu^{2}}{F} -  \, \frac{\beta^{2}}{F^{2}},
\label{19dez2}
\end{equation}
\begin{equation}
p = - \,\frac{5 \alpha^{2}}{3}\, F^{2} + \frac{1}{12} \, F - \frac{7
\mu^{2}}{3} \, \frac{1}{F} + \frac{11 \beta^{2}}{3} \,
\frac{1}{F^{2}}. \label{pter}
\end{equation}
As we saw in Sect.\ref{bmu}, for any Lagrangian that is a polynomial in $F$,
$$
\mathscr{H} = \mathscr{H}_{0} \, a^{-2}.
 $$
As discussed in \cite{aline},
the combined system of equations of the FLRW metric
and the magnetic field described by General Relativity and NLED,
are such that the
negative energy density contributions coming from
${\cal L}_{1}$ and ${\cal L}_{4}$ never overcome the positive terms arising
from ${\cal L}_{2}$ and ${\cal L}_{3}.$ Before reaching undesirable
negative energy density values,
the universe bounces (for very large values of the field) and
bounces back (in the other extreme, that is, for very small values)
to precisely avoid this difficulty. These events occur at the values
$\rho_{B} = \rho_{TA} = 0$, which follow from Friedmann's equation
in the case $\epsilon =0$.
Notice that this is not an extra condition imposed by hand
but a direct consequence of the dynamics described by ${\cal L}_{T}.$

Let us now turn to the generic
conditions needed for the universe to have a bounce and a phase of
accelerated expansion.
From Einstein's equations, the acceleration of the universe is
related to its matter content by
\beq 3 \frac{\ddot a}{a} = - \half
(\rho + 3 p). \label{acc} \eeq
In order to have an accelerated
universe, matter must satisfy the constraint $(\rho + 3 p)<0$, which translates into
\beq {\cal L}_F >
\frac{\cal L}{4\mathscr{H}^2}.\label{23set14}
\eeq
It follows that any nonlinear
electromagnetic theory that satisfies this inequality yields
accelerated expansion. In the present model, the terms
${\cal L}_{2}$ and ${\cal L}_{4}$
produce negative acceleration  and ${\cal L}_{1}$ and
${\cal L}_{3}$ yield inflationary regimes $(\ddot{a} > 0).$
%
Raychaudhuri's
equation shows that :
imposes
further restrictions on $a(t)$ at a bounce.
Indeed, the existence of a minimum (or a maximum) for the scale
factor implies that at the bounce point $B$ the inequality
$(\rho_{B} + 3 p_{B})<0$ (or, respectively, $(\rho_{B} + 3 p_{B})>
0 )$  must be satisfied. Note that, as already mentioned, at any extremum (maximum or
minimum) of the scale factor the energy density is zero.
Four distinct
periods can be identified according to the dominance of each term of the energy
density. The early regime (driven by the $F^{2}$ term); the
radiation era (where the equation of state $p = 1/3\rho$ controls
the expansion); the third accelerated evolution (where the 1/F
term is the most important one) and finally the last era where the
$1/F^{2}$ dominates and in which the expansion stops, the universe
bounces back and starts to collapse.
The bounce (for an Euclidean section)
was discussed in Sect.\ref{bmu}.
The standard, Maxwellian term
dominates in the intermediate regime. Due to the dependence on
$a^{-2}$ of the field, this phase is defined by $\mathscr{H}^{2} >> \mathscr{H}^{4}$
yielding the approximation
\begin{eqnarray}
\rho & \approx & \frac{\mathscr{H}^{2}}{2} \nonumber \\
p & \approx & \frac{\mathscr{H}^{2}}{6} \, \label{3Maio11}
\end{eqnarray}
When the universe becomes larger, negative powers of $F$
dominate and the energy density becomes typical of an
accelerated universe, that is:
\begin{eqnarray}
\rho & \approx & \frac{1}{2} \, \frac{\mu^{8}}{\mathscr{H}^{2}}
\nonumber \\
p & \approx & -\frac{ 7}{6} \, \frac{\mu^{8}}{\mathscr{H}^{2}}
\label{4Maio11}
\end{eqnarray}
In the regime between the radiation and the acceleration eras,
the energy content is described by
$$
\rho = \frac{\mathscr{H}^2}{2} + \frac{\mu^2}{2} \frac{1}{ \mathscr{H}^2},
$$
or, in terms of the scale factor,
  \beq \rho = \frac{\mathscr{H}_0^2}{2}\;
\frac{1}{a^4} + \frac{\mu^2}{2\mathscr{H}_0^2}\; a^4. \label{density} \eeq
For small $a$ it is the ordinary
radiation term that dominates. The $1/F$ term takes over only
after $a=\sqrt{\mathscr{H}_0}/\mu$, and grows without bound
afterwards.
Using this matter density in Eqn.(\ref{acc}) gives
$$
3 \frac{\ddot a}{a} +  \frac{\mathscr{H}_0^2}{2}\; \frac{1}{a^4} - \frac 3 2
\frac{{\mu^8}}{\mathscr{H}_0^2}\; a^4=0.
$$
To get a regime of accelerated expansion, we must have
$$
 \frac{\mathscr{H}_0^2}{a^4} -  3
\frac{{\mu^8}}{H_0^2}\; a^4 < 0,
$$
which implies that the universe will accelerate for $a>a_c$, with
$$
a_c = \left(\frac{H_0^4}{3\mu^8}\right)^{1/8}.
$$
%
%
For very large values of the scale factor, the energy density can
be approximated by
\begin{equation}
\rho \approx \frac{\mu^{2}}{F} - \frac{\beta^{2}}{F^{2}}
\end{equation}
and the model goes from an accelerated regime to a phase in which the
acceleration is negative. When the field attains the value $F_{TA}
= 16 \alpha^{2} \mu^{2}$ the universe stops expanding and turns to a
collapsing phase. The scale factor attains its maximum value
$$a^{4}_{max}
\approx \frac{\mathscr{H}_{0}^{2}}{8 \alpha^{2} \mu^{2}}.$$
%
%
Analytic forms for the scale factor in each regime can be found in
\cite{aline}.

\subsubsection{Cyclic universes in loop quantum gravity}

There are realizations
of cyclic models in the effective equations
for loop quantum gravity
(some features of which have been presented in
Sect.\ref{slqg}).
As discussed in Sect.\ref{slqg},
the
Klein-Gordon
equation for a scalar field under the influence of a potential, the Friedmann and
Raychaudhuri's equations in the semiclassical regime
are modified due to quantum gravity effects (see Eqns.(\ref{effkg}-\ref{effray})).
It was shown in \cite{mul} that
positively curved universes sourced by a massless scalar field can
undergo repeated expansions and contractions due to the
modifications described above. This was achieved by rewriting Eqns.
(\ref{effkg}-\ref{effray})
in the form of the classical FLRW model with the
addition of matter described by an effective equation of state,
given by
$$
\omega \equiv
\frac{p_{eff}}{\rho_{eff}}=-1+\frac{2\dot\phi^2}{\dot\phi^2+2DV}\left(
1- \frac 1 6 \frac{d\ln D}{d\ln a}\right).
$$
A violation of the null energy condition, leading to a bounce, is
accomplished when $\omega <-1$, which amounts to $d\ln D/d\ln a
>6$, or $a<0.914a_*$ \cite{mul}, with
$$
D(q)=\left(\frac{8}{77}\right)^6\;q^{3/2}
\{7[(q+1)^{11/4}-|q-1|^{11/4}]-11q[(q+1)7/4\}-|q-1|^{7/4}{\rm
sgn}(q-1)]\}^6,
$$
with $q=a^2/a^2_*$ and $a_*^2=\gamma {\ell}_{Pl}^2 j /3$, where
$\gamma\approx 0.13$, and $j$ is
a quantization parameter, which takes half-integer values.
When $V=0$, $\omega$ is
independent of the kinetic energy of the field,  and an
oscillatory behaviour follows. The addition of a potential leads
to the interruption of the cycles as soon as the potential
dominates the motion (in analogy to what was discussed in Sect.\ref{targ}
for the cosmological constant), and a period of inflation may
follow \cite{mul}. This analysis was later extended to the case of
spatially flat universes, with both negative an positive
potentials \cite{mul2, gregory}.

Yet another realization of a cyclic universe in this scenario is the
so-called \textit{emergent universe from a loop} \cite{ellis2}. As
mentioned in Sect.\ref{eu}, the Einstein universe is unstable, so
perturbations drive the universe away from this state. This
situation partially changes when loop quantum gravity corrections
are considered. Using a phase-space analysis,  it was shown in
\cite{ellis2} that a new static solution appears in the
semiclassical regime ($a<a_*$) for positive potentials (for $V<0$
this is the only solution). This new solution (called loop static,
LS) is stable, and the universe oscillates around it, for $V<V_*$,
with $V_*=39/(136\pi l^2_{Pl}a_*^2)$, while for $V>V_*$, the
equilibrium point corresponding to LS merges with that of the
Einstein universe. So in the model proposed in \cite{ellis2},  the
universe is initially at, or in the neighbourhood of the static
point LS, with $\phi$ in the plateau region of the potential with
$\dot\phi_i>0$. After undergoing a series of non-singular
oscillations in a (possibly) past-eternal phase, while the field
evolves monotonically along the potential, the cycles are eventually
broken as the magnitude of the potential increases, and the universe
enters an inflationary epoch. For this model to work, the potential
must be such that $dV/d\phi\rightarrow 0$ for $\phi\rightarrow
-\infty$, and increase monotonically to exit the cycles
\footnote{Other constraints are imposed by succesful reheating.}. An
example of a suitable potential is given by
$$
V=\alpha \left[ \exp (\beta\phi/\sqrt 3)-1\right]^2,
$$
where $\alpha$ and $\beta$ are parameters that may be constrained by the CMB spectrum.
As in the case of the classical emergent universe discussed in Sect.\ref{eu}, there are some
fine-tuning issues: the scalar field must start in the asymptotically low-energy region of $V$.

\subsubsection{The cyclic universe based on the ekpyrotic universe}
\label{ekpyrotico}

The starting point of the ekpyrotic scenario
\cite{ek}
is five-dimensional
heterotic M-theory \cite{witten}, where the fifth dimension
terminates at two boundary $Z_2$ branes, one of which is
identified with the visible universe. There are two different
versions of the ekpyrotic scenario, the old
\cite{ekold}, where there is a bulk brane between the boundary
branes and the new \cite{eknew}, where only the
boundary branes are present \cite{thesis}. The initial state in
both cases is supposed to be the vacuum state, where the
branes are flat, parallel and empty. The branes are drawn together
by the action of an attractive potential, and collide inelastically over cosmological times. Part of
the kinetic energy is transferred to the branes and used to create
matter and radiation. After the collision, the universe enters a
``standard'' big bang phase, until dark energy domination at the
end of the matter era, which causes an accelerated expansion,
diluting the content of the universe.
The whole process can be described by a 4-d effective theory, with the action (in the
Einstein frame) given by
$$
S_E=\int d^4x\;\sqrt{-g}\left(\half R-\half (\nabla\varphi)^2-V(\varphi )\right),
$$
plus higher-order corrections, where the conveniently-tailored
potential $V(\varphi )$ is responsible for the main features of the
model. The potential is slightly positive for $\varphi >
0$, and goes to zero as $\varphi\rightarrow -\infty$. For $\varphi <0$,
the potential has a minimum and is very steep and negative. The
minimum corresponds to the close approach of the branes, which
happens at such short distances that quantum gravity effects
are relevant. The field $\varphi$ moves rapidly through the
minimum, and the branes collide as $\varphi\rightarrow -\infty$.
Both the old and the new model were shown to
have problems due to excessive fine-tuning
\cite{linde}, so a cyclic version was introduced \cite{ekcyclic}.

In the cyclic ekpyrotic model,
it is assumed that the
interbrane potential is the same before and after collision (instead of being zero,
as in the non-cyclic model). After the branes
bounce and fly apart, the interbrane potential ultimately causes them to draw
together and collide again. To ensure cyclic behavior the
potential must vary from negative to positive values \cite{ekcyclic}.
The model may be adjusted in such a way that,
at distances corresponding to the present-day separation
between the branes, the inter-brane potential energy density is positive
and corresponds to the currently observed dark energy, providing roughly
70\% of the critical density today.
The cosmic acceleration restores
the Universe to a nearly vacuous state and as the brane separation
decreases, the interbrane potential becomes negative.
As the branes approach one another, the scale factor of the Universe,
in the conventional Einstein description, changes from expansion to contraction.
When the branes collide and bounce, matter and radiation are produced
and there is a second reversal transforming contraction to expansion so a new
cycle can begin \cite{ekcyclic}. Figure \ref{potek}
shows a plot of several forms of the potential
that allow for a cyclic universe in this scenario \cite{jus}.
\begin{figure}[htb]
\begin{center}
\includegraphics[width=0.8\textwidth,
]
{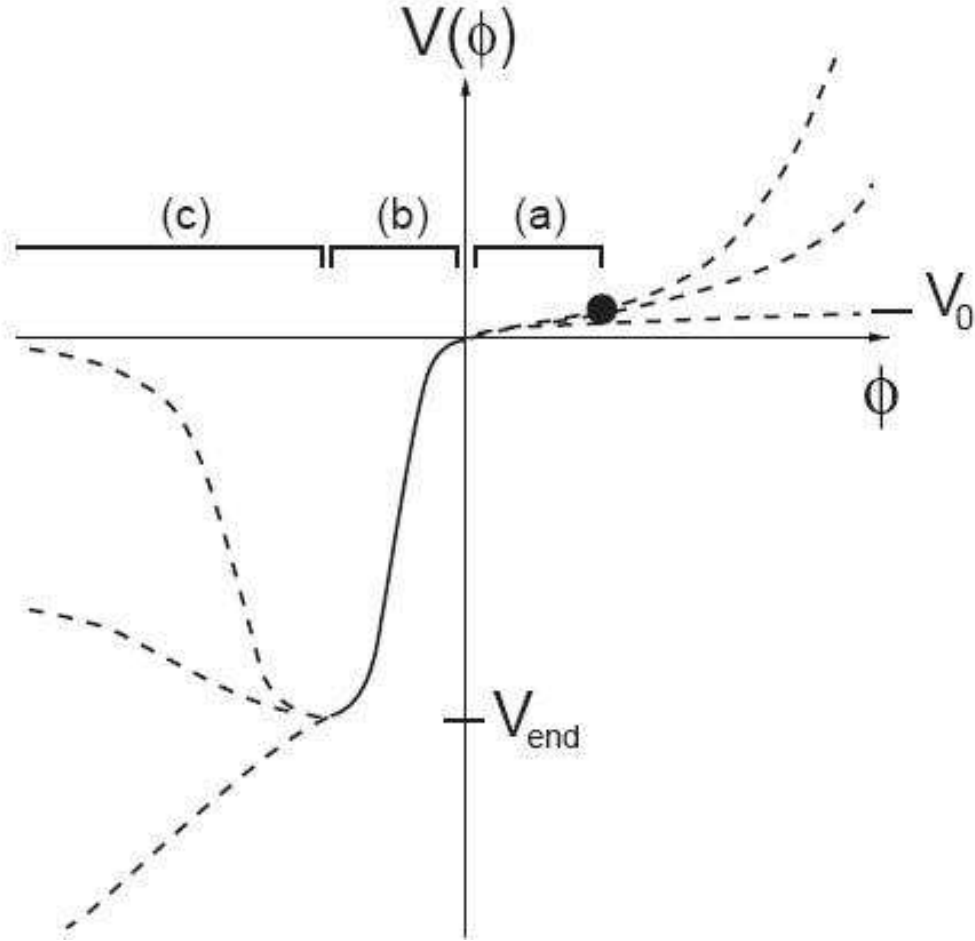}
\caption{Plot of several allowed forms for $V(\varphi )$ .}
\end{center}
\label{potek}
\end{figure}
A qualitative description of the model can be given in terms of this figure as
follows.
Currently,
the field is in region (a), at the point indicated with a dark circle, where the
potential is flat and drives cosmic acceleration. Eventually, the field
rolls towards negative values of $V$ (region b),
where cosmic expansion stops and the universe
(being nearly vacuous as a consequence of the acceleration phase)
enters a phase of slow
contraction, where the spectrum of density perturbations is
generated from quantum fluctuations in $\varphi$.
In region (c) the kinetic energy of $\varphi$ dominates
the energy density.
At the bounce, part of this kinetic energy is
converted into matter and radiation, while the perturbations in $\varphi$
are imprinted
as density fluctuations in the matter/radiation fluid. Meanwhile the field
quickly returns
back to (a) where it comes to a stop, and the universe enters the
radiation-dominated era, so commencing the next cycle.
As recognized by its authors, the
model presents two weak points (as is the case with many cyclic
models): the passage through
the would-be singular point, and the propagation of perturbations
\footnote{This second problem will be discussed in Sec.\ref{cobs}.}.
It is difficult to achieve the bounce without passing from the
semi-classical regime to the high-energy fully quantum
regime, where our use of the effective 4-dimensional theory
breaks down. The problem is that the kinetic energy
and the Hubble rate typically reach Planckian scale as
the branes approach.
In fact,
in the semi-classical regime where
loop corrections can be applied, brane collision may be prevented
\footnote{Some other problems of the model were discussed by Linde
\cite{lindec}.}.

Recently, a ''new ekpyrotic cosmology'' was presented in
\cite{eug}, where a NEC-violating ghost condensate was merged
with an ekpyrotic phase to generate a non-singular bouncing cosmology.
The authors claim to obtain a pre-bounce scale-invariant spectrum
using the
mechanism of entropy perturbation generation \cite{engen}.
This is accomplished by having two ekpyrotic scalar fields rolling down
their respective negative exponential potentials, and having its own
higher-derivative kinetic function. Notice however that the results
of
this model have been challenged in \cite{cha}.

\subsubsection{Oscillatory universe from the Steady State model}
\label{oqss}

The Steady State model \cite{ssm} was proposed as alternative to
the Big Bang model, and has fallen into disfavor because the
observations of the CMB. However, its authors have advanced a new
scenario, called the quasi-steady state model (QSSC, see
\cite{hoyle, sachs, burdos, burtres}). In this model, the
singularity is avoided by the action of a scalar field $C(x)$,
which creates matter in compliance with the Weyl postulate and the
cosmological principle, and has negative energy and stresses.
%
The cyclic solutions in the QSSC can be expected from physical
grounds as follows \cite{sachs}. To create a particle, $C(x)$ must
have energy-momentum equal or larger than that of the particle.
When $C$ is above the threshold, it creates particles and fuels
the spacetime expansion (since it has negative stresses). To this
overall expansion an oscillation is superimposed. The creation of
particles and the expansion set $C$ below the threshold, slowing
down the number of created particles, and the expansion. Here, the
cosmological constant takes control and causes contraction. The
contraction rises the background level of the $C$ field, and the
cycle starts again. As shown in \cite{sachs}, a solution to the
EOM of this theory in the FLRW setting that oscillates in this way
is given by
$$
a(t) = e^{t/P}\left(1+\eta\cos\theta (t)\right),
$$
with $\theta (t)\approx 2\pi t/Q$, where $P$ is the long term
''steady state'' time scale of expansion,  $Q$ is the period of a
single oscillation (with $P>>Q$), and $\eta$ is a parameter .

\subsubsection{Other models}

Due to the recently discovered dark energy component of the universe,
several forms for the dependence of the EOS parameter with the redshift
have been analyzed \cite{chinue}.
In fact, some data
suggest that $\omega (z)$ evolved from a value larger that $-1$ to
a value smaller that $-1$ at some recent redshift. One of the
models that describes this crossing is the quintom model
\cite{chinos1},  where $\omega$ is parameterized as
\footnote{Constraints on this form of dark energy were studied in
\cite{linder}.}
\beq \omega(\ln a) = \omega_0 + \omega_1
\cos[A\ln(a/a_c)],
\label{qmatter}
\eeq with $\omega_0$, $\omega_1$, $A$, and
$a_c$ to be fitted by observations \footnote{Similar ideas were
studied in \cite{undulant}.}. It was shown in \cite{chinos2} that
for a certain choice of the parameters, a universe filled with
quintom matter (that is, matter wit $\omega$ given by
Eqn.(\ref{qmatter})
plus radiation and normal matter expands and
contracts cyclically, yielding an inflationary period at the
beginning of each cycle,  and an acceleration period at the end.

Perhaps it is convenient at this point to remind that
a closed universe has not been
discarded by observation yet (and in fact, cannot be discarded
with certainty due to the errors inherent to any experiment),
though theoretical prejudice and observation tend
to favor $\Omega = 1$. As we saw in Chapter \ref{visco}, a nonzero bulk
viscosity $\zeta$ modifies the fluid pressure according to
$$
p=p_0-3\zeta H,
$$
where $p_0$ is the equilibrium pressure. The asymmetry in the
pressure depending on the sign of $H$ causes the increment in
energy and entropy, leading to ever-increasing cycles. It was
shown in \cite{kanekar} that a similar asymmetry can be caused by
scalar fields in a pure non-dissipative setting. Starting from a
FLRW setting plus a scalar field under the influence of a potential
which displays a minimum, an asymmetry in the pressure, given by
$p\approx -\rho$ for $H>0$, and $p\approx \rho$ for $H<0$
is generated by the oscillations of the field around the minimum
\cite{kanekar}. By imposing appropriate conditions to force a
bounce ($a\rta a$, $\dot a \rta -\dot a$, $\phi\rta \phi$, $\dot
\phi\rta \dot\phi$), it was shown that there is an in increment in
the maximum radius of expansion of the universe in each cycle,
due to conversion of work, done during expansion,
into expansion
energy. The flatness problem is gradually
ameliorated in this model, since the universe becomes
considerably long-lived and more
flat after each expansion.

To close this section, other models of a cyclic evolution for the universe
are listed next:

\begin{itemize}

\item String
theory-inspired cyclic universes inspired,
starting from the property that there exists a minimal
length, ${\ell}_{Pl}$. See \cite{ponjas}.

\item Classical spinor field under the influence of
a quartic potential in a FLRW background was discussed in
\cite{picon}. It was shown that $V=\lambda_\psi +
m\psi\bar\psi-\lambda(\bar\psi\psi)^2$ gives rise to oscillations
in the scale factor, for certain choices of the parameters.

\item A cyclic scenario that takes into account matter and radiation evolution
if the proton has a finite lifetime was studied in \cite{dicus}.

\end{itemize}

\subsection{Issues of the cyclic models}
\label{contraarg}
Cyclic universes are not free of problems. As was put forward in
\cite{penrose}, during a matter-dominated cycle, black holes with
masses ranging from stellar to galactic will form.  During the
contracting phase they will coalesce into a ``monster black hole''
with mass equal to the mass of the universe. Its entropy can be
estimated by
$$
S=\half A = 2\pi R^2 = 8\pi M^2 \gtrsim 10^{124},
$$
where the mass within one Hubble volume ($\approx 10^{23}M_\odot$)
was used. However, the entropy of the radiation in the present
Hubble volume is $\approx 10^{87}$, in such a way that black hole
formation in a previous cycle would lead to a huge excess of
entropy generation. In this scenario, the excess must have somehow
been eliminated by the bounce. But there are some ways out of this
problem. Sikkema and Israel \cite{israel} have suggested that the
inner horizon
of the monster Kerr black hole absorbs strongly
blue-shifted gravitational radiation emitted during the last
moments of the collapse. This radiation increases the mass of the
core of the black hole by a huge amount, rapidly reaching
Planckian values, and correspondingly greatly
reduces its specific entropy. If quantum effects produce a bounce,
this process would allow
the expansion to begin from a state of relatively low disorder
%
Durrer and Laukenmann \cite{durrer} have proposed another solution
to Penrose's problem. They have remarked that black hole
thermodynamics is valid only in asymptotically flat exteriors (a
fact which was also noted in \cite{israel}). They also noted that
the entropy in the radiation we observe today is actually due to
the \emph{previous} matter cycle, which may have had shorter
duration than the current cycle, leading to less clumping and
consequently less entropy production \footnote{Gravitational
perturbations were also studied in \cite{durrer}
}.

Another issue of cyclic models was raised in \cite{ave}, where the
evolution of a cosmic string network was considered in a bouncing
universe. It was shown that the string network displays an
asymmetric behaviour between the contraction and expansion epochs.
In particular, while during expansion a cosmic string network will
quickly evolve towards a linear scaling regime, in a phase of
collapse it would asymptotically behave like a radiation fluid. A
cosmic string network will add a significant contribution, in the
form of radiation, to the energy (and hence also entropy) budget
of a contracting universe, which will become ever more important
as the contraction proceeds. Hence it establishes the need for a
suitable entropy dilution mechanism. This process will also
operate, \textit{mutatis mutandis}, for other stable topological
defects. Conversely, if direct evidence is found for the presence
of topological defects (with a given energy scale) in the early
universe, their existence alone will impose constraints on the
existence and characteristics of any previous phases

\section{Perturbations in bouncing universes}
\label{cobs}

As discussed in the Introduction, inflation can solve many of the shortcomings
of the SCM, but it also has problems of its own. Bouncing models may provide
an alternative (or perhaps a complement) to standard inflation,
since in principle the problems
of the SCM come from a
``shortage of time'' for things to happen early after the big bang
\cite{vene}.
The arguments in Sect.\ref{sint} show that
an accelerated contraction
has the necessary features to solve
the problems of the SCM
\cite{pbbinfl}. Let us recall that if in the contracting phase the Hubble
radius decreases faster than the physical wavelength corresponding
to fixed comoving scales, quantum fluctuations on microscopic
scales can be stretched to scales which
are cosmological at the present time, exactly as it happens in
inflationary models (see for instance \cite{vene}).
Figure \ref{bran01} shows a sketch of the  structure
of a space-time in which standard inflation starts at $t_i$ and ends at $t_R$.
During inflation, the Hubble radius $H^{-1}(t)$ is constant,
and it grows linearly afterwards, while the
physical length corresponding to a fixed co-moving scale increases
exponentially during the period of inflation, and then grows less fast
than $H^{-1}(t)$. The figure shows that for a given $k$, the fluctuation
can be (causally) produced well inside the Hubble radius, ''leave ''
$H^{-1}(t)$, and ''re-enter'' in an appropriate way to describe
the structures we observe today.

Figure \ref{bran02} shows a universe that undergoes a contracting phase, a bounce,
and then enters an expanding epoch, assumed to be that of the SCM.
In this case,
the Hubble radius decreases relative
to a fixed comoving scale during the contracting phase, and increases faster in the
expanding phase. Fluctuations of cosmological interest today are generated sub-
Hubble but propagate outside the Hubble radius for a long time interval.
\begin{figure}[htb]
\begin{center}
\includegraphics[width=0.4\textwidth]{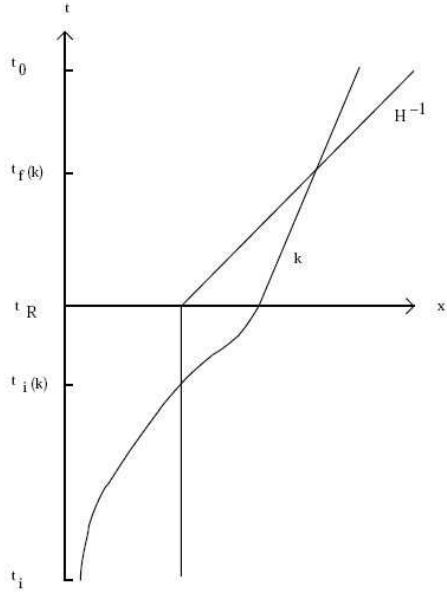}
\caption{Behavior of the comoving scale $k$ and of the Hubble radius
$H^{-1}$
as a function of time in inflation. Taken from \cite{breview}.}
\end{center}
\label{bran01}
\end{figure}
\begin{figure}[htb]
\begin{center}
\includegraphics[width=0.4\textwidth]{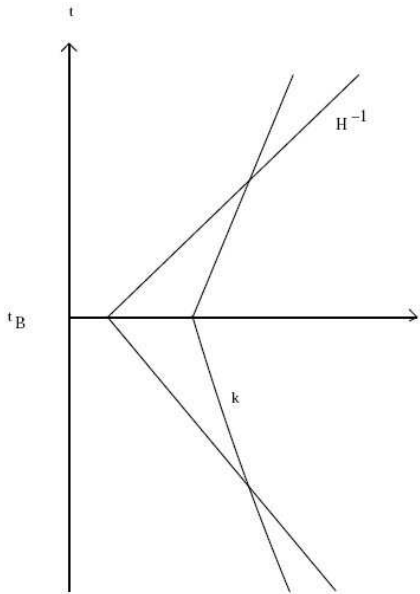}
\caption{Behavior of the comoving scale $k$ and of the Hubble radius
$H^{-1}$
as a function of time in a bouncing universe. Taken from \cite{breview}.}
\end{center}
\label{bran02}
\end{figure}
There is however,
one main difference with respect to the standard
inflationary scenario. In the latter the curvature scale
$R\propto H^2$
is (almost) constant,
while in the
former, it grows until it reaches a maximum and then decreases
\footnote{This assertion
is valid in models in which quantum effects intervene in such a way that
$R_{max}\propto \lambda_{min}^{-2}$, which is the case of loop quantum gravity
for instance, where
$\lambda_{min}\propto\ell_{Pl}$. For the models in which $H$ reaches a null value, $H^2$
can be replaced by $\ell_c=\sqrt{a^3/a''}$, see Eqn.(\ref{zero}.}.
This difference may lead to observational consequences \footnote{See \cite{inflgasp}
for a qualitative discussion of these consequences in the case of string pre-big-bang
cosmology.}, particularly regarding
the generation of a
primordial spectrum of inhomogeneities through parametric amplification
of
the quantum fluctuations of the background fields
in their vacuum state \cite{revmuk}.
These, when decomposed in
Fourier modes, satisfy a canonical Schrodinger-like equation, whose effective potential is
determined by the so-called ``pump field'', which depends in its turn on the background
geometry.
There are then two properties of the
background in a bouncing universe that can affect the final
form of the perturbation spectra \cite{inflgasp}:
(1) the growth of the curvature scale, and
(2) the fields which, together with the gravitational field,
determine the background.
Property (1) has two important consequences. The first,
is that
bouncing scenarios may lead to  ''blue'' ({\em i.e.} growing with frequency)
metric perturbation spectra,
instead of being flat, or decreasing (''red''), as in standard inflation.
A growing spectrum leads to the formation of relic backgrounds
whose amplitude is higher at higher frequency, hence
more easily detectable. A typical example is that of gravitational waves in SPPB
\cite{inflgasp} (see Eqn.(\ref{sgw})). The second is that
the growth of the curvature may also force the comoving amplitude of perturbations to
grow (instead of being frozen) outside the horizon
(see \cite{brustein} for this effect in the SPBB)
\footnote{Consequently, special attention must be taken in the application of
linear perturbation theory, see \cite{brustein}.}.

Regarding Property 2, one of the interesting consequences is the
amplification of the fluctuations of the EM field, due for instance
to the non-minimal coupling with a
scalar field (such as the dilaton, or the scalar field in WIST,
see Sect.\ref{bcreation}). A relic background of scalar
particles is also generated, which may be related to dark matter \cite{dilatongasp}.

There is yet another salient
feature of the perturbations in a bouncing universe. Since
in
the far past of this type of models the universe is assumed to be almost
flat,
one can impose vacuum initial conditions for the perturbations based on simple
quantum field theory in flat space \cite{nelsonrev}, instead of
having to set initial conditions in a high-curvature regime.

It must be remarked that solving for the
perturbations in bouncing models is in principle a
nontrivial task, since there are potential ambiguities that may arise
at the bounce, not present in standard inflation \footnote{For instance, at the
bounce the comoving Hubble scale diverges. Hence all scales are
inside the Hubble scale, at least for an instant. However, there are some
issues common to both scenarios, such as the transplanckian problem
(see for instance \cite{joras}).}. Two
views can be taken to tackle the study of perturbations in such a
scenario. The first one is to devise first a detailed model of the
bounce, and then study the properties of the post-bounce
perturbations. The problem in this case is that
total control of the high-energy physics involved in the bounce is
needed, which is not always achieved. It may also happen that the
bouncing solution under scrutiny is quite artificial from the physical point of
view, as for instance if it is not embedded in any fundamental theory.
But in any case some lessons may be extracted from the examples,
as we shall see in Sect.\ref{exact}.

A second attitude is to make some simplifying assumptions and try
to work out predictions that are independent of the UV physics
that most surely governs the bounce. This possibility has led
to a great debate \cite{debate}. In particular, in order to avoid
the specification of the details near the high-curvature regime,
matching conditions are used, leading to ambiguities.
The dependence of the post-bounce spectrum on the matching
conditions has been addressed by many authors, as will be
discussed in Sect.\ref{match}.

At this point, it is perhaps necessary to say that there are at least two
alternative procedures to deal with gravitational perturbations in
a relativistic setting. Since Lifshitz\rq s original paper
\cite{Lifshitz}, it has been a common practice to start the
examination of the theory of perturbations of
General Relativity by considering variations of non-observable
quantities, such as $\delta g_{\mu\nu}$. The main drawback of this
procedure is that it mixes true perturbations and arbitrary
(infinitesimal) coordinate transformations, which are unphysical.
As shown in \cite{bardeen}, \cite{Jones},
\cite{BrandenbergerKhan}, \cite{revmuk}, this problem can be
solved by adopting gauge-independent combinations of the perturbed
quantities expressed in terms of the metric tensor and its
derivatives. The dynamics of these gauge-independent variables is
then provided by the EE.

A second method exists, based on the quasi-Maxwellian (QM)
formulation of Einstein\rq s equations. The advantage of this method
is that it is gauge-independent from the
start, thus dealing only with observable quantities
\cite{Hawking2}, \cite{Olson}, \cite{Novello1,Novello2,Novello3}
\cite{Vishniac}.
We shall briefly review both methods in Sect.\ref{apert},
including a summary of the relation between them.

In the next sections we shall discuss examples of the two
approaches.
From an observational point of view, the crucial question is whether
bouncing models can furnish a
nearly-scale invariant spectrum of adiabatic scalar perturbations after the
bounce,
as demanded by the measurements of the WMAP \cite{wmap},
Sloan survey \cite{sdss}, and 2df \cite{2df}.
It is also of interest to see if bouncing solutions lead to
observable consequences that are markedly different
from those of inflation (see Sect.
\ref{conc}).
%

\subsection{Regular models}
\label{exact}

In the previous chapters, we have seen that it is possible to generate
bouncing models in a wide choice of scenarios, essentially by any of the
mechanisms presented in Sect.\ref{defsing}.
Obviously, the
outcome is very dependent on the choice, but specific models
can be sometimes useful in the hope of extracting tendencies of a more
general behaviour. In this sense, scalar, vector, and tensor perturbations
have been studied in many exact backgrounds displaying a bounce. An incomplete list includes
the following:
\bi
\item General relativity with
radiation and a free scalar field having negative energy
\cite{nelson1},

\item Two scalar fields \cite{kim, wands,marco}.

\item A 5d Randall-Sundrum model with radiation, in which
the extra dimension is timelike \cite{batte}.

\item Two perfect fluids \cite{fpp}.

\item A nonlinear EM Lagrangian \cite{novsal}.

\item A scalar field with higher-order corrections from string
theory, with an exponential potential (this case covers the SPBB
and the first version of the Ekpyrotic universe) \cite{tbf}.

\item A non-canonical scalar field, with Lagrangian
${\cal L}=p(X,\phi)$, where $X=1/2g^{\mu\nu}\partial_\mu\phi \partial_\nu\phi$
\cite{ap}.

\item Bounce
due to quantum cosmological effects
using Bohmian solutions of the canonical Wheeler-de Witt
equation \cite{pinho}.

\item Non-local dilaton potential stemming from string theory
\cite{ggv}.
\ei


We shall present next a short discussion of
scalar, tensor, and vector perturbations in some of these scenarios.

\subsubsection{Scalar perturbations}
\label{sp}
The evolution of scalar perturbations through a bounce has been a
subject of intense debate (see references in \cite{bozzadef}). A
consensus
for the case of
a two-component bouncing model in GR
seems to have been reached after \cite{bozzadef}.
This
model is described by a flat FLRW metric,
and one of the components has negative energy
density (to produce the bounce) and is important only near the
bounce. The components interact only gravitationally, and the
component that dominates away from the bounce has an intrinsic
isocurvature mode, in order to describe scalar fields or perfect
fluids.
The result obtained in \cite{bozzadef} is
that the spectrum of the growing mode of the Bardeen
potential in the pre-bounce is transferred to a decaying mode in
the post-bounce
\footnote{These result is supported by the references cited in
\cite{bozzadef} and also by the results in \cite{fpp}.},
\footnote{Notice that
mode-mixing is possible with $\epsilon =1$, as for instance in
\cite{pertposit}.}.

Since the
phenomenology associated to the decaying mode is
known to differ from observation \cite{ame}, we may ask what can
be done to lift the negative result of \cite{bozzadef}. One
possibility is to allow the fluids to interact. Another one is to
incorporate in the background solution the decay of the normal component to radiation
\footnote{See Sect.\ref{matcre} and \cite{nilton} for an exact
solution that has this feature.}. Yet another possibility is to
consider higher-order corrections. This has been done in several
string-inspired models \footnote{The string pre-big-bang model
without corrections furnishes a highly blue-tilted spectrum
$n_s=4$
of scalar perturbations \cite{brustein}.}, in the gravi-dilaton regime
by exploring regular backgrounds (such as those presented
in Sect.\ref{ssst}), as in
\cite{carhwang},\cite{tsuji2},\cite{tbf},
\cite{cartieralpha}. The results presented in these articles
show that although it may be possible to generate a nearly scale-invariant
spectrum in the pre-bounce phase, it corresponds to the decaying mode in the expanding
phase \footnote{The SPBB model may yield the right spectrum when axion
fluctuations are considered \cite{curvaton}.}. An exception is the model
presented in Sect.\ref{sp}.
Another exception may be the
ekpyrotic model,
where there are results indicating that
a scale-invariant spectrum may be obtained in the post-bounce phase
\cite{pbf} \footnote{See ref.\cite{abb} for another model
in which the growing mode in the contracting phase
goes over into the dominant mode in the post-bounce phase.}.

Another set of models comes from the quantum evolution of the universe.
As discussed in Sect.\ref{onto}, bouncing solutions are possible (without the need of
a ''phantom'' field)
in the context of the WdW equation, when the Bohm-de
Broglie interpretation
is used in the mini-superspace approach. A feature of this scenario is that
a full quantum treatment of both background
and perturbations is possible \cite{ppp,pinho2}.
The model analyzed in \cite{pinho} is GR plus a perfect fluid, in which the scalar
perturbations
can be described in terms of a single degree of freedom, related to the
Bardeen potential $\Phi$ (see Appendix).
The Bohmian quantum trajectory for the scale factor
is given by
\beq
a(T)=a_0\left[a+\left(\frac{T}{T_0}\right)^2\right]^\frac{1}{3(1-\omega)},
\eeq
with
$p=\omega\rho$.
The normal modes of the scalar perturbation satisfy the equation
\beq
v''_k+\left(\omega k^2-\frac{a''}{a}\right) v_k=0,
\eeq
where a prime means derivative wrt conformal time.
Following the usual procedure of expanding the modes for large (negative and positive)
values of $T$, matching the expansions, and then transforming to the
Bardeen potential, the power spectrum defined by
\beq
P_\Phi=\frac{2k^3}{\pi^2}|\Phi^2|\propto k^{n_s-1},
\eeq
yields for the post-bounce phase \cite{pinho}
\beq
n_s=1+\frac{12\omega}{1+3\omega}.
\eeq
An analogous calculation for the tensor modes gives
\beq
n_T=\frac{12\omega}{1+3\omega}.
\label{sps}
\eeq
Notice that a scale-invariant spectrum follows both for the scalar and the tensor perturbations
for the case of dust ($\omega = 0$), which is the fluid supposed to dominate
the evolution at the time of the
matching of the solutions (not necessarily the same governing at the time of the
bounce) \cite{pinho}.
An important lesson that follows from this example
and the one presented in \cite{fpp} (see Sect.\ref{tp}) is that
the spectral index is quite insensitive to the details of the bounce, being determined mostly by
the dominant component. The example also shows that
the bounce is important in the mixing of the
modes, which is relevant for the amplitude of the modes in the post-bounce phase.

\subsubsection{Vector perturbations in a contracting background}

It is a well-known result of perturbation theory that vector
perturbations (VPs) only exhibit decreasing solutions in the
context of an expanding Universe (see for instance \cite{revmuk})
\footnote{Another interesting result is that the simplest models of inflation
do not produce VPs, see for instance \cite{lindeu}.}.
However, as shown in \cite{battecont}, VPs
can increase in a contracting flat background, with a perfect fluid as source.
Hence, they might provide a signature of a bounce.
As shown in the Appendix,
the relevant equations are
$S^i_{\; k}=C^i_{\; k}/a^2$, where $C$ is a constant, and
\beq
V^i_k\propto \frac{k^2C^i_{\; k}}{a^{1-3\omega}}.
\eeq
Note that $V^i_k$ increases for $\omega =0$, and stays constant for radiation, but
$S^i_{\; k}$ always increases for decreasing $a$.
As argued in \cite{battecont}, VPs cannot be neglected in the SPBB scenario, in such a way that
near
a bounce, the metric perturbations may become too large for the use of linear theory (depending
on the value of the $C^i_{\; k}$) \footnote{Quantum corrections to the evolution of vector modes
were studied in the context of loop quantum gravity in \cite{bojovec}}.
Related results were presented in
\cite{giovavec}, where it was also shown that the growing vector mode
matches with a decaying mode after the curvature bounce, in the context of a
low-energy flat gravi-dilaton model \footnote{This is not nececssarily so in multidimensional
cosmological models, also analyzed in \cite{giovavec}.}.

Since many bouncing models are generated by
a scalar field, a relevant question is whether VP are
important in this type of scenarios. One important point is that VPs are not
supported by a scalar field at first order.
At second order, the scalar, vector, and tensor modes couple, and VPs
can be generated by scalar-scalar mode couplings \cite{mena}.
Considering exponential
potentials and power-law solutions, the ratio of the amplitudes
of second order vector perturbations in contracting and expanding phases was studied in
\cite{mena}.  The relative
magnitudes of the second order vector perturbations in the two phases depend on the
scaling solutions chosen, but at least in one of the examples studied
(dust-like collapse, \cite{fibra}),
the observable
differences between the collapsing models and the inflationary scenario could be large,
assuming that
the transition between the two phases
does not significantly alter the ratio.

\subsubsection{Tensor perturbations}
\label{tp}

The spectrum of gravitational waves can be a very powerful tool to discriminate between different
models of the universe, since gravitational waves decouple
very early from matter and travel undisturbed,
as opposed to EM waves. In particular, in the context of the SPBB
scenario, the amplification of tensor perturbations
is greatly enhanced wrt the standard inflationary scenario for
large comoving wavenumber $k$ \cite{gasper}.
This result was confirmed in \cite{brustein}, with a gravi-dilaton background solution of
the EOM
\beq
G_{\mu}^{\; \nu} =\half \left( \partial_\mu\varphi\partial^\nu\varphi-
\half\delta_\mu^{\; \nu}\partial_\alpha\varphi\right),
\eeq
\beq
\Box\varphi =0,
\eeq
given by
$$
a(\eta)=(-\eta)^{1/2}, \;\;\;\;\;\;\;\varphi (\eta) = \frac{-3-\sqrt{3}}{1+\sqrt{3}}\ln (-\eta)+{\rm const.},
$$
the typical amplitude for the normalized vacuum tensor
fluctuations outside of the horizon over a scale $k^{-1}$ is given by \cite{brustein}
\beq
|\delta_{h_k}(\eta )|\approx\left(\frac{H_1}{M_{Pl}}\right)(k\eta_1)^{3/2}\ln |k\eta|,
\label{sgw}
\eeq
where $H_1\approx (a_1\eta_1)^{-1}$ is the final contraction scale
\footnote{Scalar perturbations of this model were also investigated in \cite{brustein},
and present amplitudes and spectra similar to the tensor perturbations.}, while the
result in the standard inflationary
expansion does not have the $\ln$ dependence (see for instance \cite{solo}).
The possible influence of the nonperturbative phase, where the curvature and the dilaton are very large,
was studied by imposing a bouncing solution in \cite{tanoss}, and
by taking into account higher-derivative $\alpha '$ and quantum corrections (see Sect.\ref{ssst})
\cite{tano3},
\cite{caco}. The results in these papers show
that
that the low frequency modes, crossing the
horizon in the low-curvature regime, are unaffected by higher-order corrections, and also that
the shape
of the spectrum of the relic graviton background, obtained in the context of the pre-Big
Bang scenario, is strongly model-dependent.

This analysis was continued in \cite{guo}, where
cosmological perturbations in the
low-energy string effective action with a dilaton coupling
$F(\phi )$ were studied, with the addition of
a Gauss-Bonnet term, a kinetic term of the type
$(\nabla\phi )^4$, and a potential $V(\phi )$. Scale-invariant spectra in the string frame
and a suppressed tensor-to-scalar ratio were obtained by imposing
slow-roll inflation in the Einstein frame. The results show that it is practically impossible
to obtain these conditions without the second-order corrections
given by Eq.(\ref{hoc}), both with and without the Gauss-Bonnet term.

Analytic
and numerical results for the tensor post-bounce spectrum have been obtained for a two-component
model defined by $p_\pm=\omega_\pm \rho_\pm$ \cite{fpp}. The flat background is given by
$$
a(\tau) = a_0\left(1+\frac{\tau^2}{\tau_0^2}\right)^\alpha,
$$with
$$
d\tau = \frac{dt}{a^\beta},\;\;\;\;\;\;\beta = \frac 3 2 (2\omega_+-\omega_-+1),
$$
$$
\alpha =\frac{1}{3(\omega_-\omega_+)},\;\;\;\;\;\;\;a_0=
\left(\frac{\gamma_-}{\gamma_+}\right)^\alpha,
\;\;\;\;\;\;\tau_0^2=\frac{4\alpha^2}{{\ell}^2_{\rm Pl}}\frac{\gamma_-}{\gamma_+^2},
$$
$\gamma_+$ and $\gamma_-$ are constants, with $\gamma_-<0$, to produce the bounce.
The tensor spectrum, assuming that $-1/3<\omega_+<1$, and that the potential
that arises from Eqn.(\ref{zero})
has only one extremum at $\tau =0$, is given by \cite{fpp}
$P_h\propto \tilde k^{n_T}$, where
$$
n_T=\frac{12\omega_+}{1+3\omega_+}.
$$
Note that the spectral index does not depend on the
EOS parameter of the ``exotic'' fluid (contrary to the case of the spectral index
for the scalar perturbations). This was to be expected
since large wavelengths are comparable to the curvature scale of the
background at a time when the universe is still far from the bounce, so
the behaviour obtained in this case can be taken as generic, \textit{{i.e.}}
independent of the details of the bounce.

Yet another example of the calculation of a tensor spectrum in a
bouncing model was presented in Sect.\ref{sp}, based on the
quantum evolution (using the Bohmian quantum trajectory) of a
universe dedcribed by GR plus a perfect fluid. The result is (see
the comments after Eqn.(\ref{sps})) \beq
n_T=\frac{12\omega}{1+3\omega}. \eeq In fact, the tensor-to-scalar
ratio in this model was estimated as $T/S\approxeq 5.2\times
10^{-3}$, and the the characteristic bounce length-scale $L_0
\approxeq  1500 {\ell_{Pl}}$, (assuming that $n_s\lesssim 1.01$)
which is a value in the range in which quantum effects are
expected to be relevant, while at the same time the Wheeler-de
Witt equation is valid (without corrections from stringy/loop
effects).

\subsection{Scalar perturbations in exact models using the quasi-Maxwellian framework}
\label{spqm}
As mentioned in the introduction of this chapter,
perturbations can also be studied using the
quasi-Maxwellian (quasi-Maxwellian) method.
In this section we apply it
to two exact bouncing solutions. The first one is
generated by the non-minimal coupling of
the electromagnetic field with gravity (see Sect.(\ref{examplensu})).
As discussed in the Appendix, in the quasi-Maxwellian formalism
the scalar perturbations are completely described by the variables $E$
and $\Sigma$, which obey the equations (\ref{e8})-(\ref{f2}):
$$
\dot{E} = - \, \frac{1 + \lambda}{2} \, \rho \, \Sigma -
\frac{1}{3} \, \theta \, E,
$$
$$
\dot{\Sigma} = \left[  \frac{6 \lambda}{1 + \lambda} \left( \epsilon +
\frac{k^{2}}{3} \right) \frac{1}{ a^{2} \, \rho} - 1 \right] \, E,
$$
with $p=\lambda \rho$, and $k$ is the wave number (the subindex $k$ in $E$ and $\Sigma$
has been omited).
Combining these, we obtain the equation for the time
evolution of the electric part of the perturbed Weyl tensor:
\begin{equation}
\ddot{E} +  \dot{E}\left( \frac{4}{3}  + \lambda \right) \, \theta
+ E X = 0, \label{15dez1}
\end{equation}
where $X$ is a function of the background functions given by
$$ X \equiv \lambda \frac{ 3\epsilon + k^{2}}{a^{2}} -
\left( \lambda + \frac{2}{3} \right)  \, \rho + \frac{2 +
3\lambda}{9} \, \theta^{2}. $$
Defining a new function $g(t)$ by  $g = E \, a^{-\sigma}$, where $ \sigma
\equiv -(4 + 3\lambda)/2$, we obtain from
Eqn.(\ref{15dez1})
\begin{equation}
\ddot{g} +  \chi(t) \, g = 0, \label{15dez2}
\end{equation}
where \footnote{Notice that, as shown in the Appendix,
this equation is actually a consequence of a transformation that
takes the variables $(E,\Sigma )$ (which are not canonically conjugated)
into a new pair of variables that are canonically conjugated.}
\begin{equation}
\chi (t) \equiv \sigma \, \frac{\ddot{a}}{a} \, - \sigma (\sigma + 1) \,
\left(\frac{\dot{a}}{a}\right)^{2} + X. \label{9mar2}
\end{equation}
In the case of the bouncing universe given by Eqn.(\ref{aex}), we have
\begin{equation}
\left( t^{2} + \alpha_0^{2} \right)^2\, \ddot{g}+ \left( \alpha \, t^{2} +
\beta \, \alpha_0^{2} \right) g = 0 \label{15dez3},
\end{equation}
where $ \alpha \equiv k^{2}/3 - 7/4$ and $ \beta
\equiv k^{2}/3- 1/2$.
With the change of variable
$ z = 1/2 - it/(2\alpha_0)$, this equation takes the
form
\begin{equation}
\frac{d^{2} g}{dz^{2}} + I(z) \, g = 0, \label{23marco1}
\end{equation}
where
\begin{equation}
I(z) = - \, \frac{\beta}{4 z^{2} \, (z - 1)^{2}} + \frac{\alpha \
( 2z - 1 )^{2}}{4 z^{2} \, (z - 1)^{2}}. \label{23marco2}
\end{equation}
After a direct calculation, Eqn.(\ref{23marco1}) can be transformed into a
hypergeometric equation
\begin{equation}
z (1 - z) \, \frac{d^{2} \omega}{dz^{2}} +  [ c - ( a + b + 1) z ]
\frac{d \omega}{dz}  - a b \omega = 0, \label{10mar3}
\end{equation}
where \begin{equation}
 a = \frac{1}{2} + \sqrt{\frac{1}{4} - \alpha},
 \label{23marco3}\end{equation}
\begin{equation}
 b = \frac{1}{2} - \sqrt{\frac{1}{4} - \alpha},
 \label{23marco4}\end{equation}
\begin{equation}
 c = \frac{5}{2}.
 \label{23marco5}\end{equation}
The solution for $g(z)$ is given by
 \begin{equation}
g(z) = z^{\frac{c}{2}} \, (z - 1)^{\frac{c - a - b - 1}{2}} \,
\omega (z)\label{23marco6}
\end{equation}
ot,  in terms of the hypergeometric function
$F(a,b,c;z)$,
\begin{equation}
g(z) = z^{\frac{5}{4}} \, (z - 1)^{- \,\frac{1}{4}} \,
F\left(\frac{1}{2} + \sqrt{\frac{1}{4} - \alpha}, \frac{1}{2} -
\sqrt{\frac{1}{4} - \alpha}, \frac{5}{2}; z\right).
 \label{23marco7}
\end{equation}
Finally, the solution for the electric part of the Weyl tensor,
is given by
\begin{equation}
E_{k} = s ( -4 \alpha_0^{2})^{-\frac{5}{4}} \, ( z - 1)^{- \,\frac{3}{2}}
\,F\left(\frac{1}{2} + \sqrt{2 - \frac{k^{2}}{3}}, \frac{1}{2} -
\sqrt{2 - \frac{k^{2}}{3}}, \frac{5}{2}; z\right) \label{10mar5}
\end{equation}
where $s$ is a constant. Restricting to $z\in\Re$,
it follows that this solution is regular for $z<1$, and
can be analitically extended for all values of $z$. Hence, the
perturbation is regular.

Notice that the power spectrum of the perturbations
can be obtained using (see Appendix)
\beq
P_k=k^{-1}|E_k|^2.
\label{esp}
\eeq

The second example we shall study in this section
is the model presented in Sect.\ref{magu}, the perturbation of wihch
was analyzed by
the quasi-Maxwellian method in \cite{novsal}.
In this model, the singularity is
avoided by the introduction of nonlinear corrections to Maxwell
electrodynamics, given by
\begin{equation}
L = -\frac{1}{4}\,F + \alpha\,F^2 + \beta\,G^2,
\end{equation}
where $ F=F_{\mu\nu}F^{\mu\nu}, G \stackrel{.}{=} 
\frac{1}{2}\eta_{\alpha\beta\mu\nu}F^{\alpha\beta}F^{\mu\nu}$,
$\alpha$ and  $\beta$ are arbitrary constants. After an average procedure (see
Sect.\ref{magu}),
the expression for the scale factor for the ''magnetic unverse'' with
$\epsilon=0$ is:
\begin{equation}
a(t)^2=\mathscr{H}_0\left[\frac{2}{3}(t^2+12\alpha)\right]^{1/2}.\label{at}
\end{equation}
The interpretation of the source as a one-component perfect
fluid in an adiabatic regime leads to
instabilities \cite{nelsonns}, which are artificial,  as will be seen next.  The sound
velocity of the fluid in this case is given by \cite{Landau}
\begin{equation}
\frac{\partial p}{\partial\rho}
=\frac{\dot{p}}{\dot{\rho}}
=-\frac{\dot{p}}{\theta(\rho +p)}.
\end{equation}
This expression, involving only background quantities, is not defined at
the points where the energy density attains an extremum given by
$\theta=0$ and $\rho+p=0$. In terms of the
cosmological time, these points are determined by $t=0$ and $t=\pm
t_c=12\alpha$. Notice that they are well-behaved regular points
of the geometry, indicating that the occurrence of a singularity is
in fact caused by an inappropriate description of the source.
This difficulty can be circumvented by splitting
the part coming from Maxwell's dynamics
from the additional non-linear $\alpha-$dependent term in the
Lagrangian. As a result, we get two
noninteracting perfect fluids:
\begin{equation}
T_{\mu\nu} = T^{(1)}_{\mu\nu} + T^{(2)}_{\mu\nu},
 \protect\label{ex1}
\end{equation}
where
\begin{equation}
\protect\label{Tmunubis1} T^{(1)}_{\mu\nu} = (\rho_{1} + p_{1})\,
v_{\mu} v_{\nu} - p_{1}\,g_{\mu\nu},
\end{equation}
\begin{equation}
\protect\label{Tmunubis2} T^{(2)}_{\mu\nu} = (\rho_{2} + p_{2})\,
v_{\mu} v_{\nu} - p_{2}\,g_{\mu\nu},
\end{equation}
and
\begin{eqnarray}
\rho_{1} &=& \frac{1}{2} \, \mathscr{H}^2
\label{rho1},\\[1ex]
\protect p_{1} &=& \frac{1}{6} \,\mathscr{H}^2, \label{P21} \\[1ex]
\protect  \rho_{2} &=&- 4\alpha \, \mathscr{H}^4,
\label{rho2}\\[1ex]
\protect p_{2} &=& -\, \frac{20}{3}\, \alpha \,\mathscr{H}^4. \label{P22}
\end{eqnarray}
From this decomposition it follows that each of the
components of the fluid satisfies the conservation equation, thus
showing that the source can be described by
two non-interacting perfect fluids with equation of states $ p_{1}
= 1/3 \,\rho_{1}$ and $ p_{2} = 5/3 \,\rho_{2}.$  This splitting
should be understood only as a
mathematical device to allow for a fluid description.

From the considerations presented in Sect.\ref{another}
we obtain \cite{novsal}:
\begin{equation}
\dot{\Sigma_{1}}= -
\left(\frac{2\lambda_{1}(3\epsilon+k^2)}{a^2(1+\lambda_{1})\rho_{1}}+1\right)E_{1},
\end{equation}
\begin{equation}
\dot{\Sigma_{2}}= -
\left(\frac{2\lambda_{1}(3\epsilon+k^2)}{a^2(1+\lambda_{2})\rho_{2}}+1\right)E_{2},
\end{equation}
\begin{equation}
\dot{E_{1}}+\frac{1}{3}\theta
E_{1}=-\frac{1}{2}\left(1+\lambda_{1}\right)\rho_{1}\Sigma_{1},
\end{equation}
\begin{equation}
\dot{E_{2}}+\frac{1}{3}\theta
E_{2}=-\frac{1}{2}\left(1+\lambda_{2}\right)\rho_{2}\Sigma_{2},
\end{equation}
where $k$ is the wave number.
As shown in \cite{Novello1}, the scalar
perturbations can be expressed in terms of the two basic
variables $E_{i}$ and $\Sigma_{i}$, and the corresponding equations
can be decoupled. The result in terms the $E_{i}$ is
\beq
\ddot{E}_{i}+\frac{4+3\lambda_{i}}{3}\theta \dot{E}_{i}+ \left\{\frac
{2+3\lambda_{i}}{9}\theta^2
 \left(\frac{2}{3}+\lambda_{i}\right)\rho_{i} \frac{1}{6}(1+3\lambda_{j})\rho_{j}
-\frac{(3\epsilon+k^2)\lambda_{i}}{a^2} \right\} E_{i}=0.
\eeq
Note that in this
expression there is no summation in the indices, and $j\neq i$ , and
$\lambda_{i}=\left(\frac{1}{3}, \frac{5}{3}\right)$. In the first
case the equation for the variable $E_{1}$ becomes
\begin{equation}
\ddot{E}_{1}+\frac{5}{3} \theta \dot{E}_{1}+\left[\frac
{1}{3}\theta^2-\rho_{1}-\rho_{2} -\frac{5 }{3 a^2}\right]
E_{1}=0.
\end{equation}
Let us analyze the behavior of the perturbations in the
neighborhood of the points where the energy density attains an
extremum ({\em i.e.} the bounce and the
point in which $\rho + p$ vanishes).
The expansion of the equation of $E_1$ in the neighborhood of the
bounce (at  $t=0$) up to second order, is given by:
\begin{equation}
\ddot{E_1}+at\dot{E_1}+(b+b_1t^2)E_1=0\protect\label{equa1},
\end{equation}
where the constants $a$ and $b$ are defined as follows
\begin{equation}
a=\frac{5}{2t_c^2},
\end{equation}
\begin{equation}
b=-\frac{k^2}{\sqrt{6} \mathscr{H}_0 t_c},
\end{equation}
\begin{equation}
b_1=-\frac{b}{2t_c^2}-\frac{3}{4t_c^4}.
\end{equation}
Defining a new function $f$ as
\begin{equation}
f(t)=E_1(t)
\exp\left\{\left(+\frac{a}{4}-\frac{i}{2}\sqrt{b_1-\frac{a^2}{4}}\right)t^2\right\},
\end{equation}
and introducing the coordinate $\xi$ by
\begin{equation}
\xi=-i t^2  \sqrt{b_1-\frac{a^2}{4}},
\end{equation}
we obtain for $f$ the confluent hypergeometric equation
\cite{Abramowitz}
\begin{equation}
\xi \ddot{f}+(1/2-\xi)\dot{f}+ e f=0,
\end{equation}
where
\begin{equation}
e=\frac{i(b-a/2)}{4(b_1-a^2/4)^{1/2}}-\frac{1}{2}.
\end{equation}
The
solution of this equation is given by
\begin{equation}
f(t)= A \;  M\left(d,1/2,-i t^2 \sqrt{b_1-\frac{a^2}{4}} \right),
\end{equation}
where $A$ is an arbitrary constant and $M(d,1/2,\xi)$ is the confluent
hypergeometric function, which is
well-behaved in the neighborhood of the bounce. Hence the perturbation
$E_{1}(t)$ is regular and given by
\begin{eqnarray}
E_1(t)&=&  A \; M\left(d,1/2,-i t^2  \sqrt{b_1-\frac{a^2}{4}}\right)\nonumber \\
&\times&\exp\left\{\left(-\frac{a}{4}+\frac{i}{2}\sqrt{b_1-\frac{a^2}{4}}\right)t^2\right\}.
\end{eqnarray}
After a similar procedure, the perturbation $E_2$ obeys, in the same neighborhood,
the following equation:
\begin{equation}
\ddot{E_2}+at\dot{E_2}+(b+b_1t^2)E_2=0.\protect\label{equa2}
\end{equation}
This is the same equation we obtained for $E_1$, with
different values
of $a, b$ and $b_1$ given in this case
by
\begin{equation}
a=\frac{9}{2t_c^2},
\end{equation}
\begin{equation}
b= \frac{3}{2t^2_c}- 5\frac{k^2}{\sqrt{6} \mathscr{H}_0 t_c},
\end{equation}
\begin{equation}
b_1=-\frac{5 k^2}{t_c^3 \mathscr{H}_0\sqrt{6}}-\frac{5}{t_c^4}.
\end{equation}
The solution is given by the real part of
\begin{eqnarray}
 E_2(t)&=& A\, M\left(d,1/2,-i  t^2 \sqrt{b_1-\frac{a^2}{4}}\right)\nonumber \\
 &\times& \exp\left\{-\left(\frac{a}{4}-\frac{i}{2}\sqrt{b_1-\frac{a^2}{4}}\right)t^2\right\},
\end{eqnarray}
so the perturbation $E_{2}(t)$ is well-behaved. At
the neighborhood of the other critical point, given by
$t=t_c$, the equation for the
perturbation $E_1$ is given by
\begin{eqnarray}
\ddot{E}_1 + a\dot{E}_1 + \left(b+b_1
t\right)E_1=0,
\protect\label{equa3}
\end{eqnarray}
with
\begin{equation}
a=\frac{5}{4t_c},
\end{equation}
\begin{equation}
b=-\frac{3}{4t_c^2}-\frac{\sqrt{3} k^2}{6 \mathscr{H}_0 t_c},
\end{equation}
\begin{equation}
b_1=\frac{\sqrt{3}}{4t_c^2}\left(
\frac{k^2}{3\mathscr{H}_0}-\frac{3}{2t_c}\right),
\end{equation}
By the following variable transformation:
\begin{equation}
E_1(t)=\exp\left(-\frac{a t}{2} w(t)\right),
\end{equation}
the differential equation goes to
\begin{equation}
\ddot{w}+\left( b-(a/2)^2+b_1t\right)w=0,
\end{equation}
and the solution
is
\begin{equation}
w(t)=  w_0\, {\rm Ai}\left(-\frac{b-(a/2)^2 + b_1
t}{b^{2\\/3}}\right).
\end{equation}
The Airy function Ai is regular near $t=t_c$,
and so is $E_1$. Finally we look for the
equation of $E_2$ at the neighborhood of $t=t_c$:
\begin{eqnarray}
\ddot{E}_2 + a\dot{E}_2 + \left(b+b_1 t\right)E_2=0,
\end{eqnarray}
where
\begin{equation}
a=\frac{9}{4t_c},
\end{equation}
\begin{equation}
b=\frac{5}{t_c}\left(\frac{5}{4t_c}-\frac{\sqrt{3} m^2}{6 \mathscr{H}_0
}\right),
\end{equation}
\begin{equation}
 b_1=\frac{5\sqrt{3}}{2t_c^2}\left(
\frac{1}{t_c}-\frac{m^2}{6\mathscr{H}_0}\right).
\end{equation}
This equation differs from Eq.(\ref{equa3}) only by the numerical
values of the parameters $a, b$, and $b_1$ so we obtain the same type of regular
solution
\begin{equation}
E_2=  \,w_0 \,{\rm Ai}
\left(-\frac{b-(a/2)^2+ b_1 t}{b_1^{2/3}}\right)\exp\left(-\frac{at}{2}\right)
\end{equation}
Hence, it was shown
by a direct analysis of a specific nonsingular universe, that
in the neighborhood of the
special points in which a change of regime occurs, all independent
perturbed quantities are well-behaved, and the model is stable with regard
to scalar perturbations.

A similar analysis has been carried out for the model described by
Eqn.(\ref{at})
in the case of tensor perturbations in \cite{erico}. The result shows differences between
gravitational waves generated near a singularity and those generated near the bounce.
While in the first case the system exhibits a a node-focus transition in the $(E,\Sigma)$ plane,
independently of the perturbation wavelength $\lambda$,
in the bouncing model
the trajectories may exhibit a focus-node-focus transition, or no transition at all, depending
on the value of $\lambda$.

\subsection{Matching}
\label{match}

As mentioned in Sect.\ref{cobs},
another approach to the description of perturbations
in a bouncing
universe uses the idea of matching a contracting with an expanding phase.
The hope here again resides in the fact that some general features can be extracted from
given examples, since the matching may
be done in such a way as to avoid a very detailed specification of
the high curvature phase. Inasmuch as the result depends on the matching conditions,
this issue was the subject of a long debate \cite{debate}. We shall present next some
examples of this technique.

The case of a scalar field with an exponential potential (inspired in the string pre-big bang and the ekpyrotic model) was studied in \cite{vernizzi}.
A matching between
a contracting, scalar
field-dominated phase and an expanding, radiation-dominated
phase (and also of the corresponding perturbations)
was done
using the Israel conditions \cite{israeljc}. It was assumed that the slice of
spacetime in which high-energy physics takes control is very thin,
and can be approximated by a spacelike surface, with a negative
surface tension (to be specified by the underlying physics) required by
the jump in the extrinsic curvature.
Neglecting possible, but subdominant,
anisotropic surface stresses \footnote{This restriction was lifted in
\cite{copematch}.}, and depending on the chosen surface, it was found that
a scale-invariant spectrum could be transferred
from the contracting to the expanding phase.
A similar model has been studied in \cite{fibra}, where
it was shown that the value $p=2/3$ of the
power law $a(t) \propto (-t)^p$ was adopted for the scale
factor generates a
scale-invariant spectrum of adiabatic curvature fluctuations in
the collapsing phase. The chosen background corresponds to a
contracting Universe dominated by cold matter with null pressure.
As a result of the glueing, the spectrum is matched at the bounce
to a scale-invariant spectrum during the expanding phase. This
model was also shown to generate a scale-invariant spectrum of
gravitational waves, as already realized in \cite{gw1}.

It is useful to assume that the physics of the bounce is encoded in
the transfer matrix $T$, defined by
\begin{equation}
\begin{pmatrix}
D_+ \cr S_+
\end{pmatrix}
=
\begin{pmatrix}
T_{11} & T_{12} \cr T_{21} & T_{22}
\end{pmatrix}
\begin{pmatrix}
D_- \cr S_-
\end{pmatrix}\, .
\label{transT}
\end{equation}
$T$ gives the degree of mixing between the dominant (D) and sub-dominant (S)
modes before and after the
bounce for a fixed comoving wave number $k$.
Several combinations are possible, such as
one for which the spectrum is initially
not scale invariant but is turned into it because of
a nontrivial $k$ dependence of the transition matrix. Due to the
fact that the bounce lasts only a short time,
it is conceivable that it does not exert any influence on the large scales that are
of astrophysical
interest today. This implies that $T$ does not depend on $k$ \cite{vernizzi},
in such a way that a scale invariant pre-bounce spectrum
is transmitted without change to the post-bounce phase.
This hypothesis has
been tested in \cite{pp1}. It was shown by way of an example (a
bouncing solution in general relativity, with positive curvature
spatial section, with a scalar field as a source, by using an
expansion of the bouncing scale factor around the $\epsilon = 1$ de
Sitter-like bouncing solution) that $T$ may depend on $k$,
provided that the null energy condition (NEC) is very close to
being violated at the bounce, hence affecting the large scale
behaviour of the scalar perturbations (see however \cite{nd, mp}).
Note however that it was shown in \cite{pp1} that the spectrum of
gravitational waves is not affected by the bounce.

The authors of \cite{chu} have obtained the most general form of
the transfer matrix respecting local causality. In particular,
they have shown that no local-causality-respecting matching
condition can lead to a scale invariant spectrum for both the
pre-big-bang and the ekpyrotic model, in agreement with the result
of \cite{cremi}. They also studied a non-local model based on
string theory and showed that under certain conditions a post-bounce SIS is
possible.

A different line of attack was pursued in \cite{bozza1}
with the central assumption that the bounce in a spatially flat
universe is governed by just one physical scale (chosen as
$\eta_B$, the cosmological time at which the bounce occurs).
Working in GR and incorporating all the eventual new physics in
the matter side of EE, the general solution to the problem of the
propagation of perturbations through the bounce was presented in
\cite{bozza1}. It was shown that the spectrum of the Bardeen
potential in the expansion phase depends critically on the
relation between the comoving pressure perturbation and the
Bardeen potential in the new physics sector of the energy-momentum
tensor. Only if the comoving pressure perturbation is directly
proportional to the Bardeen potential (rather than its Laplacian,
as for any known form of ordinary matter), the pre-bounce growing
mode of the Bardeen potential persists in the post-bounce constant
mode. This would open the door to models with a scale-invariant
spectrum (hence in agreement with observations) for those cases in
which there is very slow contraction in the pre-bounce. This
result is supported by numerical analysis of a toy model in which
$\delta p \propto \Psi$ \cite{bozza1}. Examples of this type of
behaviour for the perturbations are given by models with spatial
curvature (which cannot be treated however with this approach) and
also by models with modifications coming from extra dimensions
(such as the one presented in \cite{patel}) \cite{bozza1}.

%
%
%

\subsection{Creation of cosmological magnetic fields}
\label{bcreation}

The origin, evolution, and structure of large-scale magnetic fields are amongst the most
important issues in astrophysics and cosmology. The standard model for the generation of this
fields is the dynamo, which amplifies a small seed field to the current observed values
of $1-{\rm few }\;\mu G$. There are several mechanisms to produce these seeds, but the
prevalent view is that they have a primordial origin \cite{grasso}. In particular, the
vacuum fluctuations of the EM field may be ``stretched'' by the evolution of the
background geometry to super-horizon scales, and they could appear today as
large-scale EM fields. For this to happen, conformal invariance of the EM equations
must be broken. This is the case in models such as dilaton electrodynamics
\cite{lemoine} and Weyl integrable spacetime (see Sect.\ref{dynor}, and \cite{miojcap} for
a list of references on the subject).

As a previous step in the details of the case of the EM field,
let us discuss the creation of massive scalar particles in a bouncing universe
with $\epsilon =-1$, following
\cite{costa}. The expansion factor is given by
$a(t)=t^2+a_0^2$, or $a(\eta) = a_0^2\cosh \eta$ in conformal time, as in the examples studied
in \cite{melni,NovelloSalim}. The EOM for the scalar field
is
$$
\Box\phi+\left(m^2+\frac 1 6 \xi R\right)\phi = 0.
$$
With the mode decomposition
$$
\phi_k(x)=a(\eta)^{-1/2}Y_\mathbf{k}(\vec x)\chi_k(\eta ),
$$
where $\mathbf{k}=(k,J,M)$ and the $Y_\mathbf{k}(\vec x)$ are given in terms of the spherical harmonics
(see \cite{birreldavis}), the function $\chi_k(\eta )$ satisfies the
modified Mathieu equation:
$$
\frac{d^2\chi_k}{d\eta^2}-(\lambda -2h^2\cosh^2\eta)\chi_k=0,
$$
where $\lambda \equiv -(k^2+\half m^2a_0^2)$, and $h\equiv \half ma_0$.
The number of created quanta in the (asymptotically flat)
future can be calculated
with the solutions of this equation that have the right asymptotic behaviour,
and following standard techniques. In the limit $h<<1$ (\ie$\:$ when the Compton wavelength
of the particle is much greater than $a_0$), the result is \cite{costa}
$$
|\beta_k|^2=\frac{1}{2\sinh^2\pi\tilde k}[1-\cos \left(4\tilde k\ln \frac h 2 \right)+ \varphi],
$$
where $\tilde k$ is the index in the
Mathieu functions $M_{-i\tilde k}(\eta , h)$, and is a complicated function
of $\lambda$ and $h$, which in the limit for small $h$ reduces to
$$
\lambda = -\tilde k^2-\frac{h^4}{2(\tilde k+1)}+O(h^8),
$$
and $\varphi$ is a phase, independent of $h$.
The expression for $|\beta |$ varies from 0 to 4$\times\exp (-2\pi\tilde k)$ for large
$k$, and shows that for a given $k$, the particle number depends on the product $ma_0$.

The creation of magnetic fields in a bouncing universe in models that break
the conformal invariance with a coupling to a scalar field was studied
in \cite{nilton,heat,miojcap}. In the latter, canonical quantization was applied to
the model given by
$$
S=\half \int\,d^{4}x \sqrt{-g}\;
 f(\omega)  F_{\alpha\beta}\; F^{\alpha\beta},
$$
where $\omega$\ is the scalar field, and $F_{\alpha\beta}$ an
abelian field, with $f(\omega ) = \exp (-2\omega )$. The modes of
the potential ${\cal A}_\mu=e^{-\omega }A_\mu$ satisfy the
equation \beq {\cal A}''^{(\sigma)}_{k\alpha}(\eta)+(k^2-V(\eta
)){\cal A}^{(\sigma)}_{k\alpha}, \label{modeqa} \eeq where $\sigma
= +,-$ designates the base of travelling waves, $\alpha =1,2$
describes the two transverse degrees of freedom, and $V(\eta ) =
-\omega ''+\omega '^2$. For the background described in
Eqn.(\ref{wistb}), \beq V(\eta ) = \frac{2\sigma \sinh(2\eta
)+\sigma ^2}{\cosh ^2(2\eta)}, \label{potni} \eeq where
$\sigma\equiv \sqrt 6/\lambda$, where $\lambda^2$ is the coupling
constant of the scalar field to gravity.
\begin{figure}[htb]
\begin{center}
\includegraphics[width=0.6\textwidth
]
{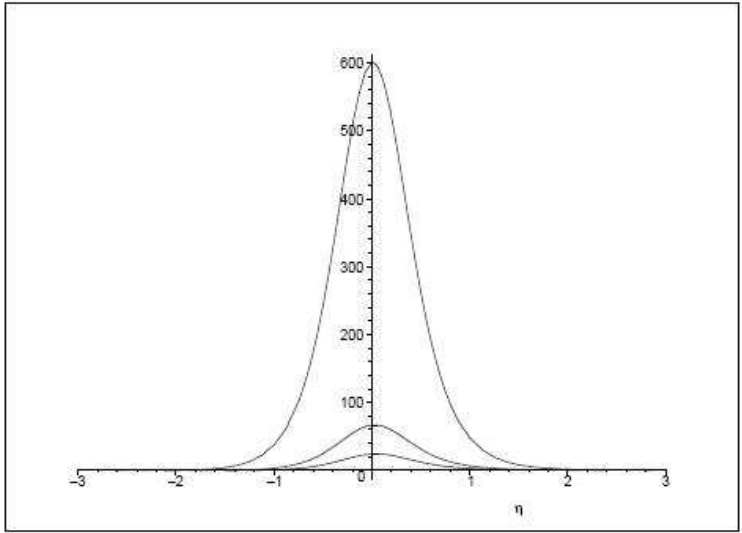}
\caption{Plot of the potential (see Eq.(\ref{potni})), for
$\lambda=0.1,0.3,0.5$ (solid, dotted, dashed line respectively).}
\label{potproko}
\end{center}
\end{figure}
The mode equation (\ref{modeqa})
admits analytical solutions in terms of hypergeometric functions, in terms of which
the Bogolubov coefficients, and the expression
for the energy density of the magnetic field $\rho_m$
can be calculated \cite{miojcap}.
The amplification factor with respect to the conformal vacuum
peaks for the modes with momenta such that $k\approx 1.31$, and is given by
\beq
\frac{\rho_m}{(\rho_m)_{cf}}\propto\exp\left(\frac{\pi\sqrt 6}{\lambda}\right),
\label{ampl}
\eeq
for $\eta >>1$. The conditions for the spectrum to be greatly amplified
today are \cite{miojcap}
$$
a_0<<ct_r,\;\;\;\;\;\;\;\;\;\lambda<<1,
$$
where $t_r$ is the time at which the scalar field is negligible, in such a way that
the EM field is free again.

At a comoving scale of about 10 kpc, the strength of conformal vacuum fluctuations
is of the order of $10^{-55}$ G. To reach the strength required to feed the galactic
dynamo, $B_{seed}\propto 10^{-20}$ G, , which is a conservative estimate,
we get from Eqn.(\ref{ampl}) that $\lambda\approx 0.1$. Taking for the comoving scale
the size of the universe ($\approx 4\times 10^6$ kpc), the amplification factor becames
$10^{46}$, and we need $\lambda \approx 0.07$. So the strength needed in both cases can be achieved
by a modest value of $\lambda$, the coupling constant of $\omega$ to gravity.

These results were obtained in a model
that did not take into account the effect of the creation of
matter by the decay of the scalar field. The solution presented in Sect.(\ref{matcre}),
namely
\beq a(\eta ) = \beta\sqrt{\cosh (2\eta )
+k_0\sinh (2\eta ) - 2k_0(\tanh \eta +1)},\eeq
with
$\beta = a_0/\sqrt{1-k_0}$, and $0<k_0<1/7$.
incorporates this feature, and its influence on the
creation of photons was discussed in \cite{nilton}.
The result, displayed in Fig.(\ref{ni}), shows that
there is a substantial increment in the number of photons
if we take into account the effect of matter creation.
\begin{figure}[htb]
\begin{center}
\includegraphics[width=0.7\textwidth
]{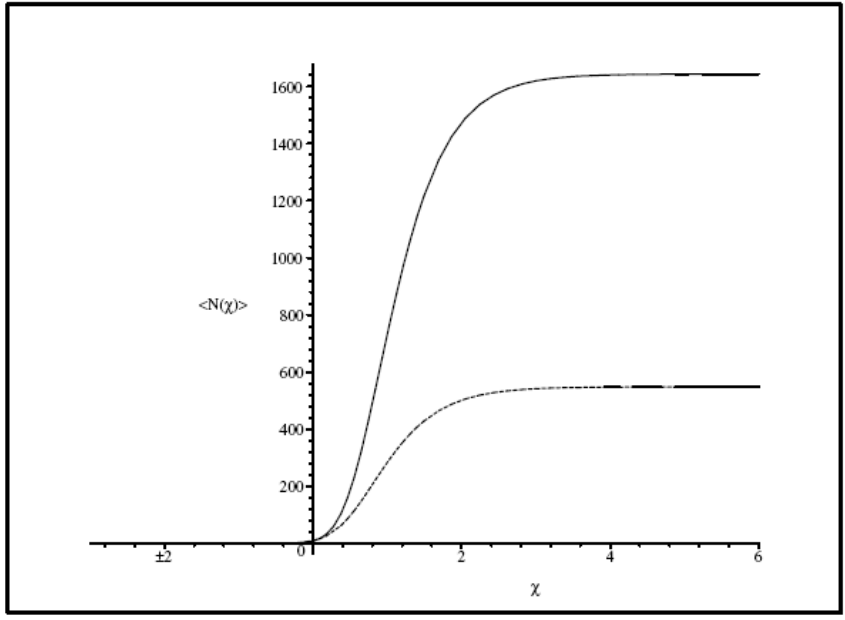}
\caption{Plot of the mean number of photons as a function of
the conformal time for $m = 20$, in the case without matter
(dashed line) and for the case with matter creation (full line),
for $\lambda = 1$.}
\end{center}
\label{ni}
\end{figure}

\subsection{Appendix}
\label{apert}
In this appendix, we give a short summary of two gauge-invariant
methods that can be applied to study the perturbations in
cosmological scenarios.
\subsubsection{Perturbations using Bardeen variables}

The fluctuations of the metric tensor can be classified by their
properties under spatial rotations into scalar, vector and
tensor perturbations. In the linear theory, their evolution is
decoupled.
In the case of scalar perturbations, the perturbed metric of a homogeneous and
isotropic spacetime can be written as
\beq
ds^2=a^2(\eta) \left\{(1+2\phi)d\eta^2-2B_{;i}d\eta\:dx^i-
[(1-2\psi)\gamma_{ij}+2E_{,i;j}dx^idx^j]\right\},
\label{scalarpert}
\eeq
where $\gamma_{ij}$ is the metric of the 3-space. We shall sketch
the case of hydrodinamical perturbations of a perfect fluid
\footnote{For other cases, such as s scalar field, see
\cite{revmuk}.} with energy-momentum tensor
\beq
T^\alpha_{\;\beta} = (\rho +p) u^\alpha u_\beta - p
\delta^\alpha_{\;\beta}.
\label{emt}
\eeq
Following \cite{bardeen}, it is convenient to build, from the four variables appearing in
(\ref{scalarpert}), two gauge-invariant quantities, given by
$$
\Phi = \phi + \frac{[(B-E')a']'}{a},
\;\;\;\;\;\;\;\;\;\Psi = \psi -\frac{a'(B-E')}{a}.
$$
In terms of these, the gauge-invariant perturbed EE are \beq -3{\cal
H}({\cal H}\Phi +\Psi ')+\nabla^2\Psi +3k\Psi = \half a^2\delta
T^{(gi)0}_{0}, \label{gi1}\eeq
\beq
({\cal H}\Phi+\Psi ')_{,i}=\half a^2\delta
T^{(gi)0}_i, \label{gi2}\eeq
\beq [(2{\cal H}'+{\cal H}^2\Phi + {\cal H}\Phi
'+\Psi ''+2{\cal H}{\Psi '}-k\Psi+\half
\nabla^2(\Phi-\Psi)]\delta^i_j-\half
\gamma^{ij}(\Phi-\Psi)_{|kj}=-\half a^2 \delta T^{(gi)i}_j,
\label{gi3}
\eeq
where the $\delta T^{(gi)\alpha}_\beta$ are gauge invariant
combinations of the $\delta T^\mu_\nu$, $B$, and $E$ (see
\cite{revmuk} for details).

In the case of
hydrodynamical matter,
the most general form of the perturbation can be written in terms of the
perturbed energy $\delta \rho$, the perturbed pressure $\delta p$,
the potential ${\cal V}$ of the 3-velocity $v^i(t,\vec x)$, and the anisotropic stress
$\sigma$ as follows \cite{bardeen}:
\begin{displaymath}
(\delta T^{\mu}_{\;\nu}) =
\left( \begin{array}{cc}
\delta\rho & -(\rho_0 +p_0)a^{-1}{\cal V}_{,i}  \\
(\rho_0 +p_0)a{\cal V}_{,i}  & -\delta p \delta_{ij}+\sigma_{ij}
\end{array} \right).
\end{displaymath}
For the case of a perfect fluid, with energy-momentum tensor given by
Eqn.(\ref{emt}), $\sigma_{ij}=0$.

The pressure perturbation can be split into its
adiabatic and entropy components as
\beq
\delta p = \left( \frac{\partial p}{\partial
\rho}\right)_S\delta\rho + \left( \frac{\partial
p}{\partial S}\right)_\rho \delta S \equiv c_s^2 \delta
\rho +\tau \delta S.
\label{ppert}
\eeq
Entropy perturbations may be important in the case of
two-component systems, such as plasma and radiation.

The gauge-invariant perturbations of the energy-momentum tensor
can be expressed in terms of the gauge-invariant energy density, pressure, and
velocity perturbation:
$$
\delta T^{(gi)0}_0=\delta\rho^{(gi)},\;\;\;\;\;\;\;\;\delta T^{(gi)0}_i=(\rho_0+p_0)
a^{-1}\delta u^{(gi)}_i,\;\;\;\;\;\;\;\;\delta T^{(gi)i}_j=-\delta p^{(gi)}\delta^i_{\;j},
$$
with
$$
\delta\rho^{(gi)}=\delta\rho +\rho_0'(B-E'),\;\;\;\;\;\;
\delta p^{(gi)}=\delta p +p_0'(B-E'),\;\;\;\;\;\;
\delta u_i^{(gi)}=\delta u_i+a(B-E')_{|i}.
$$
From Eqns.(\ref{gi1})-(\ref{gi3}) applied to this case, it follows that
$\Phi = \Psi$. Using Eqn.(\ref{ppert}), the system can be written as
\beq \Phi''+3{\cal
H}(1+c_s^2)\Phi '- c_s^2\nabla^2\Phi +[2{\cal H}'+(1+3c_s^2
)({\cal H}^2-k)]\Phi = \half a^2\tau \delta S. \label{pee} \eeq
\beq
(a\Phi)'_{,i}=\half a^2(\rho_0+p_0)\delta u_i^{(gi)}.
\label{peu2}
\eeq
For adiabatic perturbations, Eqn.(\ref{pee}) yields $\Phi$,
which determines
$\delta \rho^{(gi)}$ through Eqn.(\ref{gi1}), and
$\delta u_i^{(gi)}$
through Eqn.(\ref{peu2}).

Eqn.(\ref{pee}) can be simplified with the change of variables
$$
\Phi = \sqrt{\half}\sqrt{\frac{{\cal H}^2-{\cal H}'+k}{a^2}}\;u,
$$
yielding
$$
u''- c_s^2\nabla^2u -\frac{\theta''}{\theta}\; u = {\cal N},
$$
with
$$
\theta = \frac 1 a
\left(\frac{\rho_0}{\rho_0+p_0}\right)^{1/2}\left(1-\frac{3 \epsilon}{a^2\rho_0}\right)^{1/2},
$$
$$
{\cal N} = a^2(\rho_0+p_0)^{-1/2}\tau \delta S.
$$
\\[0.3cm]
\noindent\textbf{Vector perturbations}

The most general perturbed metric including only vector perturbations is given by
\footnote{The results quoted in this section are taken from \cite{revmuk}.}
\begin{displaymath}
(\delta g_{\mu\nu}) =
\left( \begin{array}{cc}
0 & -S^i  \\
-S^i & F^i_{\;,j}+F^j_{\;,i}
\end{array} \right),
\end{displaymath}
where the vectors $S$ and $F$ are divergenceless. From their transformation properties,
it can be shown that
$$
\sigma^i=S^i+\dot F^i
$$
(where the dot means derivative w.r.t. conformal time) is a gauge invariant quantity.
For the perturbations of the stress-energy tensor, we have
\begin{displaymath}
(\delta T^\alpha_{\beta}) =
\left( \begin{array}{cc}
0 & -(\rho + p) V^i  \\
(\rho +p)(V^i+S^i) & p(\pi^i_{\;,j}+\pi^j_{\;,i})
\end{array} \right),
\end{displaymath}
where $V^i$ and $\pi^i$ are divergenceless.
$V^i$ is related to the perturbation of the 4-velocity bu
\begin{displaymath}
(\delta u^\mu) =
\left( \begin{array}{c}
0   \\
V^i/a
\end{array} \right).
\end{displaymath}
Adopting the Newtonian gauge (in which $F=0$), from the perturbed EE we get
\beq
-\frac{1}{2a^2}\triangle S^i=(\rho + p) V^i,
\label{lapl}
\eeq
\beq
-\frac{1}{2a^4}\nabla_t(a^2(S^j_{\;,i}+S^i_{\;,j}))=p(\pi^i_{\;,j}+\pi^j_{\;,i}),
\label{pieq}
\eeq
where $\triangle$ is the spatial Laplacian.
From Eqn.(\ref{lapl}) we get
\beq
V^i_k=\frac{1}{2a^2(\rho +p)}k^2 S^i_k,
\label{vs}
\eeq
for the Fourier modes of $V$ and $S$.
Assuming that there are no anisotropic stresses, as in the case of
presureless dust, we get from Eqn.(\ref{pieq}),
$$
\nabla_t(a^2S^i_{\;k})=0.
$$
Hence $S^i_{\; k}=C�_{\; k}/a^2$, where $C$ is a constant.
From this and Eqn.(\ref{vs}), we get
\beq
V^i_k\propto \frac{k^2C�_{\; k}}{a^{1-3\omega}}.
\label{va}
\eeq
Note that $V^i_k$ increases for $\omega =0$, and stays constant for radiation, but
$S^i_{\; k}$ always increases for decreasing $a$.
\\[0.3cm]
\noindent\textbf{Tensor perturbations}

These perturbations are built using a symmetric 3-tensor
$h_{ij}$ which satisfies the constraints
$$
h^i_{\;i}=0\;\;\;\;\;\;\;\;\;\;h_{ij}^{|j}=0,
$$
in such  way that the metric for tensor perturbations is
\begin{displaymath}
(\delta g_{\mu\nu}^{(t)}) =
-a^2(\eta ) \left(  \begin{array}{cc}
0 & 0  \\
0 & h_{ij}
\end{array} \right).
\end{displaymath}
From the perturbed EE we find (see for instance \cite{mubook}
$$
h_{ij}''+2{\cal H}h_{ij}'-\triangle h_{ij}=2a^2\delta T^{(gi)T}_{ij},$$
where $\delta T^{(gi)T}_{ij}$ is the gauge-invariant
``pure tensor'' part of $\delta T_{\mu\nu}$.
In Fourier space, and introducing the rescaled variable
$h_{ij}=e_{ij}v/a$, we have
\beq
v_k''+\left(k^2-\frac{a''}{a}\right) v_k=0.
\label{zero}
\eeq

\subsubsection{The quasi-Maxwellian method}
\label{another}

The QM method it has its roots
in the formulation of Jordan and his collaborators \cite{jordan}
and uses the Bianchi identities to propagate initial
conditions.
The basic idea is to identify gauge invariant quantities from the
beginning, using Stewart's lemma \cite{stewart}, which guarantees
that the perturbation of an object $Q$ is gauge-invariant if $Q$
is either constant or a linear combination of $\delta^\mu_{\;\nu}$
with constant coefficients. In conformally flat models, the Weyl
tensor (defined below) is identically zero, so its perturbation is
a true perturbation, and not a gauge artifact. We shall see below
how to obtain a minimum set of variables to completely
characterize a perturbation, along with their evolution equations.
\\[0.3cm]
{\bf Definitions and notation}

The Weyl conformal tensor is defined by means of the expression
$$
W_{\alpha\beta\mu\nu} = R_{\alpha\beta\mu\nu} -
M_{\alpha\beta\mu\nu} + \frac{1}{6} R g_{\alpha\beta\mu\nu},
$$
where
\begin{equation}
g_{\alpha\beta\mu\nu} \equiv g_{\alpha\mu} g_{\beta\nu} -
g_{\alpha\nu} g_{\beta\mu} \protect\label{d2},
\end{equation}
and
\begin{equation}
2 M_{\alpha\beta\mu\nu} = R_{\alpha\mu} g_{\beta\nu} +
R_{\beta\nu} g_{\alpha\mu} - R_{\alpha\nu} g_{\beta\mu} -
R_{\beta\mu} g_{\alpha\nu}. \protect\label{d3}
\end{equation}
The 10 independent components of the Weyl tensor can be separated
in the electric and magnetic parts, defined (in analogy with the
electromagnetic field) as:
\begin{equation}
E_{\alpha\beta} = - W_{\alpha\mu\beta\nu} v^{\mu} v^{\nu},
\protect\label{d4}
\end{equation}
\begin{equation}
H_{\alpha\beta} = - W^{\ast}_{\alpha\mu\beta\nu} v^{\mu} v^{\nu}.
\protect\label{d5}
\end{equation}
{}The dual operation was performed with the completely
skew-symmetric Levi-Civita tensor $\eta_{\alpha\beta\mu\nu}$. From
the symmetry properties of the Weyl tensor it follows that the operation of
taking
the dual is independent on the pair in which it is applied.

It follows from these definitions that the tensors $E_{\mu\nu}$
and $H_{\mu\nu}$ are symmetric, traceless and belong to the
three-dimensional space orthogonal to the observer with 4-velocity
$v^{\mu}$, that is:
\beq E_{\mu\nu} = E_{\nu\mu},\;\;\;\;\;\;\;\;\;\; E_{\mu\nu}
v^{\mu} = 0 ,\;\;\;\;\;\;\;\;\;\; E_{\mu\nu} g^{\mu\nu} = 0,
\protect\label{d6}
\end{equation}
and similar relations for $H_{\mu\nu}$.
The metric $g_{\mu\nu}$ and the vector $v^{\mu}$ (tangent to a
timelike congruence of curves $\Gamma$) induce a projector tensor
$h_{\mu\nu}$ which separates any tensor in terms of quantities
defined along $\Gamma$ plus quantities defined on the
3-dimensional space
orthogonal to $v^{\mu}$. The tensor $h_{\mu\nu}$, defined on this
3-dimensional space is symmetric and a true projector, that is
\begin{equation}
h_{\mu\nu} h^{\nu\lambda} = {\delta_{\mu}}^{\lambda} - v_{\mu}
\hspace{0.1cm} v^{\lambda} = {h_{\mu}}^{\lambda}.
\protect\label{d8}
\end{equation}
We shall work with the FLRW geometry written in the standard
Gaussian coordinate system:
\begin{equation}
ds^{2} = dt^{2} + g_{ij} dx^{i} dx^{j} \protect\label{d9}
\end{equation}
where $g_{ij} = - a^{2}(t) \gamma_{ij}(x^{k})$. The 3-dimensional
geometry has constant curvature and thus the corresponding
Riemannian tensor $^{(3)}R_{ijkl}$ can be written as
\begin{displaymath}
^{(3)}R_{ijkl} = \epsilon \gamma_{ijkl}.
\end{displaymath}
The covariant derivative in the 4-dimensional space-time will be
denoted by the symbol ``;'' and the 3-dimensional derivative will be
denoted by ``${\|}$''.

The irreducible components of the covariant derivative of
$v^{\mu}$ are given in terms of the expansion scalar ($\theta$),
shear ($\sigma_{\alpha\beta}$), vorticity ($\omega_{\mu\nu}$) and
acceleration ($A_{\alpha}$) by the standard definition:
\begin{equation}
v_{\alpha ;\beta} = \sigma_{\alpha\beta} + \frac{1}{3} \theta
h_{\alpha\beta} + \omega_{\alpha\beta} + A_{\alpha} v_{\beta},
\protect\label{d10}
\end{equation}
where
\begin{equation}
\begin{array}{ll}
\sigma_{\alpha\beta} = \frac{1}{2} h^{\mu}_{(\alpha}
h_{\beta)}^{\nu}
v_{\mu ;\nu} - \frac{1}{3} \theta h_{\alpha\beta} ,\\ \\
\theta = {v^{\alpha}}_{;\alpha}, \\ \\
\omega_{\alpha\beta} = \frac{1}{2} h_{[\alpha}^{\mu} h_{\beta
]}^{\nu}
v_{\mu ;\nu} ,\\ \\
A_{\alpha} = v_{\alpha ;\beta} v^{\beta}.
\end{array}
\protect\label{d11}
\end{equation}
We also define
\begin{equation}
\theta_{\alpha\beta} \equiv \sigma_{\alpha\beta} + \frac{1}{3}
\theta h_{\alpha\beta}. \protect\label{d11b}
\end{equation}
\\[0.3cm]
{\bf Quasi-Maxwellian equations of
gravity and their perturbation}

We shall present in this subsection a sketch of the deduction of
the equations that govern the perturbations in the
quasi-Maxwellian formalism. The details of the calculations in
this section can be found in \cite{Novello1}. Using Einstein\rq s
equations and the definition of Weyl tensor, Bianchi identities
can be written in an equivalent form as
\begin{eqnarray}
{W^{\alpha\beta\mu\nu}}_{;\nu} & = & \frac{1}{2}R^{\mu[\alpha
;\beta]} -
\frac{1}{12}g^{\mu[\alpha}R^{,\beta]} \nonumber \\
& = & - \frac{1}{2}T^{\mu[\alpha ;\beta]} +
\frac{1}{6}g^{\mu[\alpha}T^{,\beta]}. \nonumber
\end{eqnarray}
The quasi-Maxwellian equations of gravity are obtained by
projecting these equations ({\em i.e.} the Bianchi identities are
taken as true dynamical equations which describe the propagation
of gravitational disturbances).
The evolution equation for the perturbations for $\delta\theta$,
$\delta\sigma_{\mu\nu}$, and $\delta\omega^\mu$, as well as 3
constraint equations, are obtained projecting and perturbing the
equation
$$
v_{\mu ;\alpha ;\beta}-v_{\mu ;\beta
;\alpha}=R_{\mu\omega\alpha\beta}v^\omega
$$
which follows from the
definition of the curvature tensor.
Finally we get two more equations by projecting the conservation
law $T^{\mu\nu}_{\;\;\;\;;\nu}=0$. Adding up, we have a set of
twelve equations which when perturbed yield (after straighforward
manipulations) the coupled differential equations needed to give a
complete description of the perturbation. In a general case, the
variables are
$$
{\cal V} = \left\{\delta E_{ij}, \delta H_{ij},  \delta
\omega_{ij}, \delta \sigma_{ij}, \delta \pi_{ij}, \delta a_{i},
\delta q_{i}, \delta\rho, \delta\theta, \delta V_0,\delta
V_k\right\}.
$$
From now on we will concentrate on the case of scalar irrotational
perturbations. As shown in \cite{Lifshitz},  it is useful to
develop the perturbed quantities in the spherical harmonics basis.
It is enough for our purposes to work only with the scalar
$Q^{(k)}(x^{i})$ (with $\partial Q^{(k)}/ \partial t = 0$) and the
vector and tensor quantities that follow from it, defined by
$ Q^{(k)}_{i} \equiv Q^{(k)}_{,i}$, $ Q^{(k)}_{ij} \equiv
Q^{(k)}_{,i;j}$.
The scalar $Q^{(k)}$ obeys the eigenvalue equation defined in the
3-dimensional background space by:
\begin{equation}
\bigtriangledown^{2}Q^{(k)} = k Q^{(k)} \protect\label{d13},
\end{equation}
where $k$ is the wave number, and the symbol
$\bigtriangledown^{2}$ denotes the 3-dimensional Laplacian:
\begin{equation}
\bigtriangledown^{2}Q \equiv \gamma^{ij} Q_{,i\| j} = \gamma^{ij}
Q_{,i;j} \protect\label{d14}.
\end{equation}
Since the modes do not mix at the linear order, we will drop the
superindex $(k)$ from $Q$. The traceless operator $\hat{Q}_{ij}$ is
defined as
\begin{equation}
\hat{Q}_{ij} = Q_{ij} +  \frac{k^{2}}{3}Q \gamma_{ij},
\protect\label{d15}
\end{equation}
and the divergence of $\hat{Q}_{ij}$ is given by
\begin{equation}
{\hat{Q}^{ij}}_{\;\;  ; j} = - 2 \left( \epsilon + \frac{k^{2}}{3}
\right) \hspace{0.1cm}Q^{i}. \protect\label{d16}
\end{equation}
%
%
Due to Stewart's lemma, the good (since they are gauge-invariant
an null in the background) objects in the list ${\cal V}$ are
$\delta E_{ij}$, $\delta \Sigma_{ij}$, $\delta \pi_{ij}$, $\delta
a_{i}$, and $\delta q_{i}$. According to causal thermodynamics the
evolution equation of the anisotropic pressure is related to the
shear through \cite{Israel}
\begin{equation}
\tau\dot{\Pi}_{ij} + \Pi_{ij} = \xi\sigma_{ij}
\protect\label{extra1}
\end{equation}
in which $\tau$ is the relaxation parameter and $\xi$ is the
viscosity parameter. For simplicity we will take the case in which
$\tau$ can be neglected and $\xi$ is a constant \footnote{In the
general case $\xi$ and $\tau$ are functions of the equilibrium
variables, for instance $\rho$ and the temperature $T$ and, since
both variations $\delta\Pi_{ij}$ and $\delta\sigma_{ij}$ are
expanded in terms of the traceless tensor $\hat{Q}_{ij}$, it
follows that the above relation does not restrain the kind of
fluid we are examining. However, if we consider $\xi$ as
time-dependent, the quantity $\delta\Pi_{ij}$ must be included in
the fundamental set ${\cal M}_{[A]}$.}.  Eq.(\ref{extra1}) then
gives
\begin{equation}
\Pi_{ij} = \xi\sigma_{ij} \protect\label{extra2},
\end{equation}
and the associated perturbed equation is
\begin{equation}
\delta\Pi_{ij} = \xi \hspace{0.1cm}\delta\sigma_{ij}.
\protect\label{m3}
\end{equation}
%
We shall decompose the four independent and gauge-invariant
perturbations as \footnote{In fact, $ \sqrt{\delta E_{ij} \delta
E^{ij}} $ is the only quantity that characterizes without
ambiguity a true perturbation of the Debever invariants
\cite{Novello1}.}
$$
\delta E_{ij} = \sum_k\;E^{(k)}(t)\hat Q^{(k)}_{ij},
$$
$$
\delta \Sigma_{ij} = \sum_k\;\Sigma^{(k)}(t)\hat Q^{(k)}_{ij},
$$
$$
\delta A_{i} = \sum_m\;\psi^{(m)}(t) Q^{(m)}_{i},
$$
$$
\delta q_{i} = \sum_m\;q^{(m)}(t) Q^{(m)}_{i}.
$$
It can be shown that $\psi$ is a function of $\Sigma$ and $E$
\cite{Novello1}. It follows that, restricting to the case $q=0$ (no energy flux)
\footnote{We further assume an equation of state relating the pressure and
the energy density, i.e. $p = \lambda \rho$, which is preserved
under arbitrary perturbations.},
%
$E(t)$ and $\Sigma (t)$ constitute the fundamental pair of
variables in terms of which the dynamics for the perturbed FLRW
geometry is completely characterized. Indeed, the evolution
equations for these two quantities (which follow from Einstein\rq
s equations) generate a dynamical system involving only $E$ and
$\Sigma$ (and background quantities) which, when solved, contains
all the necessary information for a complete description of all
remaining perturbed quantities of the FLRW geometry.

The evolution equations are given by \cite{Novello1}
\begin{equation}
\dot{\Sigma} = -E - \frac{1}{2}\;\xi \hspace{0.1cm}\Sigma - k^{2}
\hspace{0.1cm} \psi, \protect\label{s3}
\end{equation}
\begin{eqnarray}
\dot{E} & = & - \frac{(1 + \lambda)}{2} \rho \hspace{0.1cm}\Sigma
- \left(\frac{\theta}{3} + \frac{\xi}{2}\right) \hspace{0.1cm}E \nonumber \\
& - & \frac{\xi}{2} \hspace{0.1cm}\left(\frac{\xi}{2} +
\frac{\theta}{3}\right) \hspace{0.1cm} \Sigma - \frac{k^{2}}{2}
\hspace{0.1cm}\xi \hspace{0.1cm}\psi. \protect\label{e2}
\end{eqnarray}
As mentioned before, $\psi$ can be expressed in terms of $E$ and
$\Sigma$ \footnote{Except when $(1 + \lambda) = 0$, see
\cite{Novello1} for this case.}:
\begin{equation}
(1 + \lambda) \hspace{0.1cm}\rho \hspace{0.1cm}\psi = 2 \left(1 +
\frac{3\epsilon}{k^{2}}\right) \hspace{0.1cm}a^{-2} \hspace{0.1cm}
[ -\lambda E + \frac{1}{2} \hspace{0.1cm}\lambda \hspace{0.1cm}\xi
\hspace{0.1cm}\Sigma + \frac{1}{3} \hspace{0.1cm}\xi
\hspace{0.1cm}\Sigma]. \protect\label{e7}
\end{equation}
Thus the set of perturbed equations reduces to a time-dependent
dynamical system in the variables $E$ and $\Sigma$:
\begin{equation}
\dot{\Sigma} = F_{1}(\Sigma, E)
,\;\;\;\;\;\;\;\;\;\;\;\;\;\;\;\;\;\;\;\; \dot{E} = F_{2}(\Sigma,
E),
\protect\label{e8}
\end{equation}
with \beq F_{1} \equiv -E - \frac{1}{2}\xi \hspace{0.1cm}\Sigma -
k^{2} \hspace{0.1cm} \Psi , \protect\label{f1} \eeq and
\begin{eqnarray}
F_{2} & \equiv & - \left(\frac{1}{3}\theta + \frac{1}{2}\xi\right)
\hspace{0.1cm}E - \frac{k^{2}}{2} \hspace{0.1cm}\xi \hspace{0.1cm}\Psi \nonumber \\
& - & \left(\frac{1}{4}\xi^{2} + \frac{(1 + \lambda )}{2}\rho +
\frac{1}{6} \xi\theta\right) \hspace{0.1cm}\Sigma
\protect\label{f2}
\end{eqnarray}
where $\Psi$ is given in terms of $E$ and $\Sigma$ by
eq.(\ref{e7}), so the system \ref{e8} can be written as
\begin{equation}
\begin{pmatrix}
\dot E \cr \dot \sigma
\end{pmatrix}
=
\begin{pmatrix}
\alpha & \beta \cr \gamma & \delta
\end{pmatrix}
\begin{pmatrix}
E \cr \Sigma
\end{pmatrix}\, ,
\label{dsys}
\end{equation}
where
$$\alpha\equiv -\frac\theta 3,\;\;\;\;\beta\equiv-\frac{1+\lambda}{2}\rho,\;\;\;\;
\delta = 0,\;\;\;\;\gamma=\frac{6\lambda}{1+\lambda}\left(\epsilon+\frac{k^2}{ 3}\right)
\frac{1}{a^2\rho}-1.
$$
Since
$$
\frac{\partial\dot E}{\partial E}+\frac{\partial\dot\Sigma}{\partial\Sigma}=-\frac{\theta}{3},
$$
the
system \ref{dsys} is not Hamiltonian due to the expansion of the universe.
Nonetheless, new variables $(Q,P)$ can be introduced in such a way that the system \ref{dsys}
is Hamiltonian in them. Defining
$$
Q\equiv a^m \sigma ,\;\;\;\;\;P=a^n E,
$$
it is easily shown that the otherwise arbitrary powers
$m$ and $n$ must satisfy
the relation $m+n=1$ for the variables $Q$ and $P$ to be canonically
conjugated. It follows that
$$
\ddot P = {\cal M}_1 P + {\cal M}_2 Q.
$$
The choice $n=3\lambda/2+2$ yields ${\cal M}_2=0$,
and $P$ satisfies the equation
$$
\ddot P +\mu (t) P=0,
$$
with
$$
\mu (t) = \left(\frac 5 4 \lambda + \frac 2 3 \right) \rho + \frac {1} {a^2}
\left[\frac{3\lambda}{2}\left(\frac{3\lambda}{2}\right)\epsilon-\lambda k^2\right],
$$
which is equivalent to Eqn. (\ref{15dez2}).

This method can be extended to vector and tensor perturbations in the FLRW model
\cite{Novello2}. In the first case, the observable quantities are
described in terms of the vorticity and the shear, while the electric and
magnetic parts of the Weyl tensor suffice for the gravitational waves
\footnote{Perturbations in the Kasner solution were studied in \cite{motta}.}.
The three types of perturbation are describable in
Hamiltonian form, thus paving the way to canonical quantization
\cite{Novello3}, which was performed for
scalar, vectorial, and tensor perturbations using
the squeezed states formalism in \cite{Novello3}.
In fact, in the case of scalar perturbations,
the Hamiltonian in terms of the $(Q,P)$ variables (with
the choice $m=0$ is given by
$$
H=\frac{h_1}{2}Q^2+\frac{h_2}{2}P^2+\frac{h_3}QP,
$$
with
$$
h_1=\frac{1+\lambda}{2}\frac\rho a,\;\;\;\;h_2=\frac{6\lambda}{1+\lambda}
\left(\epsilon+\frac {k^2}{ 3}\right)\frac{1}{a\rho}-a,\;\;\;\;h_3=0.
$$
\subsection{Relation between the two methods}

The Bardeen variables $(\Phi , \Psi)$ are related to the quasi-Maxwellian
variables $(E, \Sigma )$. For instance, in the case of
scalar perturbations the relation between $E$ and $\Phi$ (for a perfect fluid)
is given by \cite{goode,tjoras}
$$
E=-k^2\Phi ,$$
from which the relation for the spectrum given in Eqn.(\ref{esp}) follows.

\section{Conclusion}
\label{conc}
The idea of a bouncing
universe has been considered since the early days of relativistic cosmology,
as shown
in this review. However,
only a few analytical solutions
describing a nonsingular universe
served as a starting point
to build a complete cosmological scenario.
%
The main reason for this neglect
by the majority of the physics community
in the last 30 years of the 20th century
was the strong influence of the
singularity theorems, which led to the belief that some sort of
singularity was inevitable in gravitational processes.
The situation should have changed with the recently discovered positive
acceleration of the universe since, in the realm of GR, the
accelerated expansion means that the matter content must satisfy
the condition $\rho + 3p <0$, which is precisely one of the
conditions needed to have a bounce in Einstein's gravity. This
violation of the SEC was already accepted in the early 80's in
order to have a phase of inflationary expansion, and
nowadays several systems are known
which do not
satisfy the inequality $\rho + 3p >0$ (see for instance
\cite{twilight}).
Hence, there is
mounting evidence against one of the
main theoretical prejudices
forbidding
bouncing universes in
GR.
Surprisingly, nonsingular models have not arisen
the interest that should be expected based on the preceding considerations
\footnote{It may be argued that this lack of interest is due to the fact that
the bounce is expected to involve scales where quantum effects render GR inapplicable.
But this is true also of the singularity theorems, as was known already in the early 70's.
Moreover, there is no evidence against the possibility of a bounce in the classical
regime \cite{melni2}, as follows from some of the models presented in
Sect.\ref{exact}, see also \cite{falciano}).}.

Almost contemporaneous to the discovery of the
accelerated expansion was the gradual advent
of a handful of cosmological models based on nonsingular solutions.
These models aimed at solving the most stringent problems of the
(pre-inflationary) cosmological standard model: the initial singularity,
the isotropy and homogeneity of the currently-observed universe,
the horizon problem,
the flatness problem\footnote{Note that the flatness problem may
in principle not be a problem in
gravitational theories other than GR (see Sect.\ref{riccis}).},
and the formation of structure \footnote{In spite of its historical
importance,
the so-called
monopole problem is not included in this list, since there is still room
for it to be
be considered as a problem of field theory first, and then (perhaps) of
the standard cosmological model, see for instance \cite{lindeb,dar}.}.
Bouncing universes have partially met these challenges.
The singularity is obviously absent, and its avoidance requires
any of the assumptions listed in Sect.\ref{defsing}, which range from
the violation of SEC (in GR) to quantum gravitational effects.

As explained in Sect.\ref{sint}, a phase of accelerated contraction
may solve the flatness problem in GR, and may also get rid of
particle horizons (see for instance \cite{vene}) \footnote{See however the concerns in
\cite{buonanno} about the efficiency of some bouncing models in erasing
possible initial inhomogeneities.}.

Finally, the amplification of primordial seeds (a problem prior to
the formation of structure) in bouncing universes
has been intensely debated recently (see Sect.\ref{cobs}).
The asymptotic behavior of these universes is markedly different
from that of the SCM or inflation. The universe at
past infinity starts to collapse from a flat empty structure-less state.
that at past infinity can be
approximated by Minkowski geometry written in terms of Milne coordinates
\footnote{We have also seen that there are eternal (non-bouncing) universes,
that start in
a de Sitter regime.}.
The transmission of the quantum fluctuations from this initial state
to the post-bounce phase is strongly model-dependent,
but there are some models which yield a scale-invariant spectrum
for the scalar perturbations in the post-bounce phase (see Sect.\ref{sp}).

An offspring of the bouncing models are the cyclic universes (see Sect.\ref{ccyclic}).
The cyclic models also attempt to solve the above-mentioned problems, and also may offer
a new view on the initial conditions: since by definition,
there is neither a beginning nor an end of
time in these models, there is no need to specify initial conditions.
Generically, cyclic universes share the problems of the universes that bounce only once.
In addition, they must assure that
the large scale structure present
in one cycle (generated by the quantum fluctuations in the
preceding
cycle) is not endangered by
perturbations or structure generated in earlier cycles, and
will not interfere with structure generated in later cycles.
One of the latest cyclic models, presented in \cite{clatest},
claims to have successfully faced these issues (however see
\cite{lindeu}).

As compelling a scenario may (or may not) seem, the
ultimate judge is observation, so we can ask if there are any
that may point to the occurrence of a bounce. As far as we know, there are
two possibilities \footnote{Some bouncing
models in GR were severely restricted in \cite{marek2},
using
SNIa data, CMB analysis, nucleosynthesis, and the age of the oldest high-redshift
objects.}:

\begin{itemize}

\item As discussed
in Sect.\ref{exact}, the tensor spectrum
of a nonsingular universe has a unique feature.
As an example, the SPBB models predict a stochastic spectrum of gravitational waves
whose amplitude increases as a function of frequency
in some frequency ranges (see Sect.\ref{tp}), hence
avoiding
the bounds due to the CMB, pulsar timing, and Doppler
tracking \cite{vuk}. The parameter space of the ``minimal'' SPBB model
\cite{maggiore}
was limited using LIGO results in \cite{vuk}. Notice also that nonsingular universes
may produce vector perturbations (see Sect.\ref{exact}).

\item The bounce may cause
oscillations, that will be superimposed
on the power spectrum of scalar perturbations. These oscillations
would also appear in the WMAP data, linked to the spectrum
through the multipole moments
which are in turn defined through the two-point correlation
function of the temperature fluctuations.
\cite{marrin}. Let us
note however, that such oscillations may be due not only to a bounce,
but also to transplanckian effects \cite{marrin} or
to non-standard initial conditions in the framework of hybrid inflation
\cite{cline}.

\end{itemize}

We would like to close by pointing out that although they do not yet
give a complete description of the universe, a better understanding
of bouncing models in classical GR should be attempted since they are inevitably
imposed upon us by the apparently
observed violation of the strong energy condition. It
must also be noted that there are at least two more reasons
to attempt this task.
First, the current
solution to the problems of the standard cosmological models (namely
inflation) is successful, but has several problems (see
Sect.\ref{sint}). Second, even if bouncing models do not succeed in
yielding a complete description of the universe (thus offering
an alternative to inflation \footnote{The
comparison of bouncing models with the inflationary
scenario has been undertaken in several articles (see for instance
Refs.\cite{vene} and \cite{lindepros}).}), they may throw light upon the
singularity problem (an issue in which inflation has nothing to
say).

Summing up, we have seen in this review that
bouncing universes have some attractive features,
but they are not complete yet:
much work is needed
to achieve a stage in which their predictions can match
those of the cosmological standard model.
Therefore, we hope this review encourages
further developments in nonsingular cosmologies.

\section{Acknowledgements}

The authors would like to thank all the participants of the \emph{Pequeno Semin\'ario}
at CBPF for interesting discussions on some parts of this review, and
J. Salim and specially N. Pinto-Neto for the reading of some chapters
and discussions. MN would like to
acknowledge support from CNPq and FAPERJ.
The authors would like to acknowledge support from ICRANet-Pescara
for hospitality during some stages of this work.


\begin{thebibliography}{999}

\bibitem{pdg} W-M. Yao et al  J. Phys. G: Nucl. Part. Phys.
\textbf{33}, 1 (2006).


\bibitem{haw67} S. Hawking, Proc. Royal Soc. (London) A \textbf{300}, 187 (1967).




\bibitem{lemaitre} G. Lema\^itre, Ann. Soc. Sci. Bruxelles, A {\bf 53}, 51 (1933).

\bibitem{penrosee} See for instance R. Penrose in \emph{General Relativity: an Einstein
centenary survey}, S. Hawking and W. Israel (eds), Cambridge U. Press (1979).

\bibitem{eins} A. Einstein  in
\emph{The principle of relativity},  eds. H. A. Lorentz, A. Einstein, H. Minkowski
and H. Weyl, Methuen, London (1950).

\bibitem{tolmanbook}
The thermodynamics of the cyclic universe was discussed in a
series of papers in the 30's by R. Tolman. The results are in his
book \emph{Relativity, Thermodynamics, and Cosmology}, Oxford, Clarendon (1934).


\bibitem{einstein} A. Einstein, Sitzungsber. Preuss. Akad. Wiss.,
Phys-Math. Kl., 235 (1931).

\bibitem{bojoi} M. Bojowald, Gen. Rel. Grav. {\bf 35}, 1877 (2003),
\texttt{gr-qc/0305069}.

\bibitem{gaspe} M. Gasperini,  Class. Quant. Grav. {\bf 17}, R1 (2000),
\texttt{hep-th/0004149}.


\bibitem{beken} J. Bekenstein, Int. J. Theor. Phys. {\bf 28}, 967
(1989).

\bibitem{beke2} J. Bekenstein, \Prl\; {\bf 46}, 623 (1981).

\bibitem{viss1} C. Molina-Paris, Matt Visser,
Phys. Lett. B {\bf 455} (1999).

\bibitem{viss2} D.
Hochberg, C. Molina-Par\'{\i}s, and M.  Visser, Phys.  Rev.  D
{\bf 59}, 044011 (1999).

\bibitem{ellis} {\em Cosmological models}, G.F.R. Ellis, Carg\`ese Lectures 1998.

\bibitem{raycha} A. Raychaudhuri, Phys. Rev. {\bf 98}, 1123 (1955).


\bibitem{ellisescola} {\em Standard Cosmology}, G.F.R. Ellis, Proceedings
of the Vth Brazilian School of Cosmology and Gravitation, World
Scientific (1987).

\bibitem{wald} See for instance {\em General Relativity}, R.  Wald, The University of
Chicago Press (1984).

\bibitem{viss3} D.  Hochberg and M.  Visser, Phys.  Rev.  D {\bf 56},
4745 (1997).


\bibitem{libromatt} {\em Lorentzian wormholes: From Einstein to Hawking}, M. Visser,
AIP Press (1996).


\bibitem{senore} J. Senovilla,
\Grg$\;$ {\bf 30}, 701 (1998). See also the discussion of a singularity theorem based on
spatial averages of matter variables
in \emph{A Singularity theorem based on spatial averages},
J. Senovilla,
Pramana special issue dedicated to A.K. Raychaudhuri,
(Naresh Dadhich, Pankaj Joshi and Probir Roy, eds),
\texttt{gr-qc/0610127}.


\bibitem{alere} H. Fuchs,
U. Kasper, D. Liebscher, V. M\"uller, and H. Schmidt, \Fp {\bf
36}, 427 (1988).


\bibitem{matt1} C. Barcelo and M. Visser,  Class. Quant. Grav. {\bf 17}, 3843
(2000).



\bibitem{bigrip} B. McInnes, JHEP \textbf{0208}, 029 (2002),
\texttt{hep-th/0112066}.

\bibitem{sing} See \emph{Singularity Theorems in General Relativity: Achievements and Open
Questions} and references therein, Jose M.M. Senovilla,
\texttt{physics/0605007}.

\bibitem{matt3} C. Catt\"oen and M. Visser,
Class. Quant. Grav. {\bf 22}, 4913 (2005), \texttt{gr-qc/0508045}.

\bibitem{twilight} See for instance C. Barcelo, M. Visser,
Int. J. Mod. Phys. D \textbf{11}, 1553 (2002),
\texttt{gr-qc/0205066}.


\bibitem{herd} C. Herdeiro and M. Sampaio, Class. Quant. Grav. \textbf{23}, 473 (2006),
\texttt{hep-th/0510052}.

\bibitem{lif} V.A. Belinsky, I.M. Khalatnikov, E.M. Lifshitz,
Adv. Phys. \textbf{19}, 525 (1970).

\bibitem{vile} A. Borde, A. Guth, A. Vilenkin,
Phys. Rev. Lett. \textbf{90}, 151301 (2003), \texttt{gr-qc/0110012}.

\bibitem{kerner3} J. Duruisseau, R. Kerner, and P. Eysseric, \Grg {\bf 15}, 797 (1983).

\bibitem{davies} See for instance P.C.W. Davies, S. Fulling, S.
Christensen and T. Bunch, \emph{Ann. Phys.} {\bf 109}, 108
(1977).

\bibitem{corrqg} See for instance
B. S. DeWitt, Phys. Rept. \textbf{19}, 295 (1975).

\bibitem{corrst} See for instance
T. Damour, A. Polyakov, Nucl. Phys. B \textbf{423}, 532 (1994),
\texttt{hep-th/9401069}, and references therein.

\bibitem{uti} R. Utiyama and B. De Witt, J. Math. Phys. {\bf 3}, 608 (1962),
see also R. Utiyama, \Prd {\bf 125}, 1727 (1962).

\bibitem{sakha} A. D. Sakharov, Sov. Phys. Doklady {\bf 12}, 1040 (1968).



\bibitem{nato} H. Nariai and K. Tomita, Progr. Th. Phys. {\bf 46}, 776 (1971).

\bibitem{toana} K. Tomita, T. Azuma and H. Nariai, Prog.
Theor. Phys. {\bf 60} 403 (1978)

\bibitem{mul1} V. M\"uller and H.-J. Schmidt, \Grg {\bf 17}, 769 (1985).

\bibitem{starobinsky} A. Starobinsky, Phys. Lett. B {\bf 91} 99 (1980).

\bibitem{fischetti} M. Fischetti, J. Hartle, and B. Hu, \Prd {\bf 20}, 1757 (1979).

\bibitem{ruz2} T. Ruzmaikina and A. Ruzmaikin, Sov. Phys. JETP {\bf 30}, 372 (1970).

\bibitem{guro} V. Ts. Gurovich, Sov. Phys. Dokladay {\bf 15}, 1105 (1971)

\bibitem{buch} H. A. Buchdahl,  Month. Notices Roy. Astr. Soc. {\bf 150}, 1 (1970).


\bibitem{bi} M. Born and L. Infeld, Proc. Roy. Soc. Lond. A \textbf{144}, 425 (1934).

\bibitem{bicknell} G. V. Bicknell, J. Phys. A {\bf 7}, 1061 (1974).

\bibitem{nariai} H. Nariai, Progr. Theor.
Phys. {\bf 46}, 433 and 776 (1971).

\bibitem{ruz} T. Ruzmaikina and A. Ruzmaikin, Sov. Phys. JETP {\bf 30}, 372 (1970).

\bibitem{gies} M. Giesswein, R. Sexl, and E. Streeruwitz, \Plb {\bf 52}, 442 (1974).

\bibitem{mac} K. Macrae and R. Riegert, \Prd {\bf 24}, 2555 (1981).

\bibitem{guro1} V. Ts. Gurovich, Sov. Phys. JETP {\bf 46}, 193 (1978).

\bibitem{bao} J. Barrow and A. Ottewill, \Jpa {\bf 16}, 2757 (1983).

\bibitem{wit} B. De Witt, Phys. Rev. {\bf 160}, 1113 and {\bf 162}, 1239 (1967).

\bibitem{gusta} V. Ts. Gurovich and A. A. Starobinskii, Sov. Phys. JETP {\bf 50}, 844 (1980).

\bibitem{murphy} G. Murphy, \Prd {\bf 8}, 4231 (1973).

\bibitem{kli} Z. Klimek, Acta Cosmologica \textbf{2}, 49 (1973).

\bibitem{prigo2}
See for instance \emph{Thermodynamic theory of structure stability and fluctuations},
I. Prigogine and P. Glansdorff, Wiley (1970).


\bibitem{kerner} R. Kerner, \Grg {\bf 14}, 453 (1982).

\bibitem{tey} P. Teyssandier and P. Tourrenc, J. Math.
Phys. \textbf{24}, 2793 (1983)

\bibitem{zia} G. Le Denmat and H. Sirousse Zia, \Prd {\bf 35}, 480 (1987).

\bibitem{maxcurv} M. Markov, JETP Lett.  {\bf 36}, 265 (1982).

\bibitem{mukh}V. Mukhanov and R.  Brandenberger, Phys.
Rev. Lett.  {\bf 68}, 1969 (1992).

\bibitem{bran1} R.  Brandenberger, V.
Mukhanov, and A.  Sornborger, \Prd {\bf 58}, 1629 (1993).

\bibitem{bd1} R.  Brandenberger, R.  Easther, and J.  Maia, JHEP {\bf 9808},
007 (1998).


\bibitem{kleinert} H. Kleinert and H-J. Schmidt,
Gen. Rel. Grav. {\bf 34}, 1295 (2002), \texttt{gr-qc/0006074}.

\bibitem{whitt} B. Whitt, Phys. Lett. \textbf{B145}, 176 (1984).


\bibitem{wands2} D. Wands, Class. Quant. Grav. {\bf 11}, 269 (1994),
\texttt{gr-qc/9307034}.

\bibitem{Nojiri} S. Nojiri and S.D. Odintsov, Phys. Lett. B {\bf 631}, 1 (2005),
\texttt{hep-th/0508049}.

\bibitem{Leigh} R. G. Leigh, Mod. Phys. Lett. \textbf{A} 4, 2767 (1989)  and
references therein.

\bibitem{ander1} P. Anderson, \Prd {\bf 28}, 271 (1983).

\bibitem{ander2} P. Anderson, \Prd {\bf 29}, 615 (1984).



\bibitem{bar} A. Barvinsky and G. Vilkovisky, Phys. Rep. {\bf 119}, 1 (1985).

\bibitem{scherk} J. Scherk and J. Schwarz, Nucl. Phys. {\bf B81}, 118 (1974).

\bibitem{odin} See for instance \emph{Introduction to modified gravity and gravitational
alternatives for dark energy}, S. Nojiri and S. Odintsov, lectures
given at 42nd Karpacz Winter School of Theoretical Physics:
Current Mathematical Topics in Gravitation and Cosmology, \texttt{hep-th/0601213}.

\bibitem{melni} V. N. Melnikov and S. V. Orlov, Phys. Lett. A \textbf{70}, 263 (1979).

\bibitem{melni2} \emph{The program of an eternal universe}, M. Novello,
Proceedings of the Vth Brazilian School of Cosmology and Gravitation,
World Scientific (1987).

\bibitem{bekesc} J. Bekenstein, Phys. Rev. D {\bf 11}, 2072
(1975).

\bibitem{turco} S.S. Bayin, F.I. Cooperstock, V. Faraoni, Astrophys. J. {\bf 428}, 439
(1994), \texttt{astro-ph/9402033}. See also S. P. Starkovich and
F. I. Cooperstock, ApJ \textbf{398}, 1 (1992).


\bibitem{bd} C. H. Brans and R. H. Dicke, Phys. Rev. {\bf 124}, 925 (1961).

\bibitem{stei} See for instance P. J. Steinhardt amd F. S. Accetta, \Prl
{\bf 64}, 2740 (1990).


\bibitem{liddle} A. R. Liddle and D. Wands, \Prd {\bf 45}, 2665 (1992).


\bibitem{kal} S. Kalyana Rama, Phys. Rev. D {\bf 56}, 6230 (1997),
\texttt{hep-th/9611223}, Phys. Rev. Lett.{\bf 78}, 1620 (1997),
\texttt{hep-th/9608026}.

\bibitem{kalo}  N. Kaloper and K. Olive,
Phys. Rev. D {\bf 57}, 811 (1998), \texttt{hep-th/9708008}.

\bibitem{serna} A. Serna and J.M. Alimi, Phys. Rev. D {\bf 53},
3074 (1996), \texttt{astro-ph/9510139}.

\bibitem{mimoso} J. P. Mimoso and D. Wands, Phys. Rev. D {\bf 51},
477 (1995), \texttt{gr-qc/9405025}.

\bibitem{diego}  See for instance D. F. Torres, H.
Vucetich, Phys. Rev. D {\bf 54}, 7373 (1996),
\texttt{gr-qc/9610022}.


\bibitem{fay} S. Fay, Class. Quant. Grav. {\bf 17}, 2663 (2000),
\texttt{gr-qc/0309087}.

\bibitem{bdwh} L.
Anchordoqui, S. Perez Bergliaffa, D. Torres, Phys. Rev. D {\bf 55},
5226 (1997).




\bibitem{borde1} A.  Borde and
A. Vilenkin, Int.  J.  Mod.  Phys. D {\bf 5}, 813 (1996).



\bibitem{barker} B. M. Barker, Astrophys. J. {\bf 219}, 5 (1978).


\bibitem{reviewssc} J. E. Lidsey, D. Wands, E. J.
Copeland, Phys. Rept. {\bf 337}, 343 (2000),
\texttt{hep-th/9909061}.

\bibitem{vene} M. Gasperini and G. Veneziano, Phys. Rept. \textbf{373}, 1 (2003),
\texttt{hep-th/0207130},
M. Gasperini and G. Veneziano,
Mod. Phys. Lett. A \textbf{8}, 3701
(1993). For an introduction see M. Gasperini,
Class. Quant. Grav. \textbf{17}, R1 (2000) \texttt{hep-th/0004149}.

\bibitem{vene2} G. Veneziano, \textit{Les Houches 1999, The primordial universe},
581-628, {\tt hep-th 0002094}.

\bibitem{constants} J. P. Uzan, Rev. Mod. Phys. {\bf 75}, 403 (2003).

\bibitem{dualities} See for instance A. A. Tseytlin, Mod. Phys. Lett. A \textbf{6}, 1721 (1991).

\bibitem{barrow1} J. D. Barrow, \Prd {\bf 48}, 3592 (1993).









\bibitem{anto} I.  Antoniadis, J.  Rizos, and K.  Tamvakis, \Np {\bf 415}, 497 (1994).


\bibitem{rizos} J.  Rizos and K.  Tamvakis, \Plb {\bf 326}, 57 (1994).

\bibitem{eas} R.
Easther and K.  Maeda, \Prd {\bf 59}, 083512 (1999).

\bibitem{easther} R. Easter and K. Maeda, \Prd {\bf 54},
7252 (1996).

\bibitem{kawai} S.  Kawai and J.  Soda, \Prd {\bf
59}, 063506 (1999).

\bibitem{yajima} H. Yajima, K. Maeda, and H. Ohkubo,
Phys. Rev. D \textbf{62}, 024020 (2000), \texttt{gr-qc/9910061}.

\bibitem{soda} S.  Kawai, M.
Sakagami, and J.  Soda, \Plb {\bf 437}, 284 (1998),
S. Alexeyev, A. Toporensky, V. Ustiansky,
Phys. Lett. B {\bf 509}, 151 (2001), \texttt{gr-qc/0009020}.

\bibitem{nowa} A. A. Al-Nowaiser,
M. Ozer, M.O. Taha, Int.  J.  Mod.  Phys.  \textbf{D} {\bf 8}, 43
(1999).




\bibitem{shinji} S. Tsujikawa, Class. Quant. Grav.
{\bf 20}, 1991 (2003), \texttt{hep-th/0302181}.

J. Khoury, B.A. Ovrut, P.J. Steinhardt and N. Turok, Phys. Rev. D64 123522
(2001), \texttt{hep-th/0105212}

\bibitem{ek} J. Khoury, B.A. Ovrut, N. Seiberg, P.J. Steinhardt and N. Turok, Phys. Rev.
D65, 086007 (2002), R.Y. Donagi, J. Khoury, B.A. Ovrut, P.J. Steinhardt
and N. Turok, JHEP 0111, 041 (2001),
J. Khoury, B.A. Ovrut, P.J. Steinhardt and N. Turok, Phys. Rev. D66
046005 (2002).


\bibitem{pauli} \emph{Theory of Relativity}, W. Pauli, Dover (1958).

\bibitem{fabris} C.P. Constantinidis, J.C. Fabris, R.G. Furtado, M. Picco,
Phys. Rev. D \textbf{61}, 043503 (2000), \texttt{gr-qc/9906122}.

\bibitem{gunzig}
E. Gunzig, A. Saa, L. Brenig, V. Faraoni, T.M. Rocha Filho, and A. Figueiredo, Phys.
Rev. D \textbf{63}, 067301 (2001), A. Saa, E. Gunzig, L. Brenig, V.
Faraoni, T.M. Rocha Filho, and A. Figueiredo, Int. J. Theor. Phys.
\textbf{40}, 2295 (2001).

\bibitem{abramo} L. Abramo, L. Brenig, E. Gunzig, A. Saa
Phys. Rev. D {\bf 67}, 027301 (2003), \texttt{gr-qc/0210069}.

\bibitem{gunzig2} E. Gunzig, A. Saa, L. Brenig,
V. Faraoni, T.M. Rocha Filho, A. Figueiredo, Phys. Rev. D \textbf{63},
067301 (2001), \texttt{gr-qc/0012085}.

\bibitem{rs} L. Randall and R. Sundrum, Phys. Rev. Lett. \textbf{\textbf{83}}, 4690 (1999).

\bibitem{bine} P. Binetruy, C. Deffayet, U. Ellwanger, D. Langlois,
Phys. Lett. B {\bf 477}, 285 (2000), \texttt{hep-th/9910219}.




\bibitem{foffa} S. Foffa, \Prd {\bf 68}, 043511 (2003), \texttt{hep-th/0304004}.


\bibitem{peloso} S. Mukherji and M. Peloso, \Plb {\bf 547},  297 (2002), hep-th/0205180.


\bibitem{sahni} Y. Shtanov and V. Sahni, \Plb \textbf{557}, 1 (2003),
\texttt{gr-qc/0208047}.

\bibitem{ekold} J. Khoury,
B.A. Ovrut, P.J. Steinhardt and N. Turok, Phys. Rev. D64 123522
(2001), \texttt{hep-th/0105212}.

\bibitem{eknew} J. Khoury, B. Ovrut, P. Steinhardt, N. Turok,
Phys. Rev. D \textbf{66}, 046005 (2002),
\texttt{hep-th/0109050}.

\bibitem{linde} R. Kallosh, L. Kofman and A. D. Linde, Phys. Rev. D \textbf{64}, 123523
(2001), \texttt{hep-th/0104073},
R. Kallosh, L. Kofman, A. D. Linde and A. A. Tseytlin, Phys.
Rev. D \textbf{64}, 123524 (2001), \texttt{hep-th/0106241}.

\bibitem{ekcyclic} P. J. Steinhardt and N. Turok, hep-th/0110030 and
Phys. Rev. D \textbf{65}, 126003 (2002), \texttt{hep-th/0111098}.


\bibitem{turok} P. J. Steinhardt and N. Turok,
\texttt{hep-th/0111030}, Phys. Rev. D 65, 126003 (2002),
\texttt{hep-th/0111098}, J. Khoury, P. J. Steinhardt
and N. Turok, Phys. Rev. Lett. 92, 031302 (2004)
\texttt{arXiv:hep-th/0307132}.

\bibitem{jus} \emph{A Briefing on the ekpyrotic / cyclic universe},
J. Khoury, \texttt{astro-ph/0401579}.


\bibitem{lindec} \textit{Inflationary Theory versus Ekpyrotic/Cyclic Scenario},
A. Linde, talk at Stephen Hawking's 60th birthday conference, Cambridge University, Jan. 2002,
\texttt{hep-th/0205259}.

\bibitem{penrose} R. Penrose, Ann. N. Y. Acad. Sci. {\bf 571}, 249 (1990).


\bibitem{israel} A. E. Sikkema and W. Israel, Nature {\bf 349}, 45 (1991).

\bibitem{robertson} H. P. Robertson, Rev. Mod. Phys. {\bf 5}, 51 (1933).



\bibitem{kardashev} N. Kardashev, Mon. Not. R. Astr. Soc. {\bf
243}, 252 (1990).

\bibitem{gibbons} G. Gibbons, \Np {\bf B292}, 784 (1987), {\bf
B310}, 636 (1988).

\bibitem{dabro} M. D\c abrowski, Annals Phys. {\bf 248}, 199 (1996),
\texttt{gr-qc/9503017}.

\bibitem{barrow} J. Barrow, Phys. Lett. {\bf B325}, 40 (1990).

\bibitem{ponjas} Y. Habaraa, H. Kawaib, and M. Ninomiya,
gr-qc/0504103, M. Fukuma, H. Kawai, M. Ninomiya, Int. J. Mod.
Phys. {\bf A19}, 4367 (2004), \texttt{hep-th/0307061}.


\bibitem{bojo} M. Bojowald, Annalen Phys. {\bf 15}, 326 (2006),
\texttt{astro-ph/0511557}.




\bibitem{bojo1} M. Bojowald, P. Singh, and A. Skirzewski,
Phys. Rev. D \textbf{70}, 124022 (2004),  \texttt{gr-qc/0408094}.

\bibitem{bojo3} M. Bojowald, G. Date,
Class. Quant. Grav. \textbf{21}, 3541 (2004), \texttt{gr-qc/0404039}.



\bibitem{mul} J. Lidsey, D. Mulryne, N. Nunes, R. Tavakol, Phys. Rev. D
\textbf{70}, 063521 (2004), \texttt{gr-qc/0406042}.



\bibitem{mul2} D. J. Mulryne, N.J. Nunes, R. Tavakol, J. E. Lidsey, Int. J. Mod. Phys.
{\bf A20}, 2347 (2005), \texttt{gr-qc/0411125}.

\bibitem{ellis2}  D. J. Mulryne, R. Tavakol, J. E.
Lidsey, G. F. R. Ellis, Phys. Rev. D {\bf 71}, 123512 (2005), \texttt{astro-ph/0502589}.

\bibitem{freese} M. G. Brown, K. Freese and W. H. Kinney,
\texttt{astro-ph/0405353}.


\bibitem{caldwell} R. Caldwell, M. Kamionkowski,  N. Weinberg,
\Prl {\bf 91}, 071301 (2003).


\bibitem{chinos1}  B. Feng, Xiu-Lian Wang, Xin-Min Zhang, Phys. Lett. B {\bf 607}, 35 (2005),
\texttt{astro-ph/0404224}.

\bibitem{chinos2} B. Feng, M. Li, Yun-Song Pia, X. Zhang, Phys. Lett. B {\bf 634}, 101 (2006),
\texttt{astro-ph/0407432}.

\bibitem{undulant} G. Barenboim, O. Mena and C. Quigg, Phys. Rev. D \textbf{71}, 063533 (2005),
\texttt{astro-ph/0412010}, G. Barenboim, O. Mena Requejo, C.
Quigg, JCAP \textbf{0604}, 008 (2006), \texttt{astro-ph/0510178}.

\bibitem{linder} E. Linder, Astropart. Phys. {\bf 25},  167 (2006),
\texttt{astro-ph/0511415}.


\bibitem{novi} \emph{The structure and evolution of the Universe},
Ya. B. Zeldovich and I. D. Novikov, U. of Chicago Press (1983).


\bibitem{hoyle} F. Hoyle, G. Burbidge, and J. V. Narlikar, Ap. J.
{\bf 410}, 437 (1993), and MNRAS {\bf 267}, 1007 (1994).

\bibitem{rees} M. Rees, The Observatory {\bf 89}, 193 (1969).

\bibitem{kanekar} N. Kanekar, V. Sahni, and Y. Shtanov,
Phys. Rev. {\bf D63}, 083520 (2001), \texttt{astro-ph/0101448}.









\bibitem{durrer} R. Durrer and J. Laukenmann, Class. Quant. Grav. {\bf 13},
1069 (1996).









\bibitem{harrison} E. Harrison, Mon. Not. R. Astr. Soc. {\bf 137},
69 (1967).


\bibitem{dabro2} J. Barrow and M. D\c abrowski, Mon. Not. R.
Astron. Soc. {\bf 275}, 850 (1995).



\bibitem{prigo} I. Prigogine,  Int. J. Theor. Phys. {\bf 9},  927 (1989),


\bibitem{de} S. De, Int. J. Theor. Phys. {\bf 32}, 1603 (1993).


\bibitem{bojo5} M. Bojowald, Phys. Rev. Lett. {\bf 89}, 261301 (2002),
\texttt{gr-qc/0206054}.

\bibitem{dadhich} N. Dadhich, A.K. Raychaudhuri, Mod. Phys. Lett. A {\bf 14}, 2135
(1999),  \texttt{gr-qc/9901081}.

\bibitem{picon} C. Armendariz-Picon, P. B. Greene, Gen. Rel. Grav. {\bf 35},  1637
(2003), \texttt{hep-th/0301129}.

\bibitem{wang} G. Yang, A. Wang, Gen. Rel. Grav. {\bf 37}, 2201 (2005),
\texttt{astro-ph/0510006}.

\bibitem{ave} P.P. Avelino, C.J.A.P. Martins, C. Santos, and E.P.S. Shellard,
Phys. Rev. D \textbf{68}, 123502 (2003), \texttt{astro-ph/0206287}.

\bibitem{marek} T. Stachowiak, M. Szydlowski,
Phys. Lett. B \textbf{646}, 209 (2007),
\texttt{gr-qc/0610121}.

\bibitem{holland} See for instance
D. Bohm and B. J. Hiley, Phys. Rep. \textbf{144}, 323 (1987), P. R.
Holland, \textit{The Quantum Theory of Motion: An Account of the
de Broglie-Bohm Causal Interpretation of Quantum Mechanichs},
Cambridge U. Press (1993).






\bibitem{bojoic} M. Bojowald, Gen. Rel. Grav. \textbf{35}, 1877 (2003),
\texttt{gr-qc/0305069}.

\bibitem{copen} See for instance
M. Jammer,
\emph{The Conceptual Development of Quantum Mechanics}, American Institute of Physics (1989).

\bibitem{lund} F. Lund, \Prd {\bf 8}, 3253 (1973).

\bibitem{thie} J. Brunnemann and T. Thiemann,
Class. Quant. Grav. {\bf 23}, 1395 (2006), \texttt{gr-qc/0505032}.


\bibitem{topo} P. Singh, A. Toporensky,
Phys. Rev. D {\bf 69}, 104008 (2004), \texttt{gr-qc/0312110}.


\bibitem{asht} A. Ashtekar, T. Pawlowski, P. Singh,
Phys. Rev. Lett. {\bf 96}, 141301 (2006), \texttt{gr-qc/0602086},
A. Ashtekar, T. Pawlowski, P. Singh,
Phys.Rev. D73 (2006) 124038,
\texttt{gr-qc/0604013}.


\bibitem{rovelli} See for instance \textit{Quantum Gravity}, C. Rovelli, Cambridge U. Press
(2004).



\bibitem{Lifshitz} E. M. Lifshitz and I.M. Khalatnikov, Adv. Phys. {\bf 12},
185 (1963).


\bibitem{Mukhanov} V. Mukhanov and R. Brandenberger, Phys. Rev,
Lett. {\bf 68}, 1969, (1992). See also R. Brandenberger, V.
Mukhanov and A. Sornborger, Phys. Rev. D {\bf 48}, 1629, (1993);
R. Moessner and M. Trodden,{\it ibid} {\bf 51}, 2801, (1995).
\bibitem{brande}
R. Brandenberger, in {\em VIII Brazilian School of Cosmology and
Gravitation}, ed.\ M. Novello, Editions Frontieres (1996).

\bibitem{hawkingellis} S. W. Hawking and G. F. Ellis,
\emph{The Large Scale
Structure of Space-Time}, Cambridge University Press (1973).
\bibitem{Novello1} M. Novello, J. M. Salim, M. C. Motta da Silva, S. E.
Jor\'as and R. Klippert, Phys. Rev. D {\bf 51}, 450 (1995).
\bibitem{Novello2} M. Novello, J. M. Salim,M.C. Motta da Silva, S. E.
Jor\'as and R. Klippert, Phys. Rev. D {\bf 52}, 730 (1995).
\bibitem{Novello3} M. Novello, J. M. Salim, M. C. Motta da Silva and R. Klippert,
Phys. Rev. D {\bf 54}, 2578 (1996).
\bibitem{Hawking2} S. W. Hawking, Ap. J. {\bf 145} (1966), 544.

\bibitem{revmuk} V. F. Mukhanov, H. A. Feldman, R. H. Brandenberger,
Phys. Rept. \textbf{215}, 203 (1992).


\bibitem{Israel} W. Israel, Ann. Phys. (NY) \textbf{100}, 310 (1976).

\bibitem{Vishniac} J.C. Hwang and E. T. Vishniac, Astrophys. J. \textbf{382}, 363
(1991).

\bibitem{NovelloSalim} M. Novello and J. Salim, \Prd {\bf 20},
377 (1979).


\bibitem{cremi} P. Creminelli, A. Nicolis, M. Zaldarriaga,
Phys. Rev. D {\bf 71}, 063505 (2005), \texttt{hep-th/0411270}.

\bibitem{fpp} \textit{Spectra of primordial fluctuations in two-perfect-fluid regular bounces},
F. Finelli, P. Peter, N. Pinto-Neto, \texttt{0709.3074 [gr-qc]}.

\bibitem{chu} Chong-Sun Chu, Ko Furuta, Feng-Li Lin
 Phys. Rev. D {\bf 73}, 103505 (2006),
\texttt{hep-th/0602148}.

\bibitem{derruelle} N. Deruelle and V. F. Mukhanov, Phys. Rev. D \textbf{52}, 5549
(1995), \texttt{gr-qc/9503050}.


\bibitem{batte}
T. J. Battefeld, G. Geshnizjani, Phys. Rev. D {\bf 73}, 064013
(2006), \texttt{hep-th/0503160}.

\bibitem{vernizzi}
R. Durrer, F. Vernizzi, Phys. Rev. D {\bf 66}, 083503 (2002),
\texttt{hep-ph/0203275}.

\bibitem{pp1} J. Martin, P. Peter, Phys. Rev. D {\bf 68}, 103517 (2003),
\texttt{hep-th/0307077}.

\bibitem{bozza1} V. Bozza JCAP \textbf{0602}, 009 (2006),
\texttt{hep-th/0512066}.

\bibitem{patel} T. J. Battefeld , S. P. Patil and R. H. Brandenberger,
\texttt{hep-th/0509043}.

\bibitem{nelson1} P. Peter, N. Pinto-Neto, Phys. Rev. D \textbf{66}, 063509 (2002),
\texttt{hep-th/0203013}.

\bibitem{kim} \emph{Evolution of linear perturbations through a bouncing world model:
Is the Harrison-Zel'dovich spectrum possible via bounce?} Han Seek
Kim, Jai-chan Hwang, \texttt{astro-ph/0607464}.

\bibitem{battecont}
T. J. Battefeld, R. Brandenberger, Phys. Rev. D \textbf{70},
121302 (2004), \texttt{hep-th/0406180}.

\bibitem{jordan} P. Jordan, J. Ehlers and R. Sachs, Akad. Wiss. Lit. Mainz Abh Math. Naturwiss. Kl. \textbf{1},
3 (1961).

\bibitem{stewart} J. Stewart and M. Walker, Proc. R. Soc. London A \textbf{341},
49 (1974).


\bibitem{Jones} B. Jones, Rev. Mod. Phys. \textbf{48}, 107 (1976).

\bibitem{BrandenbergerKhan} R. Brandenberger, R. Khan, and W. Press,
Phys. Rev. D \textbf{28}, 1809 (1983).

\bibitem{Olson} D. Olson, \Prd \textbf{14}, 327 (1976).


\bibitem{wands} L. Allen, D. Wands, Phys. Rev. D \textbf{70}, 063515 (2004),
astro-ph/0404441.

\bibitem{nd} \textit{Bouncing universes and their perturbations: Remarks on a toy model},.
N. Deruelle, \texttt{gr-qc/0404126}, N. Deruelle and A. Streich,
Phys. Rev. D \textbf{70}, 103504 (2004), \texttt{gr-qc/0405003}.

\bibitem{mp} \textit{Reply to 'Bouncing universes and their perturbations: Remarks on a toy model'}, J. Martin and P. Peter, \texttt{gr-qc/0406062}.

\bibitem{fibra} F. Finelli, R. Brandenberger, Phys. Rev. D \textbf{65}, 103522
(2002), \texttt{hep-th/0112249}.


\bibitem{xi}
D. H. Lyth, Phys. Rev. D \textbf{31}, 1792 (1985).

\bibitem{gw1} D. H. Lyth and E. D. Stewart, Phys. Lett. B \textbf{274}, 168
(1992), D. Wands, Phys. Rev. D \textbf{60}, 023507 (1999), \texttt{gr-
qc/9809062}.

\bibitem{klippert} V. De Lorenci, R. Klippert, M. Novello, J. M. Salim,
Phys. Rev. D \textbf{65}, 063501 (2002).




\bibitem{deSitter}W. de Sitter, Proc. K. Ned. Akad. Wet. \textbf{19}, 1217,
(1917).


\bibitem{goe} H.F.M. Goenner, Found. Phys. \textbf{14}, 865 (1984).







\bibitem{NovelloS2}M. Novello, L. A. R. Oliveira, J. M. Salim and
E. Elbaz, Int. J. Mod. Phys. D \textbf{1}, 641 (1993).


\bibitem{Murphy}G. L. Murphy, Phys. Rev. D \textbf{8}, 4231 (1973); J. M.
Salim and H. P. de Oliveira, Acta Phys. Pol. B \textbf{19}, 649, (1988).

\bibitem{Veneziano}G. Veneziano, hep-th/0002094, 2000; R.
Klippert, V. A. De Lorenci, M. Novello and J. M. Salim, Phys.
Lett. B \textbf{472}, 27 (2000).


\bibitem{NovelloM} V. A. De Lorenci, R. Klippert, M. Novello and J. M. Salim,
Phys. Rev. D {\bf 65}, 063501 (2002).

\bibitem{Cartier}C. Cartier, R. Durrer, E. Copeland,
Phys. Rev. D \textbf{67}, 103517 (2003).

\bibitem{novellomg8} M. Novello in \textit{Proceedings of the 8th Marcel
Grossmann Meeting on General Relativity}, Ed. Tsvi Piran, World
Scientific (1999).

\bibitem{Kolb} {\em The Early Universe}, E. W. Kolb and M. S. Turner,
Addison Wesley, California (1990).


\bibitem{Dunne}
G. Dunne and T. Hall, Phys. Rev. D {\bf 58}, 105022
(1998); G. Dunne, Int. J. Mod. Phys. A {\bf 12} (6), 1143
(1997).
\bibitem{Tajima}
T. Tajima, S. Cable, K. Shibata, and R. M. Kulsrud, Astrophys. J. {\bf 390}, 309 (1992);
M. Giovannini and M. Shaposhnikov, Phys. Rev. D {\bf 57}, 2186 (1998).
\bibitem{Campos}
A. Campos and B. L. Hu, Phys. Rev. D {\bf 58}, 125021
(1998).
 \bibitem{Hawking}S. W. Hawking, Ap. J. {\bf 145}, 544 (1966).


\bibitem{Ellis} G.F.R.Ellis and M.Bruni, Phys. Rev. D \textbf{40}, 1804 (1989).
\bibitem{Landau} \textit{Fluid Mechanics}, L. Landau and E. M. Lifshitz,
Pergamon Press (1982).










\bibitem{Wei} Wei Chen and Yong-Shi Wu, Phys Rev D \textbf{41}, 695 (1990).






\bibitem{Capozzielo} S. Capozzielo and R. de Ritis,
\texttt{astro-ph/9605070}.













\bibitem{einstein1} A.  Einstein, Berl.  Ber.  {\bf 235} (1931).


%







\bibitem{staro} A. Starobinksy, \Plb {\bf 91}, 99 (1980).

























%


 \bibitem{kerner2} J. P. Duruisseau and R.
 Kerner, \Cqg {\bf 3}, 817 (1986).




















%



 \bibitem{Born} M. Born and L. Infeld, Nature \textbf{133}, 63 (1934).

\bibitem{breton}
R, Garcia-Salcedo and N. Breton, Class. Quant. Grav. {\bf 22}, 4783 (2005),
\texttt{gr-qc/0410142}. See also
Int. J. Mod. Phys. \textbf{A15}, 4341 (2000),
\texttt{gr-qc/0004017}.

\bibitem{ademir} C.S. Camara, M.R. de Garcia Maia, J.C. Carvalho, Jose Ademir Sales Lima
Phys. Rev. D {\bf 69}, 123504 (2004), \texttt{astro-ph/0402311}.

\bibitem{Robertson} H. Robertson, Rev. Mod. Phys. \textbf{5}, 62 (1963).

\bibitem{Lif} E. M. Lifshitz and I. M. Khalatnikov, Adv. Phys. \textbf{12},
185 (1963).

\bibitem{rohr} See for instance
\emph{Classical charged particles; foundations of their theory}, F. Rohrlich,
Addison-Wesley (1965).

\bibitem{weylbook} \textit{Space, Time, and Matter}, H. Weyl, Dover (2003).



\bibitem{Thompson} C. Thompson and O. Blaes, \Prd {\bf 57}, 3219 (1998).

\bibitem{Jedamzik} K. Jedamzik, V. Jatalinic, and A. Olinto, \Prd {\bf 57},
3264 (1998).

\bibitem{smolin} \emph{The Fate of black hole
singularities and the parameters of the standard models of particle physics and cosmology},
L. Smolin, \texttt{gr-qc/9404011}.


\bibitem{magueijo} J. Barrow, D. Kimberly, J. Magueijo, Class. Quant. Grav. \textbf{21}, 4289 (2004), \texttt{astro-ph/0406369}.

\bibitem{sandvik} H. Sandvik, J. Barrow, J. Magueijo, Phys. Rev. Lett.
\textbf{88},  031302 (2002), \texttt{astro-ph/0107512}.

\bibitem{magueijo2} See for instance
J. Magueijo, Rept. Prog. Phys. \textbf{66}, 2025 (2003),
\texttt{astro-ph/0305457}.

\bibitem{ellisc} G. Ellis, J.-P. Uzan, Am. J. Phys. \textbf{73}, 240 (2005),
\texttt{gr-qc/0305099}.

\bibitem{clifton} T. Clifton, John D. Barrow, Phys. Rev. D \textbf{75},
043515 (2007), \texttt{gr-qc/0701070}.

\bibitem{rose} B. Rose,  \Cqg \textbf{3}, 975 (1986).

\bibitem{wei} Wei Chen and Yong-Shi Wu, \Prd \textbf{41}, 695 (1990).


\bibitem{arbab} A. Arbab and A. Abdel-Rahman, \Prd \textbf{50}, 7725 (1994).

\bibitem{abdalla} Abdalla and Abdel-Rahman, \Prd \textbf{46}, 5675 (1992).

\bibitem{abdel} A-M.M. Abdel-Rahman, \Prd \textbf{45}, 3497 (1992).

\bibitem{bronnikov} K. Bronnikov, A. Dobosz and I. Dymnikova,
Class. Quantum Grav. \textbf{20}, 3797 (2003).

\bibitem{sachs} R. Sachs, J. Narlikar, and F. Hoyle, Astron. Astrophys. \textbf{313},
703 (1996).

\bibitem{burdos} J. V. Narlikar, G. Burbidge, R.G. Vishwakarma,
J. Astrophys. Astron.  \textbf{28}, 67 (2007),
\texttt{arXiv:0801.2965 [astro-ph]}.

\bibitem{burtres} J. V. Narlikar, G. Burbidge, R.G. Vishwakarma,
J. Astrophys. Astron. \textbf{28}, 67 (2007),
\texttt{0801.2965 [astro-ph]}.




\bibitem{novaraujo} M. Novello and R. Araujo, \Prd$\;$ \textbf{22}, 260 (1980).

\bibitem{causal} See for instance W. Israel, Ann. Phys. \textbf{100}, 310 (1976),
W. Israel and J. Stewart, \emph{ibid} \textbf{118}, 341 (1978),
D. Pav\'on, D. Jon, and J. Casas-V\'asquez, Ann. Inst. Henri Poincar\'e
\textbf{1}, 79 (1982).

\bibitem{hensalim} H. P. de Oliveira , J. M. Salim, Acta
Phys. Polon. B \textbf{19}, 649 (1988).

\bibitem{joao} M. Novello, H. P. de Oliveira, J. M. Salim
and J. Torres, Acta
Phys. Polon. \textbf{B} 21, 571 (1990).





\bibitem{schiffer} M. Schiffer, Int. J. Theor. Phys. {\bf 30}, 410 (1991).

J. Moffat, Int. J. Mod. Phys. A \textbf{20}, 1155 (2005),
\texttt{gr-qc/0404066},
\textit{New Ekpyrotic Cosmology.},
E. Buchbinder, J. Khoury, B. Ovrut, \texttt{hep-th/0702154}.

\bibitem{sling} \textit{A Stringy Alternative to Inflation: The Cosmological Slingshot Scenario},.
C. Germani, N. Grandi, A. Kehagias, \texttt{hep-th/0611246}.





\bibitem{taha2} M. Ozer, M.O. Taha, Nucl. Phys. B \textbf{287}, 776
(1987).

\bibitem{kir} Kirzhnits and Linde, Ann. Phys. (NY) \textbf{101}, 195 (1976),
A. Linde, Rep. Prog. Phys. \textbf{42}, 861 (1979).

\bibitem{carv} J. Carvalho, J. Lima, and I. Waga, \Prd \textbf{46}, 2404 (1992).

\bibitem{ademir2} J. Lima and J. Maia, \Prd \textbf{49}, 5597 (1994).

\bibitem{friedmann} A. Friedmann,
Zeitschrift f\"ur Physik, \textbf{10}, 377 (1922), Zeitschrift f\"ur Physik, \textbf{21}, 326 (1924).

\bibitem{lem} G. Lema\^itre, Ann. Soc. Sci. Bruxelles, A {\bf 47}, 29 (1927).

\bibitem{robe} H. P. Robertson, Proc. Nat. Acad. Sci. \textbf{15}, 822 (1929).

\bibitem{walker} A. G. Walker, Q. J. Math. (Oxford), \textbf{6}, 81 (1935).

\bibitem{duncan} M. J. Duncan and L. G. Jensen, Nuc. Phys. B \textbf{328}, 171 (1989).

\bibitem{hawking5} S. W. Hawking and R. Laflamme, \Plb \textbf{209}, 39 (1988).

\bibitem{barcelos} M. Novello, J. Barcelos-Neto, J.M. Salim,
Class. Quant. Grav. \textbf{18}, 1261 (2001),
\texttt{hep-th/0006061}.

\bibitem{trodden} J. A. S. Lima and M. Trodden, Phys. Rev. D \textbf{53}, 4280 (1996),
\texttt{astro-ph/9508049}.

\bibitem{ssm} H. Bondi, T. Gold, Mon. Not. Roy. Astron. Soc.
{\bf 108}, 252 (1948),
F. Hoyle, \emph{ibid}, 372,
F. Hoyle and J. Narlikar, Proc. Roy. Soc. A \textbf{270}, 334 (1962).

\bibitem{dac} See \emph{A different approach to Cosmology}, F.
Hoyle, G. Burbidge, and J. V. Narlikar, Cambridge U. Press
(2005), and references therein.

\bibitem{weinberg} See for instance \textit{Gravitation and Cosmology}, S. Weinberg,
Wiley (1972).

\bibitem{wdw} B.S. De Witt, Phys. Rev D \textbf{160}, 1113 (1967),
J. A. Wheeler, in \emph{Batelle Recontres}, eds. C. DeWitt and J.A.Wheeler, Benjamin Press: New York,
(1968).


\bibitem{entropybounds} \emph{Cosmological Entropy Bounds}, Ram Brustein,
\texttt{hep-th/0702108}.

\bibitem{acacio} J. Acacio de Barros, N. Pinto-Neto, Int. J. Mod. Phys. D \textbf{7},
201 (1998).

\bibitem{nelsonbarros}J. Acacio de Barros, N. Pinto-Neto and M. A.
Sagioro-Leal, Phys. Lett. A \textbf{241}, 229 (1998).

\bibitem{nelson} P. Peter and N. Pinto-Neto, Phys. Rev. D \textbf{65},
023513 (2001).

\bibitem{colis} R. Colistete, Jr., J. C. Fabris, and N. Pinto-Neto, \Prd \textbf{62},
083507 (2000).

\bibitem{biswas}T. Biswas, A. Mazumdar, and W. Siegel, \texttt{hep-th/0508194}.

\bibitem{komar} A. Komar, \Prd \textbf{104}, 544 (1956).

\bibitem{clifton2} \textit{Exact Friedmann Solutions in Higher-Order Gravity Theories}
T. Clifton, \texttt{gr-qc/0703126}.




\bibitem{tamva} P. Kanti and K. Tamvakis, \Prd \textbf{68}, 024014 (2003).

\bibitem{kanti} P.Kanti, J.  Rizos, and K. Tamvakis, \Prd {\bf 59}, 083512 (1999).

\bibitem{luminet} See for instance
\emph{The rise of Big Bang Models, from Myth to Theory and
Observations}, J-P Luminet, \texttt{arXiv:0704.3579 [astro-ph]}.


\bibitem{frampton} L. Baum and P. Frampton, Phys. Rev. Lett. \textbf{98}, 071301
\textbf{2007}, \texttt{hep-th/0610213}.


\bibitem{frampton2} L. Baum, P. Frampton, \texttt{hep-th/0703162}.


\bibitem{coleslucchin} See for instance P. Coles and F. Lucchin,
\emph{Cosmology:
The Origin and Evolution of Cosmic Structure}, John Wiley \& Sons (2002).



\bibitem{earman} For a discussion of these theorems and of the concept of
singularity see for instance \emph{Bangs, Crunches, Whimpers and Shrieks:
Singularities and Acausalities in Relativistic Spacetimes}, J. Earman,
Oxford University Press, USA (1995).


\bibitem{rendall} \textit{The Nature of spacetime singularities},
A. Rendall, \texttt{gr-qc/0503112}.

\bibitem{bennet}C. Bennett, Nature \textbf{440}, 1126 (2006).

\bibitem{saka} C. Germani, W. Nelson, M.
Sakellariadou, Phys. Rev. D \textbf{76}, 043529 (2007), \texttt{gr-qc/0701172},
\emph{The Measure Problem in Cosmology},
G.W. Gibbons, N. Turok, \texttt{hep-th/0609095}.


\bibitem{narpam} J. Narlikar, T. Padmanabhan,
Annual Review of Astronomy and Astrophysics. \textbf{29}, 325
(1991), S. Hollands, R. Wald, Gen. Rel. Grav. \textbf{34}, 2043 (2002),
\texttt{gr-qc/0205058}.


\bibitem{dicus} D. A. Dicus, J. R. Letaw,  D. C. Teplitz, and V. L.Teplitz,
Astrop. J. \textbf{252}, 1 (1982), and
\textit{The Future of the Universe}, Scientific American \textbf{248}, 90 (1983).

\bibitem{ghosts} T. Chiba, JCAP 0503:008 (2005),
\texttt{gr-qc/0502070}, A. Nunez, S. Solganik, Phys. Lett. B \textbf{608},
189 (2005), \texttt{hep-th/0411102}.

\bibitem{kamen} I. Khalatnikov, A. Kamenshchik, Phys. Rep. 513
\textbf{288}, 513 (1997).

\bibitem{barrowsudden} J. D. Barrow, Class. Quant. Grav. \textbf{21}, L79 (2004),
\texttt{gr-qc/0403084}.


\bibitem{mamu} M. Markov and V. Mukhanov, Nuovo Cim. B \textbf{86}, 97 (1985),
J. Polchinski, Nucl. Phys. B \textbf{325}, 619 (1989).

\bibitem{padma} For a review see T. Padmanabhan, Phys. Rept. \textbf{380},  235 (2003),
\texttt{hep-th/0212290}.

\bibitem{zeld} Y.B. Zel'dovich, JETP Letters \textbf{6}, 316 (1967); Soviet Physics Uspekhi \textbf{11}, 381
(1968).

\bibitem{novellocc} M. Novello, R.P. Neves, Class. Quant. Grav. \textbf{20}, L67 (2003).

\bibitem{maty} J. Matyjasek, \Prd \textbf{51}, 4154 (1995).

\bibitem{over} J.M. Overduin and F.I. Cooperstock, Phys. Rev. D \textbf{58}, 043506 (1998).

\bibitem{mtw} \textit{Gravitation}, C. Misner, K. Thorne, and J. Wheeler, W. H. Freeman ed. (1973).

\bibitem{nossou}  M. Novello, E. Goulart, J.M. Salim, S.E. Perez Bergliaffa,
Class. Quantum Grav. \textbf{24}, 3021 (2007),
\texttt{gr-qc/0610043}.

\bibitem{hartle} See for instance J. J. Halliwell, in \textit{Quantum Cosmology and Baby Universes}, edited by S. Coleman,
J. B. Hartle, T. Piran, and S. Weinberg (World Scientific, Singapore,
1991), p. 159.


\bibitem{demaret} J. Demaret, Nature \textbf{277}, 199 (1979).

\bibitem{bojobscg} For a detailed analysis of the assumptions of the singularity theorems,
and conclusions following from them see \textit{Singularities and Quantum Gravity},
M. Bojowald, Lectures given at 12th Brazilian School of Cosmology and Gravitation (XII BSCG), Rio de Janeiro, Brazil, 10-23 Sep 2006,
\texttt{gr-qc/0702144}, to be published by AIP.





\bibitem{assingh} Abhay Ashtekar, Tomasz Pawlowski, Parampreet Singh,
Phys. Rev. D \textbf{74}, 084003 (2006), \texttt{gr-qc/0607039}.

\bibitem{vander} K. Vandersloot, Phys. Rev. D \textbf{75}, 023523 (2007),
\texttt{gr-qc/0612070}.

\bibitem{manyw} H. Everett, III, Rev. Mod. Phys. \textbf{29}, 454 (1957).

\bibitem{nlqm} See for instance G. C. Ghirardi,  A. Rimini
and T. Weber, Phys. Rev. D \textbf{34}, 470 (1986)

\bibitem{decoh} R. B. Griffiths, J. Stat. Phys. \textbf{36}, 219 (1984).

\bibitem{eli} \emph{Estrutura e processos de cria\c c\~ao do universo nos mitos
e na consci\^encia}, Elisabeth C. Cotta Mello, PhD thesis UFRJ (2005).

\bibitem{primeval} See for instance \textit{Primal Myths: Creation Myths
around the World}, Barbara C. Sprowl, Harper (1991).

\bibitem{bir} \emph{A Survey of Modern Algebra}, G. Birkhoff and S. Mac
Lane, 5th ed. New York: Macmillan, (1996)

\bibitem{sstrings} See for instance \emph{String theory and M-theory: A modern introduction},
K. Becker, M. Becker, J.H. Schwarz, Cambridge U. Press (2007).


\bibitem{madden} See for instance
R. Brustein, R. Madden, Phys. Rev. D \textbf{57}, 712 (1998),  \texttt{hep-th/9708046}.

\bibitem{cartier} C. Cartier, E. J. Copeland, R. Madden, JHEP 0001:035,2000,
\texttt{hep-th/9910169}.

\bibitem{kaloper} N. Kaloper, R. Madden and K.A. Olive, Nucl. Phys. B \textbf{452}, 677 (1995) .

\bibitem{dresden} M. Dresden, Physica A \textbf{110}, 1
(1982).

\bibitem{belinsky} V. A. Belinsky and I. M. Khalatnikov,
Sov. Phys. JETP \textbf{42}, 205 (1976).

\bibitem{olival} M. Novello and F. d'Olival,
Acta Physica Polonica vol \textbf{B11}, 3 (1980).

\bibitem{ligia} M. Novello and L. Rodrigues, Lett. Nuovo Cim. \textbf{40}, 317 (1984).

\bibitem{bendix} See for instance
\emph{Theory of bifurcations of dynamical systems on a plane}, A. Andronov, E. Leontovish, I. Gordon, and
A. Maier, Israel Program for Scientific Translations (Jerusalem, 1971).

\bibitem{veresh} G. Vereshkov, Yu. Gushkan, S. Ivanov, V. Nesterenko, and A. Poltavtsev,
Sov. Phys. JETP, \textbf{46}, 1041 (1977).

\bibitem{z1} Ya. B. Zeld'ovich, Sov. Phys. JETP Lett., \textbf{12}, 307 (1970),
L. Halpern, Ark. Phys. \textbf{34}, 539 (1973).

\bibitem{marioluiz} M. Novello and L. A. Oliveira, Phys. Lett. \textbf{109}, 454 (1985).

\bibitem{markov} M. A. Markov, Sov. Phys. JETP \textbf{24}, 584,
(1967))

\bibitem{nelson4} \textit{\textit{}Dynamical theories of Brownian motion}, E. Nelson, Princeton U. Press (1967).


\bibitem{confor} J. D. Barrow and S. Cotsakis, Phys.Lett.
B \textbf{214}, ,515 (1988), K. Maeda, Phys. Rev. D \textbf{39}, 3159 (1989), \textit{A Note on Wavemap-Tensor Cosmologies},
S. Cotsakis and J. Miritzis,
\texttt{gr-qc/0107100}.

\bibitem{makler} See for example
M. Novello, M. Makler, L.S. Werneck, and C.A. Romero, Phys. Rev. D \textbf{71}, 043515 (2005),
\texttt{astro-ph/0501643}.

\bibitem{nosso} M. Novello, S. E. Perez Bergliaffa,
J. Salim, Phys. Rev. D \textbf{69}, 127301 (2004),
\texttt{astro-ph/0312093}.

\bibitem{nilton} J. Salim, S. Perez Bergliaffa, N. Souza,
Class. Quant. Grav. \textbf{22}, 975 (2005),
\texttt{astro-ph/0410423}.


\bibitem{massimo} M. Giovannini, Class. Quantum Grav. \textbf{21}, 4209 (2004).


\bibitem{chamb} H. A. Chamblin and H. S. Reall, Nucl. Phys. B \textbf{562}
133 (1999), \texttt{hep-th/9903225}.

\bibitem{maartens} R. Maartens,  Living Rev. Rel. \textbf{7}, 7 (2004),
\texttt{gr-qc/0312059}.

\bibitem{kiritsis} A. Kehagias, E. Kiritsis, JHEP \textbf{9911}, 022 (1999),
\texttt{hep-th/9910174}.


\bibitem{gregory} P. Singh, K. Vandersloot, G. Vereshchagin, Phys. Rev. D \textbf{74},
043510 (2006), \texttt{gr-qc/0606032}.

\bibitem{menace} R. Caldwell, Phys. Lett. B \textbf{545}, 23 (2002),
\texttt{astro-ph/9908168}.

\bibitem{romero} See M. Novello in  \emph{Vth Brazilian Schol of Cosmology and
Gravitation}, World Scientific (1987).

\bibitem{pavon} D. Pavon, D. Jou, and J. Casas-Vazquez, Ann. Inst. H. Poincar\'e,
vol. XXXVI, 79 (1982).

\bibitem{ellismaartens} G. Ellis, R. Maartens, Class. Quant. Grav. \textbf{21}, 223 (2004),
\texttt{gr-qc/0211082}.

\bibitem{murugan} G. Ellis, J. Murugan, C. Tsagas, Class. Quant. Grav. \textbf{21},  233 (2004),
\texttt{gr-qc/0307112}.

\bibitem{hartlehaw} J.B. Hartle, S.W. Hawking,
Phys. Rev. D \textbf{28}, 2960 (1983).

\bibitem{Vilenkin} A. Vilenkin, Phys. Rev. D \textbf{33}, 3560 (1986).

\bibitem{bojobou} M. Bojowald, Phys. Rev. D \textbf{64}, 084018 (2001),
\texttt{gr-qc/0105067}.



\bibitem{pbbinfl}
M. Gasperini and G. Veneziano,
Phys. Rev. D \textbf{50}, 2519 (1994).



\bibitem{inflgasp} For a discussion of this topics in the context of
String Pre-Big-Bang cosmology, see M. Gasperini, \Cqg {\bf 17}, R1 (2000).

\bibitem{dilatongasp}
See for instance
M. Gasperini, \emph{Relic dilatons in string cosmology}, in
\emph{Proc. of the 12th Italian Conference on Gen.
Rel. and Grav. Phys.} (Roma 1996),
eds. M. Bassan et al. (World Scientific, Singapore, 1997),
p. 181.


\bibitem{pertposit} Hwang J and Noh H, 2002 Phys. Rev. D 65, 124010
Gordon C and Turok N, 2003 Phys. Rev. D 67, 123508
Martin J and Peter P, 2003 Phys. Rev. D 68, 103517
Martin J and Peter P, 2004 Phys. Rev. Lett. 92, 061301
Martin J and Peter P, 2004 Preprint gr-qc/0406062
[18] Deruelle N, 2004 Preprint gr-qc/0404126
Deruelle N and Streich A, 2004 Phys. Rev. D 70, 103504

\bibitem{debate} See for instance
Lyth D, 2002 Phys. Lett. B 524, 1
Brandenberger R and Finelli F, 2001 JHEP 0111, 056
Lyth D, 2002 Phys. Lett. B 526, 173
Hwang J, 2002 Phys. Rev. D 65, 063514; Tsujikawa S, 2002 Phys. Lett. B 526, 179
Martin J, Peter P, Pinto-Neto N and Schwarz D J, 2002 Phys. Rev. D 65, 123513
Hwang J and Noh H, 2002 Phys. Lett. B 545, 207
Martin J, Peter P, Pinto-Neto N and Schwarz D J, 2003 Phys. Rev. D 67, 028301
Tsujikawa S, Brandenberger R and Finelli F, 2002 Phys. Rev. D 66, 083513
Cartier C, Durrer R and Copeland E, 2003 Phys. Rev. D 67, 103517
Peter P, Pinto-Neto N and Gonzalez D A, 2003 JCAP 0312, 003,
Durrer R and Vernizzi F, 2002 Phys. Rev. D 66, 083503.

\bibitem{tbf} Tsujikawa S, Brandenberger R and Finelli F, 2002 Phys. Rev. D \textbf{66}, 083513,
\texttt{hep-th/0207228}.


\bibitem{novsal}  \textit{The stability of a bouncing universe},
M. Novello, J.M. Salim,
Annals of the Tenth Marcel Grossmann Meeting, p. 770,
World Scientific (Singapore),
(2005).


\bibitem{merlau} See for instance J. Merlau-Ponty, \textit{Cosmologie do XXe sie\`cle},
Paris, Gallimard (1965).


\bibitem{bardeen} J.M. Bardeen, Phys. Rev. D \textbf{22}, 1882 (1980).

\bibitem{bozzadef} V. Bozza, G. Veneziano, Phys. Lett. B \textbf{625}, 177 (2005),
hep-th/0502047, V. Bozza, G. Veneziano, JCAP 0509:007,2005, \texttt{gr-qc/0506040},
V. Bozza, JCAP 0602, 009 (2006), \texttt{hep-th/0512066}.

\bibitem{ggv} M. Gasperini, M. Giovannini, G. Veneziano, Phys. Lett. B
569, 113 (2003).

\bibitem{ame} L. Amendola, F. Finelli, Phys. Rev. Lett. \textbf{94}, 221303 (2005),  \texttt{astro-ph/0411273}.

\bibitem{cartieralpha} \textit{Scalar perturbations in an alpha regularized cosmological bounce},
C. Cartier, \texttt{hep-th/0401036}.


\bibitem{carhwang} C. Cartier, Jai-chan Hwang, E.J. Copeland,
Phys. Rev. D \textbf{64}, 103504 (2001),
\texttt{astro-ph/0106197}.

\bibitem{tsuji2} S. Tsujikawa, Phys. Lett. B \textbf{526}, 179 (2002),
\texttt{gr-qc/0110124}.


\bibitem{pbf} A.J. Tolley, N. Turok, P.J. Steinhardt, Phys. Rev. D \textbf{69},
106005 (2004), hep-th/0306109, T.J. Battefeld, S.P. Patil, R.H. Brandenberger,
Phys. Rev. D \textbf{73}, 086002 (2006), \texttt{hep-th/0509043}.

\bibitem{ppp} P. Peter, E. Pinho, and N. Pinto-Neto, JCAP \textbf{07}, 014
(2005).

\bibitem{pinho} P. Peter, E.J.C. Pinho, N. Pinto-Neto, Phys. Rev. D \textbf{75},
023516 (2007), \texttt{hep-th/0610205}.

\bibitem{pinho2} P. Peter, E.J.C.
Pinho, N. Pinto-Neto, Phys. Rev. D \textbf{73}, 104017 (2006), \texttt{gr-qc/0605060}.

\bibitem{ap} \emph{K-Bounce}, L.R. Abramo, P. Peter, \texttt{0705.2893 [astro-ph]}.

\bibitem{curvaton} K. Enqvist and M. Sloth, Nucl. Phys. B \textbf{626}, 395  (2002),
D. Lyth and D. Wands, Phys. Lett. B \textbf{524}, 5  (2002),
V. Bozza, M. Gasperini, M. Giovannini, and G. Veneziano, Phys. Lett. B \textbf{543}, 14  (2002),
V. Bozza, M. Gasperini, M. Giovannini and G. Veneziano, Phys. Rev. D \textbf{67}, 063514 (2003).

\bibitem{abb} \emph{On the Transfer of Adiabatic Fluctuations through a Nonsingular Cosmological Bounce},
S. Alexander, T. Biswas, R.H. Brandenberger,
\texttt{arXiv:0707.4679 [hep-th]}.

\bibitem{marco} F. Di Marco, F. Finelli, R. Brandenberger, Phys. Rev. D \textbf{67},
063512 (2003), \texttt{astro-ph/0211276}.

\bibitem{mena} F.C. Mena, D.J. Mulryne, R. Tavakol
Class. Quant. Grav. \textbf{24}, 2721 (2007),
\texttt{gr-qc/0702064}.

\bibitem{giovavec} M. Giovannini, Phys. Rev. D \textbf{70}, 103509 (2004)
\texttt{hep-th/0407124}.

\bibitem{nelsonrev} N. Pinto-Neto, Int. J. Mod. Phys. D \textbf{13},  1419 (2004),
\texttt{hep-th/0410225}.



\bibitem{bojovec} M. Bojowald, G. Mortuza Hossain, Class. Quant. Grav. \textbf{24}, 4801 (2007), \texttt{0709.0872 [gr-qc]}.

\bibitem{indianos} B. S. Sathyaprakash, P. Goswami, K. P. Sinha, Phys. Rev. D 33, 2196 (1986).

\bibitem{altshu} B. L. Altshuler, Class. Quantum Grav. {\bf 7}, 189 (1990).


\bibitem{frogun} V. Frolov, E. Gunzig, and N. Van den Bergh,
Class. Quantum Grav. {\bf 8}, L125 (1991).

\bibitem{costa} I. Costa, N. Deruelle, M. Novello, e N. F. Svaiter,
\Cqg {\bf 6}, 1893 (1989).

\bibitem{birreldavis} \textit{Quantum Fields in Curved Space},
N. D. Birrell, P. C. W. Davies, Cambridge U. Press (1984).

\bibitem{grasso} D. Grasso and H. Rubinstein, Phys. Rept. {\bf 348}, 163 (2001).

\bibitem{lemoine}
Lemoine D., Lemoine M., Phys. Rev. {\bf D 52}, (1995),
1955, Gasperini M., Giovannini M. and Veneziano G. Phys. Rev.
Lett. {\bf V 75}, (1995), 3796.

\bibitem{miojcap} J. Salim, N. Souza, S. E. Perez Bergliaffa, T. Prokopec,
JCAP \textbf{0704}, 011 (2007), \texttt{astro-ph/0612281}.

\bibitem{rosen} N. Rosen, Ap. J. {\bf 297}, 347 (1985).

\bibitem{pachner} J. Pachner, Mon. Not. Astr. Soc. {\bf 131}, 173 (1965).

\bibitem{rosen2} N. Rosen, Int. J. Theor. Phys. {\bf 2}, 189 (1969).

\bibitem{hoyleold} F. Hoyle and J. Narlikar, Proc. Roy. Soc. A \textbf{278},
465 (1964).

\bibitem{carloni} S. Carloni, P. Dunsby, D. Solomons,
Class. Quant. Grav. \textbf{23}, 1913 (2006), \texttt{gr-qc/0510130}.


\bibitem{goheer} \textit{Dynamical systems analysis of anisotropic cosmologies in $R^n$-gravity},
N. Goheer, J. Leach, P. Dunsby, \texttt{arXiv:0710.0814 [gr-qc]}.

\bibitem{cycling} D. Easson, R. Gregory, G. Tasinato, I. Zavala,
JHEP \textbf{0704}, 026 (2007),
\texttt{hep-th/0701252}.

\bibitem{gasper} M. Gasperini and M. Giovannini, Phys. Rev., D \textbf{47}, 1519 (1993), \texttt{gr-qc/9211021},
Phys. Lett. B \textbf{282} , 36 (1992).


\bibitem{caco} C. Cartier, Edmund J. Copeland, M. Gasperini, Nucl. Phys. \textbf{B}
607, 406 (2001),
\texttt{gr-qc/0101019}.

\bibitem{brustein} R. Brustein,
M. Gasperini, Massimo Giovannini, V. F. Mukhanov, G. Veneziano, Phys.Rev. D \textbf{51}, 6744 (1995), \texttt{hep-th/9501066}.

\bibitem{tanoss} M. Galluccio, M. Litterio, F. Occhionero,
Phys. Rev. Lett. \textbf{79}, 970 (1997), \texttt{gr-qc/9608007}.

\bibitem{tano3} M. Gasperini, Phys. Rev. D \textbf{56}, 4815 (1997),  \texttt{gr-qc/9704045}.




\bibitem{nelsonns} P. Peter and N. Pinto-Neto, Phys. Rev D vol 65,
023513 (2001)

\bibitem{Abramowitz} M. Abramowitz and Irene A. Stegun, \textit{Handbook
of Mathematical Functions}, Dover, page 504 (1974).


\bibitem{marrin} J. Martin, C. Ringeval, Phys. Rev. D \textbf{69}, 083515
(2004), astro-ph/0310382.



\bibitem{mubook} \emph{Physical foundations of cosmology}, V. Mukhanov,
Cambridge U. Press (2005).

\bibitem{siete} L. Anchordoqui, J. Edelstein, C. Nunez, S. Perez Bergliaffa, M. Schvellinger, M. Trobo, Fabio Zyserman, Phys. Rev. D \textbf{64}, 084027 (2001), \texttt{hep-th/0106127}.

\bibitem{breview} Lectures on the theory of cosmological perturbations.
Robert H. Brandenberger,
Lect. Notes Phys.646:127-167,2004,
\texttt{hep-th/0306071}.

\bibitem{wmap} D. N. Spergel et al., Astrophys. J. Suppl. 148, 175 (2003);
H. V. Peiris et al., Astrophys. J. Suppl. 148, 213 (2003),
D. N. Spergel et al., \texttt{arXiv: astro-ph/0603449}.

\bibitem{sdss} M. Tegmark et al., Phys. Rev. D 69, 103501 (2004);
K. Abazajian et al., Astron. J. 128, 502 (2004).

\bibitem{2df} W. J. Percival et al., Mon. Not. Roy. Astron. Soc. 327,
1297 (2001); S. Cole et al., Mon. Not. Roy. Astron. Soc.
362, 505 (2005).

\bibitem{deflation} M. Gasperini, G. Veneziano,
Mod. Phys. Lett. A \textbf{8}, 3701 (1993), \texttt{hep-th/9309023}.

\bibitem{solo} L. P. Grishchuk and M. Solokin, Phys. Rev. D \textbf{43}, 2556 (1991).

\bibitem{guo} Zong-Kuan Guo, Nobuyoshi Ohta, Shinji Tsujikawa,
Phys. Rev. D \textbf{75}, 023520 (2007), \texttt{hep-th/0610336}.

\bibitem{joras} R. Brandenberger, S.E. Joras, J. Martin,  Phys. Rev. D \textbf{66},
083514 (2002), \texttt{hep-th/0112122}.

\bibitem{erico} \emph{Gravitational Waves in Singular and Bouncing FRW
Universes}, V. Antunes, E. Goulart, and M. Novello, submitted for publication.

\bibitem{israeljc} W. Israel, Nuovo Cim. B
44S10, 1 (1966) [Erratum-ibid. B 48, 463 (1967 NUCIA,B44,1.1966)].

\bibitem{copematch} E. J. Copeland, D. Wands, JCAP 0706:014,2007,  \texttt{hep-th/0609183}.

\bibitem{heat} M. Giovannini, Class. Quant. Grav. \textbf{21}, 4209 (2004), \texttt{hep-th/0406098}.

\bibitem{loll} R. Loll, Phys. Rev. Lett. \textbf{75}, 3048 (1995),
\texttt{gr-qc/9506014}.

\bibitem{motta} M. Novello, J.M. Salim, M.C. Motta da Silva, R. Klippert,
Phys. Rev. D \textbf{61}, 124025 (2000).

\bibitem{felipe} F. Falciano, N. Pinto-Neto, E.Sergio Santini,
Phys. Rev. D \textbf{76}, 083521 (2007), \texttt{arXiv:0707.1088
[gr-qc]}.

\bibitem{aranha} R.F. Aranha, I. Damiao Soares, H.P. de Oliveira, E.V. Tonini,
JCAP 0710:008 (2007).

\bibitem{witten} P. Ho\u rava and E. Witten, Nucl. Phys. B
\textbf{460}, 506 (1996), \texttt{hep-th/9510209}.

\bibitem{thesis} For an introduction see for instance
\emph{A primer on the ekpyrotic scenario}, S. Rasanen,
\texttt{astro-ph/0208282}.

\bibitem{eug} E. Buchbinder, J. Khoury, B. Ovrut,
Phys. Rev. D \textbf{76}, 123503 (2007), hep-th/0702154.


\bibitem{engen} J. L. Lehners, P. McFadden, N. Turok and P. J. Steinhardt, \emph{Generating ekpyrotic
curvature perturbations before the big bang}, \texttt{hep-th/0702153}.

\bibitem{cha} \emph{The New Ekpyrotic Ghost},
R. Kallosh, Jin U. Kang, A. Linde, V. Mukhanov,
\texttt{arXiv:0712.2040 [hep-th]}.

\bibitem{coleman} S. Coleman, Phys. Rev. {\bf 15}, 2929 (1977).

\bibitem{vile2} A. Vilenkin, Phys. Lett. B \textbf{117}, 25 (1982).

\bibitem{copenu}E. Copeland, A. Mazumdar, N. Nunes, Phys. Rev. D \textbf{60}, 083506
(1999), \texttt{astro-ph/9904309}.


\bibitem{medved} A.J.M. Medved,
Class. Quant. Grav. \textbf{21}, 2749 (2004),
\texttt{hep-th/0307258}.

\bibitem{myers} J. L. Hovdebo, R. C. Myers,
JCAP 0311:012 (2003),
\texttt{hep-th/0308088}.

\bibitem{aline} M. Novello, Aline N. Araujo and J. M. Salim,
sent for publication.

\bibitem{chinue} See for instance Hao Wei, Shuang Nan Zhang,
Phys. Lett. B \textbf{644}, 7 (2007), \texttt{astro-ph/0609597},
and references therein.

\bibitem{lindepros} A. Linde,
Phys. Scripta T \textbf{117}, 40-48 (2005),
\texttt{hep-th/0402051}.

\bibitem{goode} S. W. Goode, Phys. Rev. D. \textbf{39}, 2882 (1989).

\bibitem{tjoras} \emph{Scalar Perturbations in the Friedman-Robertson-Walker Universe}
S. Joras, MSc Thesis, CBPF 1994/02 J82 (in Portuguise).

\bibitem{lindeb} \emph{Particle physics and inflationary cosmology}, A. Linde,
\texttt{hep-th 0503203}.

\bibitem{falciano} \emph{A classical bounce: constraints and consequences},
F. Falciano, M. Lilley, P. Peter, \texttt{0802.1196 [gr-qc]}.


\bibitem{dar} See references in S. Dar, Q. Shafi, A. Sil,
Phys. Rev. D \textbf{74}, 035013 (2006),
\texttt{hep-ph/0607129}.

\bibitem{cline} C. P. Burgess, J. M. Cline, F. Lemieux and R. Holman,
JHEP 0302, 048 (2003), \texttt{hep-th/0210233}.

\bibitem{marek2} M. Szydlowski, W. Godlowski,
A. Krawiec, J. Golbiak,
Phys. Rev. D \textbf{72}, 063504 (2005),
\texttt{astro-ph/0504464}.

\bibitem{buonanno} A. Buonanno, T. Damour, Phys. Rev. D \textbf{64}, 043501
(2001), gr-qc/0102102.

\bibitem{clatest} J. K. Erickson, S. Gratton, P. J. Steinhardt and N. Turok,
\emph{Cosmic perturbations through
the cyclic ages}, \texttt{hep-th/0607164}.

\bibitem{lindeu} \emph{Inflationary Cosmology},
A. Linde, \texttt{arXiv:0705.0164}.

\bibitem{maggiore} M. Gasperini, M. Maggiore, and G. Veneziano, Nucl.
Phys. B \textbf{494}, 315 (1997).

\bibitem{vuk}
V. Mandic, A. Buonanno,
Phys. Rev. D \textbf{73}, 063008 (2006), \texttt{astro-ph/0510341}


\end{thebibliography}
\end{document}